\documentclass[10pt,journal,compsoc]{IEEEtran}

\usepackage[caption=false,font=footnotesize,labelfont=sf,textfont=sf]{subfig}

\usepackage[english]{babel}

\usepackage[numbers,sort]{natbib}

\usepackage{todonotes}
\usepackage{booktabs}
\usepackage{tabularx}
\usepackage{lscape}
\usepackage{multirow}
\usepackage[edges]{forest}
\usetikzlibrary{arrows.meta}
\usepackage{subfig}
\usepackage[acronym]{glossaries}
\glsdisablehyper
\usepackage[inline]{enumitem}
\usepackage{makecell}
\usepackage{algorithm}
\usepackage{algpseudocode}
\usepackage{float}
\usepackage{hyperref}
\usepackage{url}

\usepackage{amsmath,amssymb,amsfonts}
\usepackage{graphicx}
\usepackage{textcomp}
\usepackage{pifont}
\newcommand{\cmark}{\text{\ding{51}}}
\newcommand{\xmark}{\text{\ding{55}}}

\makeatletter
\newcommand{\setID}[1]{%
    \phantomsection
    #1\def\@currentlabel{\unexpanded{#1}}\label{#1}%
}
\makeatother

\newcommand{\setAcronym}[2]{\newacronym{#1}{#1}{#2}}

\setAcronym{IaaS}{Infrastructure as a Service}
\setAcronym{PaaS}{Platform as a Service}
\setAcronym{FaaS}{Function as a Service}
\setAcronym{VM}{Virtual Machine}
\setAcronym{SLA}{Service Level Agreement}
\setAcronym{SLO}{Service Level Objective}
\setAcronym{ITIL}{Information Technology Infrastructure Library}
\setAcronym{PCA}{Principal Component Analysis}
\setAcronym{QoS}{Quality of Service}
\setAcronym{GDPR}{General Data Protection Regulation}
\setAcronym{HPC}{High Performance Computing}

\newcommand{\customlabel}[1]{\label{#1}}

\newcounter{observation}
\newcommand{\obs}[1]{\noindent\refstepcounter{observation}\textbf{O\theobservation:}\customlabel{o:#1}}
\newcommand{\obsref}[1]{O\ref{o:#1}}

\newcounter{finding}
\newcommand{\fin}[1]{\noindent\refstepcounter{finding}\textbf{IF\thefinding:}\customlabel{if:#1}}
\newcommand{\finref}[1]{IF\ref{if:#1}}

\newcounter{mainfinding}
\newcommand{\mainfin}[1]{\refstepcounter{mainfinding}\item[MF\themainfinding:\customlabel{mf:#1}]}

\newcommand{\designref}[1]{%
 \begin{tikzpicture}[baseline=(char.base)]
   \node[draw,circle,inner sep=0.5pt, fill=black, text=white] (char){\small #1};
 \end{tikzpicture}%
 }

\newcommand{\iconWidth}{0.4cm}
\newcommand{\iconRep}{\includegraphics[width=\iconWidth]{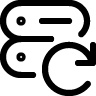}}
\newcommand{\iconExp}{\includegraphics[width=\iconWidth]{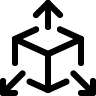}}
\newcommand{\iconVol}{\includegraphics[width=\iconWidth]{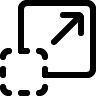}}
\newcommand{\iconVel}{\includegraphics[width=\iconWidth]{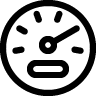}}
\newcommand{\iconHor}{\includegraphics[width=\iconWidth]{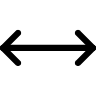}}
\newcommand{\iconVer}{\includegraphics[width=\iconWidth]{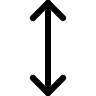}}
\newcommand{\iconHom}{\includegraphics[width=\iconWidth]{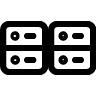}}
\newcommand{\iconHet}{\includegraphics[width=\iconWidth]{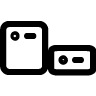}}
\newcommand{\iconEmpty}{\includegraphics[height=\iconWidth]{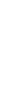}}
\newcommand{\emptySpace}{\hspace{0.4cm}}

\newcommand{\iconTextHeight}{0.275cm}
\newcommand{\iconRepInline}{\includegraphics[width=\iconTextHeight]{figures/icons/rep.png}}
\newcommand{\iconExpInline}{\includegraphics[width=\iconTextHeight]{figures/icons/exp.png}}
\newcommand{\iconVolInline}{\includegraphics[width=\iconTextHeight]{figures/icons/vol.png}}

\newcommand{\iconHorInline}{\includegraphics[width=\iconTextHeight]{figures/icons/hor.png}}
\newcommand{\iconVerInline}{\includegraphics[width=\iconTextHeight]{figures/icons/ver.png}}
\newcommand{\iconHomInline}{\includegraphics[width=\iconTextHeight]{figures/icons/hom.png}}
\newcommand{\iconHetInline}{\includegraphics[width=\iconTextHeight]{figures/icons/het.png}}

\newcommand{\simulator}{OpenDC}
\newcommand{\provider}{Solvinity}

\newcommand{\toolname}{Capelin}

\newenvironment{myitemize}
{
\setlength{\leftmargini}{10pt} 
\setlength{\parskip}{0pt} 
\begin{enumerate}  
  \setlength{\itemindent}{0pt} 
  
  \setlength{\itemsep}{0pt} 
  \setlength{\parsep}{0pt} 
  \setlength{\parskip}{0pt} 
}
{ \end{enumerate}
}

\newcommand{\vcutS}{\vspace*{-0.15cm}}
\newcommand{\vcutM}{\vspace*{-0.25cm}}
\newcommand{\vcutL}{\vspace*{-0.5cm}}

\begin{document}

    \title{Capelin: Data-Driven Capacity Procurement for\\ Cloud Datacenters using Portfolios of Scenarios}
    
    \author{[Technical Report on the TPDS homonym article]\\
        Georgios~Andreadis,
        Fabian~Mastenbroek,
        Vincent~van~Beek,
        and~Alexandru~Iosup%
        \IEEEcompsocitemizethanks{%
            \IEEEcompsocthanksitem G. Andreadis, F. Mastenbroek, V. van Beek, and A. Iosup are with Electrical Engineering, Mathematics \& Computer Science, Delft University of Technology, 2628 CD Delft, Netherlands.
            \IEEEcompsocthanksitem V. van Beek is also with Solvinity, 1100 ED Amsterdam, Netherlands.
            \IEEEcompsocthanksitem A. Iosup is also with Computer Science, Vrije Universiteit Amsterdam, 1081 HV Amsterdam, Netherlands.
        }%
    }
    

    

\IEEEtitleabstractindextext{%
\begin{abstract} 
Cloud datacenters provide a backbone to our digital society.
Inaccurate capacity procurement 
for cloud datacenters
can lead to significant performance degradation, denser targets for failure, and unsustainable energy consumption.
Although this activity is core to improving cloud infrastructure, 
relatively few comprehensive approaches and support tools exist for mid-tier operators,
leaving 
many
planners with merely rule-of-thumb judgement.
%
%
We derive requirements from a unique survey of experts in charge of diverse datacenters in several countries.
We propose Capelin, a data-driven, scenario-based capacity planning system for mid-tier cloud datacenters.
%
%
Capelin introduces the notion of portfolios of scenarios, which it leverages in its probing for alternative capacity-plans.
At the core of the system, a trace-based, discrete-event simulator
enables the exploration of
different possible topologies, with support for scaling the volume, variety, and velocity of resources, and 
for horizontal (scale-out) and vertical (scale-up) scaling.
%
%
Capelin compares alternative topologies and for each gives detailed quantitative operational information, which could facilitate human decisions of capacity planning.
%
%
%
%
%
We implement and open-source Capelin, and show through comprehensive trace-based experiments it can aid practitioners.
%
%
The results give evidence that reasonable choices can be worse by a factor of 1.5-2.0 than the best, in terms of performance degradation or energy consumption. 
%

\end{abstract}

\begin{IEEEkeywords}
Cloud, procurement, capacity planning, datacenter, practitioner survey, simulation
\end{IEEEkeywords}}

    \maketitle
    \IEEEdisplaynontitleabstractindextext

    \IEEEraisesectionheading{\section{Introduction}\label{sec:introduction}}
\vcutS{}

\IEEEPARstart{C}{loud} datacenters are critical for today's increasingly digital society~\cite{tr:idc19cloud,tr:flexera20,release:gartner19cloud}.
Users have come to expect near-perfect availability and high quality of service, at low cost and high scalability.
Planning the capacity of cloud infrastructure is a critical yet non-trivial optimization problem that could lead to significant service improvements, cost savings, and environmental sustainability~\cite{book/BarrosoHR18}.
This activity includes short-term capacity planning, which includes the process of provisioning and allocating resources from the capacity already installed in the datacenter, and {\it long-term capacity planning}, which is the process of {\em procuring} machines that form the datacenter capacity.
This work focuses on the latter, which is a process involving large amounts of resources and decisions that are difficult to reverse.

\begin{figure}[t]
    \centering
    \includegraphics[width=\columnwidth]{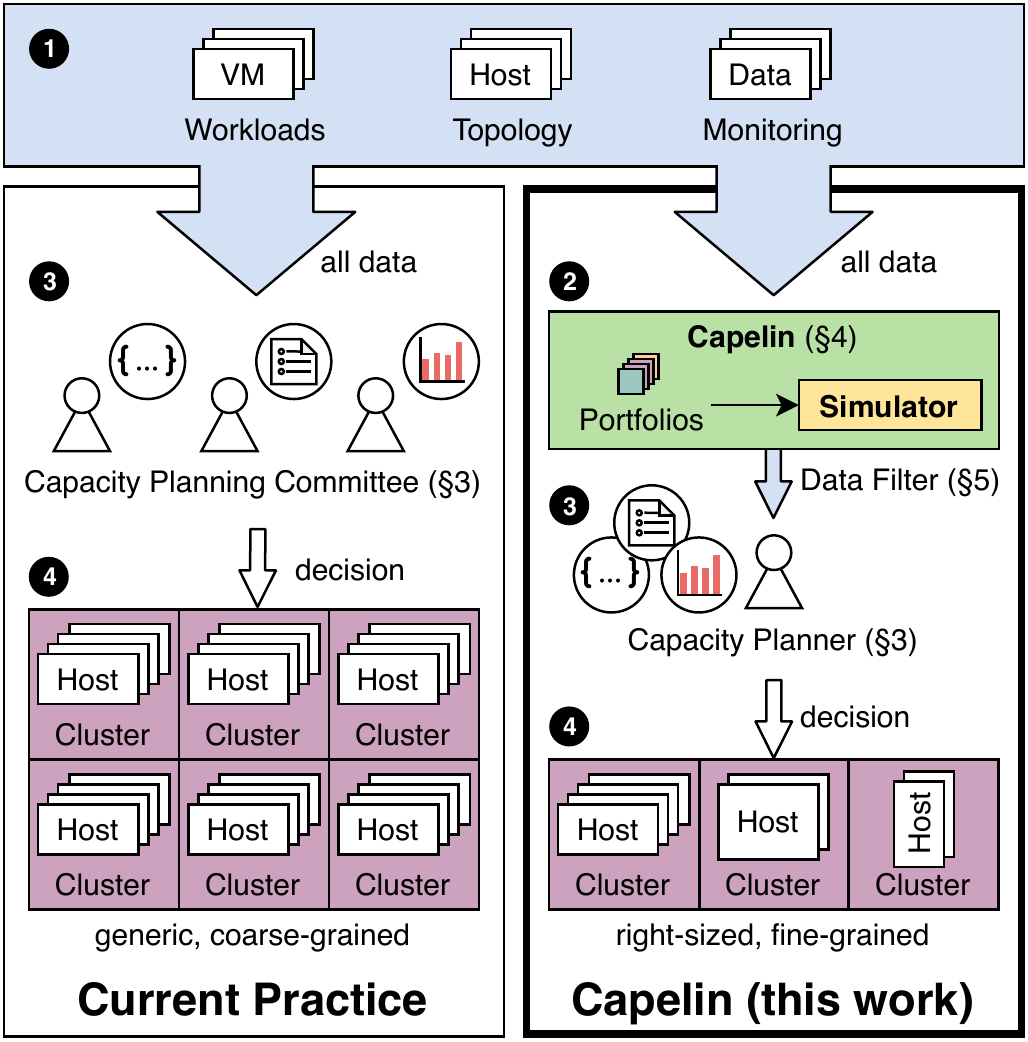}
    \vcutL{}
    \caption{\toolname{}, a new, data-based capacity planning process for datacenters, compared against the current approach.}
    \label{fig:capelin-in-context}
    \vspace*{-0.5cm}
\end{figure}

Although many approaches to the long-term capacity-planning problem have been published~\cite{Zhang2007,Rolia2005,Carvalho2017}, companies use much rule-of-thumb reasoning for procurement decisions.
To minimize operational risks, many such industry approaches currently lead to significant overprovisioning~\cite{Glanz2012Sep}, or miscalculate the balance between underprovisioning and overprovisioning~\cite{Nazareth2017}.
In this work, as Figure~\ref{fig:capelin-in-context} depicts, 
we approach the problem of capacity planning for mid-tier cloud datacenters with a semi-automated, specialized, data-driven tool for decision making.

We focus in this work mainly on {\it mid-tier providers} of cloud infrastructure that operate at the low- to mid-level tiers of the service architecture, ranging from IaaS to PaaS. Compared to the extreme-scale operators Google, Facebook, and others in the exclusive GAFAM-BAT group, the mid-tier operators are small-scale. However, they are both numerous and they are responsible for much of the datacenter capacity in modern, service-based and knowledge-driven economies. 
This work addresses
four main capacity planning challenges for mid-tier cloud providers. 
First, 
the {\bf lack of published knowledge about the current practice of long-term cloud capacity planning}. 
For a problem of such importance and long-lasting effects, it is surprising that the only studies of how practitioners make and take long-term capacity-planning decisions are either over three decades old~\cite{Lam1987} or focus on non-experts deciding how to externally procure capacity for IT services~\cite{Bauer2017}. 
A survey of expert capacity planners could reveal new requirements. 

%
%
%

Second, we observe the {\bf need for a flexible instrument for long-term capacity planning, one that can address various operational scenarios}.
State-of-the-art tools~\cite{vmwarecapacityplanner,ibmcapacityplanner,HPCapAdvisor2014} and techniques~\cite{Tang2017,Ghosh2014,Carvalho2017} for capacity-planning operate on abstractions that match only one vendor or focus on simplistic problems. 
Although single-vendor tools, such as VMware's Capacity Planner~\cite{vmwarecapacityplanner} and IBM's Z Performance and Capacity Analytics tool~\cite{ibmcapacityplanner}, can provide good advice for the cloud datacenters equipped by that vendor, they do not support real-world cloud datacenters that are heterogeneous in both software~\cite{DBLP:conf/usenix/AmvrosiadisPGGB18}\cite[\S{2.4.1}]{book/BarrosoHR18} and hardware~\cite{DBLP:conf/usenix/DuplyakinRMWDES19,DBLP:journals/ijhpca/BolzeCCDDJJLLMMNPQRTT06}\cite[\S{3}]{book/BarrosoHR18}.
Yet, to avoid vendor lock-in and licensing costs, 
cloud datacenters 
acquire heterogeneous hardware and software from multiple sources and
could, for example, combine VMware's, Microsoft's, and open-source OpenStack+KVM virtualization management technology, and complement it with container technologies. 
Although linear programming~\cite{Xu2015}, game theory~\cite{Tang2017}, stochastic search~\cite{Ghosh2014}, and other optimization techniques work well on simplistic capacity-planning problems, 
they do not address the multi-disciplinary, multi-dimensional nature of the problem. 
As Figure~\ref{fig:capelin-in-context} (left) depicts, 
without adequate capacity planning tools and techniques, 
practitioners need to rely on rules-of-thumb calibrated with casual visual interpretation of the complex data provided datacenter monitoring.
This state-of-practice likely results in overprovisioning of cloud datacenters, to avoid operational risks~\cite{DBLP:journals/ccr/GreenbergHMP09}. Even then, evolving customers and workloads could make the planned capacity insufficient, leading to risks of not meeting Service Level Agreements~\cite{book/BeyerJPM16,DBLP:conf/sosp/AlipourfardGKHV19}, inability to absorb catastrophic failures~\cite[p.37]{book/BarrosoHR18}, and even unwillingness to accept new users.



Third, we identify the {\bf need for comprehensive evaluations of long-term capacity-planning approaches, based on real-world data and scenarios.}
Existing tools and techniques have rarely been tested with real-world scenarios, and even more rarely with real-world operational traces that capture the detailed arrival and execution of user requests. 
Furthermore, for the few thus tested, the results are only rarely peer-reviewed~\cite{Rolia2005,DBLP:conf/sosp/AlipourfardGKHV19}. 
%
%
We advocate comprehensive experiments with real-world operational traces and diverse scaling scenarios to test capacity planning approaches. 

Fourth and last, we observe the {\bf need for publicly available, comprehensive tools for long-term capacity planning}.
However, 
and in stark contrast with the many available tools for short-term capacity planning, 
few procurement tools are publicly available, and even fewer are open-source. 
From the available tools, none can model all the aspects needed to analyze cloud datacenters from \S\ref{sec:model}.

We propose in this work \toolname{}, a data-driven, scenario-based alternative to current capacity planning approaches.
Figure~\ref{fig:capelin-in-context} visualizes our approach (right column of the figure) and compares it to current practice (left column).
Both approaches start with inputs such as workloads, current topology, and large volumes of monitoring data (step~\designref{1} in the figure).
From this point on, the two approaches diverge, ultimately resulting in qualitatively different solutions. 
The current practice expects a committee of various stakeholder to extract meaning from all the input data~(\designref{3}), which is severely hampered by the lack of decision support tools. 
Without a detailed understanding of the implications of various decisions, the final decision is taken by committee, and it is typically an overprovisioned and conservative approach~(\designref{4}). 
In contrast, 
\toolname{} adds and semi-automates a data-driven approach to 
data analysis and decision support~(\designref{2}), and 
enables capacity planners
to take fine-grained decisions based on curated and greatly reduced data~(\designref{3}). 
With such support, even a single capacity planner can make a tailored, fine-grained decision on topology changes to the cloud datacenter~(\designref{4}).
More than a purely technical solution, this approach can change organizational 
processes.
Overall, our main contribution is:
%


\begin{myitemize}
\item We design, conduct, and analyze community interviews on capacity planning in different cloud settings~(Section~\ref{sec:interviews}).
    We use broad, semi-structured interviews, from which 
    we identify new, real-world requirements. 
    
    \item We design \toolname{}, a semi-automated, data-driven approach for long-term capacity planning in cloud datacenters~(Section~\ref{sec:design}).
    At the core of \toolname{} is an abstraction, the capacity planning portfolio, which expresses sets of ``what-if'' scenarios. Using simulation, \toolname{} estimates the consequences of alternative decisions. 
    
    
    \item We demonstrate \toolname{}'s ability to support capacity planners through experiments based on real-world operational traces and scenarios~(Section~\ref{sec:evaluation}).
    We implement a prototype of \toolname{} as an extension to \simulator{}, an open-source platform for datacenter simulation~\cite{Iosup2017}.
    We conduct diverse trace-based experiments.
    Our experiments cover four different scaling dimensions, and workloads from both private and public clouds.
    They also consider different operational factors such as the scheduler allocation policy, and phenomena such as correlated failures and performance interference~\cite{VanBeek2019,Koh2007,Zhai2017}.
    
    \item We release our prototype of \toolname{}, consisting of extensions to \simulator{}~2.0~\cite{DBLP:conf/ccgrid/MastenbroekAI+21}, as Free and Open-Source Software~(FOSS), for practitioners to use.
    \toolname{} is engineered with professional, modern software development standards and produces reproducible results.
    
\end{myitemize}
\section{A System Model for DC Operations}
\label{sec:background}\label{sec:model}

\begin{figure}[t]
    \centering
    \includegraphics[width=\columnwidth]{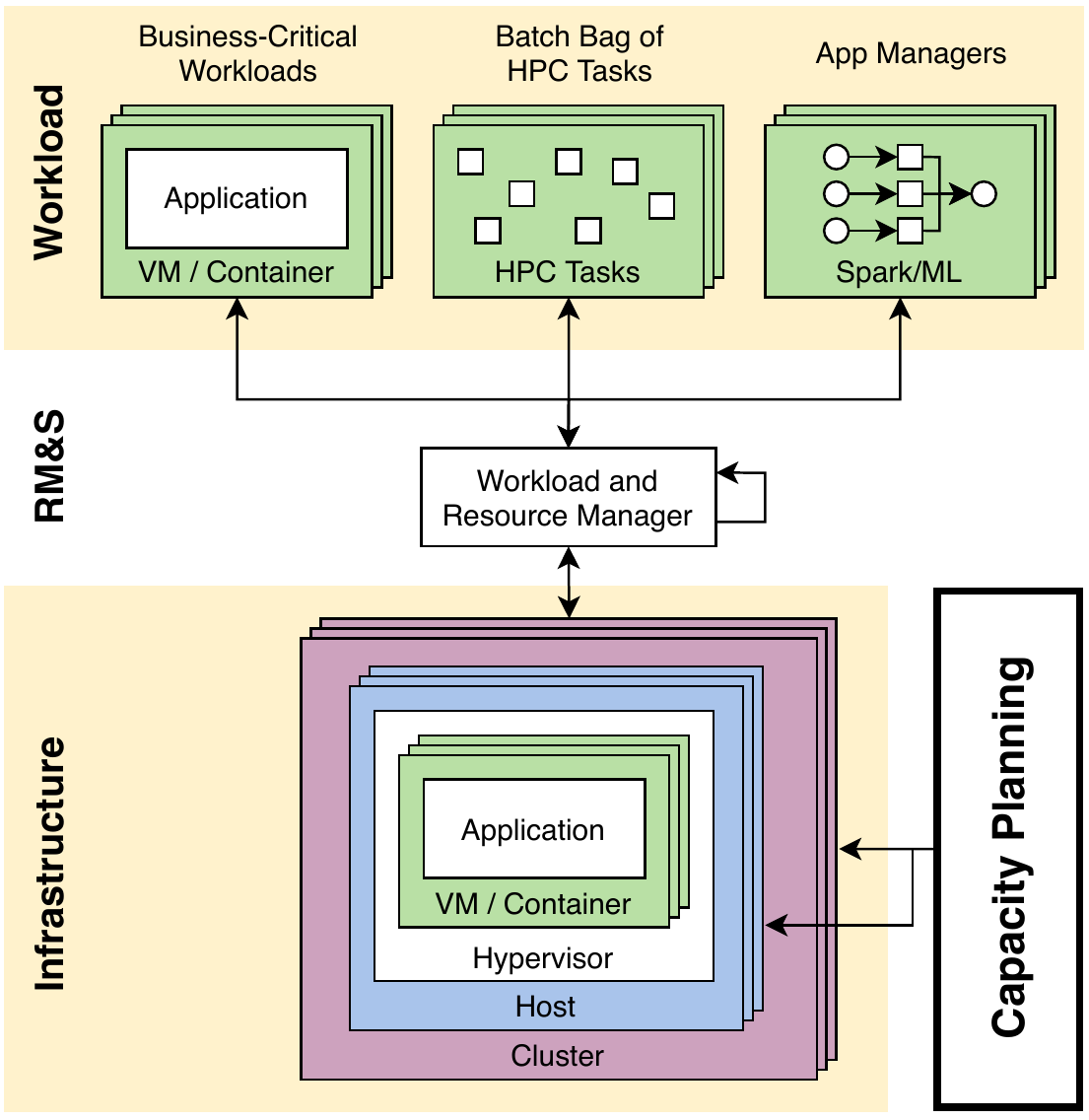}
    \vspace*{-0.75cm}
    \caption{Generic model for datacenter operation.} 
    \label{fig:system-model}
    \vspace*{-0.5cm}
\end{figure}

In this work we assume the generic model of cloud infrastructure and its operation depicted by Figure~\ref{fig:system-model}~(next page).

{\bf Workload:}
The workload consists of applications executing in {\em \glspl{VM}} and {\em containers}.
The emphasis of this study is on {\it business-critical workloads}, which are long-running, typically user-facing, and back-end enterprise services at the core of an enterprise's business~\cite{Shen2011,Shen2015}.
Their downtime, or even just low \gls{QoS}, can incur significant and long-lasting damage to the business.
We also consider virtual {\em public cloud workloads} in this model, submitted by a wider user base. 

The 
business-critical
workloads we consider also include 
virtualized \gls{HPC} parts. 
These 
are primarily comprised of conveniently~(embarrassingly) parallel tasks, e.g., Monte Carlo simulations, forming {\em batch bags-of-tasks}. Larger \gls{HPC} workloads, such as scientific workloads from healthcare, also fit in our model.

Our system model also considers app managers, such as the big data frameworks Spark and Apache Flink, and machine learning frameworks such as TensorFlow, which orchestrate virtualized workflows and dataflows. 

{\bf Infrastructure:}
The workloads described earlier run on physical datacenter infrastructure.
Our model views datacenter infrastructure as a set of physical clusters of possibly {\it heterogeneous hosts} (machines), each host being a node in a datacenter rack.
A host can execute multiple \gls{VM} or container workloads, managed by a {\em hypervisor}.
The hypervisor allocates computational time on the CPU between the workloads that request it, through {\em time-sharing} (if on the same cores) or {\em space-sharing} (if on different cores).

We model the CPU usage of applications for discretized time slices.
Per slice, all workloads report requested CPU time to the hypervisor and receive the granted CPU time that the resources allow.
We assume a generic memory model, with memory allocation constant over the runtime of a \gls{VM}.
As is common in industry, we allow overcommission of CPU resources~\cite{Baset2012}, but not of memory resources~\cite{Shen2015}. 

{\bf Infrastructure phenomena:} Cloud datacenters are complex hardware and software ecosystems, in which complex phenomena emerge.
We consider in this work two well-known operational phenomena, performance variability caused by performance interference between collocated VMs~\cite{VanBeek2019,Koh2007,Krebs2014} and correlated cluster failures~\cite{Gallet2010,DBLP:conf/dsn/BirkeGCWE14,DBLP:conf/icdcs/El-SayedZS17}. 

{\bf Live Platform Management}~(RM\&S in Figure~\ref{fig:system-model}){\bf :}
We model a workload and resource manager that performs management and control of all clusters and hosts, and is responsible for the lifecycle of submitted \gls{VM}s, including their placement onto the available resources~\cite{Andreadis2018}.
The resource manager is configurable and supports various {\em allocation policies}, defining the distribution of workloads over resources.
The devops team monitors the system and responds to incidents that the resource management system cannot self-manage~\cite{book/BeyerJPM16}.

{\bf Capacity Planning:}
Closely related with infrastructure and live platform management is the activity of {\em capacity planning}.
This activity is conducted periodically and/or at certain events by a capacity planner (or committee). 
The activity typically consists of first {\em modeling} the current state of the system (including its workload and infrastructure)~\cite{Menasce2001}, {\em forecasting} future demand~\cite{Chamness2011}, deriving a capacity {\em decision}~\cite{Zhang2009}, and finally {\em calibrating and validating} the decision~\cite{Kejariwal2017}. The latter is done for \gls{QoS},  possibly expressed as detailed \glspl{SLA} and \glspl{SLO}.
In Section~\ref{sec:interviews} we analyze the current state of practice and in Section~\ref{sec:related-work} we discuss existing approaches in literature.

{\bf Which cloud datacenters are relevant for this model?}
We focus in this work on capacity planning for {\em mid-tier cloud infrastructures}, characterized by relatively small-scale capacity, temporary overloads being common, and a lack of in-house tools or teams large enough to develop them quickly.  
In Section~\ref{sec:interviews} we analyze the current state of the capacity planning practice in this context and in Section~\ref{sec:related-work} we discuss existing approaches in related literature.

{\bf Which tools support this model?}
We are not aware of analytical tools that can cope with these complex aspects. 
Although tools for \gls{VM} simulation exist~\cite{Calheiros11,Hirofuchi18,Nunez12}, few support CPU over-commissioning and none outputs detailed VM-level metrics; the same happens for infrastructure phenomena.
From the few industry-grade procurement tools who published details about their operation, none supports the diverse workloads and phenomena considered here.

\begin{table*}[!t]
    \centering
    \caption{Summary of interviews. (Notation: TTD = Time to Deploy, CP = Cloud Provider, DC = Datacenter, M = Monitoring, m/y = month/year, NIT = National IT Infrastructure Provider, SA = Spreadsheet Analysis.)}
    \vcutM{}
    \begin{tabularx}{\linewidth}{rXccccccc}
        \toprule
        Int. & Role(s) & Backgr. & Scale & Scope & Tooling & Workload Comb. & Frequency & TTD \\
        \midrule
        1 & Researcher              & CP     & rack & multi-DC & M & combined & 3m, ad-hoc & ? \\
        2 & Board Member   & NIT & iteration & multi-DC & -- & combined & 4--5y & 12--18m \\
        3 & Manager, Platform Eng.  & CP     & rack & multi-DC & M & combined & ad-hoc & 4--5m \\
        4 & Manager                 & NIT & iteration & per DC & M & benchmark & 6--7y & 18m \\
        5 & Hardware Eng.           & NIT & iteration & per DC & M & benchmark & 6y & 18m \\
        6 & Researcher              & NIT & rack & multi-DC & M & separate & 6m & 12m \\
        7 & Manager                 & NIT & iteration & multi-DC & M, SA & combined & 5y & 3.5-4y \\
        \bottomrule
    \end{tabularx}
    \label{tab:interviews:results}
    \vcutL{}
\end{table*}

\section{Real-World Experiences with Capacity Planning in Cloud Infrastructures}
\label{sec:interviews}

Real-world practice can deviate significantly from published theories and strategies.
In this section, we conduct and analyze interviews with 8 practitioners from a wide range of backgrounds and multiple countries, to assess whether this is the case in the field of capacity planning.

\vcutS{}
\subsection{Method}
\label{sec:interviews:method}
\vcutS{}

Our goal is to collect real-world experiences from practitioners systematically and without bias, yet also leave room for flexible, personalized lines of investigation.

\vcutS{}
\subsubsection{Interview type}
\vcutS{}
The choice of interview type is guided by the trade-off between the systematic and flexible requirements.
A text survey, for example, is highly suited for a systematic study, but generally does not allow for low-barrier individual follow-up questions or even conversations.
An in-person interview without pre-defined questions allows full flexibility, but can result in unsystematic results.
We use 
the \emph{general interview guide approach}~\cite{Turner2010}, 
a semi-structured type of interview that ensures certain key topics are covered but permits deviations from the script.
We conduct in-person interviews with a prepared script of ranked questions, and allow the interviewer the choice of which scripted questions to use and when to ask additional questions.

\vcutS{}
\subsubsection{Data collection}
\vcutS{}

Our data collection process involves three steps.
Firstly, we {\it selected and contacted} a broad set of prospective interviewees representing various kinds of datacenters, with diverse roles in the process of capacity planning, and with diverse responsibility in the decisions. 

Secondly, we {\it conducted and recorded the interviews}. 
Each interview is conducted in person and digitally recorded with the consent of the interlocutor.
Interviews last between 30 and 60 minutes, depending on availability of the interlocutors and complexity of the discussion.
To help the interviewer select questions and fit in the time-limits imposed by each interviewee,
we rank questions by their importance and group questions broadly into 5 categories: (1) introduction, (2) process, (3) inside factors, (4) outside factors, and (5) summary and followup.
The choice between questions is then dynamically adjusted to give precedence to higher-priority questions and to ensure each category is covered at least briefly.
The script itself is listed in Appendix~\ref{sec:interview-script}.

Thirdly, the recordings are {\it manually transcribed} into a full transcript to facilitate easy analysis.
Because matters discussed in these interviews may reveal sensitive operational details about the organisations of our interviewees, all interview materials are handled {\it confidentially}.
No information that could reveal the identity of the interlocutor or that could be confidential to an organization's operations is shared without the explicit consent of the interlocutor.
In addition, all raw records 
will be destroyed directly after this study.


\vcutS{}
\subsubsection{Analysis of Interviews}
\vcutS{}
Due to the unstructured nature of the chosen interview approach, we combine a question-based aggregated analysis with incidental findings.
Our approach is inspired by the Grounded Theory strategy set forth by \citeauthor{Coleman2007}~\cite{Coleman2007}, and has two steps.
First, for each transcript, we {\it annotate} each statement made based on which questions it is relevant to.
This may be a sub-sentence remark or an entire paragraph of text, frequently overlapping between different questions.
We augment this systematic analysis with more general findings, including comments on unanticipated topics.

Secondly, we traverse all transcripts for each question and form {\it aggregate observations for each question} in the transcript.
Appendix~\ref{sec:interview-results} details the full findings.
From these, we synthesize \toolname{} requirements~(\S\ref{sec:design:requirements}).

\vcutS{}
\subsection{Observations from the Interviews}
\label{sec:interviews:results}
\vcutS{}

Table~\ref{tab:interviews:results} summarizes the results of the interviews. 
In total, we transcribed 
over 35,000 words in 3 languages,
which is a very large amount of raw interview data. 
We conducted 7 interviews with
practitioners from commercial and academic datacenters, with roles ranging from capacity planners, to datacenter engineers, to managers.
%
We summarize here our main observations:

\obs{typical-process}
A majority of practitioners find that the process involves a {\em significant amount of guesswork and human interpretation} (see detailed finding (\finref{typical:guessing}) in App.~\ref{sec:interview-results}).
Interlocutors managing commercial infrastructures emphasize multi-disciplinary challenges such as lease and support contracts, and personnel considerations~(\finref{typical:commercial:lease-support-important}, \finref{typical:commercial:financial-vs-human}).

\obs{absence-of-tooling}
In all interviews, we notice the {\em absence of any dedicated tooling} for the capacity planning process~(\finref{tools:none}).
Instead, the surveyed practitioners rely on visual inspection of data, through monitoring dashboards~(\finref{tools:format}).
We observe two main reasons for not using dedicated tooling:
(1) tools tend to under-represent the complexity of the real situation, and
(2) have high costs with many additional, unwanted features~(\finref{tools:issues}).

\obs{type-of-services}
The organizations using these capacity planning approaches provide a {\em range of digital services}, ranging from general IT services to specialist hardware hosting~(\finref{services:general}).
They run
\gls{VM} workloads, in both commercial and scientific settings, and batch and \gls{HPC} workloads, mainly in scientific settings~(\finref{services:per-domain}).

\obs{main-factors}
A {\em large variety of factors} are taken into account when planning capacity~(\finref{factors:plurality}).
The three named in a majority of interviews are (1) the use of historical monitoring data, (2) financial concerns, and (3) the lifetime and aging of hardware~(\finref{factors:majority}).

\begin{figure*}[!t]
    \centering
    \includegraphics[width=0.8\textwidth]{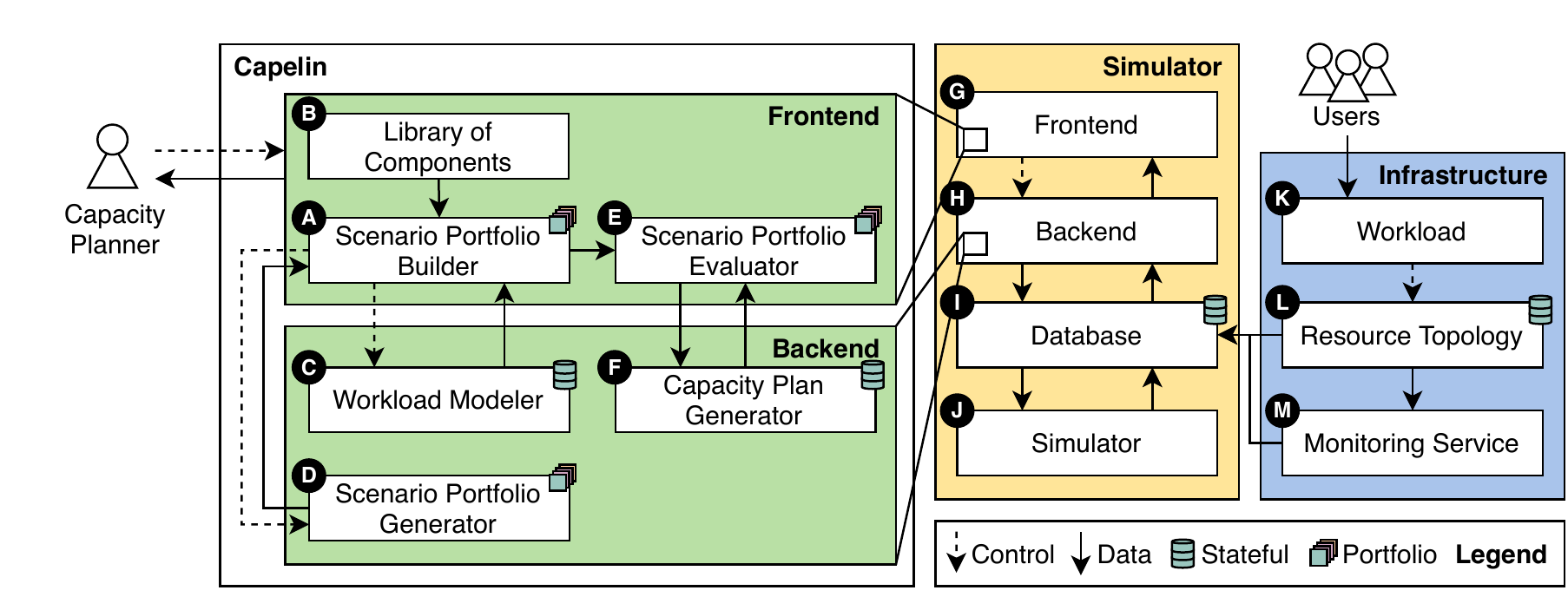}
    \vcutM{}
    \caption[Overview of the architecture of \toolname{}.]{An overview of the architecture of \toolname{}. \toolname{} is provided information on the current state of the infrastructure and assists the capacity planner in making capacity planning decisions. Labels indicate the order of traversal by the capacity planner (e.g., the first step is to use component \designref{A}, the scenario portfolio builder).}
    \label{fig:architecture}
    \vcutM{}
\end{figure*}

\obs{definition-of-success}
{\em Success and failure} in capacity planning are underspecified. 
Definitions of success differ:
two interviewees see the use of new technologies as a success~(\finref{success:new-tech}), and 
one interprets the absence of total failure events 
as a success~(\finref{success:absence-of-failures}).
Challenges include chronic underutilization~(\finref{failures:underutilization}), increasing complexity~(\finref{failures:complexity}), and small workloads~(\finref{failures:small-deployments}). 
Failures include decisions taking long~(\finref{failures:long-decisions}), misprediction~(\finref{failures:mispredictions}), and new technology having unforeseen consequences~(\finref{failures:new-tech}).

\obs{frequency}
{\it The frequency of capacity planning processes seems correlated with the duration of core activities using it:}
commercial clouds deploy within 4-5 months from the start of capacity planning, whereas scientific clouds take 1--1.5 years~(\finref{frequency:frequency}, \finref{frequency:time-to-deployment}). 

\obs{financial}
We found three {\em financial and technical factors} that play a role in capacity planning: (1) funding concerns, (2) special hardware requests, and (3) the cost of new hardware~(\finref{financial:overall}).
In two interviews, interlocutors state that financial considerations prime over the choice of technology, such as the vendor and model~(\finref{financial:choice-of-tech}).

\obs{human}
The {\em human aspect} of datacenter operations is emphasized in 5 of the 7 interviews~(\finref{human:very-involved}).
%
The datacenter administrators need training~(\finref{human:focus}), and wrong decisions in capacity planning lead to stress within the operational teams~(\finref{human:stress}).
Users also need training, to leverage heterogeneous or new resources~(\finref{human:focus}).

\obs{wishes}
We observe a {\em wide range of requirements and wishes expressed by interlocutors} about custom tools for the process.
Fundamentally, the tool should help manage the increasing complexity faced by capacity planners~(\finref{ideal:complexity}). 
A key requirement for any tool is interactivity: practitioners want to be able to interact with the metrics they see and ask questions from the tool during capacity planning meetings~(\finref{ideal:interactivity}).
The tool should be affordable and usable without needing the entire toolset of the vendor~(\finref{ideal:availability}). 
One interviewee asks for support for infrastructure heterogeneity, to support scientific computing~(\finref{ideal:heterogeneity}).

\obs{what-ifs}
Two interviewees detail {\em ``what-if'' scenarios} they would like to explore with a tool, using several dimensions~(\finref{ideal:what-if}):
(1) the {\em topology}, in the form of the computational and memory capacity needed, or new hardware arriving;
(2) the {\em workload}, and especially emerging kinds; 
and
(3) the {\em operational phenomena}, such as failures and the live management of the platform (e.g., scheduling and fail-over scenarios).

\section{Design of \toolname{}: a Capacity Planning System for Cloud Infrastructure}
\label{sec:design}

In this section, we synthesize requirements and design around them a capacity planning approach for cloud infrastructure.
We propose \toolname{}, a scenario-based capacity planning system that helps practitioners understand the impact of alternatives. 
Underpinning this process, we propose as core abstraction the {\it portfolio of capacity planning scenarios}.


\vcutM{}
\subsection{Requirements Analysis}
\label{sec:design:requirements}
\vcutS{}

In this section, 
from the results of Section~\ref{sec:interviews},
we synthesize the core functional requirements addressed by \toolname{}.
Instead of aiming for full automation -- a future objective that is likely far off for the field of capacity planning -- the emphasis here is on human-in-the-loop decision support~\cite[P2]{Iosup2018}.

\begin{description}
    \item[(\setID{FR1})] \textbf{Model a cloud datacenter environment}~(see \obsref{absence-of-tooling}, \obsref{type-of-services}, \obsref{financial}, and~\obsref{wishes}):
    The system should enable the user to model 
    the datacenter topology and virtualized workloads introduced in Section~\ref{sec:model}.
    
    \item[(\setID{FR2})] \textbf{Enable expression of what-if scenarios}~(see \obsref{absence-of-tooling}, \obsref{what-ifs}):
    Users can express what-if scenarios with diverse topologies, failures, and workloads.
    The system should then execute the what-if scenario(s), and produce and justify a set of user-selected \gls{QoS} metrics. 
    
    \item[(\setID{FR3})] \textbf{Enable expression of \gls{QoS} requirements},
    in the form of \glspl{SLA}, consisting of several \glspl{SLO}~(see \obsref{absence-of-tooling}, \obsref{definition-of-success}, \obsref{wishes}).
    These requirements are formulated as thresholds or ranges of acceptable values for user-selected metrics. 
    
    \item[(\setID{FR4})] \textbf{Suggest a portfolio of what-if scenarios}, 
    based on user-submitted workload traces, given topology, and specified \gls{QoS} requirements~(see \obsref{absence-of-tooling}, \obsref{what-ifs}).
    This greatly simplifies identifying meaningful scenarios. 
    
    \item[(\setID{FR5})] \textbf{Provide and explain a capacity plan},  
    optimizing for minimal capacity within acceptable \gls{QoS} levels, as specified by \ref{FR4}~(see \obsref{absence-of-tooling}, \obsref{wishes}).
    The system should explain and 
    visualize the data sources it used to make the plan. 
\end{description}

\subsection{Overview of the \toolname{} Architecture}
\label{sec:design:architecture}
\vcutS{}

On the previous page,
Figure~\ref{fig:architecture} depicts an overview of the \toolname{} architecture.
\toolname{} extends 
\simulator{}, an open-source, discrete event simulator with multiple years of development and operation~\cite{Iosup2017}.
We now discuss each main component of the \toolname{} architecture, taking the perspective of a capacity planner.
We outline the abstraction underpinning this architecture, the capacity planning portfolios, in \S\ref{sec:design:portfolio}.

\vcutS{}
\subsubsection{The \toolname{} Process}
\vcutS{}
The frontend and backend of \toolname{} are embedded in \simulator{}.
This enables \toolname{} to leverage the simulator's existing platform for datacenter modeling and allows for inter-operability with other tools as they become part of the simulator's ecosystem.
The capacity planner interacts with the frontend of \toolname{}, starting with the {\em Scenario Portfolio Builder}~(component \designref{A} in Figure~\ref{fig:architecture}), addressing~\ref{FR2}.
This component 
enables the planner to 
construct scenarios, 
using pre-built components from the {\em Library of Components}~(\designref{B}).
The library contains workload, topology, and operational building blocks, facilitating fast composition of scenarios.
If the (human) planner wants to modify historical workload behavior or anticipate future trends, the {\em Workload Modeler}~(\designref{C}) can model
workloads and synthesize custom loads.

The planner might not always be aware of the full range of possible scenarios. 
The {\em Scenario Portfolio Generator}~(\designref{D}) suggests customized scenarios 
extending the given base-scenario~(\ref{FR4}).
The portfolios built in the builder can be explored and evaluated in the {\em Scenario Portfolio Evaluator}~(\designref{E}).
Finally, based on the results from this evaluation, the {\em Capacity Plan Generator}~(\designref{F}) suggests plans to the planner~(\ref{FR5}). 

\vcutS{}
\subsubsection{The Datacenter Simulator}
\vcutS{}


In Figure~\ref{fig:architecture}, the {\em Frontend}~(\designref{G}) acts as a portal, through which infrastructure stakeholders interact with its models and experiments. 
The {\em Backend}~(\designref{H}) responds to frontend requests, acting as intermediary and business-logic between frontend, 
and database and simulator. 
The {\em Database}~(\designref{I}) manages the state, 
including topology models, historical data, simulation configurations, and simulation results.
It receives inputs from the real-world topology and monitoring services, in the form of workload traces.
The {\em Simulator}~(\designref{J}) evaluates the configurations stored in the database and reports the simulation results 
back to the database.

\simulator{}~\cite{Iosup2017,DBLP:conf/ccgrid/MastenbroekAI+21} is the simulation platform backing \toolname{}, enabling the capacity planner to model~(\ref{FR1}) and experiment~(\ref{FR5}) with the cloud infrastructure, interactively.
The software stack of this platform is composed of a web app frontend, a web server backend, a database, and a 
discrete-event simulator. This kind of simulator offers a good trade-off between accuracy and performance, even at the scale of mid-tier datacenters and with long-term workloads. 

\vcutS{}
\subsubsection{Infrastructure}
\vcutS{}
The cloud infrastructure is at the foundation of this architecture, forming the system to be managed and planned.
We consider three components within this infrastructure: The {\em workload}~(\designref{K}) submitted by users, the (logical or physical) {\em resource topology}~(\designref{L}), and a {\em monitoring service}~(\designref{M}).
The infrastructure follows the system model described in Section~\ref{sec:background}.


\vcutS{}
\subsection{A Portfolio Abstraction for Cap. Planning}
\label{sec:design:portfolio}
\vcutS{}

\begin{figure}[!t]
    \centering
    \includegraphics[width=0.75\columnwidth]{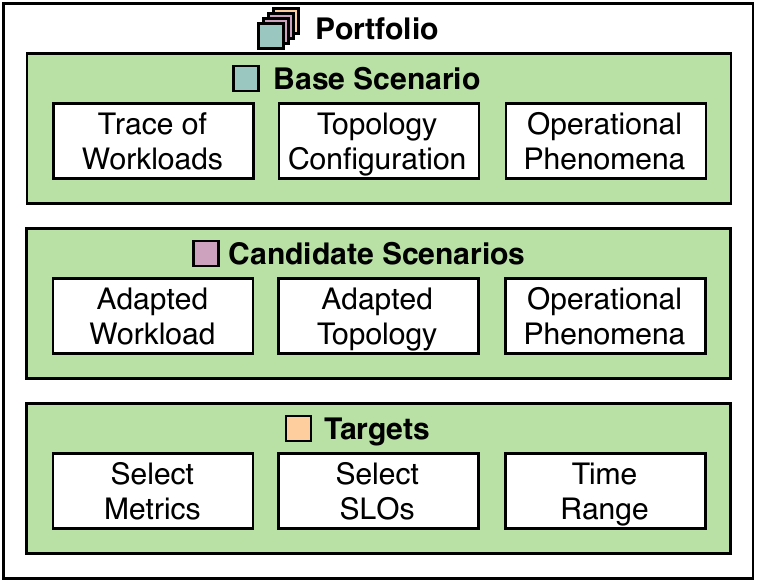}
    \vcutM{}
    \caption{Abstraction of a capacity planning portfolio, consisting of a base scenario, a number of candidate scenarios, and comparison targets.}
    \label{fig:portfolio}
    \vcutM{}
\end{figure}

In this section, 
we propose a new abstraction,  
which organizes multiple scenarios into a {\em portfolio}~(see Figure~\ref{fig:portfolio}).
Each portfolio includes a base scenario, a set of candidate scenarios given by the user and/or suggested by \toolname{},
and a set of targets to compare scenarios.
In contrast, most capacity planning approaches in published literature are tailored towards a {\em single} scenario---a single potential hardware expansion, a single workload type, one type of service quality metrics.
This approach does not 
cover the complexities that capacity planners are facing (see Section~\ref{sec:interviews:results}). 
Our portfolio reflects the multi-disciplinary and multi-dimensional nature of capacity planning by 
including {\em multiple} scenarios and a {\it set} of targets. We describe them, in turn.
%
%
%


\vcutS{}
\subsubsection{Scenarios}
\vcutS{}
A scenario represents a point in the capacity planning~(datacenter design) space to explore. 
It consists of a combination of workload, topology, and a set of {\it operational phenomena}.
Phenomena can include correlated failures, performance variability, security breaches, etc., %
allowing the scenarios to more accurately capture the real-world operations. 
Such phenomena are often hard to predict intuitively during capacity planning, due to emergent behavior that can arise at scale.

The baseline for comparison 
in a portfolio is the \emph{base scenario}.
It represents the status quo of the infrastructure or, when planning infrastructure from scratch, it consists of very simple base workloads and topologies.

\begin{table*}[!t]
	\centering
	\caption{Experiment configurations. A legend of topology dimensions is provided below. (Notation: PI = Performance Interference, pub = public cloud trace, pri = private cloud trace.)}
	\label{tab:experiment-overview}
	\vcutM{}
	\begin{tabularx}{\linewidth}{lXccccccccc}
		\toprule
		& & \multicolumn{4}{c}{Candidate Topologies} & \multicolumn{2}{c}{Workloads} & \multicolumn{2}{c}{Op. Phenomena} & \\
		\cmidrule(lr){3-6}
        \cmidrule(lr){7-8}
        \cmidrule(lr){9-10}
		Sec. & Focus & Mode & Quality & Direction & Variance & Trace & Loads & Failures & PI & Alloc. Policy \\ \midrule
		\S\ref{sec:experiments:hor-ver} & Hor. vs. Ver. & \iconRep{} \iconExp{} & \iconVol{} \emptySpace{} & \iconHor{} \iconVer{} & \iconHom{} \iconHet{} & pri & sampled & $\cmark$ & $\cmark$ & active-servers \\
		\S\ref{sec:experiments:more-vel} & Velocity & \iconRep{} \iconExp{} & \emptySpace{} \iconVel{} & \emptySpace{} \iconVer{} & \iconHom{} \iconHet{} & pri & sampled & $\cmark$ & $\cmark$ & active-servers \\
		\S\ref{sec:experiments:phenomena} & Op. Phen.  & \iconEmpty{} -- \iconEmpty{} & -- & -- & -- & pri & original & $\xmark$ / $\cmark$ & $\xmark$ / $\cmark$ & all \\
		\S\ref{sec:experiments:composite} & Workloads & \iconExp{} \emptySpace{} & \iconVol{} \iconVel{} & \iconHor{} \iconVer{} & \iconHom{} \emptySpace{} & pri / pub & sampled & $\cmark$ & $\xmark$ & active-servers \\
		\bottomrule\noalign{\vskip 2mm}  
		\multicolumn{11}{@{}c}{
		    \includegraphics[width=\textwidth]{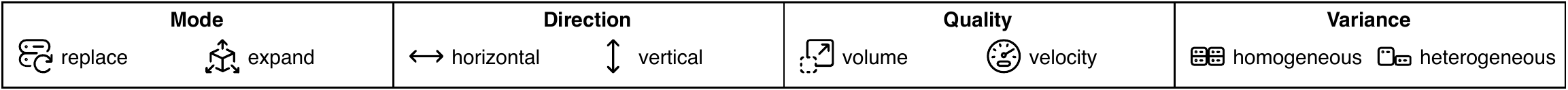}
		}
	\end{tabularx}
	\vcutM{}
\end{table*}

The other scenarios in a portfolio, called \emph{candidate scenarios}, represent changes to the configuration that the capacity planner could be interested in.
Changes can be effected in one of the following four dimensions: 
(1) \emph{Variety:} 
qualitative changes to the workload or topology (e.g., different arrival patterns, or resources with more capacity); 
(2) \emph{Volume:} 
quantitative changes to the workload or topology (e.g., more workloads or more resources); 
(3) \emph{Velocity:} 
speed-related changes 
to workload or topology (e.g., faster resources); and 
(4) \emph{Vicissitude} combines (1)-(3) over time.

This approach to derive candidate scenarios is systematic, and although abstract it allows approaching many of the practical problems discussed by capacity planners. 
For example, an ongoing discussion 
is horizontal scaling~(scale-out) vs. vertical~(scale-up)~\cite{DBLP:journals/internet/SalihogluO18}.
Horizontal scaling, which is done by adding clusters and commodity machines, contrasts to vertical scaling, which is done by acquiring more expensive, ``beefy'' machines. 
Horizontal scaling is typically cheaper for the same performance, and offers a broader failure-target~(except for cluster-level failures). 
Yet, vertical scaling could lower operational costs, due to fewer per-machine licenses, fewer switch-ports for networking, and smaller floor-space due to fewer racks.
Experiment~\ref{sec:experiments:hor-ver} explores this dichotomy.

\vcutS{}
\subsubsection{Targets}
\vcutS{}
A portfolio also has a set of targets that prescribe on what grounds the different scenarios should be compared.
Targets include the metrics that the practitioner is interested in and their desired granularity, along with relevant \glspl{SLO}~(\ref{FR3}).
Following the taxonomy defined by the performance organization SPEC~\cite{Herbst2018}, 
we support both {\it system-provider metrics} (such as operational risk and resource utilization) and {\it organization metrics} (such as \gls{SLO} violation rates and performance variability).
The targets also include a time range over which these metrics should be recorded and compared.

\section{Experiments with \toolname{}}
\label{sec:evaluation}\label{sec:exp}

In this section, we explore how \toolname{} can be used to answer capacity planning questions.
We conduct extensive experiments using \toolname{} and data derived from operational traces collected long-term from private and public cloud datacenters.

\vcutS{}
\subsection{Experiment Setup}\label{sec:exp:setup}
\vcutS{}

We implement a prototype of \toolname{}~(\S\ref{sec:exp:setup:sw}), and verify the reproducibility of its results and that it can be run within the expected duration of a capacity planning session~(\S\ref{sec:exp:setup:exec}).
All experiments use 
{\it long-term, real-world traces as input}. 

Our experiment design, which Table~\ref{tab:experiment-overview} summarizes, is comprehensive and addresses key questions such as:
Which input workload~(\S\ref{sec:exp:setup:wl})? 
Which datacenter topologies to consider~(\S\ref{sec:exp:setup:topology})?
Which operational phenomena~(\S\ref{sec:exp:setup:phenomena})? Which allocation policy~(\S\ref{sec:exp:setup:policies})? 
Which user- and operator-level performance metrics to use, to compare the scenarios proposed by the capacity planner~(\S\ref{sec:exp:setup:metrics})?


The most important decision for our experiments is which scenarios to explore.
Each experiment takes in a capacity planning portfolio~(see Section~\ref{sec:design:portfolio}), starts from a base scenario, and aims to extend the portfolio with new candidate scenarios and its results. 
{\it The baseline is given by expert datacenter engineers, and has been validated with hardware vendor teams.}
\toolname{} creates new candidates by modifying the base scenario along dimensions such as variety, volume, and velocity of any of the scenario-components. 
In the following, we experiment systematically with each of these.





\vcutS{}
\subsubsection{Software prototype}\label{sec:exp:setup:sw}
\vcutS{}
We extend the open-source \simulator{} simulation platform~\cite{Iosup2017} with 
capabilities 
for modeling and simulating the {\it virtualized workloads} prevalent in modern clouds.
We 
model the CPU and memory usage of each \gls{VM} along with hypervisors deployed on each managed node.
Each hypervisor implements a fair-share scheduling model for \glspl{VM}, granting each \gls{VM} at least a fair share of the available CPU capacity, but also allowing them to claim idle capacity of other \glspl{VM}.
The scheduler permits {\it overprovisioning of CPU resources, but not of memory resources}, as is common in industry practice.
We also
model a workload and resource manager that controls the deployed hypervisors and decides based on configurable allocation policies~(described in \S\ref{sec:exp:setup:policies}) to which hypervisor to allocate a submitted \gls{VM}.
Our experiments and workload samples are orchestrated by \toolname{}, which is written in Kotlin~(a modern JVM-based language), and processed and analyzed by a suite of tools based on Python and Apache Spark.
More detail about the software implementation is given in Appendix~\ref{sec:software-implementation}.

{\it We release our extensions of the open-source \simulator{} codebase and the analysis software artifacts on GitHub}\footnote{\url{https://github.com/atlarge-research/opendc}}, as part of release~2.0~\cite{DBLP:conf/ccgrid/MastenbroekAI+21}.
We conduct thorough validation and tests of both the core \simulator{} and our additions, as detailed in Section~\ref{sec:simulator-validation}.

\vcutS{}
\subsubsection{Execution and Evaluation}\label{sec:exp:setup:exec}
\vcutS{}
Our results are fully reproducible, 
regardless of the physical host running them.
All setups are repeated 32 times.
The results, in files amounting to hundreds of GB in size due to the large workload traces involved, are evaluated statistically and verified independently.
Factors of randomness (e.g., random sampling, policy decision making if applicable, and performance interference modeling) are seeded with the current repetition to ensure deterministic outcomes, and for fairness are kept consistent across scenarios.

\toolname{} could be used during capacity planning meetings. 
A single evaluation takes 1--2 minutes to complete, enabled by many technical optimizations we added to the simulator.
The full set of experiments is conveniently parallel and takes around 1 hour and 45 minutes to complete, on a ``beefy'' but standard machine with 64 cores and 128GB RAM; parallelization across multiple machines would reduce this to minutes.

\vcutS{}
\subsubsection{Workload}\label{sec:exp:setup:wl}
\vcutS{}
We experiment with a business-critical workload trace from \provider{}, a {\it private cloud provider}. 
The anonymized version of this trace has been published in a public trace archive~\cite{Iosup2008}.
We were provided with the full, deanonymized data artifacts of this trace, which consists of more than 1,500 \glspl{VM} along with information on which physical resources where used to run the trace and which \glspl{VM} were allocated to which resources.
We cannot release these full traces due to confidentiality, but release the summarized results. 

The {\it full trace} includes a range of \gls{VM} resource-usage measurements, aggregated over 5-minute-intervals over three months.
It consumes 3,063 PFLOPs~(\textit{exascale}), with the mean CPU utilization on this topology of 5.6\%.
This low utilization is in line with industry, where utilization levels below 15\% are common~\cite{Vasan2010}, and reduce the risk of not meeting \glspl{SLA}.

For all experiments, we consider the full trace, and further generate three other kinds of workloads as samples (fractions) of the original workload. These 
workloads are 
sampled from the full trace, resulting, in turn, to 306 PFLOPs~(0.1 of the full trace), 766~(0.25), and 1,532~(0.5).
To {\it sample}, \toolname{} takes randomly \glspl{VM} from the full trace and adds their entire load, until the resulting workload has enough load.
We illustrate this in pseudocode, in Algorithm~\ref{alg:vm-sampling}.

\begin{algorithm*}
\caption{Sampling procedure for the \glspl{VM} in \provider{} trace (as described in \S\ref{sec:exp:setup:wl}).}\label{alg:vm-sampling}
\begin{algorithmic}[1]
\Procedure{SampleTrace}{\textit{vms}, \textit{fraction}, \textit{totalLoad}}
    \State $\textit{selected} \gets \emptyset$ \Comment{The set of selected VMs}
    \State $\textit{load} \gets 0$ \Comment{Current total load (FLOP)}
    \While{$|\textit{vms}| > 0$}
        \State $\textit{vm} \gets \text{Randomly removed element from }\textit{vms}$
        \State $\textit{vmLoad} \gets \text{Total load of }\textit{vm}$
        \If{$\frac{\textit{load} + \textit{vmLoad}}{\textit{totalLoad}} > \textit{fraction}$}
            \State \Return \textit{selected}
        \EndIf
        \State $\textit{load} \gets \textit{load} + \textit{vmLoad}$
        \State $\textit{selected} \gets \textit{selected} \cup \{ \textit{vm} \}$
    \EndWhile
    \State \Return \textit{selected}
\EndProcedure
\end{algorithmic}
\end{algorithm*}



For the \S\ref{sec:experiments:composite} experiment, we further experiment with a public cloud trace from Azure~\cite{Cortez2017}.
We use the 
most recent release of the trace.
The formats of the Azure and the \provider{} traces are very similar, indicating 
a de facto standard has emerged across the 
private and public cloud communities.
One difference in the level of anonymity of the trace requires an additional assumption.
Whereas the \provider{} trace expresses CPU load as a frequency (MHz), the Azure trace expresses it as a utilization metric ranging from 0 to the number of cores of that \gls{VM}.
Thus, for the Azure trace, 
in line with Azure \gls{VM} types on offer we assume a maximum frequency of 3 GHz and scale each utilization measurement by this value.
The Azure trace is also shorter than \provider{}'s full trace, so we shorten the latter to Azure's length of 1 month.

We combine for the \S\ref{sec:experiments:composite} experiment the two traces and investigate possible phenomena arising from their interaction.
We disable here performance interference, because we can only derive it for the \provider{} trace~(see \S\ref{sec:exp:setup:phenomena}).
To combine the two traces, we first take a random sample of 1\% from the (very large) Azure trace, which results in 26,901~\glspl{VM} running for one month.
We then further sample this 1\%-sample, using the same method as for \provider{}'s full trace.
The full procedure is listed in Algorithm~\ref{alg:composite-sampling}.

\begin{algorithm*}
\caption{Sampling procedure for combining the private and private traces (as described in \S\ref{sec:exp:setup:wl}).}\label{alg:composite-sampling}
\begin{algorithmic}[1]
\Procedure{SampleMultipleTraces}{\textit{vmsPri}, \textit{fractionPri}, \textit{vmsPub}, \textit{fractionPub}}
    \State $\text{Ensure VMs in }\textit{vmsPri}\text{ and }\textit{vmsPri}\text{ have same length}$
    \State $\textit{vmsPub} \gets \text{Randomly sample 0.01 of all VMs in }\textit{vmsPub}$
    \State $\textit{totalLoad} \gets \text{Total CPU load of the private trace}$
    \State $\text{vmsPriSelected} \gets \textsc{SampleTrace}(vmsPri, fractionPri, totalLoad)$
    \State $\text{vmsPubSelected} \gets \textsc{SampleTrace}(vmsPub, fractionPub, totalLoad)$
    \State \Return $\textit{vmsPriSelected} \cup \textit{vmsPubSelected}$
\EndProcedure
\end{algorithmic}
\end{algorithm*}




\begin{table}
    \centering
    \caption{Aggregate statistics for both workloads used in this study. (Notation: AP = \provider{}.)}
    \label{tab:workload-stats}
    \vcutM{}
    \begin{tabularx}{\linewidth}{Xrrr}
        \toprule
        Characterization & & AP & Azure  \\
        \midrule
        \multirow{2}{*}{\makecell[cl]{VM submissions\\per hour}} & Mean ($\times 10^{-3}$) & 31.836 & 4.547 \\
        & CoV & 134.605 & 17.188 \\
        \midrule
        \multirow{2}{*}{VM duration [days]} & Mean & 20.204 & 2.495 \\
        & CoV & 0.378 & 3.072 \\
        \midrule
        \multirow{2}{*}{CPU load [TFLOPs]} & Mean ($\times 10^2$) & 9.826 & 64.046 \\
        & CoV & 
2.992 & 4.654 \\
        \bottomrule
    \end{tabularx}
    \vcutM{}
\end{table}


\vcutS{}
\subsubsection{Datacenter topology}\label{sec:exp:setup:topology}
\vcutS{}
As explained at the start of~\S\ref{sec:exp:setup}, 
for all experiments we set
the topology that ran \provider{}'s original workload (the full trace in~\S\ref{sec:exp:setup:wl}) as the base scenario's topology.
This topology is very common for industry practice. 
It is a subset of the complete topology of the \provider{} when the full trace was collected, but we cannot release the exact topology or the entire workload of \provider{} due to confidentiality. 


From the base scenario, \toolname{} derives candidate scenarios as follows.
First, it creates a temporary topology by choosing
half of the clusters in the topology, consisting of average-sized clusters and machines, compared to the overall topology. 
Second, it varies the temporary topology, in four dimensions:
\begin{enumerate*}[label=(\arabic*)]
    \item the {\it mode of operation}: replacement (removing the original half and replacing it with the modified version) and expansion (adding the modified half to the topology and keeping the original version intact);
    
    \item the {\it modified quality}: volume (number of machines/cores) and velocity (clock speed of the cores);
    
    \item the {\it direction of modification}: horizontal (more machines with fewer cores each) and vertical (fewer machines with more cores each); and 
    
    \item the {\it kind of variance}: homogeneous (all clusters in the topology-half modified in the same way) and heterogeneous (two thirds in the topology-half being modified in the designated way, the remaining third in the opposite way, on the dimension being investigated in the experiment).
    
\end{enumerate*}
Each 
dimension
is varied to ensure cores and machine counts multiply to (at least) the same total core count as before the change, in the modified part of the topology.
For volume changes,
we differentiate between a horizontal mode, where machines are given 28 cores (a standard size for machines in current deployments), and vertical modes, where machines are given 128 cores (the largest CPU models we see being commonly deployed in industry).
For velocity changes, we differentiate between the clock speed of the base topology and a clock speed that is roughly 25\% higher.
Because we do not investigate memory-related effects, the total memory capacity is preserved.

Last, due to confidentiality, 
we can describe the base and derived 
topologies only in relative terms. 


\vcutS{}
\subsubsection{Allocation policies}\label{sec:exp:setup:policies}
\vcutS{}
We consider several policies for the placement of \gls{VM}s on hypervisors: 
\begin{enumerate*}[label=(\arabic*)]
    \item prioritizing by available memory~({\tt mem}),
    \item by available memory per CPU core~({\tt core-mem}),
    \item by number of active \gls{VM}s~({\tt active-servers}),
    \item mimicking the original placement data~({\tt replay}), and
    \item randomly placing \gls{VM}s on hosts~({\tt random}).
\end{enumerate*}
Policies 1-3 are actively used in production datacenters~\cite{VanBeek2015}.

For each policy we use two variants, following the Worst-Fit strategy~(selecting the resource with the {\it most} available resource of that policy) and the Best-Fit strategy~(the inverse, so selecting the {\it least} available, labeled with the postfix~{\tt -inv} in \S\ref{sec:experiments:phenomena}).


\vcutS{}
\subsubsection{Operational phenomena}\label{sec:exp:setup:phenomena}
\vcutS{}

Each capacity planning scenario can include operational phenomena. In these experiments, we consider two such phenomena, 
\begin{enumerate*}[label=(\arabic*)]
    \item performance variability caused by performance interference between collocated \glspl{VM}, and \item correlated cluster failures.
\end{enumerate*}
Both 
are enabled, 
unless otherwise mentioned. 

We assume a common model~\cite{VanBeek2019,Koh2007} of performance interference, with a {\it score} from 0 to 1 for a given set of collocated workloads, with 0 indicating full interference between \glspl{VM} contending for the same CPU, and 1 indicating non-interfering \glspl{VM}.
We derive the value 
from the \emph{CPU Ready} fraction of a \gls{VM} time-slice: the fraction of time a \gls{VM} is ready to use the CPU but is not able to, due to other \glspl{VM} occupying it.
We mine the placement data of all \glspl{VM} running on the base topology and collect the set of collocated workloads along with their mean score, 
defined as the mean CPU ready time fraction subtracted from 1, conditioned by the total host CPU load at that time, rounded to one decimal.
At simulation time, this 
score is then 
activated if a \glspl{VM} is collocated with at least one of the others in the recorded set and the total load level on the system is at least the recorded load.
The 
score is then applied to each collocated \glspl{VM} 
with probability $1/N$, where $N$ is the number of collocated \glspl{VM}, 
by multiplying its requested CPU cycles with the score and granting it this (potentially lower) amount of CPU time.

The second phenomenon we model are cluster failures, which are based on a common model for space-correlated failures~\cite{Gallet2010} where a failure may trigger more failures within a short time span; these failures form a {\it group}.
We consider in this work only hardware failures that crash machines~(full-stop failures), with subsequent recovery after some duration.
We use a lognormal model with parameters for failure inter-arrival time, group size, and duration, as listed in Table~\ref{tab:failure-parameters}. 
The failure duration is further restricted by a minimum of 15 minutes, since faster recoveries and reboots at the physical level are rare.
The choice of parameter values is inspired by GRID'5000~\cite{Gallet2010}~(public trace also available~\cite{DBLP:journals/jpdc/JavadiKIE13}) and Microsoft Philly~\cite{DBLP:conf/usenix/JeonVPQXY19}, 
scaled to 
\provider{}'s topology. 

\begin{table}
    \centering
    \caption{Parameters for the lognormal failure model we use in experiments. We use the normal logarithm of each value.} 
    \label{tab:failure-parameters}
    \vcutM{}
    \begin{tabularx}{\linewidth}{Xrr}
        \toprule
        Parameter [Unit] & Scale & Shape  \\
        \midrule
        Inter-arrival time [hour] & $24 \times 7$ & $2.801$ \\
        Duration [minute] & $60$ & $60 \times 8$ \\
        Group size [machine-count] & $2$ & $1$ \\
        \bottomrule
    \end{tabularx}
    \vcutM{}
\end{table}

\vcutS{}
\subsubsection{Metrics}\label{sec:exp:setup:metrics}
\vcutS{}
In our article, we use the following metrics:
\begin{enumerate}[label=(\arabic*)]
    \item the total requested CPU cycles (in MFLOPs) of all \glspl{VM},
    \item the total granted CPU cycles (in MFLOPs) of all \glspl{VM},
    \item the total overcommitted CPU cycles (in MFLOPs) of all \glspl{VM}, defined as the sum of CPU cycles that were requested but not granted,
    \item the total interfered CPU cycles (in MFLOPs) of all \glspl{VM}, defined as the sum of CPU cycles that were requested but could not be granted due to performance interference,
    \item the total power consumption (in Wh) of all machines using a linear model based on machine load~\cite{Blackburn2008}, with an idle baseline of 200~W and a maximum power draw of 350~W,
    \item the number of time slices a \gls{VM} is in a failed state, summed across all \glspl{VM}.
    \item the mean CPU usage (in MHz), defined as the mean number of granted cycles per second per machine, averaged across machines,
    \item the mean CPU demand (in MHz), defined as the mean number of requested cycles per second per machine, averaged across machines,
    \item the mean number of deployed \gls{VM} images per host,
    \item the maximum number of deployed \gls{VM} images per host,
    \item the total number of submitted \glspl{VM},
    \item the maximum number of queued \glspl{VM} in the system at any point in time,
    \item the total number of finished \glspl{VM},
    \item the total number of failed \glspl{VM}.
\end{enumerate}

\textit{Note on the model for power consumption:} The current model, i.e., linear in the server load with offsets, is based on a peer-reviewed model and common to other simulators commonly used in practice, such as CloudSim and GridSim, and produces in general reasonable results for CPU power consumption. More accurate energy models appear for example in GreenCloud and in CloudNetSim++, which model the dynamic  energy-performance trade-off when using the DVFS technique, and in iCanCloud’s E-mc2 extension and in DISSECT-CF, which model every power state of each resource. 

\vcutS{}
\subsubsection{Listing of Full Results}\label{sec:exp:setup:full-results}
\vcutS{}
In the subsections below, we highlight a small selection of the key metrics.
For full transparency, we present the entire set of metrics for each experiment in the appendices.
Appendix~\ref{sec:full-visual-results} visualizes the full results for all metrics and Appendix~\ref{sec:full-tabular-results} lists the full results for the two most important metrics in tabular form.

\vcutS{}
\subsection{Horizontal vs. Vertical Resource Scaling}
\label{sec:experiments:hor-ver}
\vcutS{}




Our main findings from this experiment are:
\begin{description}
    
    \mainfin{complex-trade-offs} \toolname{} enables 
    the exploration of
    a complex trade-off portfolio of multiple metrics and capacity dimensions.
    
    \mainfin{vertical} Vertically scaled topologies can improve power consumption (median lower by 1.47x-2.04x) but can lead to significant performance penalties (median higher by 1.53x-2.00x) and increased chance of \gls{VM} failure (median higher by 2.00x-2.71x, which is a high risk{\bf !})
    
    \mainfin{failures-important} \toolname{} reveals how correlated failures impact various topologies. Here, 147k--361k VM-slices fail.
    
\end{description}



\begin{figure}[!t]
    \centering
    \includegraphics[width=\columnwidth]{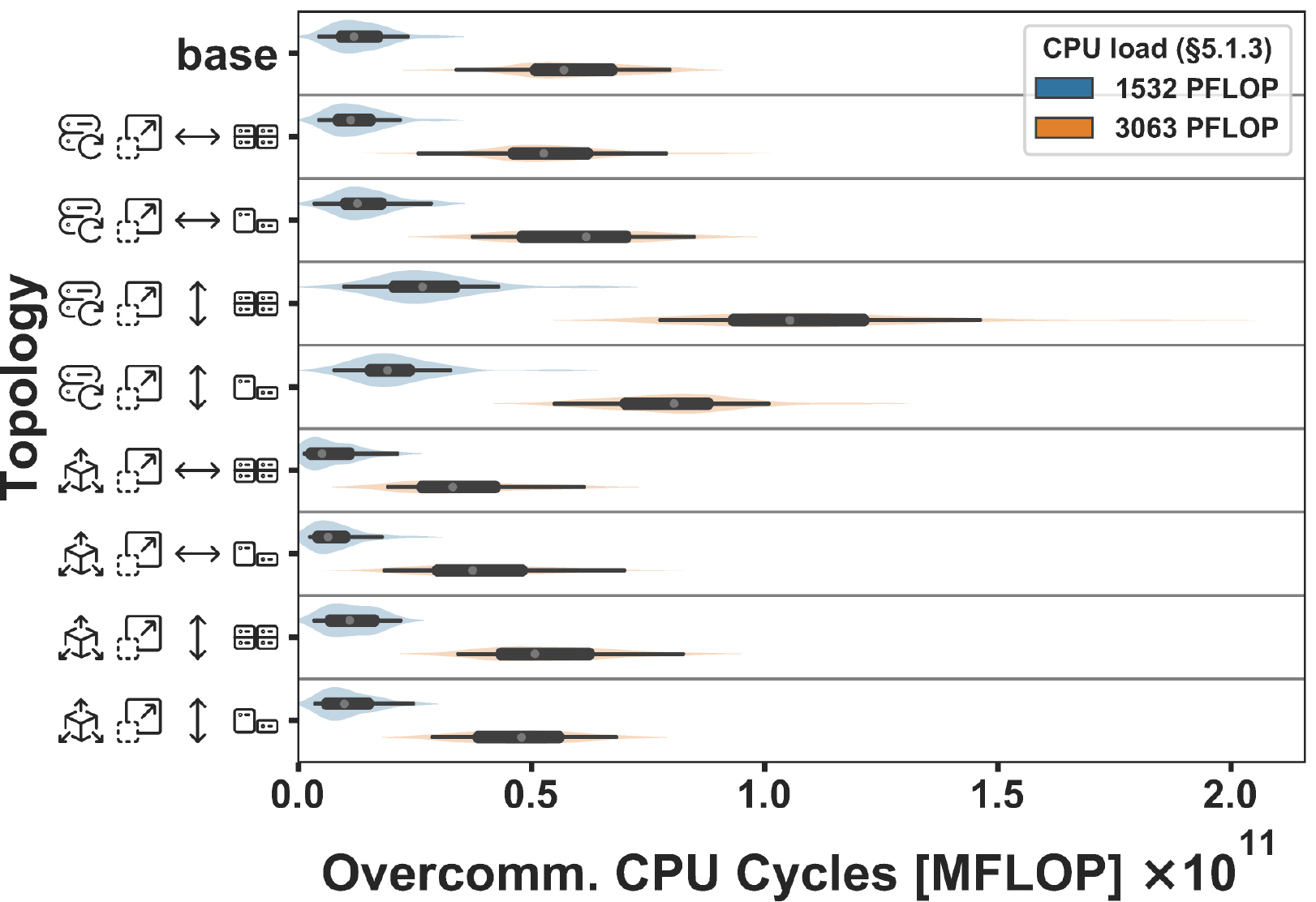}\vspace*{0.25cm}
    \includegraphics[width=\columnwidth]{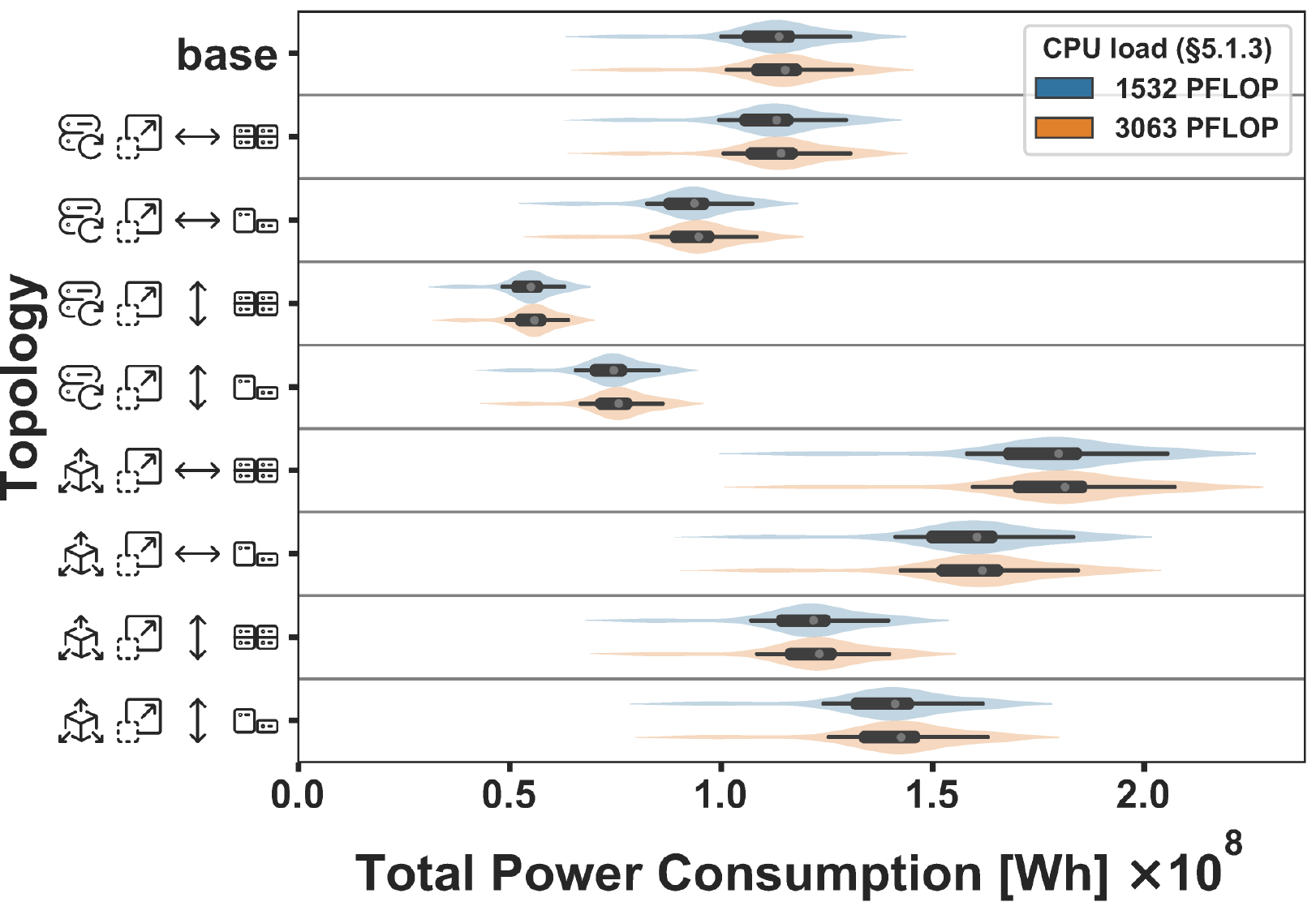}\vspace*{0.25cm}
    \includegraphics[width=\columnwidth]{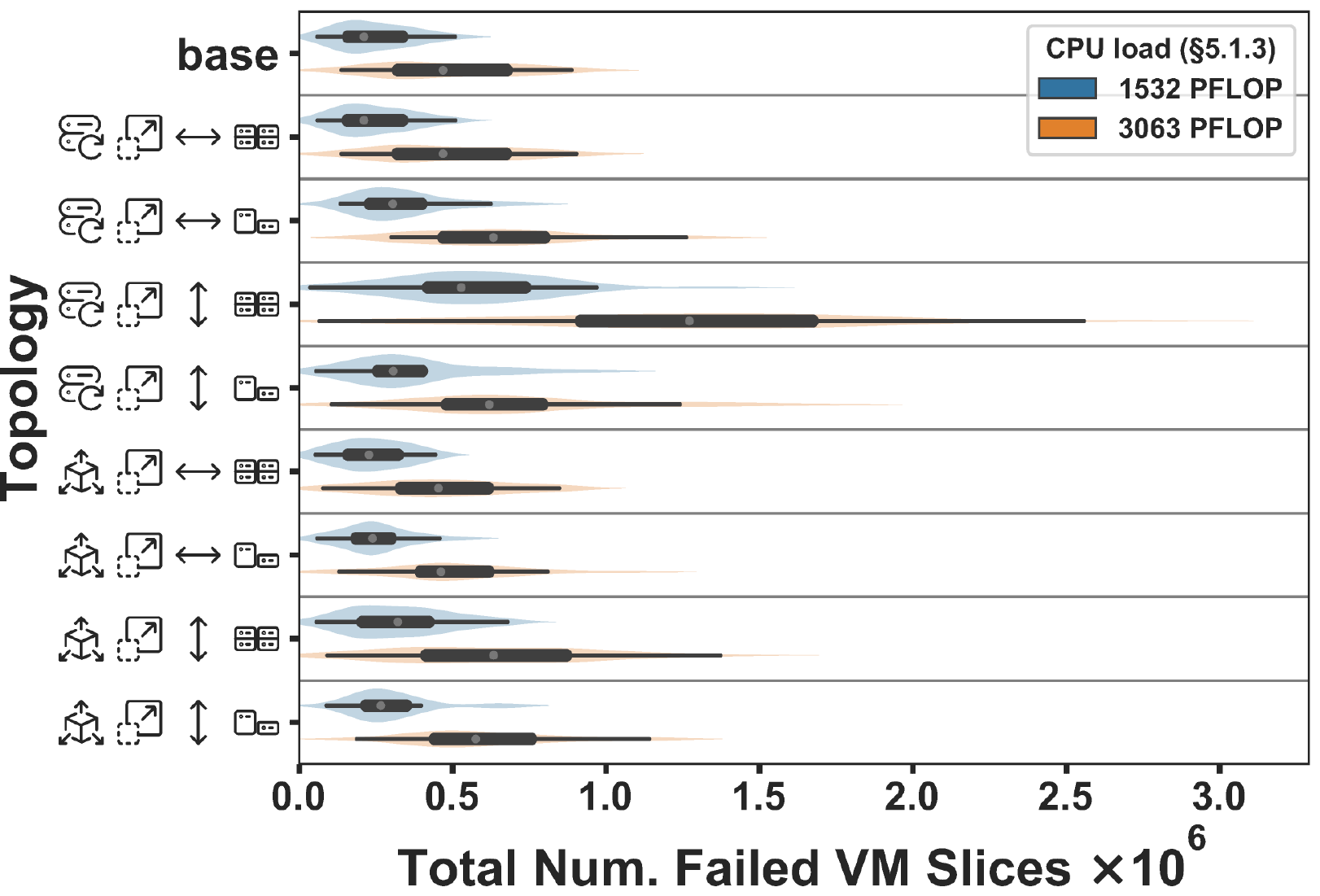}
    \caption{Results for a portfolio of candidate topologies and different workloads(\S\ref{sec:experiments:hor-ver}): 
    (top) overcommitted CPU cycles, 
    (middle) total power consumption,
    (bottom) total number of time slices in which a \gls{VM} is in a failed state. 
    Table~\ref{tab:experiment-overview} describes the symbols used to encode the topology.}
    \label{fig:hor-ver}
    \label{fig:hor-ver:overcommitted}
    \label{fig:hor-ver:power}
    \label{fig:hor-ver:failures}
    \vcutL{}
\end{figure}

The scale-in vs. scale-out decision has historically been a challenge across the field~\cite{DBLP:journals/internet/SalihogluO18}\cite[\S1.2]{book/HarcholBalter13}.
We investigate this decision in a portfolio of scenarios centered around horizontally~(symbol \iconHorInline{}) vs. vertically~(\iconVerInline{}) scaled resources~(see~\S\ref{sec:exp:setup:topology}).
We also vary: (1) the decision mode, by replacing the existing infrastructure~(\iconRepInline{}) vs. expanding it~(\iconExpInline{}), and (2) the kind of variance, homogeneous resources~(\iconHomInline{}) vs. heterogeneous~(\iconHetInline{}). 
On these three dimensions, \toolname{} creates candidate topologies by {\it increasing} the volume~(\iconVolInline{}) and compares their performance using four workload intensities, two of which are shown in this analysis.
We consider three metrics for each scenario: 
Figure~\ref{fig:hor-ver:overcommitted}~(top) depicts the overcommitted CPU cycles,
Figure~\ref{fig:hor-ver:power}~(middle) depicts the power consumption, and
Figure~\ref{fig:hor-ver:failures}~(bottom) depicts the number of failed \gls{VM} time slices.

\begin{figure*}
    \centering
    \subfloat[Requested CPU cycles\label{fig:full:hor-ver:requested}]{\includegraphics[width=0.5\linewidth]{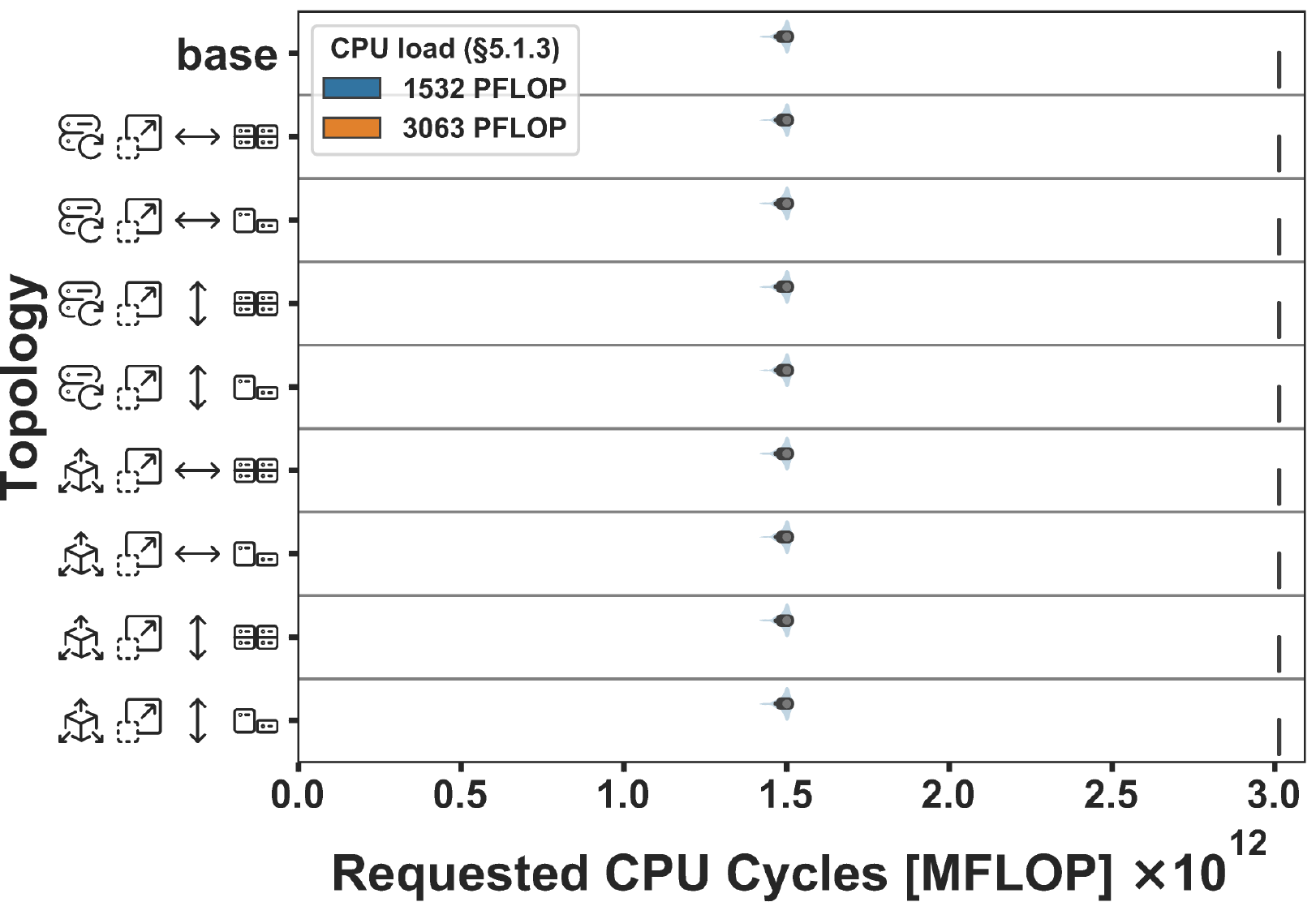}}%
    \subfloat[Granted CPU cycles\label{fig:full:hor-ver:granted}]{\includegraphics[width=0.5\linewidth]{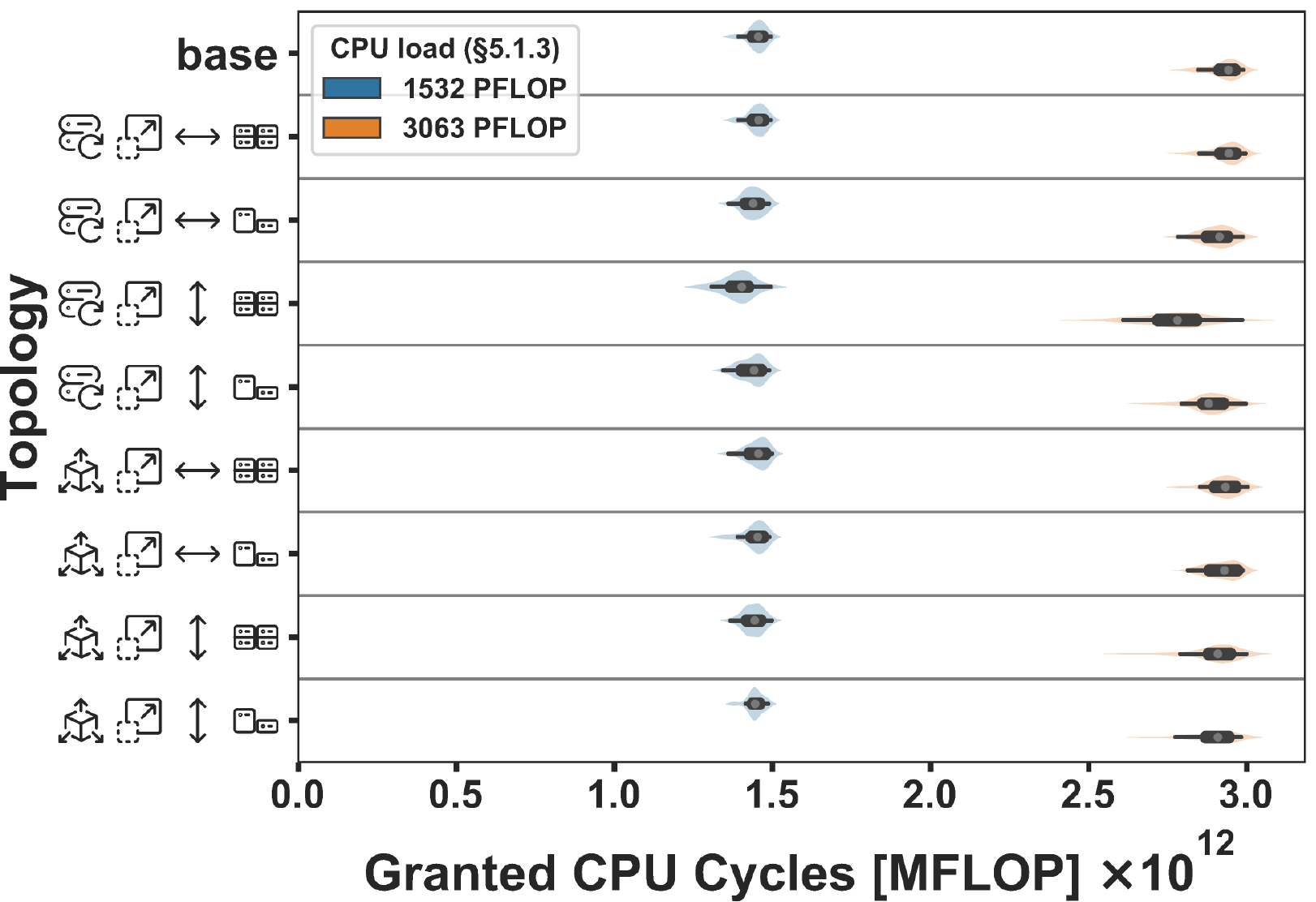}}\\
    \subfloat[Overcommitted CPU cycles\label{fig:full:hor-ver:overcommitted}]{\includegraphics[width=0.5\linewidth]{figures/plots/horizontal_vs_vertical_total_overcommitted_burst.pdf}}%
    \subfloat[Interfered CPU cycles\label{fig:full:hor-ver:interfered}]{\includegraphics[width=0.5\linewidth]{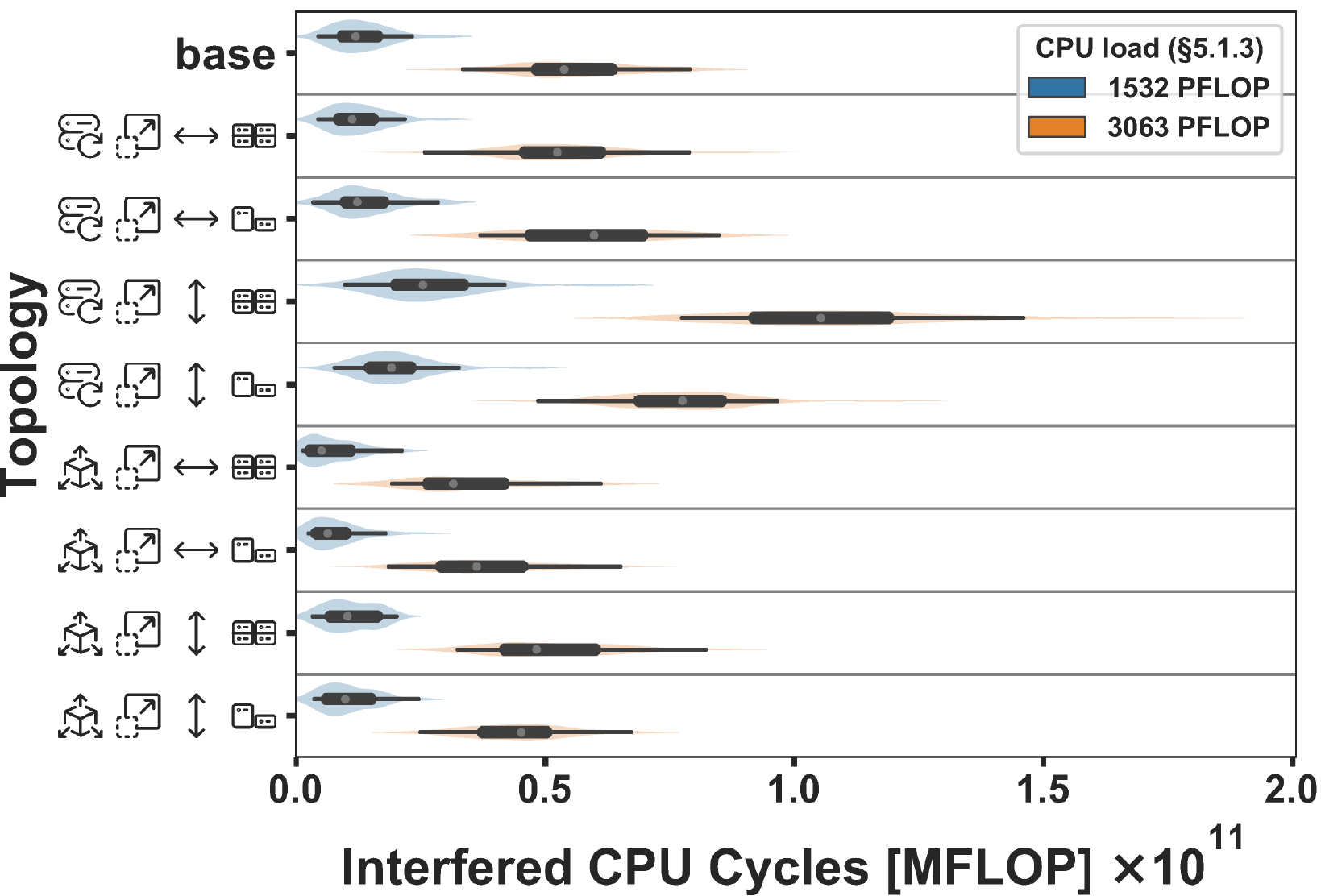}}\\
    \subfloat[Total power consumption\label{fig:full:hor-ver:power}]{\includegraphics[width=0.5\linewidth]{figures/plots/horizontal_vs_vertical_total_power_draw.pdf}}%
    \subfloat[Total number of time slices in which a \gls{VM} is failed, aggregated across \glspl{VM}\label{fig:full:hor-ver:failures:vms}]{\includegraphics[width=0.5\linewidth]{figures/plots/horizontal_vs_vertical_total_failure_vm_slices.pdf}}\\
    \caption{Performance of different horizontally and vertically expanded topologies, compared across workloads. For a legend of topologies, see Table~\ref{tab:experiment-overview}. Continued in Figure~\ref{fig:full:hor-ver:2}.}
    \label{fig:full:hor-ver:1}
\end{figure*}

Our key {\it performance} indicator is overcommitted CPU cycles, that is, the count of CPU cycles requested by \glspl{VM} but not granted, either due to collocated \glspl{VM} requesting too many resources at once, or due to performance interference effects taking place.
We observe 
in Figure~\ref{fig:hor-ver:overcommitted}~(top)
that vertically scaled topologies~(symbol \iconVerInline{}) have significantly higher overcommission~(lower performance) than their horizontally scaled counterparts~(\iconHorInline{}, the other three symbols identical).
The median value is higher for vertical than for horizontal scaling, for both replaced~(\iconRepInline{}) and expanded~(\iconExpInline{}) topologies, by a factor of 1.53x--2.00x (calculated as the ratio between medians of different scenarios at full load). 
This is a large factor, suggesting that vertically scaled topologies are 
more susceptible to overcommission, and thus lead to higher risk of performance degradation.
The decrease in performance observed in this metric is mirrored by the granted CPU cycles metric in Figure~\ref{fig:full:hor-ver:granted} (Appendix~\ref{sec:full-visual-results}), which decreases for vertically scaled topologies.
Among replaced topologies~(all combinations including~\iconRepInline{}), the horizontally scaled, homogeneous topology~(\iconRepInline{}\iconVolInline{}\iconHorInline{}\iconHomInline{}) yields the best performance, and in particular the lowest median overcommitted CPU.
We also observe that expanded topologies~(\iconExpInline{}) have lower overcommission than the base topology, so adding machines is worthwhile.
We observe all these effects strongly for the full trace (3,063 PFLOPs), but less pronounced for the lower workload intensity (1,531 PFLOPs).

But performance is not the only criterion for capacity planning. 
We turn to {\it power consumption}, as a proxy for cost analysis and environmental concerns.
We see here that vertically scaled topologies~(\iconVerInline{}) drastically improve power consumption, for median values by a factor of 1.47x--2.04x,
contrasting their worse performance compared to horizontal scaling~(\iconHorInline{}).
As expected, all expanded topologies~(\iconExpInline{}), which have more machines, incur higher power-consumption than replaced topologies~(\iconRepInline{}).
Higher workload intensity~(i.e., for the 3,063 PFLOPs results) incurs higher power consumption, although less pronounced than earlier. 

\begin{figure*}
    \subfloat[Mean CPU usage\label{fig:full:hor-ver:cpu-usage}]{\includegraphics[width=0.5\linewidth]{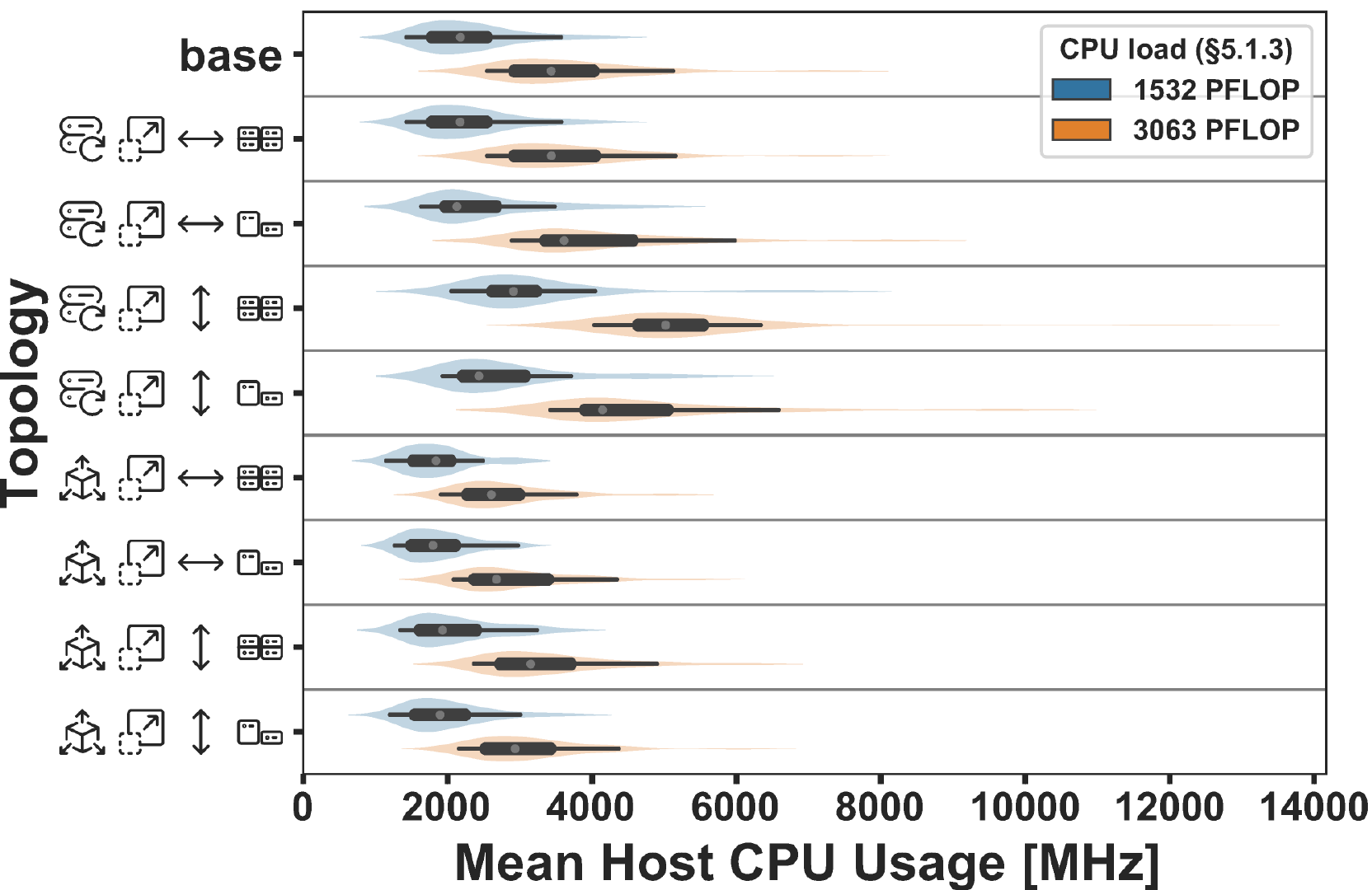}}%
    \subfloat[Mean CPU demand\label{fig:full:hor-ver:cpu-demand}]{\includegraphics[width=0.5\linewidth]{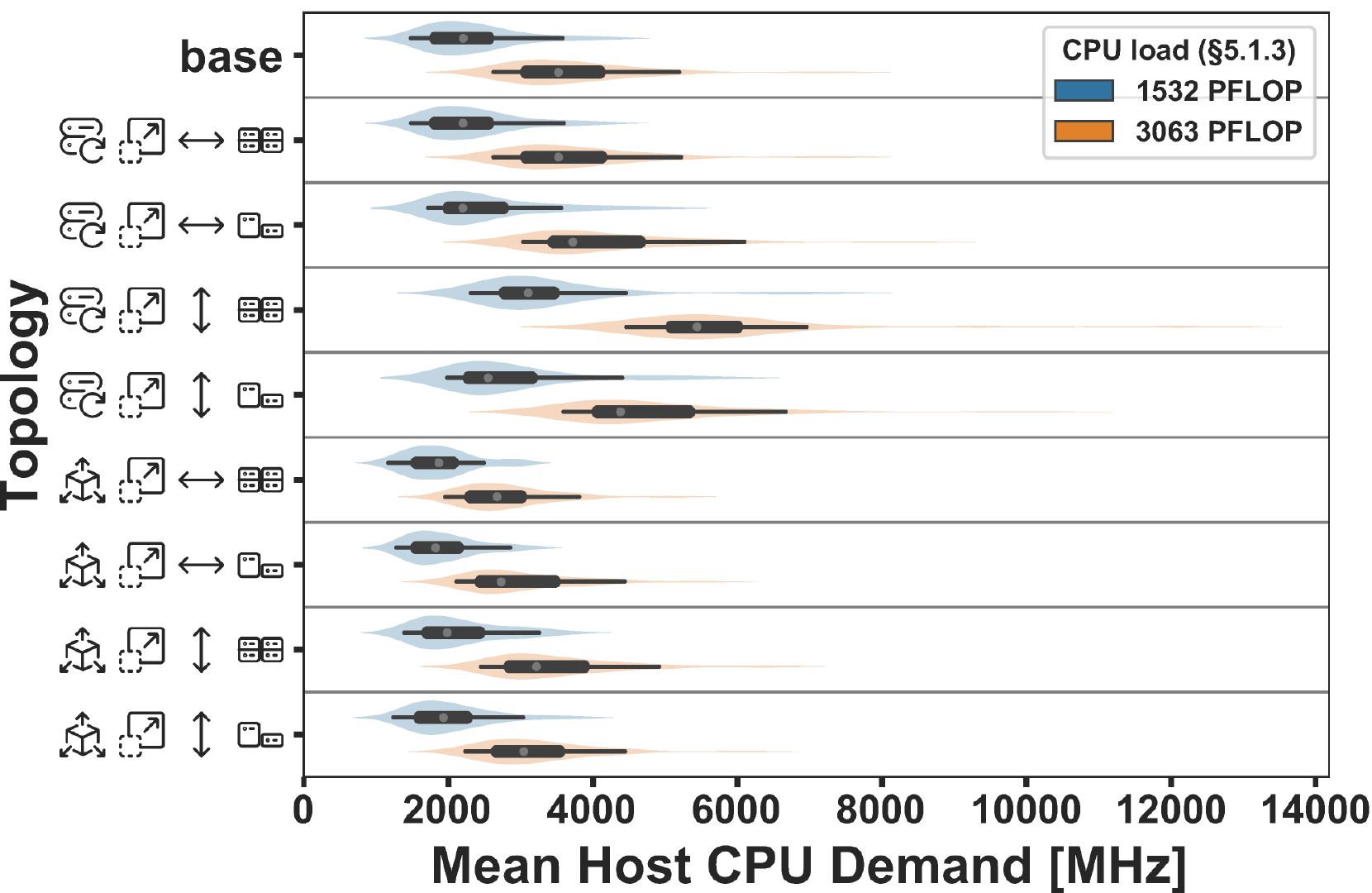}}\\
    \subfloat[Mean number of \glspl{VM} per host\label{fig:full:hor-ver:mean-vm-count}]{\includegraphics[width=0.5\linewidth]{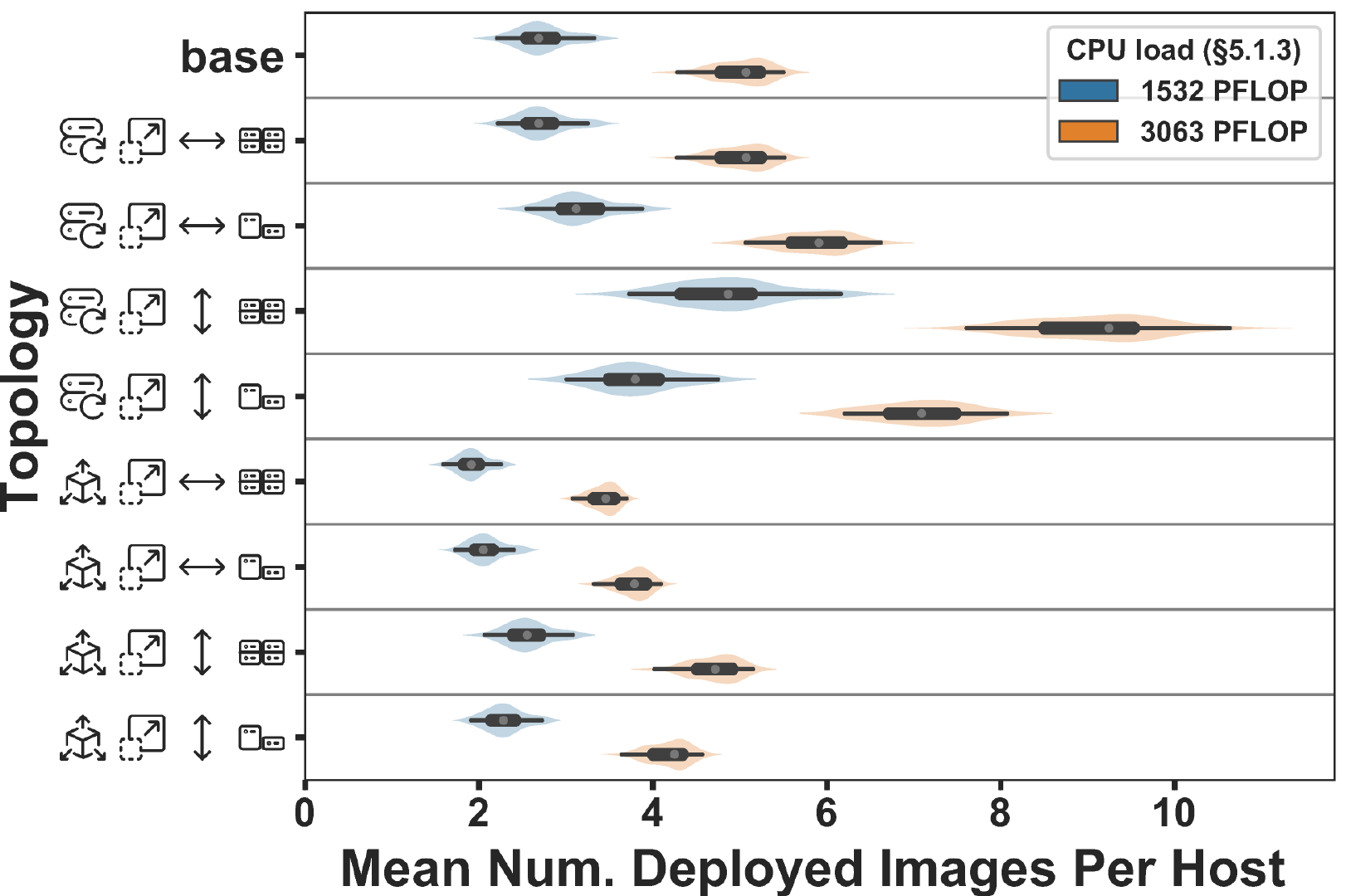}}%
    \subfloat[Max number of \glspl{VM} per host\label{fig:full:hor-ver:max-vm-count}]{\includegraphics[width=0.5\linewidth]{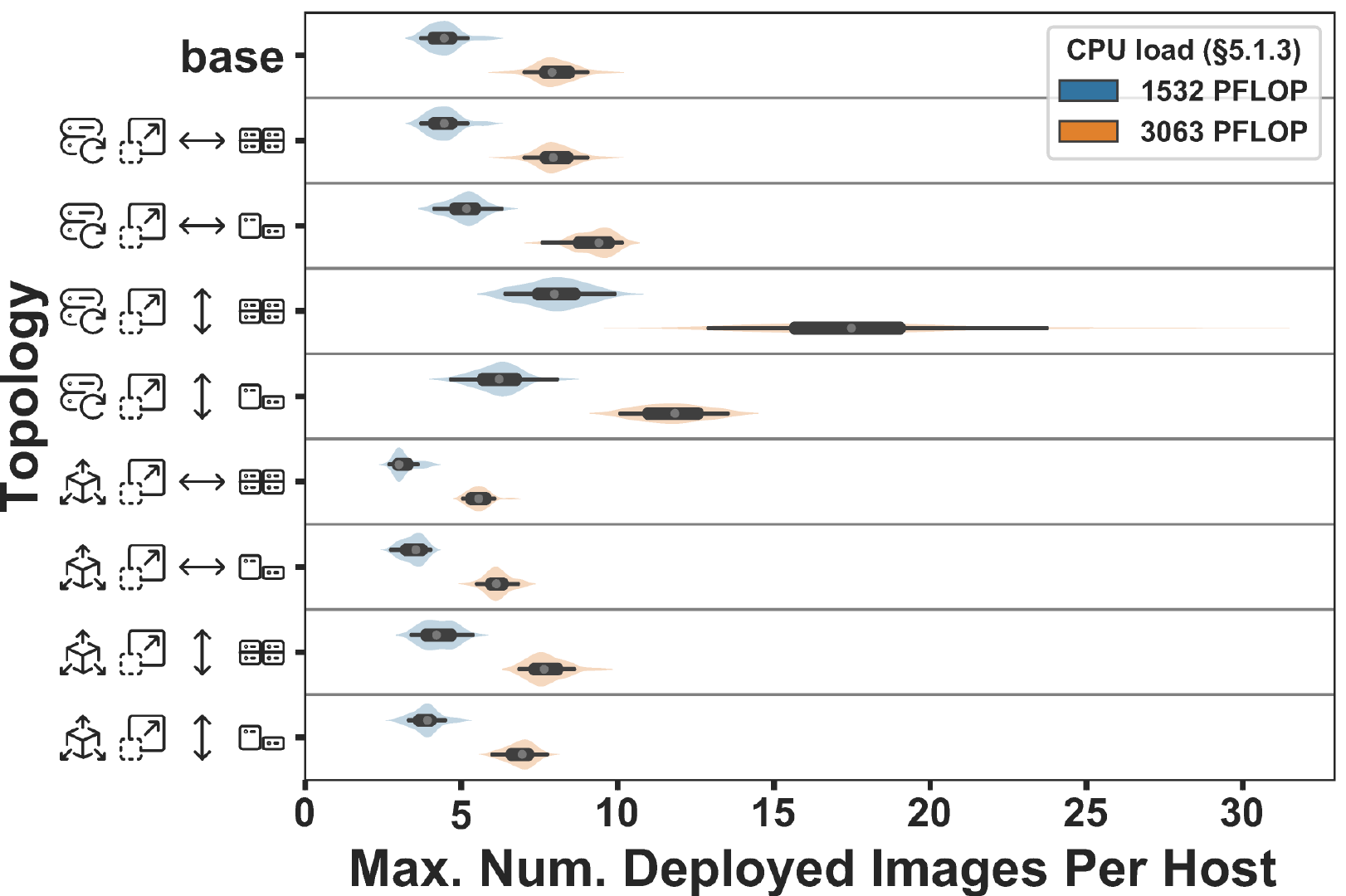}}\\
    \caption{Performance of different horizontally and vertically expanded topologies, compared across workloads. For a legend of topologies, see Table~\ref{tab:experiment-overview}. Continued in Figure~\ref{fig:full:hor-ver:3}.}
    \label{fig:full:hor-ver:2}
\end{figure*}

We also consider the {\it amount of failed \gls{VM} time-slices}. 
Each failure here is full-stop~(\S\ref{sec:exp:setup:phenomena}), which typically escalates an alarm to engineers. Thus, this metric should be minimized.
We observe significant differences here: the median failure time of a homogeneous vertically scaled topology~(\iconVerInline{}\iconHomInline{}) is between 2.00x--2.71x higher than the base topology.
This metric shows similarities qualitatively with the overcommitted CPU cycles.
Vertical scaling is correlated not only with worse performance, but also with higher failure counts.
We see that vertical scaling leads to a significant increase in the maximum number of deployed images per physical host (Figure~\ref{fig:full:hor-ver:max-vm-count}), which leads to larger failure domains and thus potentially higher failure counts.
The effect is less pronounced when making heterogeneous compared to homogeneous procurement.

Our findings show that \toolname{} gives practitioners the possibility to {\em explore a complex trade-off portfolio} of dimensions such as power consumption, performance, failures, workload intensity, etc.
Optimization questions surrounding horizontal and vertical scaling can therefore be approached {\it with a data-driven approach}.
We find that decisions including heterogeneous resources can provide meaningful compromises between more generic, homogeneous resources; they also lead to different decisions related to personnel training~(not shown here).
We show significant differences between candidate topologies in all metrics, translating to very different power costs, long-term.
We conclude that {\em \toolname{} can help test intuitions and support complex decision making}.

\subsection{Expansion: Velocity}
\label{sec:experiments:more-vel}

Our main findings from this experiment are:
\begin{description}
    
    \mainfin{velocity} \toolname{} enables exploring a range of resource dimensions frequently considered in practice, such as component velocity.
    
    \mainfin{velocity-impact} Increasing velocity can reduce overcommitted CPU cycles by 3.3\%.
    
    \mainfin{velocity-expansion} Expanding a topology by velocity can improve performance by 1.54x, compared to expansion by volume.\vcutS{}
    
\end{description}

\begin{figure}
    \centering
    \includegraphics[width=\columnwidth]{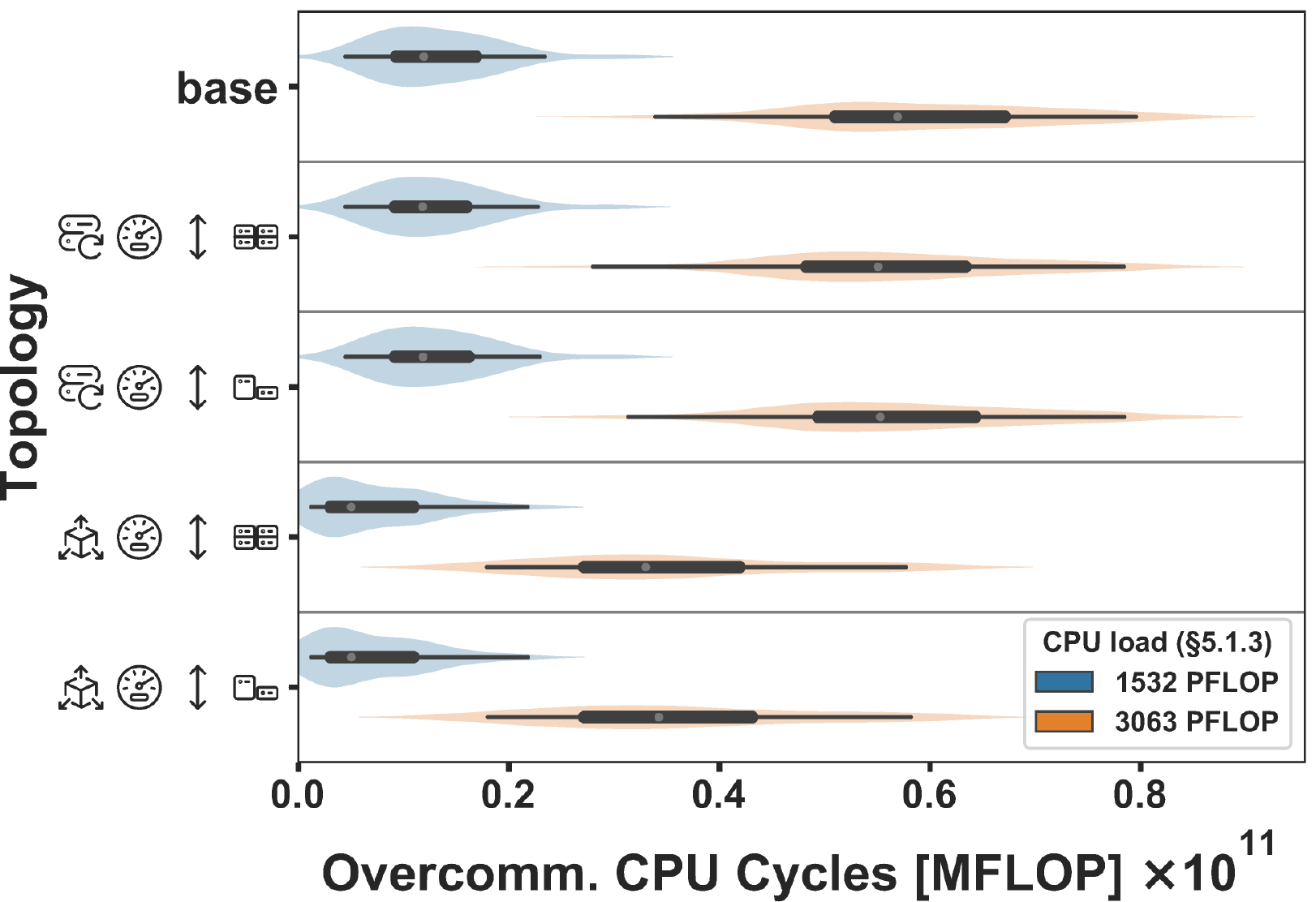}
    \caption{Overcommitted CPU time for a portfolio of candidate topologies and different workloads, for Experiment~\ref{sec:experiments:more-vel}.}
    \label{fig:more-vel:overcommitted}
\end{figure}

In vertical horizontal scaling, practitioners are also faced with the decision of which qualities to scale.
This experiment varies the velocity of resources both homogeneously and heterogeneously, while replacing or expanding the existing topology.
Figure~\ref{fig:more-vel:overcommitted} depicts the explored scenarios and their performance, in the form of overcommitted CPU cycles.

We find that in-place, homogeneous vertical scaling of machines with higher velocity leads to slightly better performance, by a percentage of 3.3\% (compared to the base scenario, by median).
In this dimension, performance varies only slightly between homogeneously and heterogeneously scaled topologies, for all metrics (see also Appendix~\ref{sec:full-visual-results}).
Expanding the topology homogeneously (\iconExpInline{}\iconVerInline{}\iconHomInline{}) with a set of machines with higher CPU frequency helps reduce overcommission more drastically, also improving it beyond the lowest overcommission reached by homogeneous vertical expansion in the previous experiment, in Figure~\ref{fig:hor-ver:overcommitted}.
When expanding, this cross-experiment comparison {\em shows an improvement of performance with a factor of 1.54x}.

\begin{figure}[!t]
    \centering
    \includegraphics[width=\columnwidth]{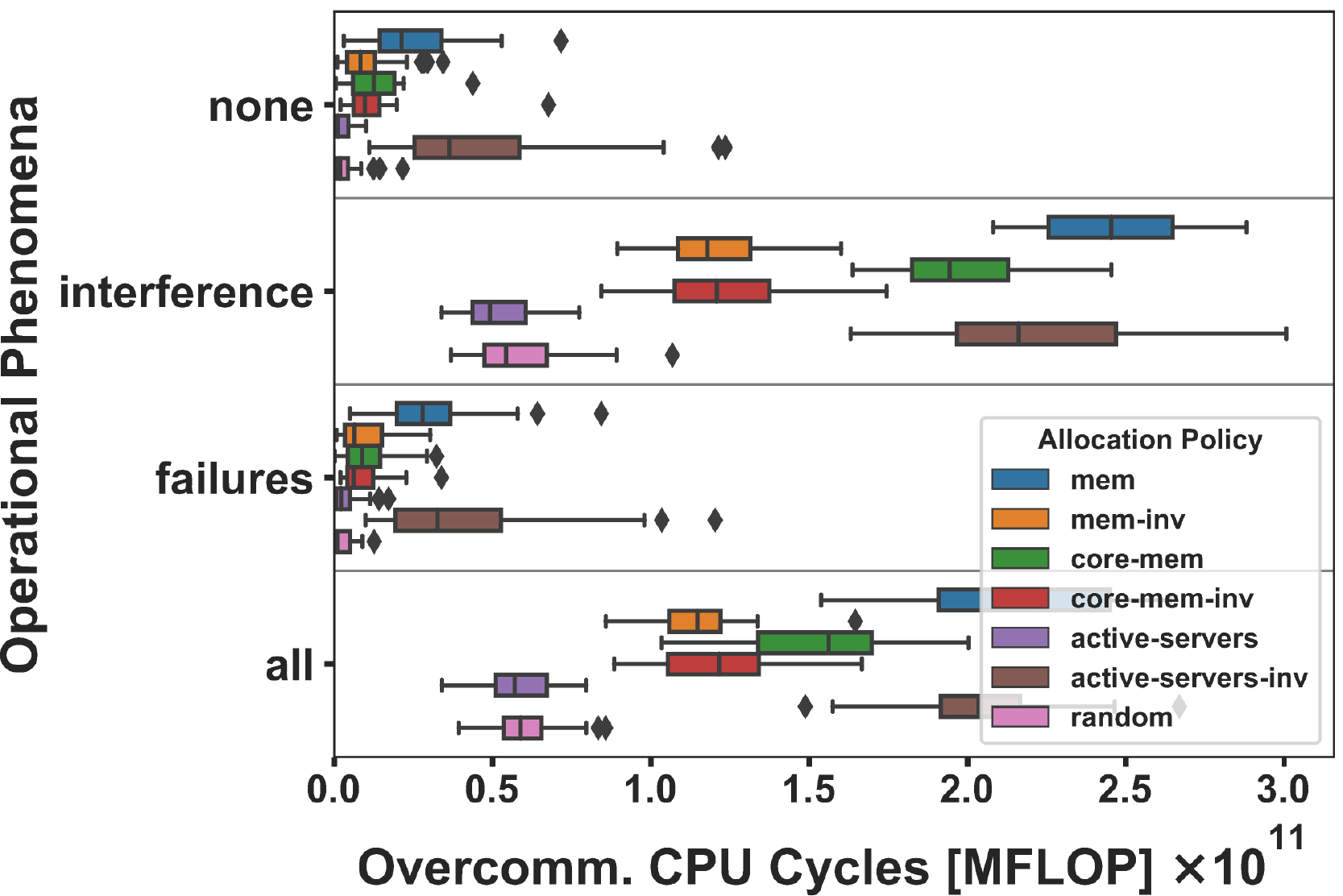}
    \vcutM{}
    \caption{Overcommitted CPU cycles for a portfolio of {\it operational phenomena} (the ``none'' 
    through ``all'' sub-plots), and {\it allocation policies} (legend), for Experiment~\ref{sec:experiments:phenomena}.} 
    \label{fig:phenomena:overcommitted}
    \vcutL{}
\end{figure}

\vcutS{}
\subsection{Impact of Operational Phenomena}
\label{sec:experiments:phenomena}
\vcutS{}


Our main findings from this experiment are:
\begin{description}
    
    \mainfin{detail-important} \toolname{} enables the exploration of
    diverse allocation policies
    and operational phenomena, both of which lead to important differences in capacity planning.
    
    \mainfin{perfinterf-important} Modeling performance interference can explain 80.6\%---94.5\% of the overcommitted CPU cycles.
    
    \mainfin{perf-values} Different allocation policies lead to different performance interference intensities, 
    and to median overcommitted CPU cycles different 
    by factors between 1.56x and 30.3x compared to the best policy---high risk{\bf !}
    
\end{description}

\begin{figure*}
    \centering
    \subfloat[Requested CPU cycles\label{fig:full:phenomena:requested}]{\includegraphics[width=0.5\linewidth]{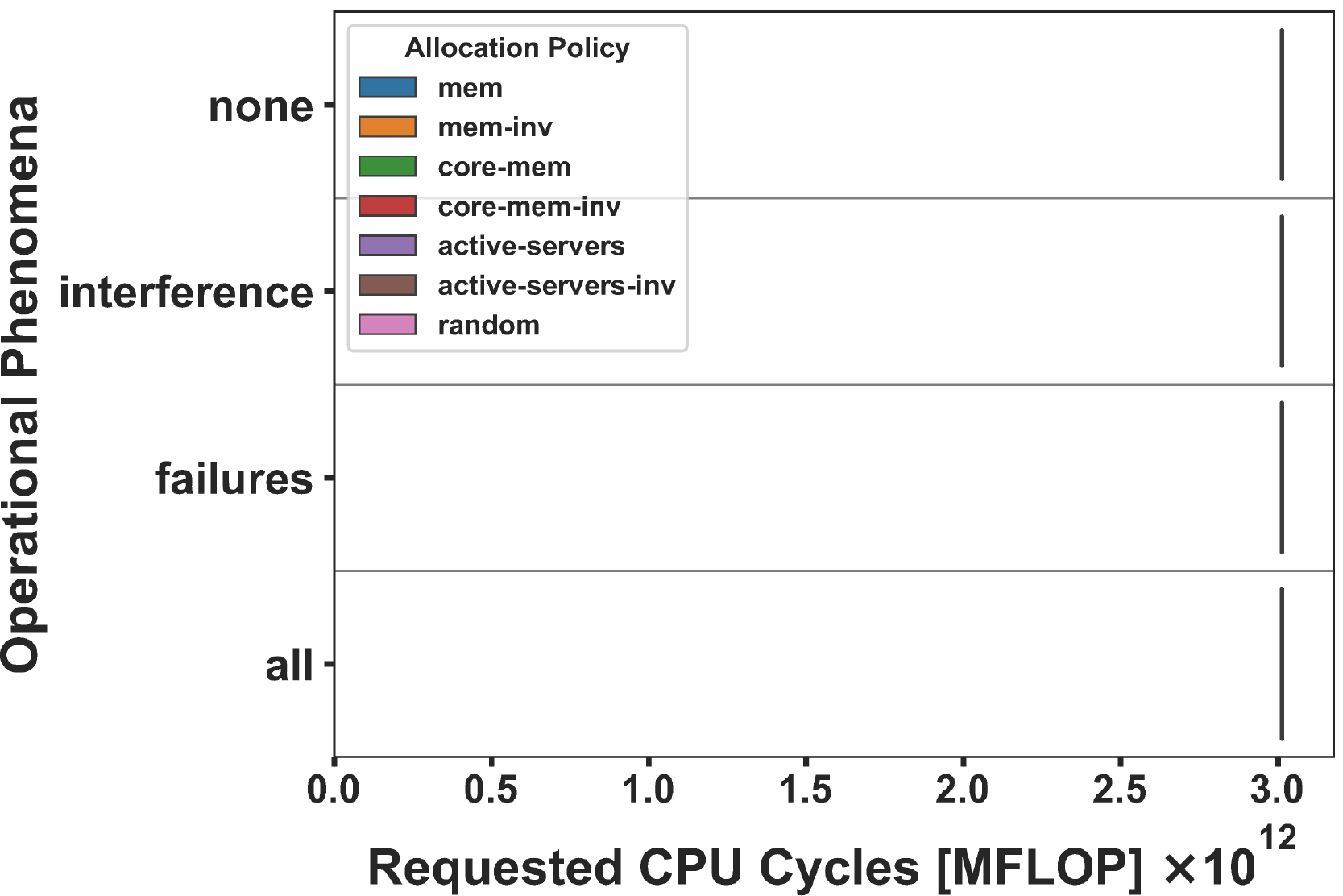}}%
    \subfloat[Granted CPU cycles\label{fig:full:phenomena:granted}]{\includegraphics[width=0.5\linewidth]{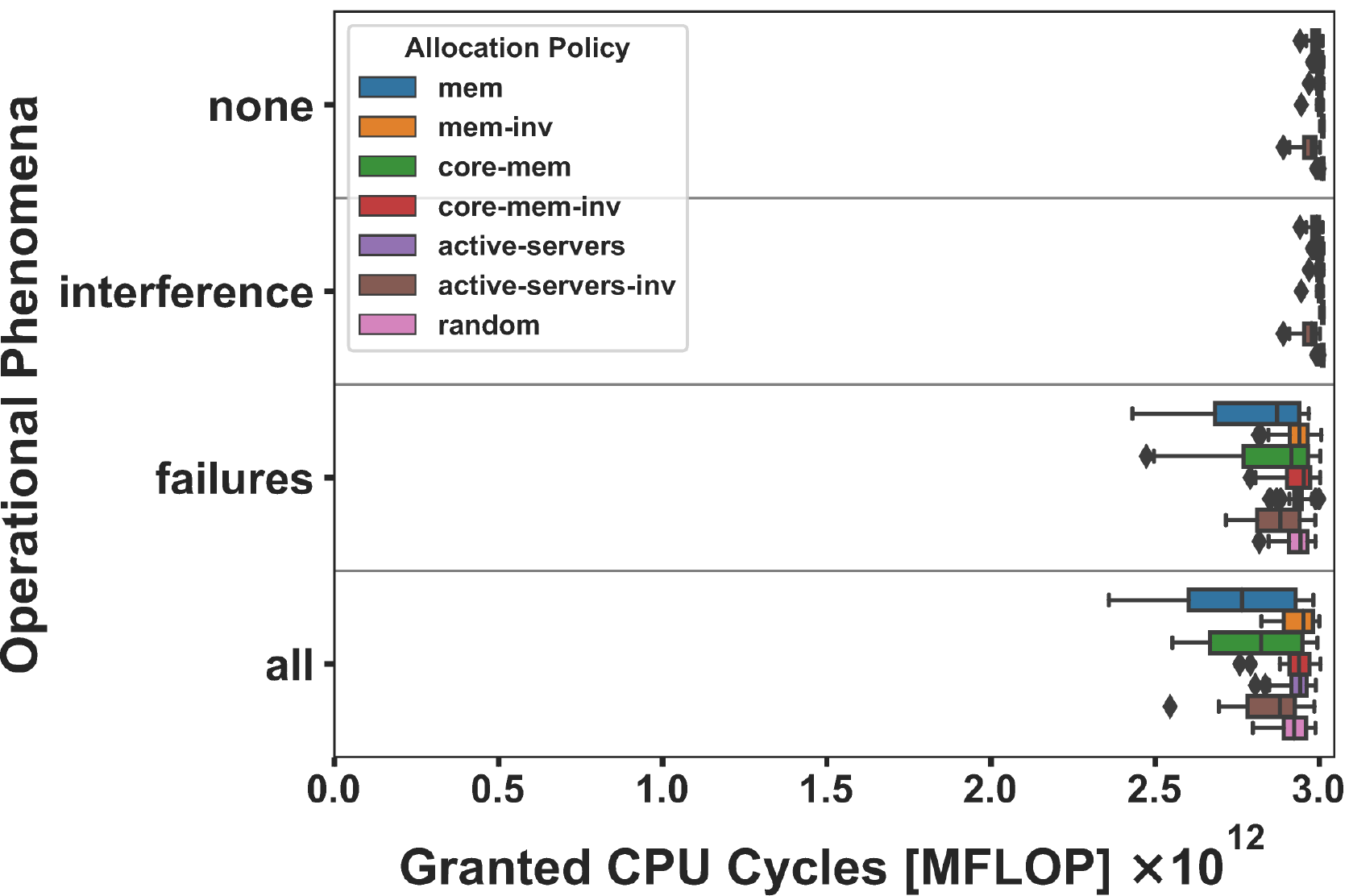}}\\
    \subfloat[Overcommitted CPU cycles\label{fig:full:phenomena:overcommitted}]{\includegraphics[width=0.5\linewidth]{figures/plots/operational_phenomena_total_overcommitted_burst.pdf}}%
    \subfloat[Interfered CPU cycles\label{fig:full:phenomena:interfered}]{\includegraphics[width=0.5\linewidth]{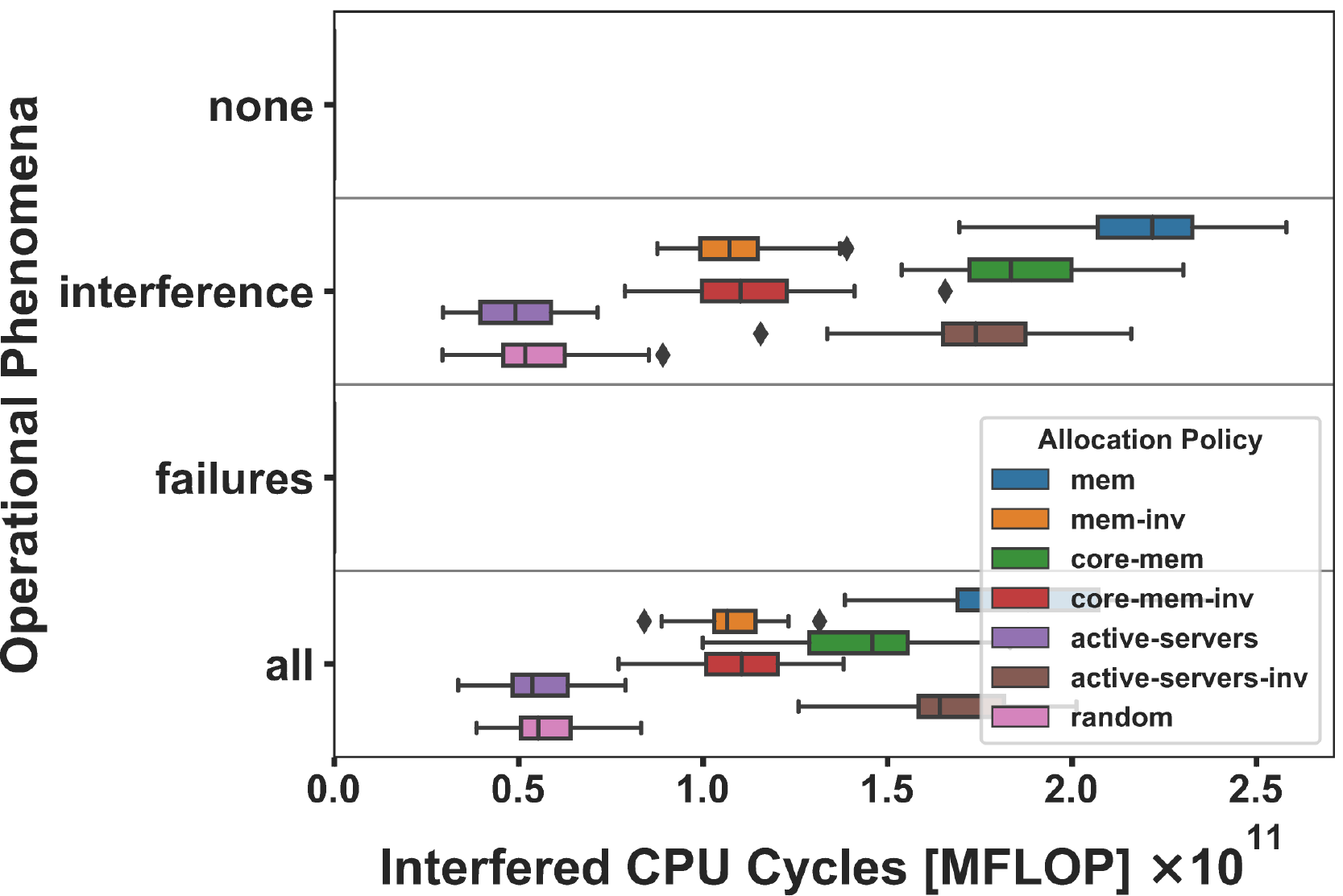}}\\
    \subfloat[Total power consumption\label{fig:full:phenomena:power}]{\includegraphics[width=0.5\linewidth]{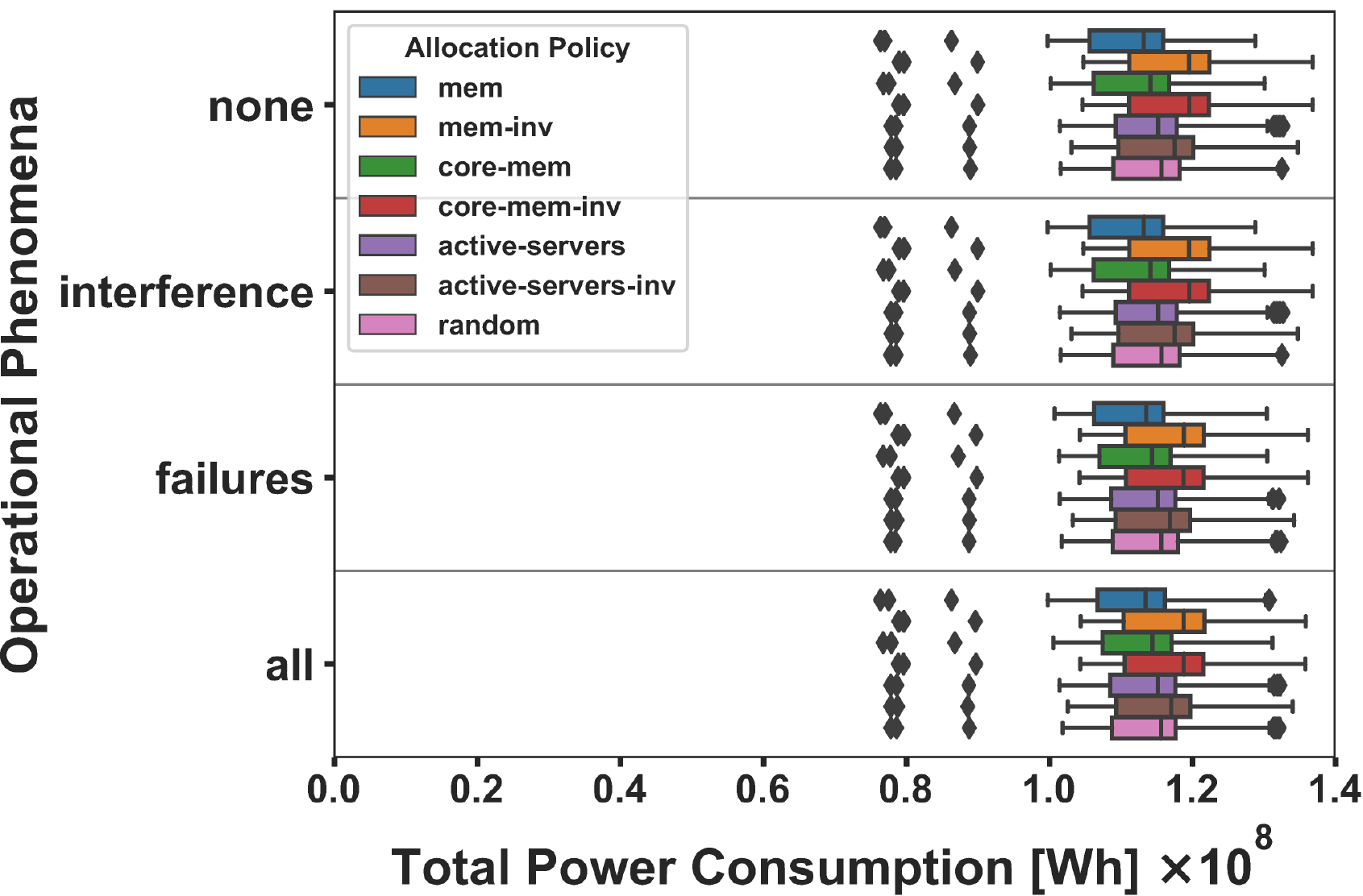}}%
    \subfloat[Total number of time slices in which a \gls{VM} is failed, aggregated across \glspl{VM}\label{fig:full:phenomena:failures:vms}]{\includegraphics[width=0.5\linewidth]{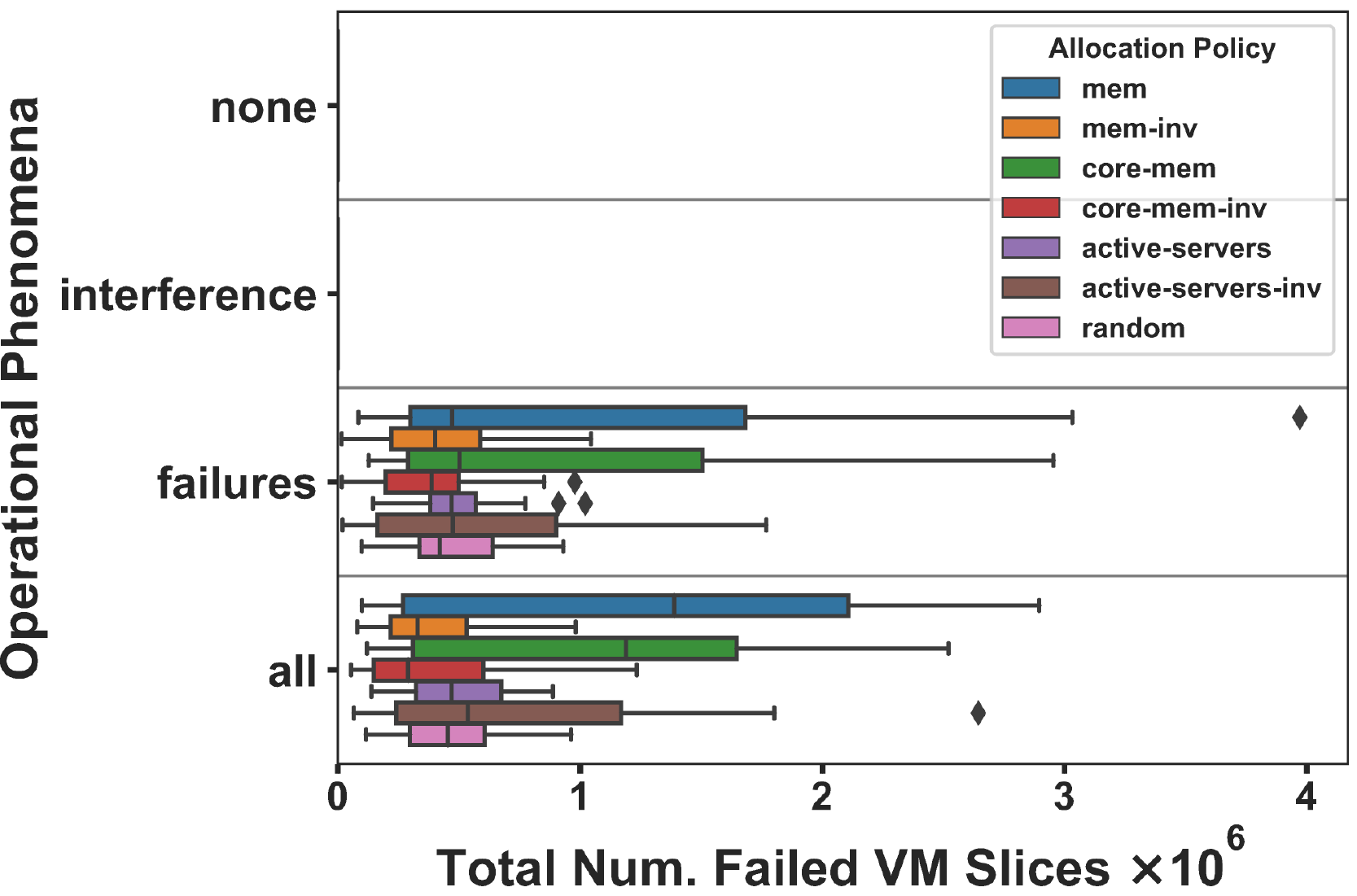}}\\
    \caption{Impact of operational phenomena and different allocation policies on the base topology. For a legend of topologies, see Table~\ref{tab:experiment-overview}. Continued in Figure~\ref{fig:full:phenomena:2}.}
    \label{fig:full:phenomena:1}
\end{figure*}

\begin{figure*}
    \subfloat[Mean CPU usage\label{fig:full:phenomena:cpu-usage}]{\includegraphics[width=0.5\linewidth]{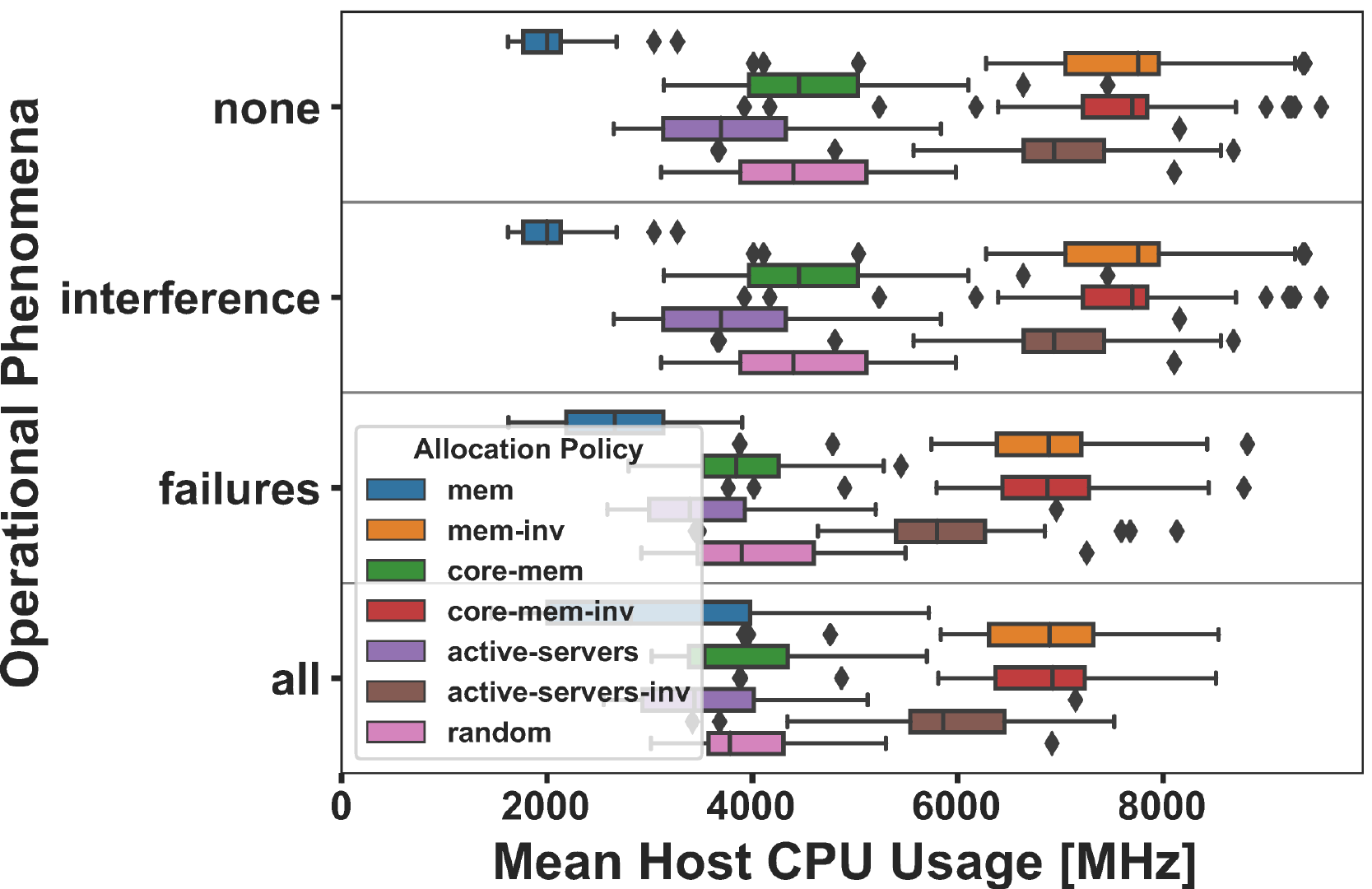}}%
    \subfloat[Mean CPU demand\label{fig:full:phenomena:cpu-demand}]{\includegraphics[width=0.5\linewidth]{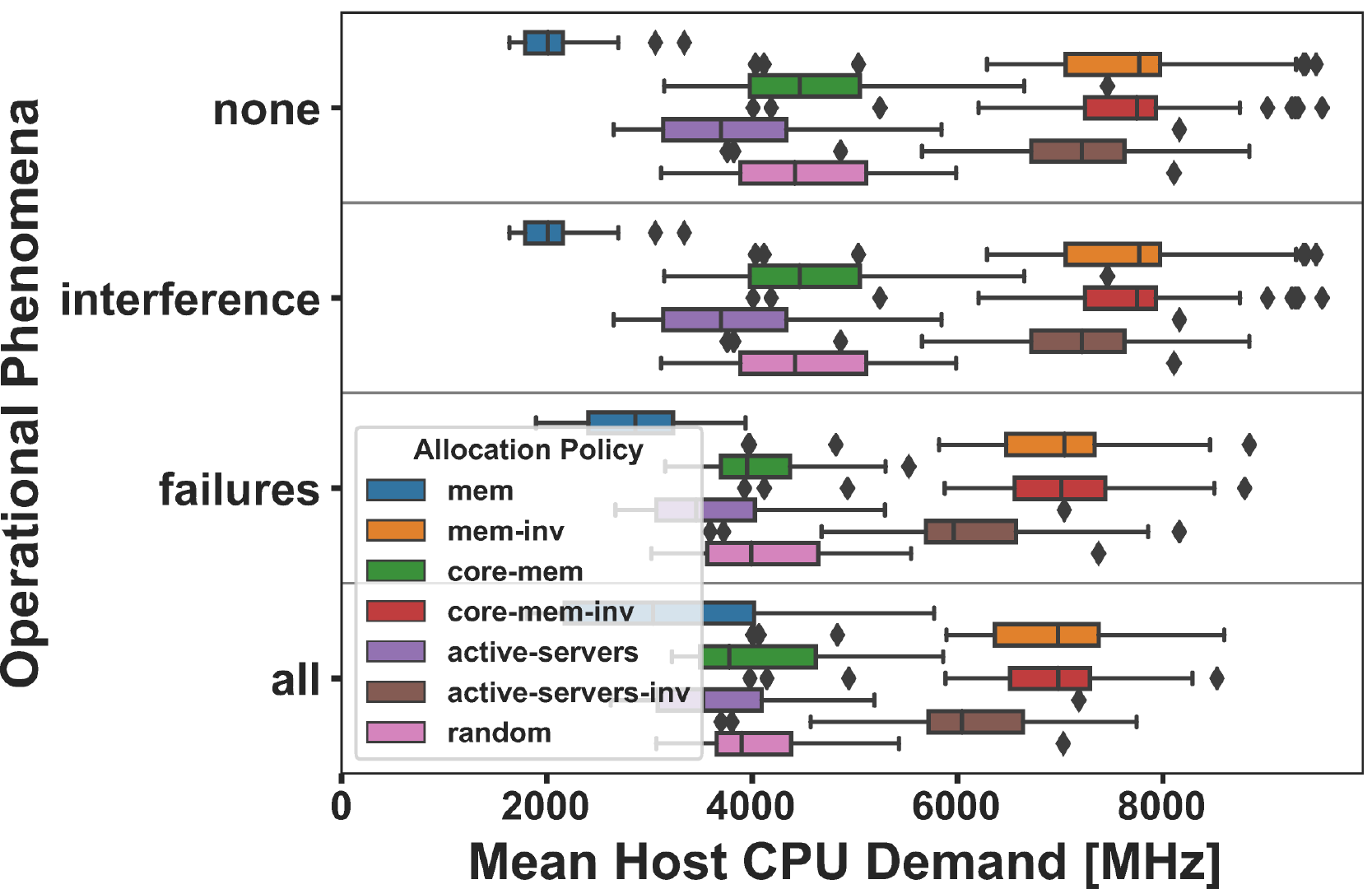}}\\
    \subfloat[Mean number of \glspl{VM} per host\label{fig:full:phenomena:mean-vm-count}]{\includegraphics[width=0.5\linewidth]{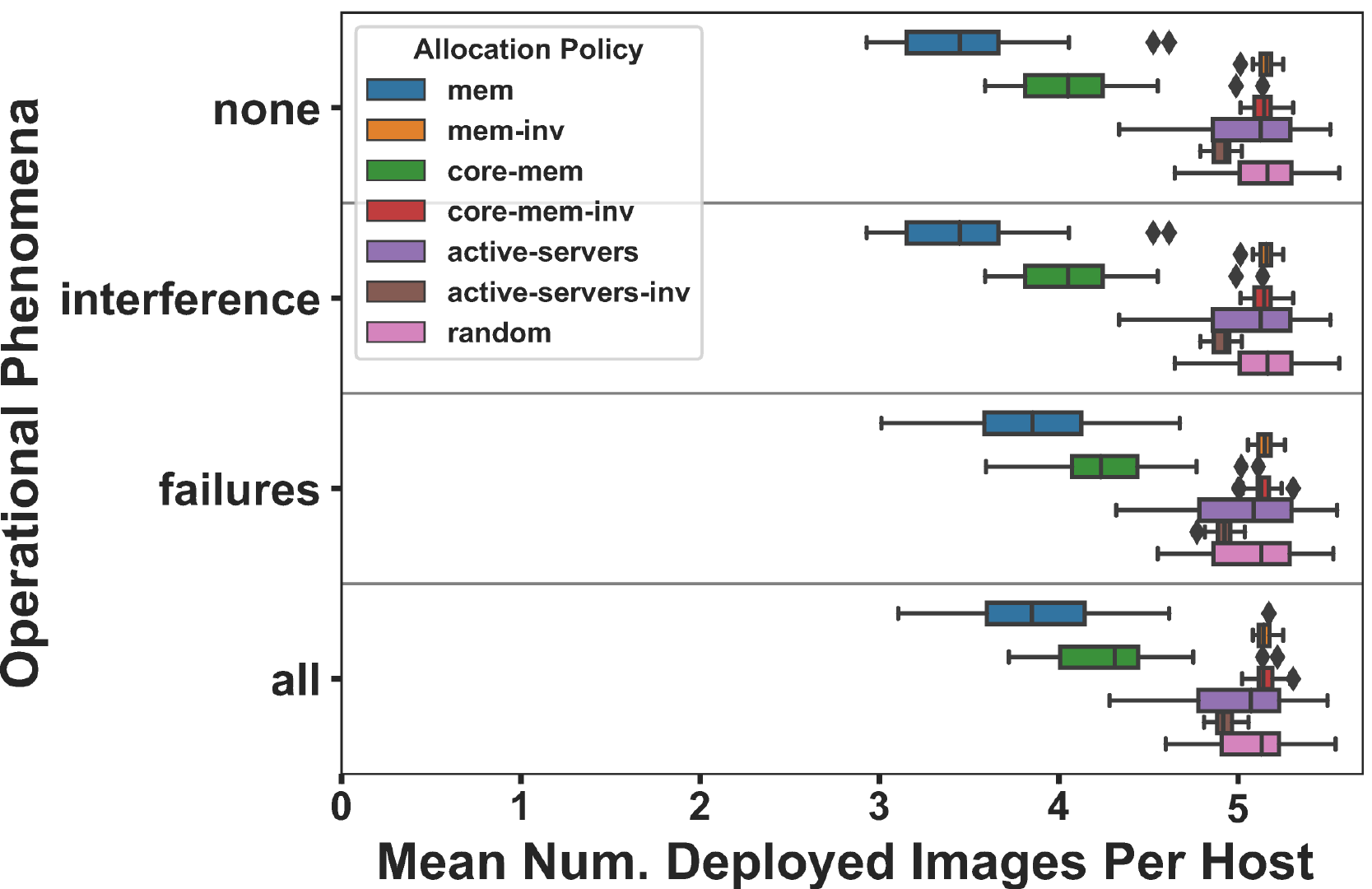}}%
    \subfloat[Max number of \glspl{VM} per host\label{fig:full:phenomena:max-vm-count}]{\includegraphics[width=0.5\linewidth]{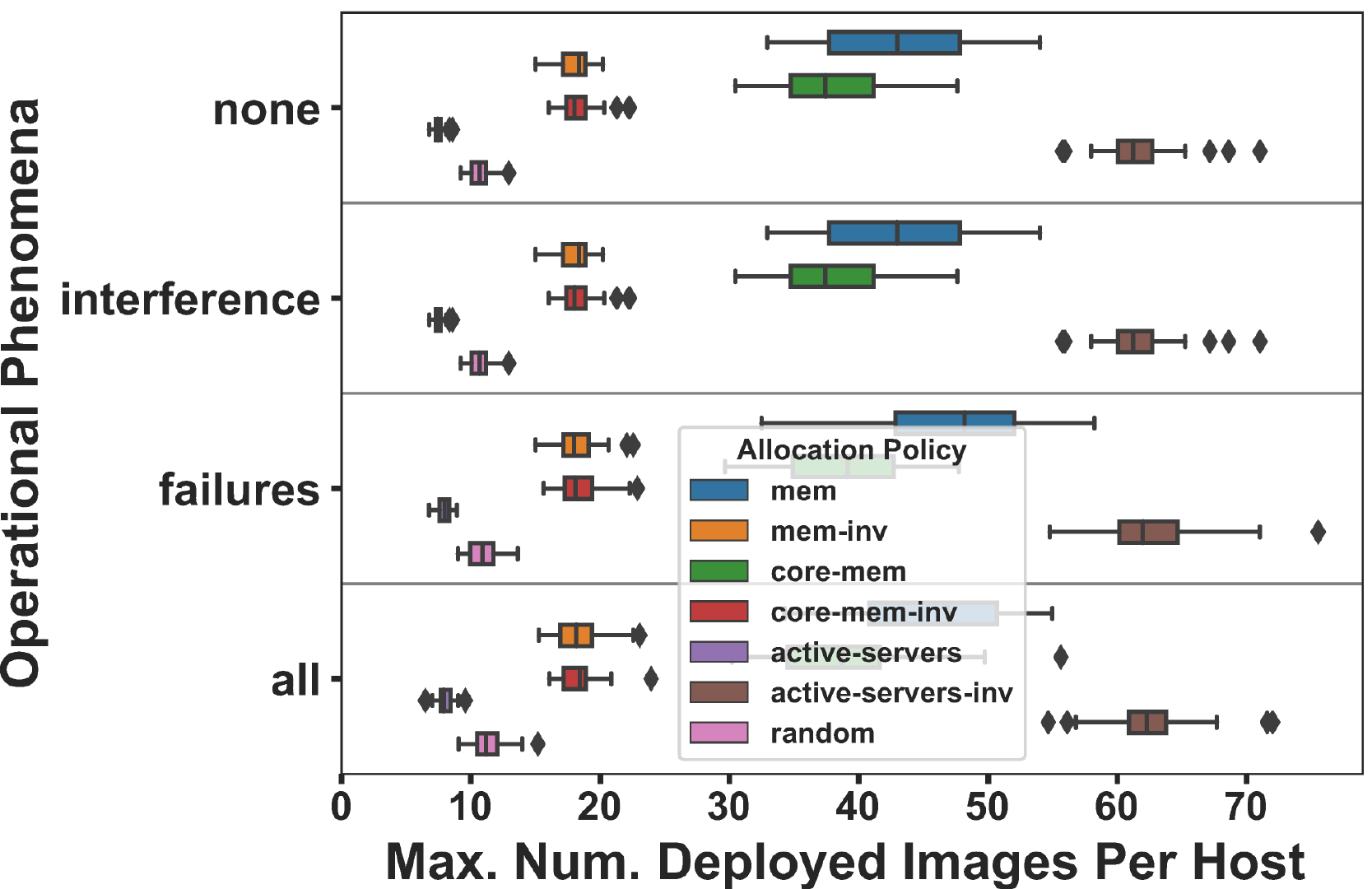}}\\
    \caption{Impact of operational phenomena and different allocation policies on the base topology. For a legend of topologies, see Table~\ref{tab:experiment-overview}. Continued in Figure~\ref{fig:full:phenomena:3}.}
    \label{fig:full:phenomena:2}
\end{figure*}

This experiment addresses operational factors 
in the capacity planning process.
We explore the impact of better handling of physical machine failures, the impact of (smarter) scheduler allocation policies, and the impact of (the absence of) performance interference on overall performance.
Figure~\ref{fig:phenomena:overcommitted} shows the impact of different operational phenomena on performance, for different allocation policies.
We observe that performance interference has a strong impact on overcommission, dominating it compared to 
the ``failures'' sub-plot, where only failures are considered, or with 
the ``none'' sub-plot, where no failures or interference are considered. 
Depending on the allocation policy, it represents between 80.6\% and 94.5\% of the overcommission recorded in simulation for the ``all'' sub-plot, where both failures and interference are considered.
This is visualized more in detail in Figure~\ref{fig:full:phenomena:interfered} (\S\ref{sec:full-visual-results}), which plots the interference itself, separately.
We also see the large impact that live resource management (in this case, the allocation policy) can have on Quality of Service.
Median ratios vary between 1.56x and 30.3x vs. the best policy, with {\em active-servers}~(see \S\ref{sec:exp:setup:policies}) generally best-performing.
Finally, we observe that enabling failures increases the colocation ratio of \glspl{VM} (see Figure~\ref{fig:full:phenomena:mean-vm-count}, \S\ref{sec:full-visual-results}).

We conclude {\em \toolname{} can help model aspects that are important but typically not considered for capacity planning.} 


\begin{figure}[!t]
    \centering
    \hspace*{-0.25cm}\includegraphics[width=1.03\columnwidth]{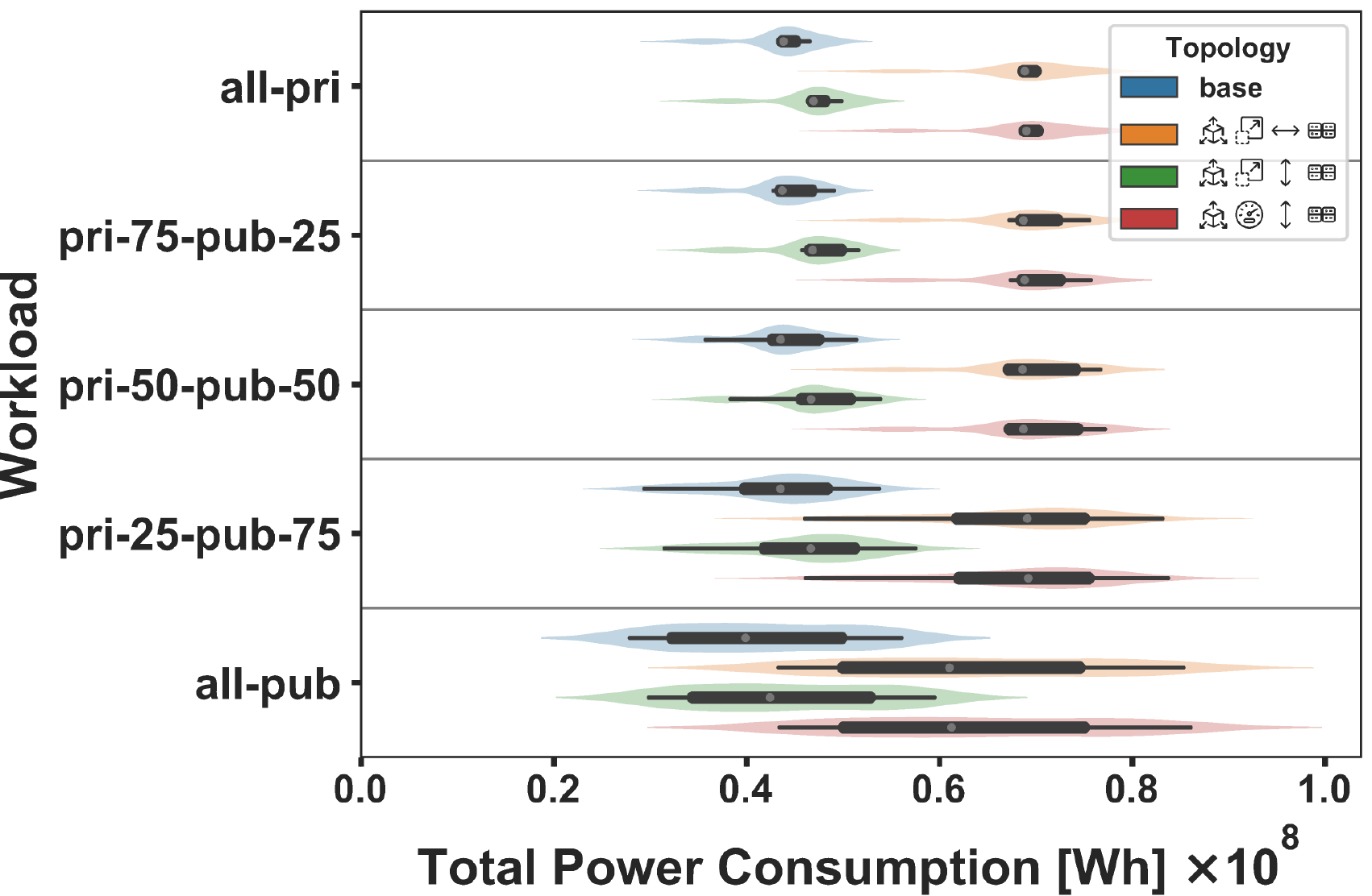}
    \vcutM{}
    \vspace*{-0.05cm}
    \caption{Total power consumption for a portfolio of candidate topologies (legend), 
    subject to {\it different workloads} (the ``all-pri'' to ``all-pub'' sub-plots), for Experiment~\ref{sec:experiments:composite}.} 
    \label{fig:composite:power}
    \vcutL{}
\end{figure}

\vcutS{}
\subsection{Impact of a New Workload}
\label{sec:experiments:composite}
\vcutS{}

Our main findings from this experiment are:
\begin{description}
    
    \mainfin{new-workload-enabled} \toolname{} enables exploring what-if scenarios that include new workloads as they become available.
    
    \mainfin{power-detail} Power consumption can vary significantly more in all-private vs. all-public cloud scenarios, with the range higher by 4.79x--5.45x.
    
\end{description}


This experiment explores the impact that a new workload type can have if added to an existing workload, an exercise capacity planners have to consider often, e.g., for new customers.
We combine here the 1-month \provider{} and Azure traces~(see \S\ref{sec:exp:setup:wl}).

Figure~\ref{fig:composite:power} shows the power consumption for different combinations of both workloads and different topologies.
We observe the unbiased variance of results~\cite[p.~32]{Devore2009} 
is positively correlated with the fraction of the workload taken from the public cloud (Azure).
Depending on topology, the variance increase with this fraction ranges from 4.78x to 5.45x.
Expanding the volume horizontally~(\iconExpInline{}\iconVolInline{}\iconHorInline{}) leads to the lowest increase in variance.
The workload statistics listed in Table~\ref{tab:workload-stats} show that the Azure trace has far fewer \glspl{VM}, with higher load per \gls{VM} and shorter duration, thus explaining the increased variance.
Last, all candidate topologies have a higher power consumption than the base topology.

\begin{figure*}
    \centering
    \subfloat[Requested CPU cycles\label{fig:full:composite:requested}]{\includegraphics[width=0.5\linewidth]{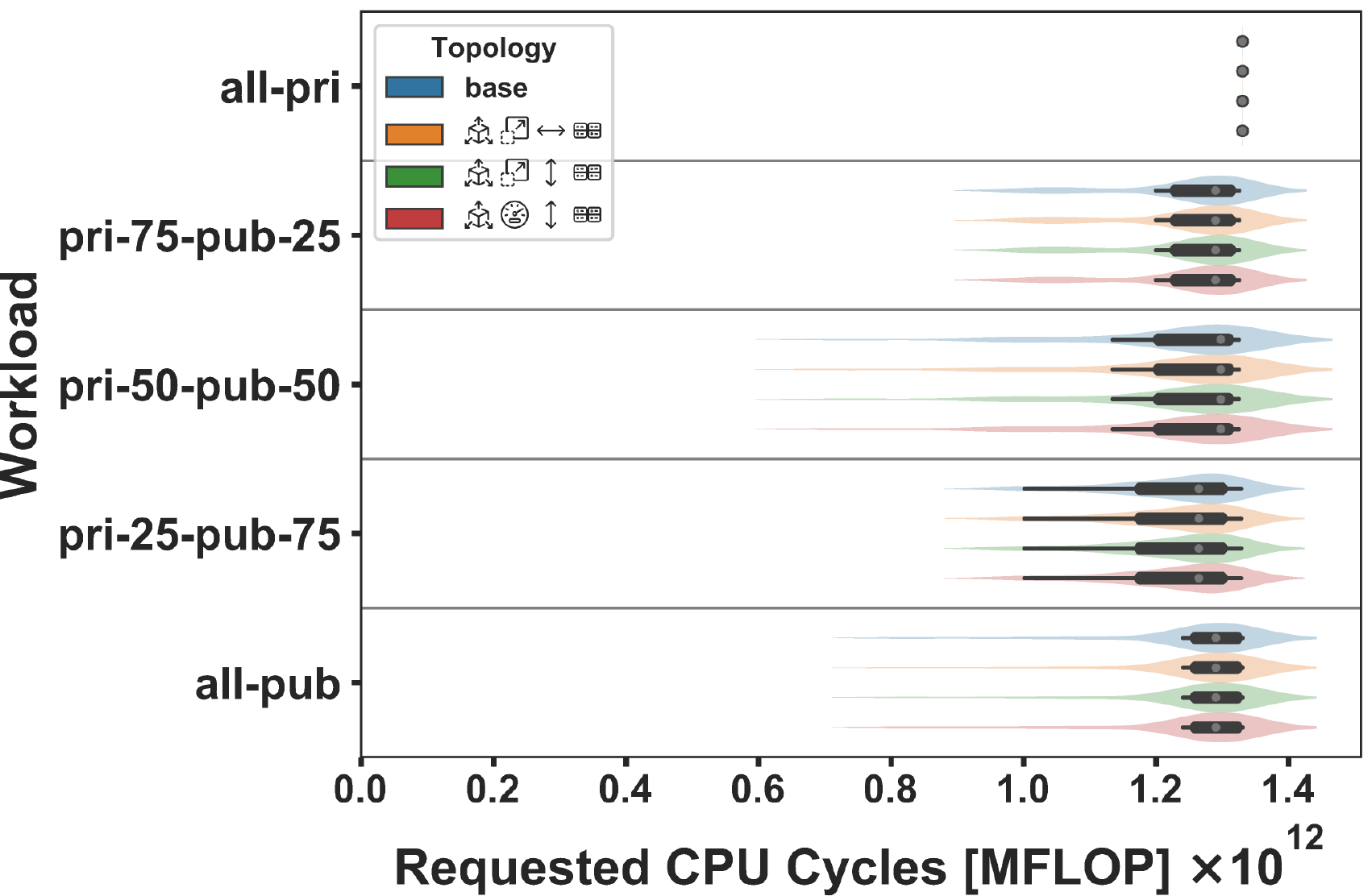}}%
    \subfloat[Granted CPU cycles\label{fig:full:composite:granted}]{\includegraphics[width=0.5\linewidth]{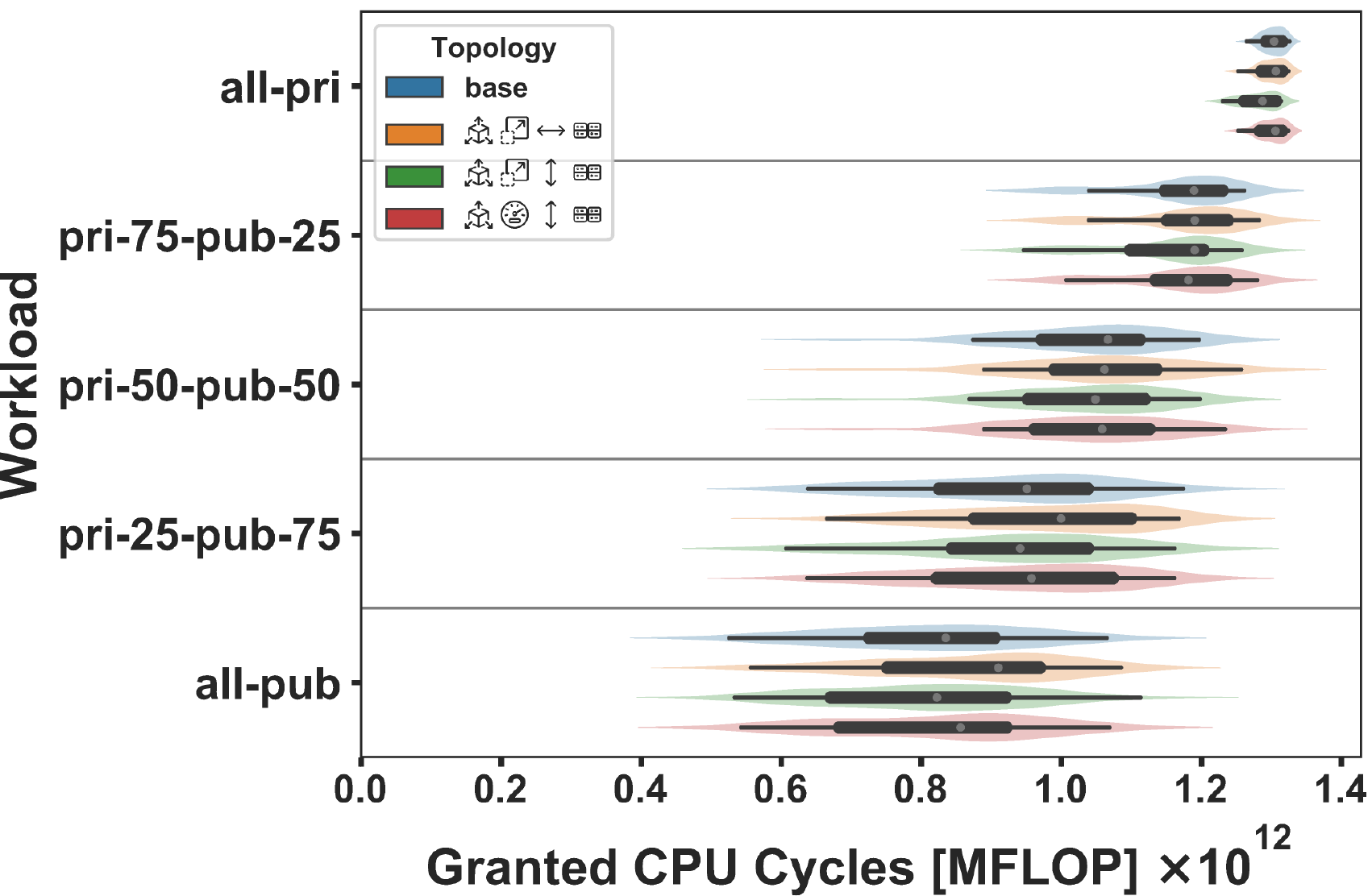}}\\
    \subfloat[Overcommitted CPU cycles\label{fig:full:composite:overcommitted}]{\includegraphics[width=0.5\linewidth]{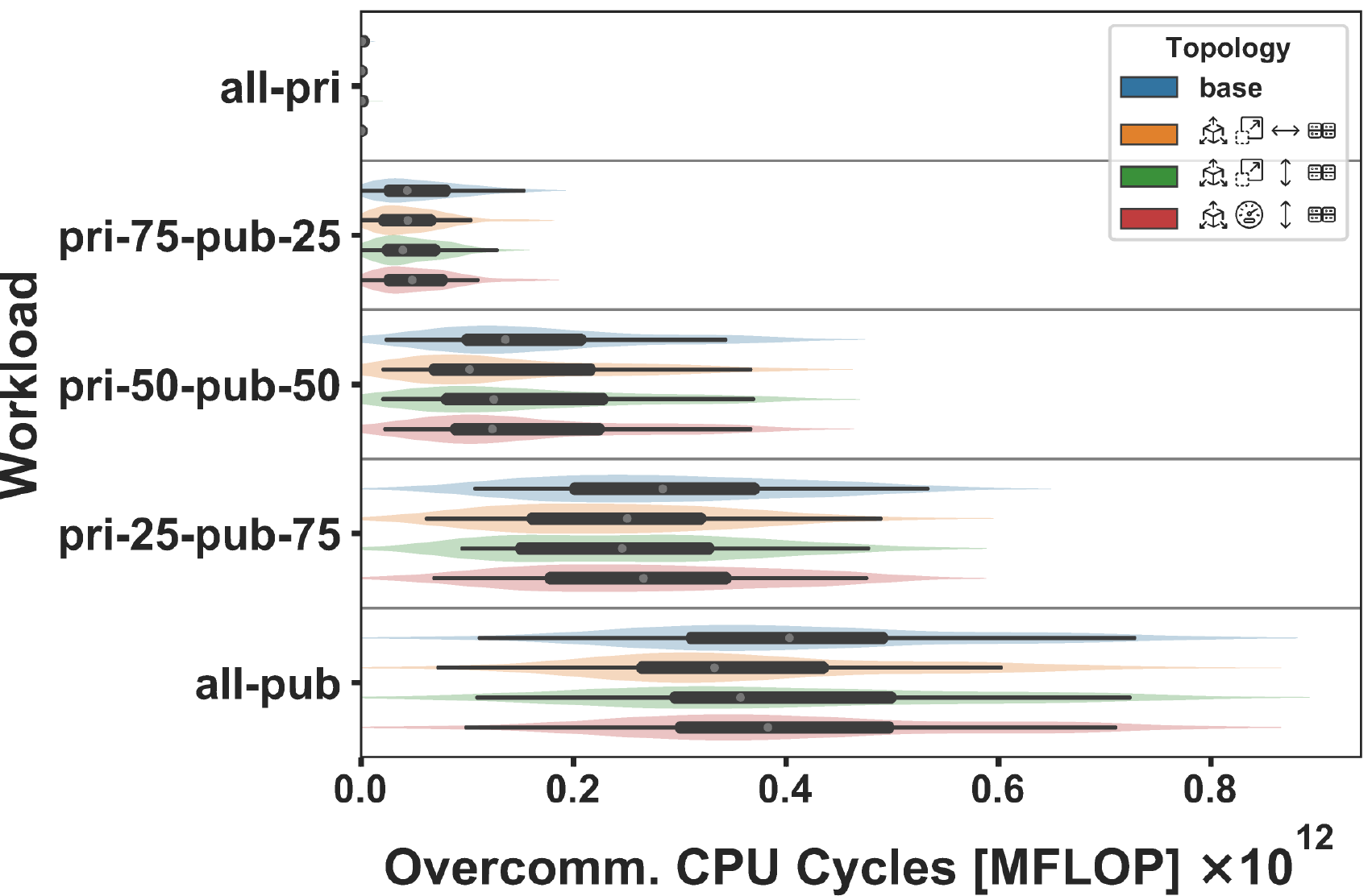}}%
    \subfloat[Interfered CPU cycles\label{fig:full:composite:interfered}]{\includegraphics[width=0.5\linewidth]{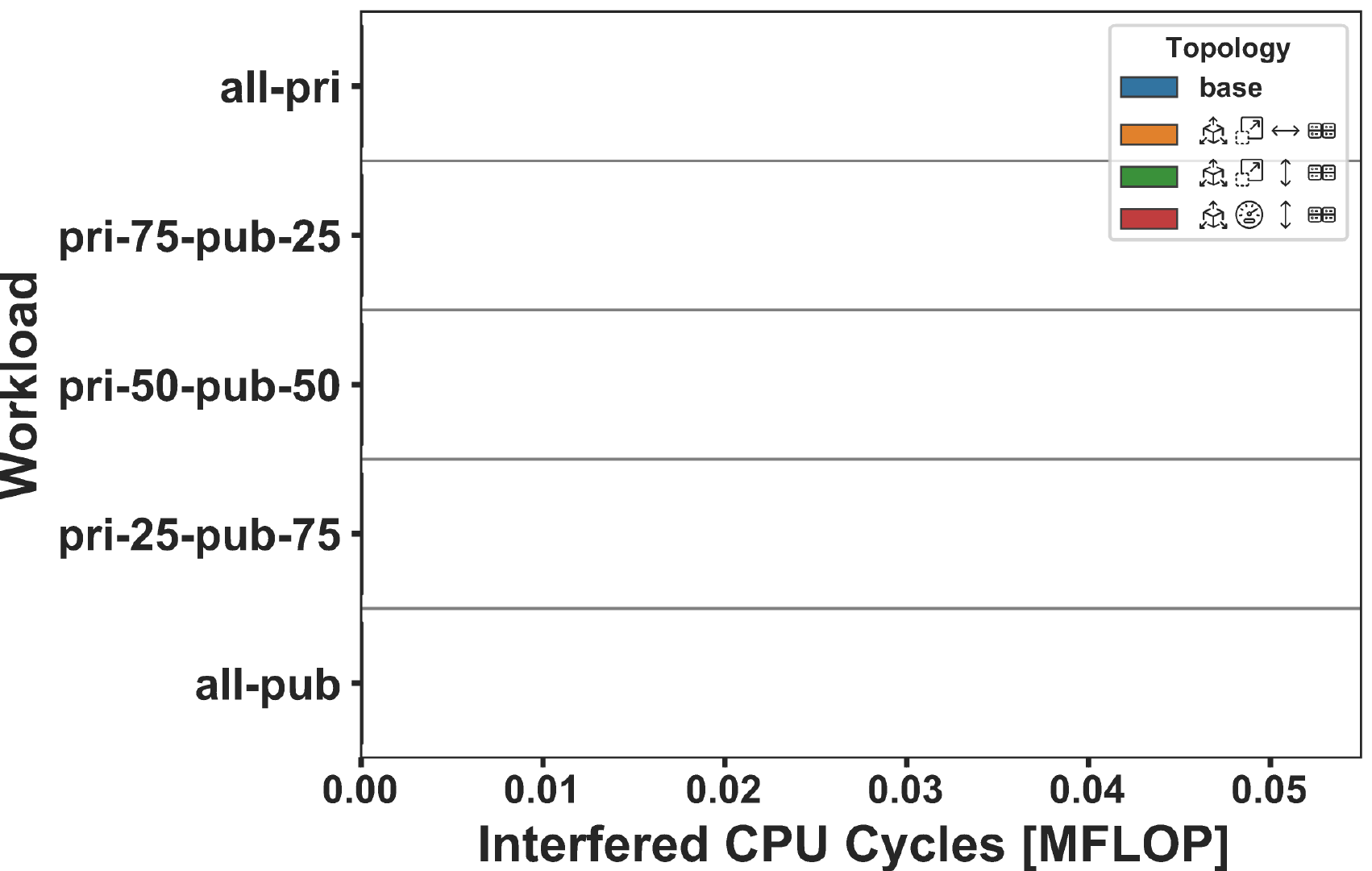}}\\
    \subfloat[Total power consumption\label{fig:full:composite:power}]{\includegraphics[width=0.5\linewidth]{figures/plots/composite_workload_total_power_draw.pdf}}%
    \subfloat[Total number of time slices in which a \gls{VM} is failed, aggregated across \glspl{VM}\label{fig:full:composite:failures:vms}]{\includegraphics[width=0.5\linewidth]{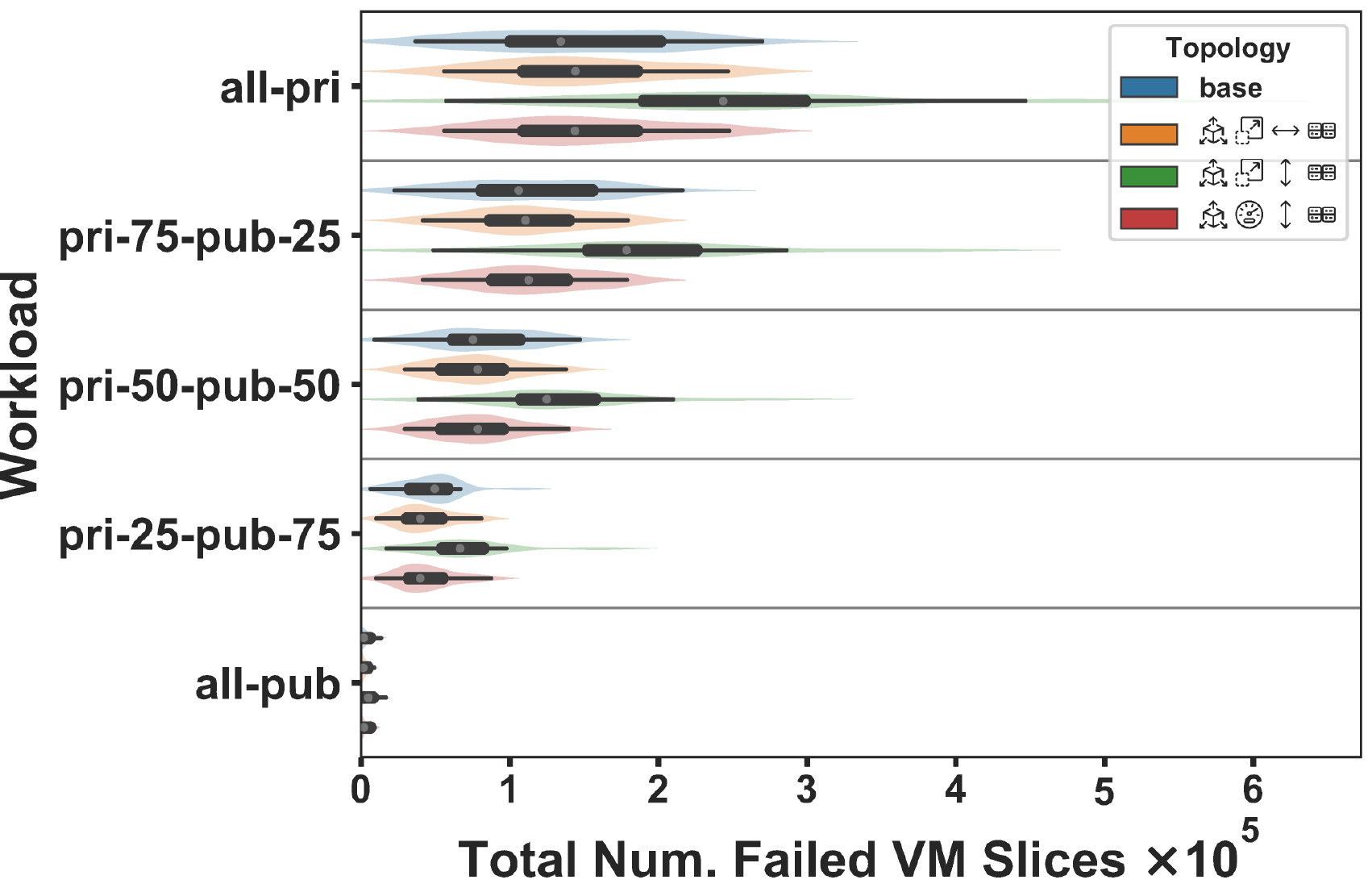}}\\
    \caption{Impact of a composite workload (consisting of private and public workloads) on different topologies. For a legend of topologies, see Table~\ref{tab:experiment-overview}. Continued in Figure~\ref{fig:full:composite:2}.}
    \label{fig:full:composite:1}
\end{figure*}

We also observe performance degrading with increasing public workload fraction (see Figure~\ref{fig:full:composite:overcommitted}, \S\ref{sec:full-visual-results}), calling for a different topology or more sophisticated provisioning policy to address the differing needs of this new workload.
We see that horizontal volume expansion (\iconExpInline{}\iconVolInline{}\iconHorInline{}) provides the best performance in the majority of workload transition scenarios.

We conclude {\em \toolname{} can support new workloads as they appear}, so before they are 
deployed. 

    \section{Validation of the Simulator}
\label{sec:simulator-validation}
We discuss in this section the validity of the outputs of the (extensions to the) simulator. 
Capelin uses datacenter-level simulation using real-world traces to evaluate portfolios of capacity planning scenarios.
Although real-world experimentation would provide more realistic outputs, evaluating the vast amount of scenarios generated by Capelin on physical infrastructure is prohibitively expensive, hard to reproduce, and cannot capture the scale of modern datacenter infrastructure, notwithstanding environmental concerns.
Alternatively, we can use mathematical analysis, where datacenter resources are represented as mathematical models (e.g., hierarchical and queuing models).
However, this approach is limited because its accuracy relies on preexisting data from which the models are derived.
Further considering the complexity and responsibilities of modern datacenters, this approach becomes infeasible.

Given that the effectiveness of Capelin depends heavily on (the correctness of) simulator outputs, we have worked very carefully and systematically to ensure the validity of the simulator.
For the validity of the simulator, we consider three main aspects: (1) validity of results, (2) soundness of results, and (3) reliability of results.
Below, we discuss for each of these aspects our approach and results.

\subsubsection*{T1. How to ensure simulator outputs are valid?}
We consider simulator outputs valid if a realistic base model (e.g., the datacenter topology) with the addition of a workload and other assumptions (e.g., operational phenomena) can reflect realistically real-world scenarios based on the same assumptions. 

We ensure validity of simulator outputs by tracking a wide variety of metrics (see Section \ref{sec:exp:setup:metrics}) during the execution of simulations in order to validate the behavior of the system. 
This selection is comprised of metrics of interest which we analyze in our experiments, but also fail-safe metrics (e.g., total requested burst) that we can verify against known values.

Moreover, we employ step-by-step inspection using the various tools offered by the Java ecosystem (e.g., Java Debugger, Java Flight Recorder, and VisualVM) to verify the state of individual components on a per-cycle basis.

\subsubsection*{T2. How to ensure simulator outputs are sound?}
While the simulator may produce valid outputs, for them to be useful, these outputs must also be realistic and applicable to users of Capelin.
That is, the assumptions that support the datacenter model must hold in the real world, for the simulator outputs to be sound and in turn be useful.
Concretely, a particular choice of scheduling policy might produce valid results, yet may not reflect reality.

To address this, we have created ``replay experiments'' that replicate the resource management decisions made by the original infrastructure of the traces, based on placement data from that time.
We do not support live migration of VMs that occurs in the placement data, since VM placements are currently fixed over time in OpenDC.
However, the majority of VMs do not migrate at all.
Capacity issues due to not supporting live migration are resolved by scheduling VMs on other hosts in the cluster based on the \texttt{mem} policy.

\begin{figure}
    \includegraphics[width=\linewidth]{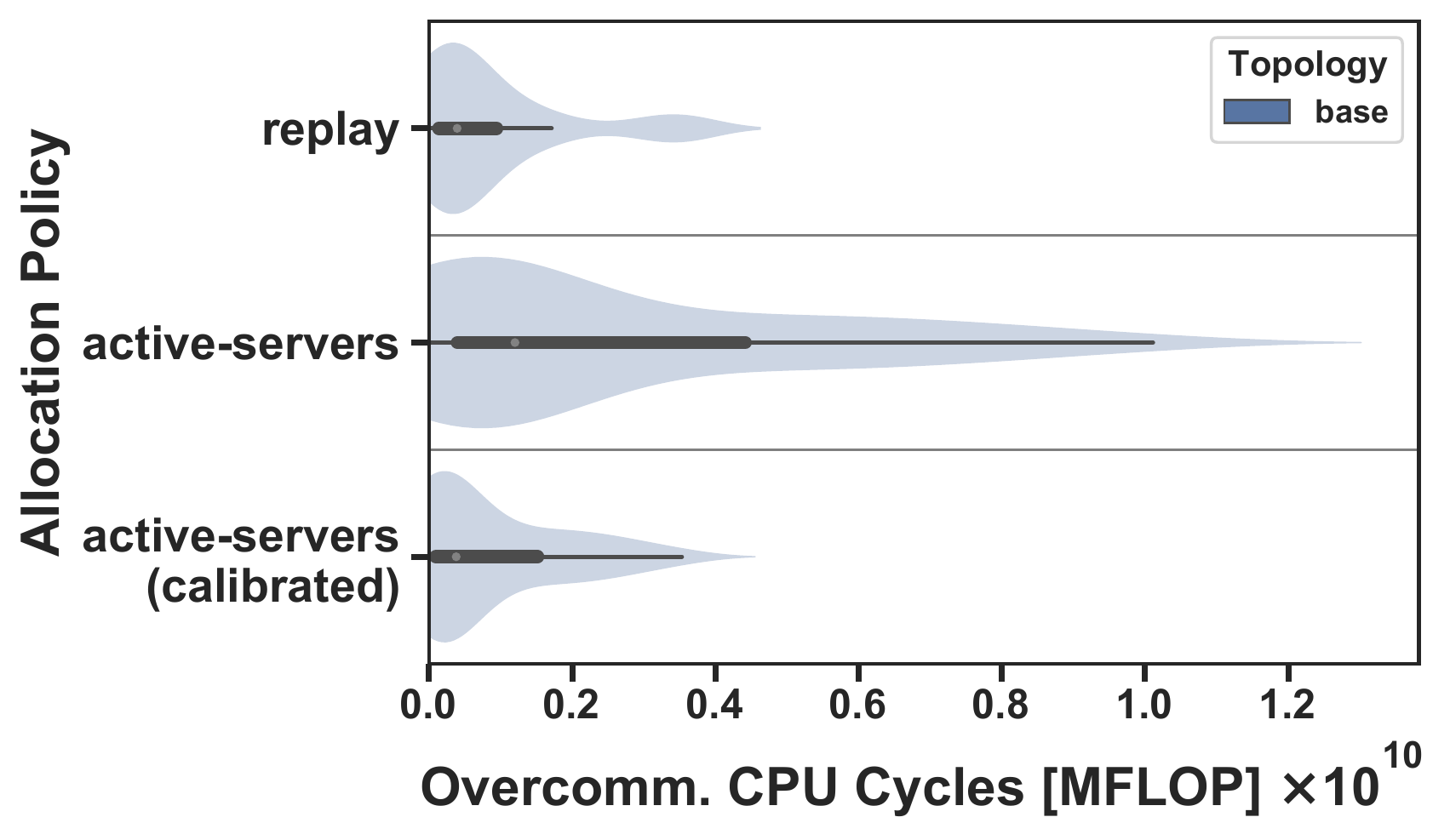}
    \caption{Validation with a replay policy, copying the exact cluster assignment of the original deployment. For a legend of topologies, see Table~\ref{tab:experiment-overview}.}
    \label{fig:validation:overcommitted-burst}
\end{figure}

\begin{figure}
    \includegraphics[width=\linewidth]{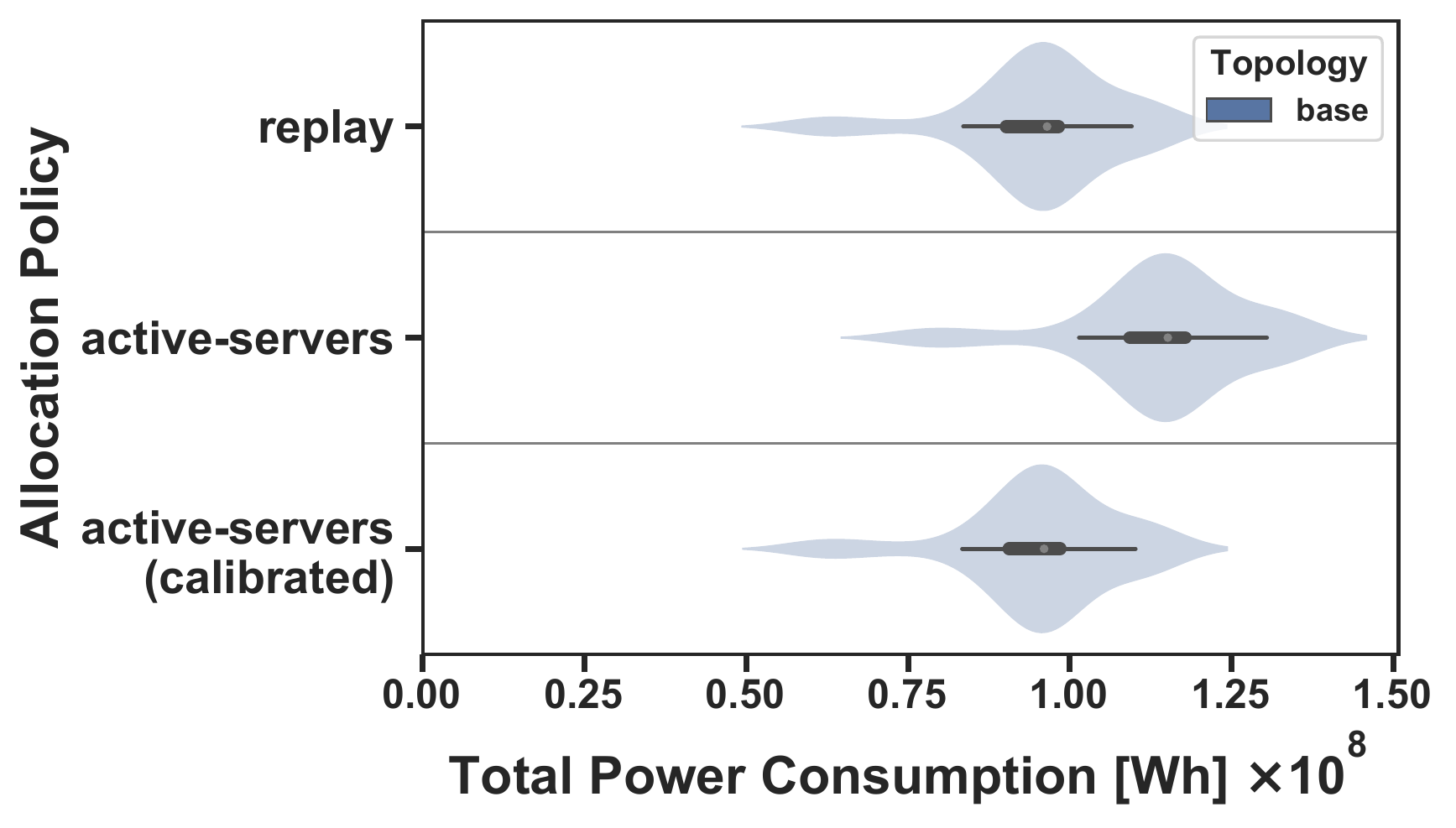}
    \caption{Validation with a replay policy, copying the exact cluster assignment of the original deployment. For a legend of topologies, see Table~\ref{tab:experiment-overview}.}
    \label{fig:validation:power-draw}
\end{figure}

The ``replay experiments'' are run in an identical setup to the experiments in Section \ref{sec:exp} and its results are compared to the \texttt{active-servers} allocation policy.
We visualize both raw results and calibrated results, obtained through only linear transformations (shifting and scaling values) to account for possible constant discrepancy factors.
We find that:
\begin{enumerate}
    \item The total overcommitted burst shows distributions that are similar in shape but differ in scale, for both policies.
    This can be explained by the fact that \texttt{active-servers} policy is not as effective as the manual placements on the original infrastructure in addition to the influence of performance interference (Figure~\ref{fig:validation:overcommitted-burst}).
    
    \item Other metrics show very similar distributions. 
    Small differences may be accounted to the number of VMs being slightly smaller in the ``replay experiments`` due to missing placement data (Figure~\ref{fig:validation:power-draw}).
    
\end{enumerate}

Furthermore, we have had several meetings with both industry and domain experts to discuss the simulator outputs in depth, validate our models and assumptions, and spot inconsistencies.
Moreover, we have had proactive communication with the experts about possible issues with the simulator that arose during development, such as unclear observations.

\subsubsection*{T3. How to ensure no regression in subsequent simulator versions?}
Although we may at one point trust the simulator to produce correct outputs, the addition or modification of functionality in subsequent versions of the simulator may inadvertently affect the output compared to previous versions. 

We safeguard against such issues by means of snapshot testing.
With snapshot testing, we capture a snapshot of the system outputs and compare it against the outputs produced by subsequent simulator versions.
For this test, we consider a downsized variant of the experiments run in this work and capturing the same metrics.
These tests execute after every change and ensure that the validity of the simulator outputs is not affected.
In case some output changes are intentional, the test failures serve as a double check. 

Furthermore, we use assertions in various parts of the simulator to validate internal assumptions.
This includes verifying that messages in simulation are not delivered out-of-order and validating that simulated machines do not reach invalid states.

Finally, we employ industry-standard development practices. 
Every change to the simulator or its extensions requires an independent code review before inclusion in the main code base. 
In addition, we automatically run for each change static code analysis tools (e.g. linting) to spot common mistakes.
    \section{Other Threats to Validity}
\label{sec:threats}

In this section, we list and address threats to the validity of our work that go beyond the validity of the simulator.

\subsection{Interview Study} \label{sec:threats:interviews}

Confidentiality limits us from sharing the {\em source transcripts of our analysis}.
Inherent in such a study is the threat to validity caused by this limitation.
To minimize this threat, the process used is meticulously described and the full findings presented.
We also point to the resonance that many of the results find with observations in other work.

The {\em limited sample size} of our study presents another threat to the validity of our interview findings.
This is difficult to address, due to the labor-intensive transcription and analysis conducted already in this study.
Follow-up studies should further address this concern by conducting a textual survey with a wider user base, requiring less time investment per interlocutor.

\subsection{Experimental Study} \label{sec:threats:experiments}

We discuss three threats related to the experimental study.

\subsubsection{Diversity of Modeled Resources}
Building topologies in practice requires consideration of many {\em different kinds of resources}.
In our study, we only actively explore the CPU resource dimension in the capacity planning process, to restrict the scope.
This could be seen 
Adding or removing CPUs to/from a machine however can relate to different types of memory or network becoming applicable or necessary.
This can have impacts on costs and energy consumption, altering the decision support provided in this study.
Nevertheless, the performance should suffer only minimal impact from this, since CPU consumption can be regarded as the critical factor in these considerations.
In addition, \toolname{} and it's core abstraction of portfolios of scenarios offers a broader framework and future extensions to \simulator{} will directly become available to planners using \toolname{}.

\subsubsection{Public Data Artifacts}
A second threat to validity could be perceived in the {\em absence of public experiment data artifacts}.
The confidentiality of the trace and topology we use in simulation prohibits the release of detailed artifacts and results.
However, an anonymized version of the trace is available in a public trace archive, which can be used to explore a restricted set of the workload.
The Azure traces used in the experiment in \S\ref{sec:experiments:composite} are public, along with our sampling logic for their use, and can therefore be locally used along with the codebase.

Last, a threat to validity could be seen in the {\em validity of the outputs of the (extensions to the) simulator itself}.
We cover this threat extensively in Section~\ref{sec:simulator-validation}.

\subsubsection{Allocation Policies}
We discuss in this section the relevance of the chosen allocation policies in this work and how they relate to allocation policies used in popular resource management tools such as OpenStack, Kubernetes, and VMWare vSphere.

The allocation policies used in this work use a ranking mechanism which orders candidate hosts based on some criterion (e.g., available memory or number of active VMs) and selects either the lowest or highest ranking host.

OpenStack uses by default the Filter Scheduler\footnote{\url{https://docs.openstack.org/nova/latest/user/filter-scheduler.html}} for placement of VMs onto hosts. For this, it uses a two step process, consisting of \textit{filtering} and \textit{weighing}. 
During the filtering phase, the scheduler filters the available hosts based on a set of user-configured policies (e.g., based on the number of available vCPUs). 
In the weighing phase, the scheduler uses a selection of policies to assign weights to the hosts that survived the filtering phase, and select the host with the highest weight.
How the weights are determined can be configured by the user, but by default the scheduler will spread VMs across all hosts evenly based the available RAM\footnote{\url{https://docs.openstack.org/nova/latest/admin/configuration/schedulers.html\#id18}}, similar to the \texttt{available-mem} policy in this work.

Kubernetes conceptually uses almost exactly the same process as OpenStack\footnote{\url{https://kubernetes.io/docs/concepts/scheduling-eviction/kube-scheduler/\#kube-scheduler-implementation}}, but by default uses more extensive weighing policies to ensure the workloads are balanced over the hosts, also taking into account dynamic information such as resource utilization. A key difference with OpenStack is that Kubernetes does not consider the memory requirements of workloads when weighing the hosts.

VMWare vSphere offers DRS (Distributed Resource Scheduler) which automatically balances workloads across hosts in a cluster based on memory requirements of the workloads. 
    
\section{Related Work}
\label{sec:related-work}

We summarize in this section the most closely related work, which we identified through a survey of the field that yielded over 75 relevant references.
Overall, our work is the first to: (1) conduct community interviews with capacity planning practitioners managing cloud infrastructures, which resulted in unique insights and requirements, 
(2) design and evaluate a data-driven, comprehensive approach to cloud capacity planning, which models real-world operational phenomena and provides, through simulation, multiple VM-level metrics as support to capacity planning decisions.
%

\subsection{Community Interviews}

Related to (1), we see two works as closely related to our {\em interview study} of practitioners.
In the late-1980s, \citeauthor{Lam1987} conducted a written questionnaire survey~\cite{Lam1987} and, mid-2010s, \citeauthor{Bauer2017} conducted semi-structured interviews~\cite{Bauer2017}.
The target group of these studies differs from ours, however, since both focus on practitioners from different industries planning the resources used by their IT department.
We summarize both related works below.

\citeauthor{Lam1987} (\citeyear{Lam1987}) conduct a written survey with 388 participants~\cite[p.~142]{Lam1987}.
The survey consists of scaled questions where practitioners indicate how frequently they use certain strategies in different stages of the capacity planning process~\cite[p.~143]{Lam1987}.
Their results indicate that very few respondents believe that they use ``sophisticated'' forecasting techniques for their capacity planning activities, with visual trending being the most popular strategy at that time.
They find that ``many companies still rely on the simplistic, rules-of-thumb, or judgmental approach'' to capacity planning~\cite[p.~8]{Lam1987}.
More importantly even, the authors believe that there is a ``significant gap between theory and practice as to the usability of the scientific and the more sophisticated techniques''.
The conclusions \citeauthor{Lam1987} draw from their survey and the relations we observe in their results are resonant with the findings of our study.
This stresses the need for a usable and comprehensive capacity planning system for today's computer systems.

\citeauthor{Bauer2017} (\citeyear{Bauer2017}) conduct 12 in-person interviews with ``IT capacity-management practitioners''~\cite{Bauer2017} in six different industries.
Similar to our interviewing style, the interviews were ``semi-structured'', guided by questions prepared in advance.
The questions range from capacity planning process questions to more managerial questions around organizational structure.
After manual evaluation of the interview transcripts, the authors find that practitioners often state that the number of capacity planning roles in organizations is decreasing, while the discipline is still very much relevant.
The practitioners also find that ``vendor-relationship management and contract management'' are playing an increasing role in the capacity planning process, as well as redundancy and multi-cloud considerations.
These results, even if for a different target group, resonate with our findings in two ways: (1) they underline our call for the need to focus on the capacity planning process as an essential part of resource management, and (2) emphasize the multi-disciplinary, complex nature of the decisions needing to be taken.

\subsection{Capacity Planning Approaches}

Related to (2), our work extends the body of related work in three key areas: (1) process models for capacity planning, (2) works related to capacity planning, and (3) system-level simulators.

\subsubsection{Process Models for Capacity Planning}

Firstly, we survey process models for capacity planning published in literature.
To enable their comparison, 
we unify the terminology and the stages proposed by these models, and create the super-set of systems-related stages summarized in Table~\ref{tab:book-model-comparison}.
We observe that the first stages (assessment and characterization) have the broadest support among models.
However, we also find significant differences in the comprehensiveness of models.
We observe that the later stages (deployment and calibration) tend to receive more attention only in more recent publications.
From a systems perspective, \toolname{} proposes the first comprehensive process.

\subsubsection{Works Related to Capacity Planning}

Secondly, we survey systematically the main scientific repositories and collect 56 works related to capacity planning.
While we plan to release the full survey at a later stage, we share key insights here.
We find that the majority of studies only consider one resource dimension, and four inputs or less for their capacity planning model.
Few are simulation-based~\cite{Rolia2005,Mylavarapu2010,Patel2013,Ostberg2017,Carvalho2017,DBLP:conf/sosp/AlipourfardGKHV19}, with the rest using primarily analytical models.
We highlight three of these works below and position them in relation to this work.

\citeauthor{Rolia2005} proposes the first trace-based approach to the problem~\cite{Rolia2005}.
Their ``Quartermaster'' capacity manager service motivates the use of what-if questions to optimize \glspl{SLO}, with the help of trace-based analysis and optimizing search for optimal capacity plan suggestions.
It's underlying simulation is restricted to replay with no additional modelling of phenomena or policies.
This severely limits the scope and coverage of the exploration, regarding only one dimension (quantity of CPUs).
The work also does not formally specify what-if scenarios, even though mentioning the wide variety of scenarios (questions) that can be formulated.

\citeauthor{Carvalho2017} uses queuing-theory models to optimize the computational capacity of datacenters~\cite{Carvalho2017}.
Their models are built from high-level workload characteristics derived from traces and include admission control policies.
The simplifying assumptions made in constructing these simulation models restrict the realism of their output.
In addition, while this work emphasizes the role that trade-offs play in the decision-making process, the trade-offs themselves are only evaluated on a single-metric scale (combining multiple metrics into one), leaving practitioners with a single output plan to accept or reject.

The notable Janus~\cite{DBLP:conf/sosp/AlipourfardGKHV19} presents a real-time risk-based planning approach for datacenter networks.
The scope of this study differs from our scope, in that it addresses networks and aims to assist in real-time, operational changes.
However, we share a focus on operational risks and involved costs, and Janus also is evaluated with the help of real-world traces.

Our scope of long-term planning (procurement) excludes more dynamic, short-term process such as Google's Auxon~\cite{Hixson2015} or the Cloud Capacity Manager~\cite{Kesavan2013}, which address the live management of capacity {\it already procured}; explained differently, Capelin (this work) helps decide on long-term capacity procurement, whereas Auxon and others like focus on the different problems of what to do with that capacity, short-term, once it is already there.
Other work investigates the dynamic management of physical components, such as CPU frequency scaling~\cite{Lucanin2016}.

\subsubsection{System-Level Simulators}

Thirdly, we survey system-level simulators, and study 10 of the best-known in the large-scale distributed systems community.
Among the simulators that support \glspl{VM} already~\cite{Calheiros11,Hirofuchi18,Nunez12} and could thus be useful for simulating cloud datacenters, few have been tested with traces at the scale of this study, few support CPU overcommissioning, none supports both operational phenomena used in this work, and none can output detailed VM-level metrics.





\begin{table}
    \centering
    \caption{Comparison of process models for capacity planning. 
    Sources: \citeauthor{Lam1987}~\cite[p.~92]{Lam1987}, \citeauthor{Howard1992}~\cite{Howard1992} (referenced by \citeauthor{Browning1994}~\cite[p.~7]{Browning1994}), \citeauthor{Menasce2001}~\cite[p.~179]{Menasce2001}, \citeauthor{Gunther2007}~\cite[p.~22]{Gunther2007}, and 
    \citeauthor{Kejariwal2017}~\cite[p.~4]{Kejariwal2017}.}
    \label{tab:book-model-comparison}
    \vcutM{}
    \setlength{\tabcolsep}{2pt}
    \begin{tabularx}{\columnwidth}{Xccccc|c}
        \toprule
        Stage & \cite{Lam1987} & \cite{Browning1994} & \cite{Menasce2001} & \cite{Gunther2007} & \cite{Kejariwal2017} & \toolname{} \\
        \midrule
        Assessing current cap. & \checkmark & & \checkmark & \checkmark & \checkmark & \checkmark \\
        Identifying all workloads & & \checkmark & & & & \checkmark \\
        Characterize workloads &\checkmark & \checkmark & \checkmark & \checkmark & & \checkmark \\
        Aggregate workloads & & \checkmark & & & & \checkmark \\
        Validate workload char. & & & \checkmark & & & \checkmark \\
        Determine resource req. & & \checkmark & & & & \checkmark \\
        Predict workload & \checkmark & & \checkmark & & \checkmark & \checkmark \\
        Characterize perf. & \checkmark & & \checkmark & & & \checkmark \\
        Validate perf. char. & \checkmark & & \checkmark & & & \checkmark \\
        Predict perf. & \checkmark & & \checkmark & & & \checkmark \\
        Characterize cost & & & \checkmark & & & \checkmark \\
        Predict cost & & & \checkmark & & & \checkmark \\
        Analyze cost and perf. & & & \checkmark & & & \checkmark \\
        Examine what-if scen. & \checkmark & & & & & \checkmark \\
        Design system & & & & \checkmark & & \checkmark \\
        Iterate and calibrate & & & & & \checkmark & \checkmark \\
        \bottomrule
    \end{tabularx}
    \vcutM{}
\end{table}

\section{Conclusion and Future Work}
\label{sec:conclusion}


Accurately planning cloud datacenter capacity is key to meeting the needs of the 2020s society whilst saving costs and ensuring environmental sustainability.
Although capacity planning is crucial, 
the current practice has not been analyzed in decades
and publicly available tools to support practitioners are scarce.
\toolname{}, a data-driven, scenario-based alternative to current planning approaches, 
addresses these problems.


In this work, we have designed, implemented, and evaluated \toolname{}.
We have conducted a guided interview with diverse practitioners from a variety of backgrounds, whose results 
led us to synthesize five functional requirements.
We have designed \toolname{} to meet them, 
including the ability to model datacenter topologies and virtualized workloads, to express what-if scenarios and \gls{QoS} requirements, to suggest scenarios to evaluate, and to evaluate and explain capacity plans. 
\toolname{} uses a novel abstraction, the capacity planning portfolio, to represent, explore, and compare 
scenarios.
Experiments based on real-world workload traces collected from private and public clouds
demonstrate \toolname{}'s capabilities. 
Results show
that \toolname{} can support capacity planning processes, exploring changes from a baseline scenario alongside four dimensions. 
We found that capacity plans common in practice could potentially lead to significant performance degradation, e.g., 1.5x--2.7x.
We also gave evidence of the important, but often discounted, role that operational choices (e.g., the allocation policy) and operational phenomena (e.g., performance interference) play in capacity planning.

We have released \toolname{} as FOSS for capacity planners to use.
We will continue to support it and, in future work, we plan to 
deepen and 
engineer \toolname{}. 
We are investigating the use of Machine Learning and conventional Artificial Intelligence search techniques to make the Capacity Plan Generator component more capable of exploring the enormous design-space.
We intend to conduct a structured survey, in the form of a textual questionnaire, to reach a larger base of capacity planning practitioners and augment the initial findings made in our interview study.
We see opportunities for research into cloud user behavior when emerging resources are deployed, a factor especially relevant in scientific clouds.
We also plan to include more workload types, such as virtualized \gls{FaaS} workloads.

    \appendices

    \bibliographystyle{plainnat}
    \interlinepenalty=10000
    \bibliography{main}

    \section{Capacity Planning Interview Script}
\label{sec:interview-script}

\newcommand{\onestar}{{\tt *}}
\newcommand{\twostar}{{\tt **}}
\newcommand{\threestar}{{\tt ***}}

In this chapter, we list the interview script used for the interviews described in Section~\ref{sec:interviews}.
To encode instructions to the reviewer, we use the following notation.
The font determines the type of instruction: Questions are written in standard fonts, emphasis indicates instructions to be read to the interlocutor by the interviewer (not necessarily verbatim), and mono-space font represents instructions for the reviewer.
The questions are numbered for cross-reference and divided into 5 category.
Each question is assigned one out of three priority levels, indicated by asterisks (\onestar, \twostar, and \threestar), with three stars indicating the highest priority.
Each category (section) of questions is allocated a number of minutes, listed between parentheses.

\subsection{Part 1: Overview (15')}
\emph{Thanks for agreeing to meet with me. 
To get the most out of this conversation, would you allow me to record the conversation for note-taking purposes? 
I will not share the recording with anyone else but you, if you want a copy, and I will delete the recording at the end of my thesis project.}

\emph{If you have any concerns about sharing this kind of information with me, let us talk quickly and openly about it. 
We hope you will share openly. 
Rest assured, we want more to learn about the issues around capacity planning than to publish on it.}

\emph{I will transcribe the recording for myself, and for any use of a snippet of your words, I will ask you specifically for approval to release, with due reference, unless you want me to keep the author anonymous, of course.}

\emph{We are interested in learning more about how businesses think about having IT infrastructure and services always available and plentiful. 
We call this capacity planning, and know we are referring here only to IT and the IT team, and not to other types of ``capacity''. 
We want to learn and share with you what processes are used, what challenges exist, and how can we help solve them.}

\begin{description}[leftmargin=5.2em,style=nextline]
    \item[(\setID{Q1}) \threestar] How important is it to have IT infrastructure and services always available and plentiful in your business?
    \item[(\setID{Q2}) \threestar] What kind of services do you provide? How important is it for your different services?
    \item[(\setID{Q3}) \threestar] Can you give us an example of a success in capacity planning? Share with us a good idea, a good process, some situation when all worked well?
    \item[(\setID{Q4}) \threestar] Can you give us an example of an insuccess in capacity planning? Share with us a mistake, an erroneous process, some situation when many things failed to worked well or took much more to get through than expected?
    \item[(\setID{Q5}) \threestar] What does the typical process for capacity planning look like at your company? You can start with an overview, or even from a concrete example, like how to get a new cluster in operation.
    \item[(\setID{Q6}) \twostar] A few yes or no questions:
    \begin{enumerate}
        \item Do you have ``what if'' scenarios?
        \item Do you consider hybrid or public clouds to be part of your capacity planning process?
        \item Would you be willing to share historical data on capacity planning?
        \item Do you consider human personnel (availability, experience, training) when planning for new capacity?
    \end{enumerate}
\end{description}

\subsection{Part 2: The Process (15')}
\begin{description}[leftmargin=5.2em,style=nextline]
    \item[(\setID{Q7}) \threestar] Who are the stakeholders? Who gets to take the decision? Who gets to give input? Is this a board-level decision? Is it left to operations?
    \item[(\setID{Q8}) \threestar] On what time and infrastructure scale are your typical decisions?
    \item[(\setID{Q9}) \threestar] What factors do you take into account when making capacity planning decisions? Does this differ per stakeholder; if so, how?
    \item[(\setID{Q10}) \twostar] What is the margin for error in the decision making process?
    \item[(\setID{Q11}) \onestar] How frequently are capacity planning decisions made? Also, how long does a decision take?
    \item[(\setID{Q12}) \onestar] What kind of data sources do you consult in your capacity planning process?
    \item[(\setID{Q13}) \onestar] What kind of tools do you use in your capacity planning process? For planning, recording, sharing information at different levels in the organization, etc.
    \item[(\setID{Q14}) \twostar] How are errors or issues about capacity planning preserved? How frequent/severe are the errors that are made? How do people learn from these issues?
\end{description}

\subsection{Part 3: Inside Factors (15')}
\begin{description}[leftmargin=5.2em,style=nextline]
    \item[(\setID{Q15}) \threestar] What kinds of IT services and infrastructure are part of your capacity planning processes?
    \item[(\setID{Q16}) \twostar] I will ask the same question about four kinds of workloads.\\
    What are your capacity planning processes for business-critical workloads? \\
    What are your capacity planning processes for big data workloads?\\
    What are your capacity planning processes for serverless workloads?\\
    What are your capacity planning processes for high performance computing workloads?
    \item[(\setID{Q17}) \threestar] How do you try to combine multiple workloads in the same capacity planning process? Shared infrastructure? Shared services? What role do hybrid or public cloud offerings play in your capacity planning process?
    \item[(\setID{Q18}) \threestar] Because serverless workloads are so new, I’d like to ask a couple more questions about them. With such fine-granularity and variable workloads, how do you reason about needs?\\
    Do you reason differently about them than about other (more traditional) workloads?\\
    How do you reason about workload isolation (performance, availability, security, etc.)?
    \item[(\setID{Q19}) \onestar] What are some typical ``what if'' scenarios?
\end{description}

\subsection{Part 4: Outside Factors (10')}
\begin{description}[leftmargin=5.2em,style=nextline]
    \item[(\setID{Q20}) \threestar] What regulatory constraints (laws, standards; e.g. \gls{GDPR}, concerning where you get the capacity) play a role in the decision process?
    \item[(\setID{Q21}) \threestar] What financial aspects (costs of resources, personnel, etc.) or technical aspects (new generation of hardware/software/IT paradigms) play a role in the decision process?
    \item[(\setID{Q22}) \threestar] How and which human factors are involved in your decision making on resource capacity planning?
    \item[(\setID{Q23}) \twostar] Do you make capacity planning decisions on a multi-datacenter level, or on a local level?
    \item[(\setID{Q24}) \threestar] Do you do capacity planning specifically for production, development, and/or test?
    \item[(\setID{Q25}) \onestar] What are some typical ``what if'' scenarios?
\end{description}

\subsection{Part 5: Summary and Follow-Up (5')}
\begin{description}[leftmargin=5.2em,style=nextline]
    \item[(\setID{Q26}) \threestar] What would your ideal process for capacity planning look like? Something that is not already there? 
    \item[(\setID{Q27}) \twostar] Which other processes do you see capacity planning linked with? For example, managing change, evolution of business requirements, etc.?
    \item[(\setID{Q28}) \twostar] What other aspects would you like to share?
\end{description}

\noindent
\textbf{Follow-Up Points}
\\
{\tt Explain what I will do with the information.}
\\
{\tt Ask if they want a summary or report related to my thesis project.}
\\
{\tt If they answered "yes" to sharing historical data, follow up here.}

    \section{Detailed Interview Results}
\label{sec:interview-results}

In this appendix, we present our detailed analysis of the interlocutors' statements, organized around the questions listed in the interview script in Appendix~\ref{sec:interview-script}.
Findings are numbered and prefixed with {\bf IF} ({\bf I}nterview {\bf F}inding).

\subsection*{1 - Q1: Importance of Availability} \label{sec:interview-results:q1}
\fin{availability:commercial-vs-science}
We observe that availability appears to be critical in commercial cloud environments.
In scientific cloud infrastructure, availability generally appears to be perceived as less important than in commercial cloud environments.

\subsection*{1 - Q2: Services} \label{sec:interview-results:q2}
\fin{services:general}
The organizations of the interlocutors provide a wide range of services, from general IT services to specialist hardware hosting.
The interlocutors themselves are mainly concerned with compute cloud services.
These can be divided into virtualized offerings (\gls{VM} hosting), batch workload services, and specialized HPC hosting.

\fin{services:per-domain}
Batch workloads and HPC hosting are only seen in the scientific infrastructure surveyed.
\gls{VM} hosting is dominant in the commercial infrastructure space and can only be found in half of the surveyed scientific infrastructures.

\subsection*{1 - Q3: Success in Capacity Planning} \label{sec:interview-results:q3}
\fin{success:scope}
The success stories being told vary from large installations with significant upfront effort to more flexible, iterative installations.
Flexibility is still often valued even in large scale designs, in the form of strategies leaving room for adaptations later-on.

\fin{success:absence-of-failures}
One interview characterizes the absence of total failure scenarios in the past as a success story for the capacity planning team.

\fin{success:new-tech}
The utilization of new hardware with beneficial features, such as competitive pricing or an increase in parallelism, is a success story recounted in two of the interviews.

\subsection*{1 - Q4: Insuccess in Capacity Planning} \label{sec:interview-results:q4}
\fin{failures:abundance}
A first observation is the abundance of failure stories, especially compared to the number of success stories.
A possible explanation is that the process could be largely taken for granted.
The practices is mainly revisited when suboptimal situations arise.

\subsubsection*{Challenges Capacity Planners Face}
We find that capacity planning practitioners face many challenges in the process.

\fin{failures:underutilization}
Most interlocutors see and are discontent with a pervasive under-utilization of their system.
This under-utilization can be caused by operational risk minimization taking precedence, newly installed resources not being directly used, and delays in the process.

\fin{failures:complexity}
We observe the challenge of increasing complexity, both in the managed resources and the tooling needed to correctly monitor them.
The heterogeneity of hardware, especially in \gls{HPC} domains, is also mentioned in accounts of this.

\fin{failures:small-deployments}
Some remark that supporting the small to medium workload deployments is much more difficult than planning for the larger deployments.
While the larger units each have a larger financial impact, the small units tend to be neglected and found in need of sufficient leftover capacity.

\subsubsection*{Failure Stories}
We summarize the most common failure types below.

\fin{failures:new-tech}
In some cases, the adoption of new technologies can have unforeseen negative consequences.
A failure story of the adoption of an (unnamed) new processor architecture ends in an entire rollback of the installation, due to users having difficulties properly utilizing the new hardware, and due to high power consumption.

\fin{failures:long-decisions}
Some capacity planning decisions take so long that technological perceptions have changed due to the rapidly changing nature of the field.
This leads to hardware choices needing to be implemented that, at the time of actual installation, are considered suboptimal.

\fin{failures:mispredictions}
A notorious challenge seems to be the prediction of future ratios between different resource types (mainly number of cores, memory units, storage capacity).
We observe a number of failure stories surrounding the misprediction of how different resource type might relate in the future, resulting in significant parts being underutilized or certain capacity dimensions running out of capacity far faster than others.
This last consequence can lead to reduced \gls{QoS}.

\fin{failures:afterthought}
We observe cases where capacity planning is only seen as an afterthought.
A representative example is the fast onboarding of a new client where available capacity or time to acquire new capacity is judged too optimistically.

\subsection*{1 - Q5: Typical Process} \label{sec:interview-results:q5}

\fin{typical:periodicity}
The typical processes we see have two shapes: a periodic process, typically centered around the lifecycle of topology resources, and a more ad-hoc process, triggered by less predictable events such as the arrival of new users. 
The former is dominant in most surveyed scientific clouds, while the latter is more common in commercial clouds.
One scientific cloud in the set has a combination of both.

\fin{typical:guessing}
A sentiment expressed in the majority of interviews is that guessing and human interpretation of monitoring data are a big part of the process.
The tooling in the area seems underutilized, further discussed in Q13.

\fin{typical:non-computational-factors}
Next to computational performance of resources, electricity and cooling also play a significant role in the equation.
This both impacts and is impacted by the choice of hardware.

\subsubsection*{Commercial Infrastructures}
We observe a difference in the typical process and challenges involved in this process between commercial and scientific infrastructures.
Below, we outline the findings for commercial infrastructures.

\fin{typical:commercial:financial-vs-human}
We see all interlocutors from commercial backgrounds facing a dilemma between combining purchases and spreading them.
The former can lead to significant cost savings but periods of more intense effort for employees, while the latter has the opposite advantages and drawbacks.
A possible generalization of this is the competition and interplay between financial and human factors, both impacting the capacity planning process in different ways.

\fin{typical:commercial:lease-support-important}
We also see some interlocutors with commercial cloud backgrounds describing lease and support contracts as being especially important in the set of factors taken into account in a typical capacity planning process.
The timing and duration of these conditions can have significant impact on the decision taken.

\subsubsection*{Scientific Infrastructures}
We now summarize the main findings for the typical process scientific infrastructures.

\fin{typical:scientific:benchmark}
Most scientific clouds seem to follow a typical public competitive dialogue process for their resource selection.
In at least half of the surveyed scientific infrastructures, this includes a benchmark-driven process.
This entails providing hardware contractors with a set of benchmark applications that need to be optimized by the contractor.
Results for applications in this benchmark are often weighted by importance or frequency of deployment.

\fin{typical:scientific:users}
We observe that most scientific clouds take their biggest users as the main indicator of the needs of a platform.
Half of scientific clouds also take into account a broader user feedback survey, reaching users regardless of size or importance.

\fin{typical:scientific:perception}
Almost all scientific cloud interlocutors perceive their process as fundamentally different from the commercial capacity planning process.
The main perceived difference is the mode of operation, which they believe to be budget-centered rather than demand-centered.
These interlocutors also consider their budget-centered to have less capacity planning efforts than commercial efforts.
Whether this assessment is accurate is difficult to objectively judge, although analysis from other questions seems to indicate that there are more aspects of ``traditional'' capacity planning in their process than commonly perceived.

\subsection*{1 - Q6: Yes/No Questions} \label{sec:interview-results:q6}

\fin{yes-no:what-if}
What-if scenarios do not seem to be established practice currently.
Some unstructured examples fitting the format (e.g. fail-over scenarios, clients arriving) are mentioned informally, but not in a structured form.

\fin{yes-no:hybrid}
Most of the scientific clouds have exploratory projects running where they investigate the possibility of offloading demands to a public cloud.
This indicates increased interest in hybrid infrastructure offerings.

\fin{yes-no:share-data}
A significant portion of interlocutors is willing to share historical data with the interviewer.
This could signal interest in academia and industry for more research being conducted in topics surrounding their capacity planning decision making.

\fin{yes-no:humans}
The majority of interlocutors considers human personnel in some form in the capacity planning process.
This topic is further analysed in Q22.

\subsection*{2 - Q7: Stakeholders} \label{sec:interview-results:q7}

\fin{stakeholders:common}
All processes surveyed seem to have an executive board at the head of the process.
This board seek out advice from experts in the domains relevant to their decision.
These can include technical advisors or scientific experts.

\fin{stakeholders:decision-level}
For all surveyed instances, the final decision seems to be at board level.
While the input of domain experts is sought out, the final decision is made by the management.

\fin{stakeholders:commercial}
The interlocutors with commercial background also have an engineer in charge of the capacity planning process, monitoring the current situation and coordinating the lifecycle-based planning process.
The process in this case also includes input from the hardware contract administration for contracts.
Occasionally, the sales department is involved, if the decision affects the shaping or pricing of services provided to customers.

\fin{stakeholders:scientific}
In scientific environments, an important part of the set of stakeholders tend to be scientific partners and (governmental) funding agencies.
Half of the surveyed processes here also take into account user input through a user questionnaire.

\subsection*{2 - Q8: Time and Infrastructure Scale} \label{sec:interview-results:q8}

\fin{scale:time}
Unlike the frequency or trigger of capacity planning processes, the time scale of the decisions made seem to be roughly uniform across interlocutors.
We observe that the aging and thus deterioration of hardware is seen as the most important factor here, with a mean of 5 years until the next decision for a specific machine/rack.
Commercial environments seem to tend towards faster replacement (3-5 years), while scientific environments seem to replace less quickly (4-7 years).

\fin{scale:infrastructure:commercial}
The infrastructure scale of decisions for commercial environments tends to be a single rack.
Making multi-rack decisions is desired, due to potential cost savings, but not always possible.

\fin{scale:infrastructure:scientific}
In scientific environments, the infrastructure scale of decisions seems to be larger, with most surveyed infrastructures working on scales of entire cluster/site iterations.
One infrastructure works on a smaller scale, making single-machine or single-rack decisions.

\subsection*{2 - Q9: Factors} \label{sec:interview-results:q9}

\fin{factors:plurality}
The number of factors is remarkable, with more than 25 distinct factors being named in the full set of interviews.

\fin{factors:majority}
Nevertheless, the factors that span across a majority of interviews are few.
Only three factors are named in more than half of the interviews: the use of historical monitoring data, financial concerns (such as budget size), and the lifetime of hardware.
These are followed up by a set of four factors mentioned in slightly less than half of the interviews: user demand, new technologies, incoming projects, and the benchmark performance of different solutions.

\fin{factors:exclusive-to-scientific}
We observe a number of factors particular to scientific infrastructures but not being mentioned in the commercial set.
The most important here are the benchmark performance of solutions (which is often required by public competitive acquisition processes) and user demands.
One surveyed infrastructure optimizes for throughput here, meaning the number of times the benchmark can be run in a certain time frame.

\fin{factors:exclusive-to-commercial}
Similarly, we observe a number of factors unique to commercial infrastructures.
The most prominent are lease contracts, current offerings that the provider has, and personnel capacities.

\subsection*{2 - Q10: Margin for Error} \label{sec:interview-results:q10}

\fin{margin}
The margin for error is difficult to objectively measure, due to the multi-faceted nature of this process.
Two main consequences of errors are mentioned by interlocutors.
First, financial losses can occur due to overestimation of the demand of a certain resource, such as specific accelerators or storage capacity, or due to underestimation, as can happen if the ratio of resource types is mispredicted.
Second, personnel can come under pressure, due to available capacity being smaller than expected, starting a search for spare capacity in any of the managed clusters.

\subsection*{2 - Q11: Frequency and Time to Deployment} \label{sec:interview-results:q11}

\fin{frequency:frequency}
While interlocutors from commercial backgrounds report a frequency of at least once per three months (depending on an ad-hoc component), counterparts from scientific infrastructures generally report a frequency upwards of four years.
There is one notable exception to this rule, with one of the scientific clouds which takes a decision twice a year.
In general, we observe a separation between fast-paced commercial planning cycles and longer cycles in scientific clouds.

\fin{frequency:time-to-deployment}
Similar to the frequency of planning events, the time from start of the event to deployment is determined largely by the background of the infrastructure.
Commercial clouds tend to finish deployment within 4--5 months, while scientific clouds tend to take 1--1.5 years to deploy.
We see a positive correlation with the frequency of planning instances, meaning that a higher frequency trends to be paired with a shorter time to deployment.

\fin{frequency:special-architectures}
In some scientific clusters, we see a part of the topology containing specialized hardware, such as accelerators, getting a special process with more rapid cycles than the rest of the architecture.
This could be due to the faster pace of evolution that these kinds of technologies experience.

\subsection*{2 - Q12: Data Sources} \label{sec:interview-results:q12}

\fin{data-sources:historic}
With the exception of one infrastructure, historic utilization data from monitoring agents is universally reported to be used in the process.

\fin{data-sources:others}
Next to historic utilization data, we see operational data such as lease contracts and maintenance periods being involved in the process.
We also observe some interlocutors explicitly mention taking global market developments into account.

\subsection*{2 - Q13: Tooling} \label{sec:interview-results:q13}

\fin{tools:none}
The main observation here is that none of the surveyed infrastructures have dedicated tooling for the capacity planning of their infrastructures.
They use monitoring tools (with dashboards) and/or spreadsheets, combined with human interpretation of the results.
Decisions are, in one infrastructure, being preserved in minutes and mails.

\fin{tools:format}
We observe that planners typically consume the data they receive from monitoring in visual formats, in plots over time.
Being able to visually investigate and interpret developments plays an important role here.

\fin{tools:monitoring-products}
The most commonly used tool for monitoring seems to be Grafana, which allows teams to build custom dashboards with the information they see as relevant.
NetApp monitoring tools are mentioned as being used by one commercial party.
One scientific infrastructure reports basing their results on custom SQL queries of monitoring data.
Another scientific infrastructure uses spreadsheets as the primary medium for analysis.

\fin{tools:issues}
We identify two key issues being raised explaining the absence of dedicated tooling.
First, tools tend to be too platform specific or work only in one layer of the hierarchy and thus return misleading results.
We see the issue being the mismatch between the complexity of the reality on the ground and the complexity that these tools assume of the topology.
Second, tools tend to have high cost and carry a number of additional features that planners reportedly do not find useful, meaning that the high price is not justified by the value the planners receive out of these tools.

\subsection*{2 - Q14: Errors} \label{sec:interview-results:q14}

\fin{errors:frequency}
We observe several occurrences of failures being mentioned, although the perceived frequency varies.
One interlocutor believes that (slightly) erroneous plans are made constantly, since it is not possible to predict accurately what will be needed in the future, while another interlocutor claims the errors made are not very frequent.
On average, the frequency is perceived as low, drawing contrast to the failures being mentioned in the rest of the interview.

\fin{errors:severity}
The severity of an error is hard to measure objectively if not actively monitored.
The (subjective) descriptions of how severe errors vary from losing potential income, to having underutilized hardware, to hitting storage limits.
This raises a different point, surrounding the definition of errors or failures in the field of capacity planning.
An underutilized new cluster may be seen as a minor error, since service is typically not affected and the only cost seems to be additional power usage and environmental footprint.

\fin{errors:reporting}
We did not observe any structured approach to recording errors in the process.
Whether they only remain tacit team knowledge or are still recorded somewhere is not clear, although our interpretation indicates the former.

\fin{errors:timing}
While most interlocutors seem to describe negative capacity planning incidents as being infrequent and having low severity, the examples being given in response to other questions tend to be from the most recent (if not one of the last) iterations.
This is partly explainable with more recent memories being more readily accessible, but also might indicate a more structural underappreciation of the possibility for failures or suboptimal choices in the process.

\subsection*{3 - Q15: Services and Infrastructure Part of Process} \label{sec:interview-results:q15}

\fin{services-in-process:meta}
We observe that the majority of interlocutors considers only one type of service as part of their capacity planning process and a minority considers two or more.

\fin{services-in-process:commercial}
In the commercial settings we survey, we see that \glspl{VM} holding business-critical workloads are most universally considered as part of the process.
One interlocutor mentions a new container platform as also being part of the process, although it is internally approximated as a \gls{VM} while planning.

\fin{services-in-process:scientific}
In the scientific settings we survey, we see that batch workloads, \gls{HPC} workloads, \gls{VM} workloads, and baremetal hosting services are equally popular.
One provider also mentions shared IT services (more general IT functionality) as also being a part of the process.

\subsection*{3 - Q16: Processes for Specific Workloads} \label{sec:interview-results:q16}

\fin{workloads:business-critical}
The instances running business critical workloads report two special aspects that they consider for these workloads: special redundancy requirements and live management concerns (primarily migration and offloading).

\fin{workloads:big-data}
We do not observe any special processes being mentioned for Big Data workloads.

\fin{workloads:serverless}
The processes for serverless workloads are still very much in a stage of infancy, as most interlocutors having container or \gls{FaaS} solutions only host them as experimental pilot projects.
In terms of capacity planning, one interlocutor points out that for the container platform they currently build they only approximate the containers with \glspl{VM} in their reasoning.
However, they acknowledge that the density and characteristics of this new workload might be very different and that they may need to have special process for this in the future.

\fin{workloads:hpc}
Two of the interlocutors reporting that \gls{HPC} is a part of their process, state that capacity planning for \gls{HPC} workloads is even more challenging than for conventional workloads, due to the increased heterogeneity in the hardware platforms needed for this domain.

\subsection*{3 - Q17: Combining Workloads} \label{sec:interview-results:q17}

\fin{combining-workloads:as-one}
All interlocutors, with one exception, consider all workloads combined in one process.
The interlocutor forming the exception states that certain different workload types in their cloud are hosted on different infrastructure and separated entirely, with no synchronization occurring between the different efforts.

\fin{combining-workloads:benchmark}
A popular approach in scientific infrastructures seems to be to combine workloads through a weighted benchmark suite.
This scores topologies by running important representatives from each workload type and combining the scores into a single score.

\fin{combining-workloads:risk}
One interlocutor with commercial background points out that there is a trade-off between combining processes, thus gaining efficiency but also increasing the risk of failure, and keeping processes separate, thus loosing efficiency but also reducing the risk.

\subsection*{3 - Q18: Serverless Workloads} \label{sec:interview-results:q18}

\fin{serverless:popularity}
We observe that, with one exception, all interviews describe introduction of (pilot) serverless programs in their services.
One interlocutor sees serverless as a fast growing business, but another interlocutor contrasts this with an observation that the demand for it is still limited.

\fin{serverless:differences}
Interlocutors see a number of differences with traditional workloads.
They observe differing usage patterns with finer granularity of execution units.
This leads to higher fluctuation of the load and faster deployment patterns.

\fin{serverless:expectations}
Three of the interviews detail expectations on how their capacity planning will change with serverless workloads becoming more prevalent.
They expect impacts on the resource needs, such as the CPU to memory ratio and the allowed overcommission ratio.
One also states that guaranteeing workload isolation is likely to become significantly more difficult.

\fin{serverless:treatment}
Currently, none of the interlocutors state having a special subprocess for serverless in their capacity planning approach.
They agree, however, that this might need to change in the future, as serverless workloads increase in popularity.

\fin{serverless:lack-of-info}
We observe two key issues hindering the specialization of (parts of) the capacity planning process towards new workloads such as serverless.
First, not yet enough information is available on this new workload type and its behavior.
This makes reasoning about its capacity needs more difficult, at least with conventional capacity planning methods.
Second, interlocutors report a lack of personnel to dedicate to research into effective and efficient hosting of this new workload type.

\subsection*{3 - Q19: What-If Scenarios} \label{sec:interview-results:q19}
\fin{inside-what-ifs:scenarios}
This question was asked infrequently due to time constraints.
One interlocutor answered that scenarios they look at indirectly are customer-based scenarios (if a certain customer needs to be hosted) and new hardware releases and acquirements (with new specifications and properties).
See also \finref{ideal:what-if} for requested what-if scenarios in tooling.

\subsection*{4 - Q20: Regulatory Constraints} \label{sec:interview-results:q20}

\fin{regulatory:general}
We gather a number of laws and standards relevant to the capacity planning process.
We conclude that regulatory constraints can definitely play a role in capacity planning.

\fin{regulatory:banking}
Financial institutions tend to have strict standards for the capacity they acquire, such as a guaranteed amount of fail-over capacity.
This requires a capacity planning process (and recorded trail of that process) that meets these standards.

\fin{regulatory:gdpr}
We observe that privacy regulations such as the \gls{GDPR} are only of limited concern in the capacity planning process.
One interlocutor managing a scientific infrastructure states that \gls{GDPR} only affects the placement and planning of privacy-critical applications on their platform, in the form of preventing public cloud offloads for these specific applications.
Another interlocutor mentions the storage of logs could be affected by \gls{GDPR}, as its introduction leads to less log storage demands and thus less storage capacity being needed for that purpose.

\fin{regulatory:competitive}
In scientific infrastructures, we observe the competitive dialogue procedure playing a big role in shaping the process.
Publicly funded institutions need to shape their acquirement processes around a public tender with competitive dialogue, which limits how and which hardware components can be selected.

\fin{regulatory:exploits}
Security standards can also steer the choice of certain technologies in the capacity planning process.
We observe one case where reported exploits in a container platform limit a quick deployment of that technology.

\fin{regulatory:special-regulations}
Special regulations that hold in the country of origin of a certain hardware vendor can also play a role.
We hear one example of a supercomputer manufacturer prohibiting its hardware being used by personnel and users having the nationality of a certain set of countries set by the government of the manufacturer.

\subsection*{4 - Q21: Financial and Technical Aspects} \label{sec:interview-results:q21}

\fin{financial:overall}
Overall, there are three most frequently cited financial and technical aspects across interviews.
The first is funding concerns, looking at the source of funds for future expansions and maintenance.
The second consists of special hardware requests from users.
The third is the cost of new hardware, in line with global cost developments on the market.

\fin{financial:commercial}
For commercial infrastructures, there are a wide range of similarly frequent financial and technical factors that are taken into consideration.
Noteworthy are the timing and costs of lease contracts, which receive special attention, and historical sale and usage models for existing services.

\fin{financial:scientific}
For scientific infrastructures, the size of publicly funded grants is a factor that was mentioned in all interviews.
Special hardware requests are the second most frequent factor here, followed up by the total cost of ownership (including electricity and cooling costs) and new hardware developments.

\fin{financial:real-variables}
One interlocutor with commercial background made an observation that we believe resonates with statements by other interlocutors, as well: The variables in the equation that are most relevant are the factors around the hardware, not the cost of the hardware itself.
Support contracts, lease contracts, personnel cost -- all these factors play a significant role.

\fin{financial:choice-of-tech}
In two interviews, we observe the point being made that their choice of technology is inferior to the financial considerations they make.
This underlines the importance of financial aspects in the process.

\fin{financial:computational-investment-relation}
One interview raises an interesting relation, between financial investment and the energy consumed by resources that can be acquired with this investment.
It observes a trend in which constant investment can lead to increasing energy costs, due to the falling cost of computational resources when compared by energy usage.

\subsection*{4 - Q22: Human Factors} \label{sec:interview-results:q22}

\fin{human:overall}
Overall, we observe two human factors being mentioned most frequently: the need to account for personnel capacity and time to install and set up hardware and software, and the usage patterns that users exhibit when using the infrastructures (each of these is mentioned in 3 interviews).

\fin{human:focus}
Strongly present in the commercial sphere is an awareness of the need for training personnel, especially when switching technologies.
In scientific infrastructures, the focus seems to rest more on end-users.
Their usage demands and their abilities (and training) are most frequently raised as factors in this category.

\fin{human:listening-to-users}
Listening to users and their demands has its limits, however, as one interlocutor points out.
They state that if administrators ask users if they would like more computing power, the answer will likely often be ``yes''.

\fin{human:stress}
One interlocutor points out that improper capacity planning can lead to stress in the team, because it can lead to short term remedial actions becoming necessary, such as gathering left-over capacity from the entire resource pool.

\fin{human:proportionality}
One interlocutor observes that personnel does not grow proportionally to the size of the managed resource pool, but that specialization and having specialized staff for certain technologies is the deciding factor.
We see this sentiment being shared in many of the interviews.

\fin{human:very-involved}
We conclude that the human factor plays a significant role in the process, for most surveyed infrastructures.
5 out of the 7 interviews place special emphasis on this.
Hiring costs and personnel hours can add up especially, as one interlocutor points out.

\subsection*{4 - Q23: Multi-Datacenter or Local Level} \label{sec:interview-results:q23}

\fin{multi-dc:majority}
We observe that the majority of surveyed clouds takes decisions on a multi-datacenter level.
The interlocutors that report single-datacenter decision making cite differences in architecture and requirements between different sites as the main cause.

\fin{multi-dc:scope-of-single-decisions}
One interlocutor points out that while the scope of decision making is multi-datacenter, the scope of single decisions still focuses on single racks.

\subsection*{4 - Q24: Production vs. Development} \label{sec:interview-results:q24}

\fin{prod-vs-dev:overall}
3 out of the 7 interviews mention capacity planning differently for development and testing resources than for production resources.
For one of these interlocutors, this involves setting aside a dedicated set of testing nodes per cluster.
For another, this means splitting workloads into different datacenters.
For a third, this involves having lower redundancy on certain development machines.

\fin{prod-vs-dev:everything-prod}
We see a set of 2 other interviews claiming that every resource in their topology is considered production, even if it is in fact used as a development or testing machine.

\subsection*{4 - Q25: What-If Scenarios} \label{sec:interview-results:q25}

\fin{aspect-what-ifs:improvement}
This question was asked infrequently due to time constraints.
One interlocutor names scenarios where costs are conditioned against certain amounts of cooling, with different cooling types.
Another interlocutor points out that their process currently does not contain any what-if scenarios, but believes that they should in the future.
This would create a better understanding of possible future outcomes, using factors such as the timing of deployments or new, unknown workload patterns.

\subsection*{5 - Q26: Ideal Process} \label{sec:interview-results:q26}

\fin{ideal:flexibility}
An aspect shared by 5 out of the 7 interviews is the call for a more flexible, fast-paced process.
Planners across academia and industry believe the current process they follow is not always able to adequately keep up with the latest hardware trends, with special mention of accelerators.
Most of them also see another issue arising from this: procured hardware is often idle for a (relatively long) time before it is utilized.
An ideal process would address these two issues by adding hardware more flexibly, i.e. in smaller batches.
This is not straightforward, due to the economies of scale that sometimes only come into effect at larger batch sizes.

\fin{ideal:faster-planning}
One interlocutor mentions that their ideal process would also require less time than it currently does.
The years of analysis and discussions should be reduced to months.

\fin{ideal:smaller-scope}
We observe two interlocutors mentioning a preference for smaller-scale decision-making, with more attention to detail.
One interlocutor mentions this with respect to topologies, having per-rack decisions replace multi-datacenter decisions.
Another interlocutor mentions this with respect to the application domain, separating different domains into different sub-processes due to the difficulty of capacity planning large heterogeneous environment.

\fin{ideal:training}
One interlocutor expresses the desire for an increased focus in the process on training of users in order to exploit the full potential of new hardware once it arrives.

\subsubsection*{Tooling}
Especially from interlocutors with commercial backgrounds, we hear a wide variety of requests for better tooling for their activities.
We list them below, grouped into categories.

\fin{ideal:interactivity}
In 3 of the 7 interviews, we observe a demand for capacity planning tools, helping the practitioners rely less on their intuition.
A key request is interactivity, with answers wanted within a maximum of two weeks.
Getting immediate answers during a meeting would be even better.
One interlocutor describes this as an ``interactive dashboard''.
One interlocutor also states that having the tool answer questions at different levels of accuracy (increasing over time) would also beneficial, to facilitate quick estimates upfront and more detailed analysis over a longer period, e.g. between two meetings.

\fin{ideal:availability}
We observe the requirement that tools should be affordable.
Interlocutors state that high prices of an existing tooling platform that might help this activity are mainly due to the many other features in that platform being packaged along with the capacity planning functionality needed.

\fin{ideal:complexity}
Interlocutors also express the need for tools that help in addressing complexities in the process.
Being able to track the details of current capacity and being able to predict needed capacity would be a first step in this direction.

\fin{ideal:heterogeneity}
One interlocutor managing a scientific infrastructure points out that any tool should for this activity should support making heterogeneous decisions, which are more difficult to make but are yet still necessary, especially in the academic domain.
The request for heterogeneous capabilities is repeated by interlocutors from commercial backgrounds.

\fin{ideal:multi-disciplinary}
Interlocutors managing commercial infrastructure call for tools that are aware of multi-disciplinary aspects in the process.
This includes lifecycle processes (such as aging and maintenance) and lease contracts.

\fin{ideal:trend-analysis}
One interlocutor also expresses the wish for workload trend analysis capabilities in any tool for this activity.

\fin{ideal:what-if}
Two interlocutors list a number of what-if scenarios that they would like to explore with a capacity planning tool.
We list the questions underlying these scenarios here, in no particular order.

\begin{enumerate}
    \item Deciding how much more capacity is needed after certain decisions are taken, given projected CPU usage, memory commission, and overbooking ratios.
    \item Seeing the impact different new kinds of workloads have before they become common.
    \item Deciding when to buy new hardware.
    \item Modeling fail-over scenarios.
    \item Deciding whether special user requests can be granted before responding to users.
    \item Choosing the best lease duration.
    \item Deciding in which cluster to place new workloads.
    \item Choosing the best overlap duration between acquiring new hardware and decommissioning old hardware.
\end{enumerate}

\subsection*{5 - Q27: Other Processes Linked With Capacity Planning} \label{sec:interview-results:q27}

\fin{links:none-direct}
One interlocutor sees no direct link between capacity planning and other processes, although an indirect link is present.
Capacity planning, according to the interlocutor, tends to be on the end of the pipeline: only after acquiring new projects is the challenge of finding capacity for these projects considered.

\fin{links:scheduling}
One interlocutor sees a close relationship between the process and resource management strategies and research.
The live management of the infrastructure can have significant impact on the needed capacity, just as the capacity can have consequences for the management strategies that should be employed.
Migration and consolidation approaches need special attention here.

\subsection*{5 - Q28: Other Aspects Shared} \label{sec:interview-results:q28}

\fin{other:initial-plans}
One interlocutor states that, no matter what one plans, the future always looks (slightly) differently.
This does eliminate the need for planning, but underlines the need to be flexible and plan for unforeseen changes down the road.

\fin{other:speed}
One interlocutor (from a scientific background) points out the difference of speed in capacity planning processes between Europe and the United States.
They believe that infrastructures in the U.S. have a faster pace of capacity planning than comparable infrastructures in Europe.
They find this disadvantageous, due to new hardware improvements being slower to arrive.

\fin{other:coherence}
Two interlocutors observe that it is easier to obtain grants for a proposal of an infrastructure addressing one coherent need or project.
The surveyed scientific infrastructure tends to serve a far more heterogeneous set of use cases, which makes acquiring sufficient funds more difficult.

    \section{Software Implementation} \label{sec:software-implementation}
We depict in the figures below the program structure and dependencies of the OpenDC simulation model and Capelin extension in the form of class diagrams.
Furthermore, we highlight the interaction between components in the simulator using sequence diagrams.

\clearpage
\onecolumn
\begin{figure}[H]
  \includegraphics[trim={0.5cm 5cm 0.5cm 4cm},clip,width=\textwidth]{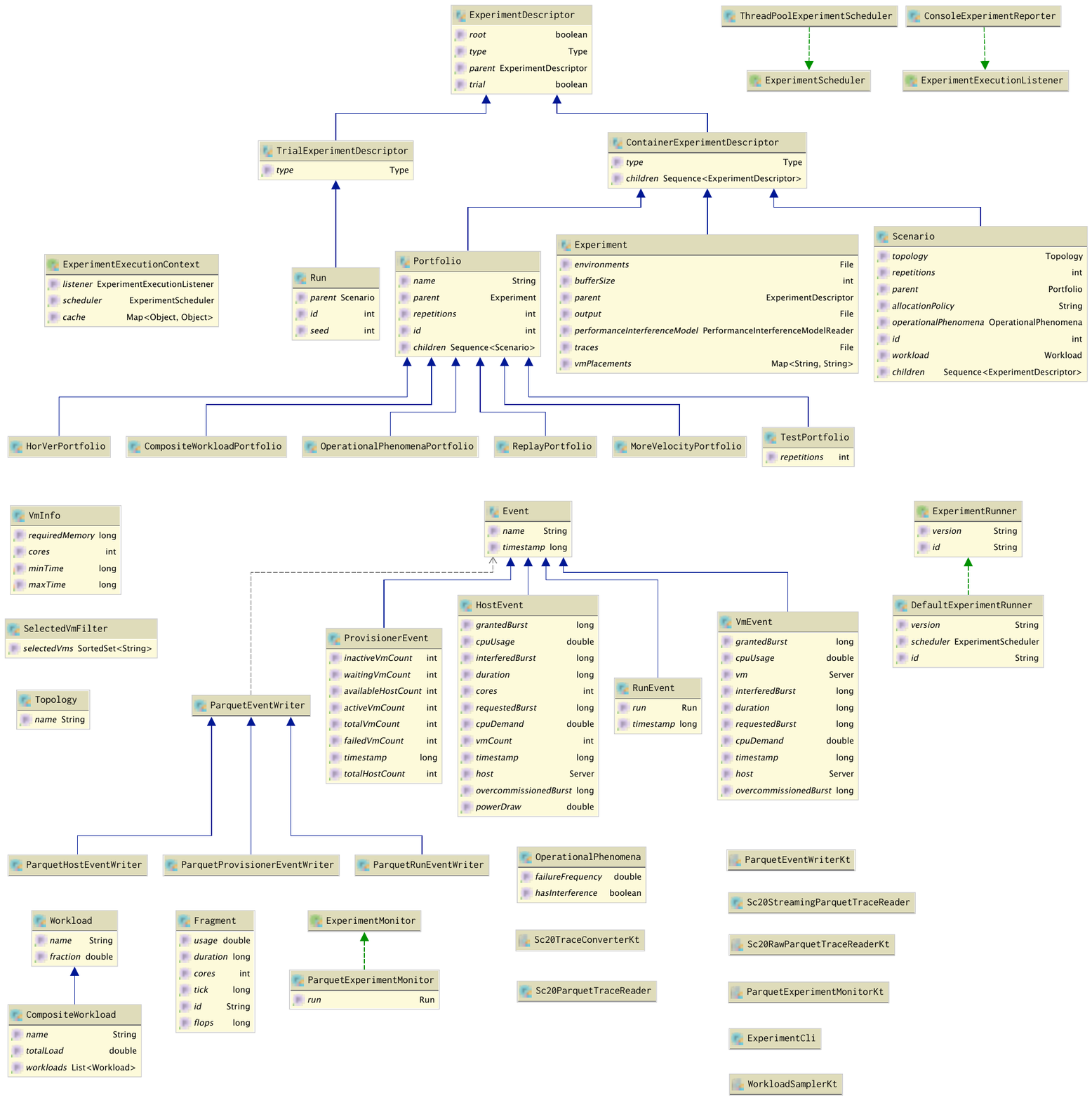}
  \caption{Class diagram for the Capelin extension in OpenDC.}
  \label{fig:class-diagram-sc20}
\end{figure}

\begin{figure}[H]
  \includegraphics[trim={0.5cm 3.4cm 0.5cm 2cm},clip,width=\textwidth]{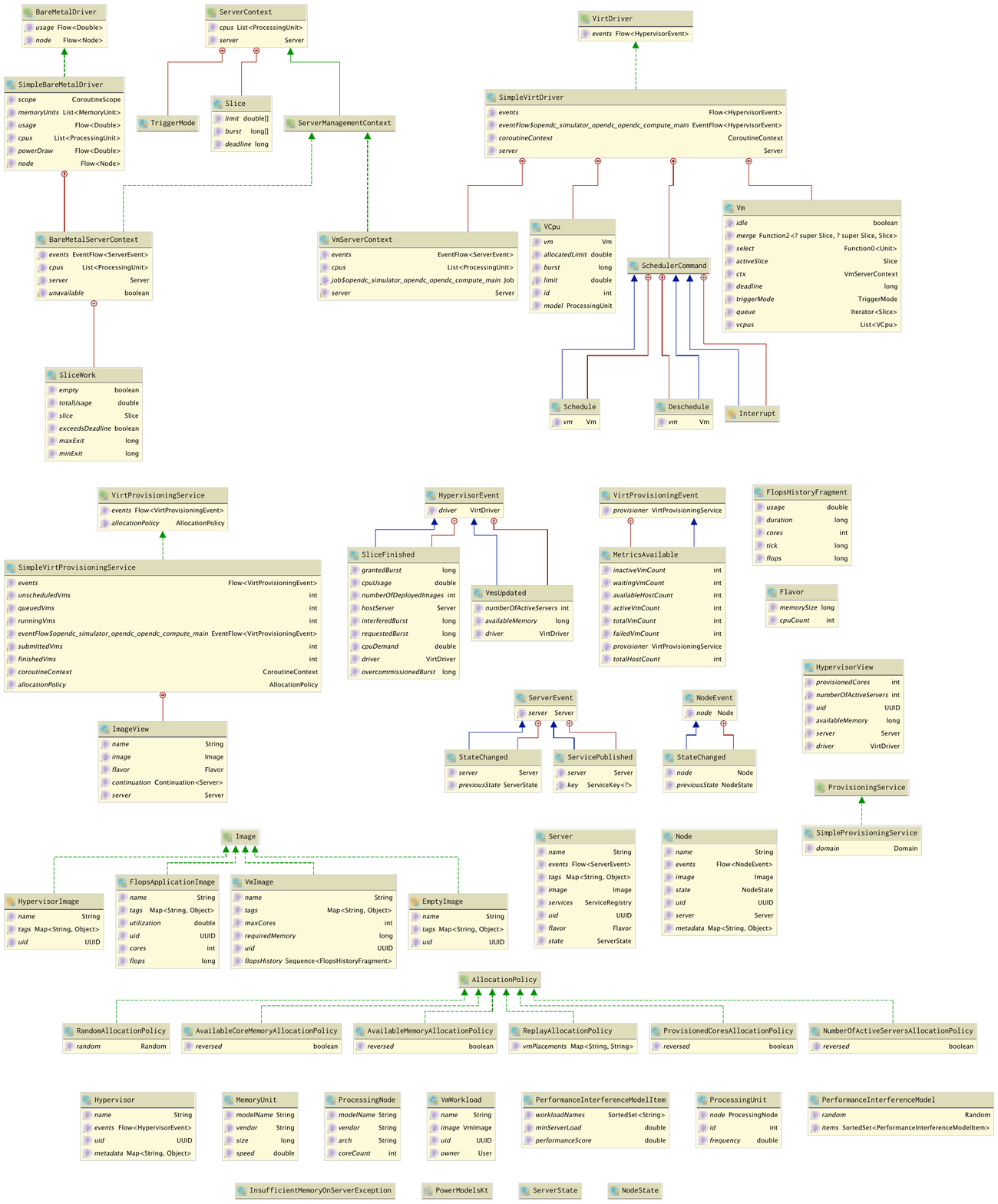}
  \caption{Class diagram for the OpenDC simulation model.}
  \label{fig:class-diagram-compute}
\end{figure}

\begin{figure}[H]
  \centering
  \includegraphics[width=0.9\textwidth]{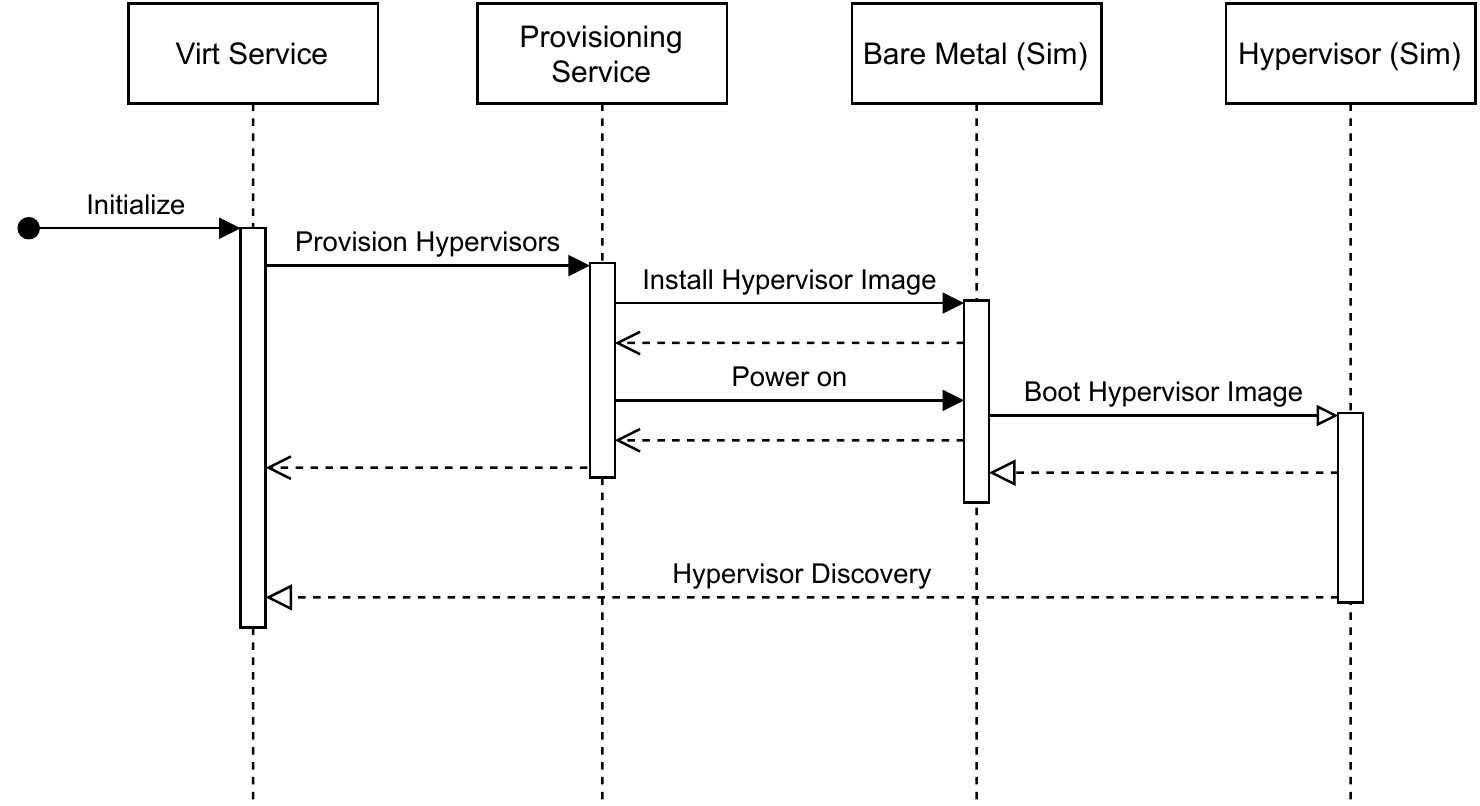}
  \caption{Interaction diagram for the initialization of the OpenDC simulation model.}
  \label{fig:interaction-diagram-init}
\end{figure}

\begin{figure}[H]
  \centering
  \includegraphics[width=0.9\textwidth]{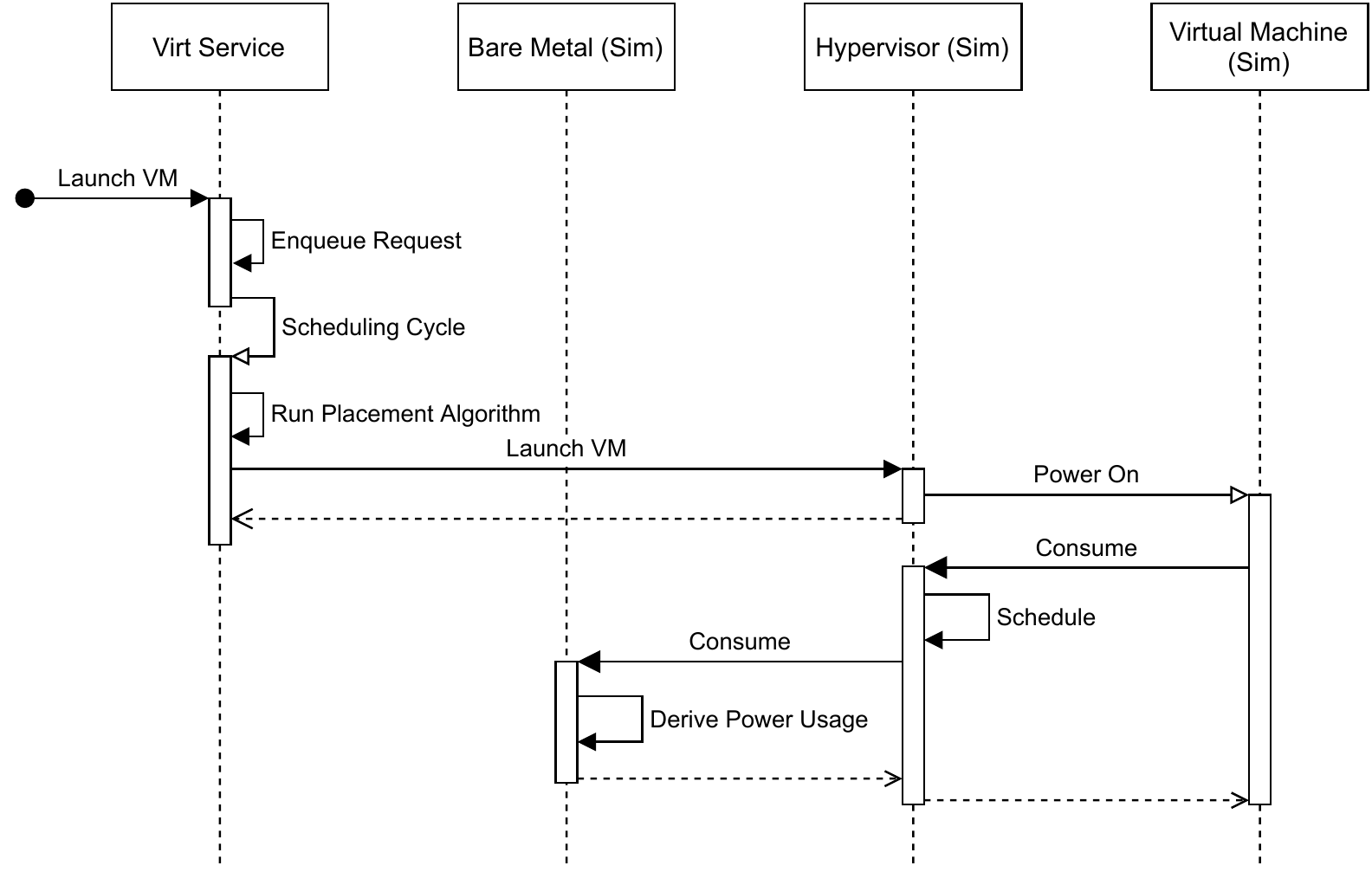}
  \caption{Interaction diagram for the launch a virtual machine in the OpenDC simulation model.}
  \label{fig:interaction-diagram-launch}
\end{figure}

\begin{figure}[H]
  \centering
  \includegraphics[width=0.9\textwidth]{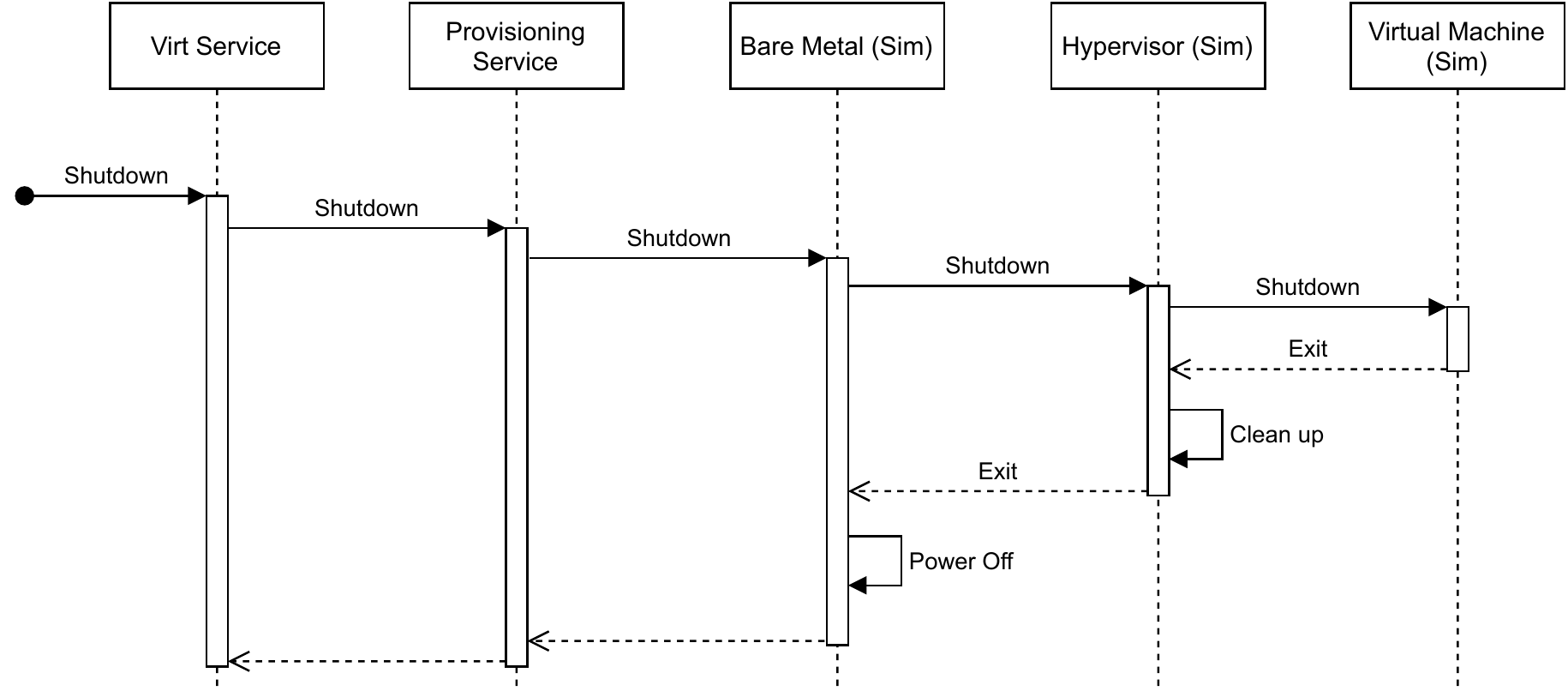}
  \caption{Interaction diagram for the finalization of the OpenDC simulation model.}
  \label{fig:interaction-diagram-exit}
\end{figure}

\clearpage
\twocolumn
    \section{Full Visual Results} \label{sec:full-visual-results}

In the figures below, we visualize the full set of metrics (\S\ref{sec:exp:setup:metrics}) for each experiment.
We differentiate between select overviews (depicting only the results of a subset of workloads) and summary overviews (aggregating over all workloads).

\begin{figure*}
    \subfloat[Total VMs Submitted\label{fig:full:hor-ver:vms-submitted}]{\includegraphics[width=0.5\linewidth]{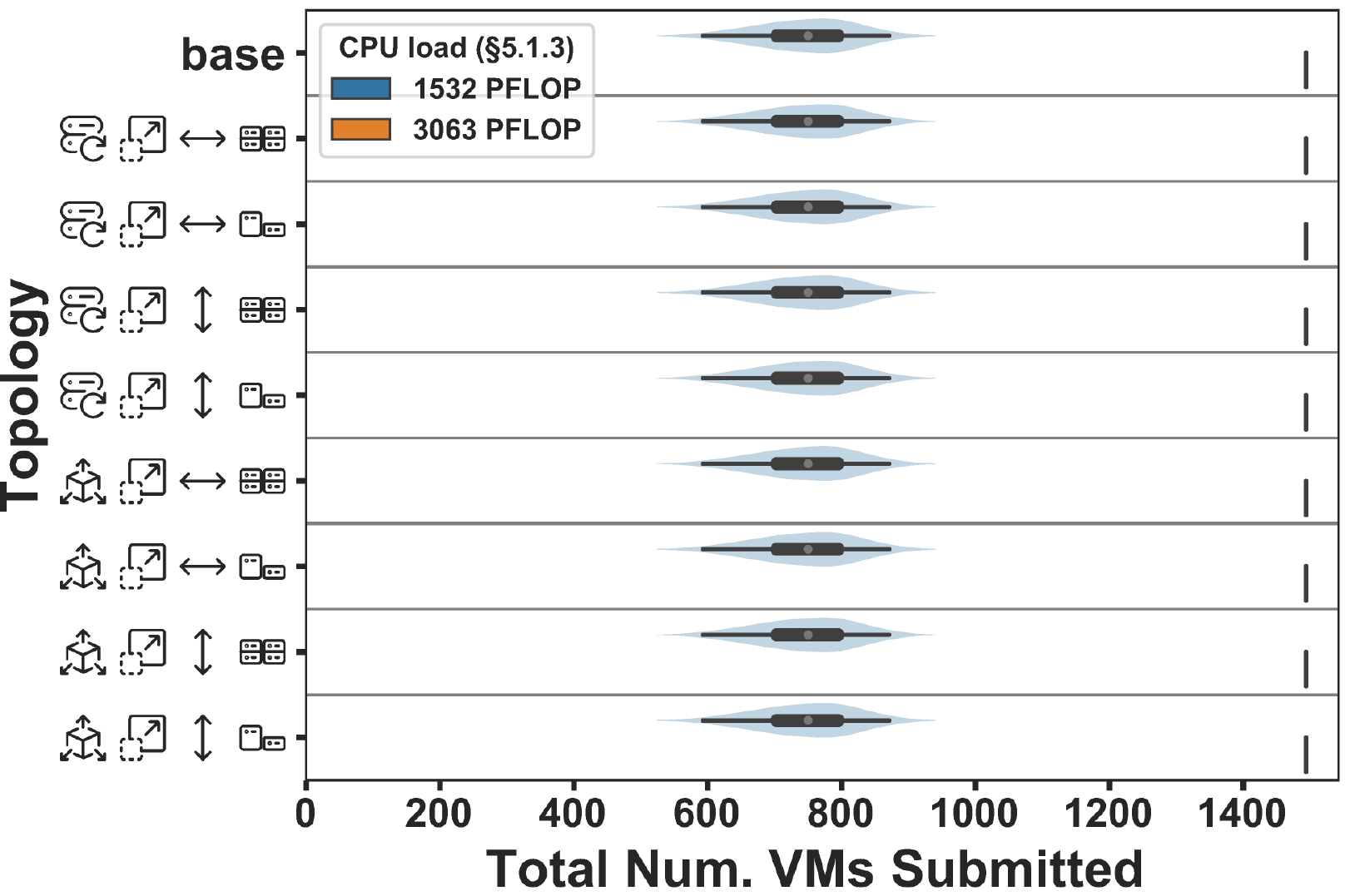}}%
    \subfloat[Total VMs Queued\label{fig:full:hor-ver:vms-queued}]{\includegraphics[width=0.5\linewidth]{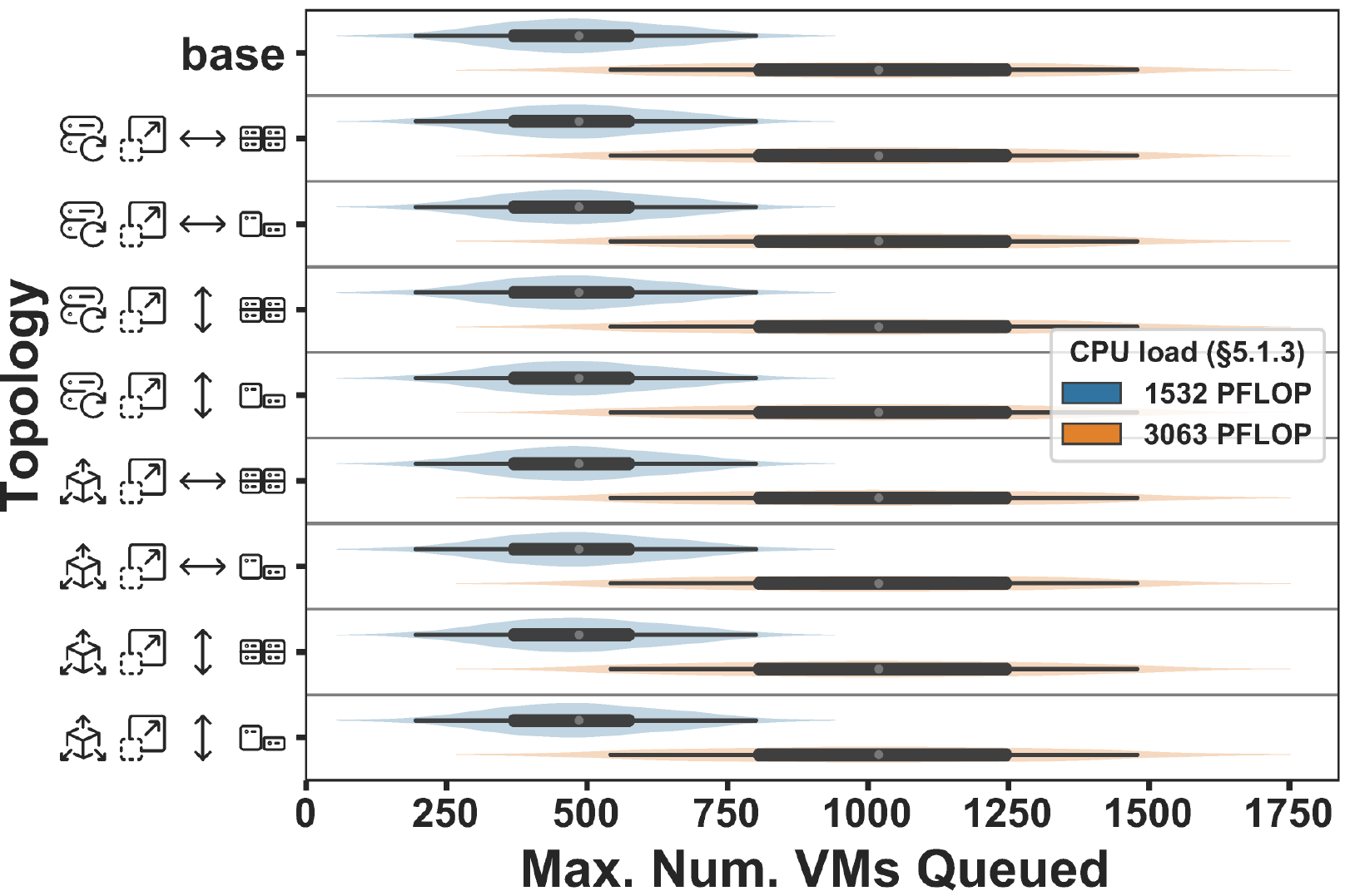}}\\
    \subfloat[Total VMs Finished\label{fig:full:hor-ver:vms-finished}]{\includegraphics[width=0.5\linewidth]{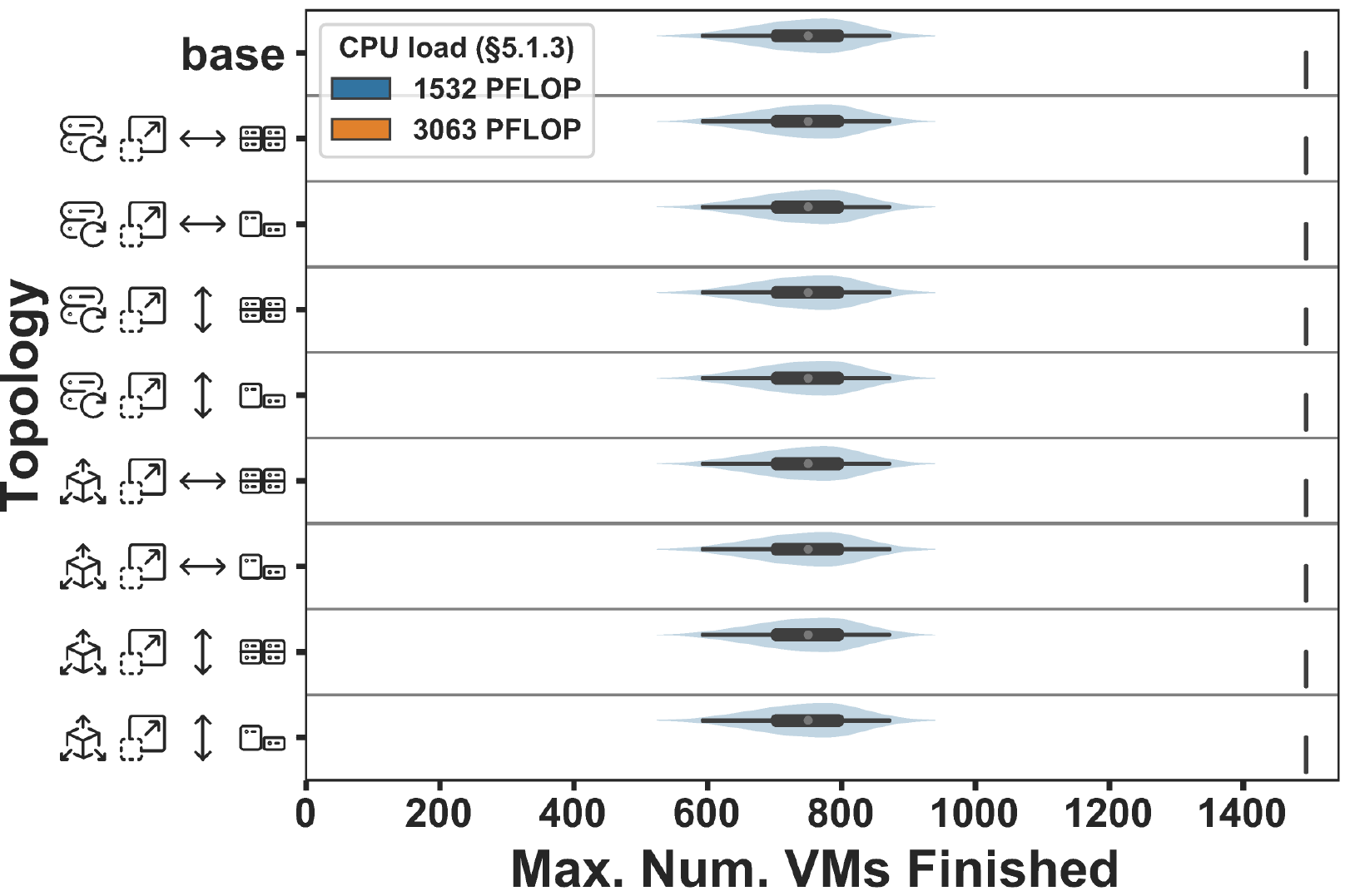}}%
    \subfloat[Total VMs Failed\label{fig:full:hor-ver:vms-failed}]{\includegraphics[width=0.5\linewidth]{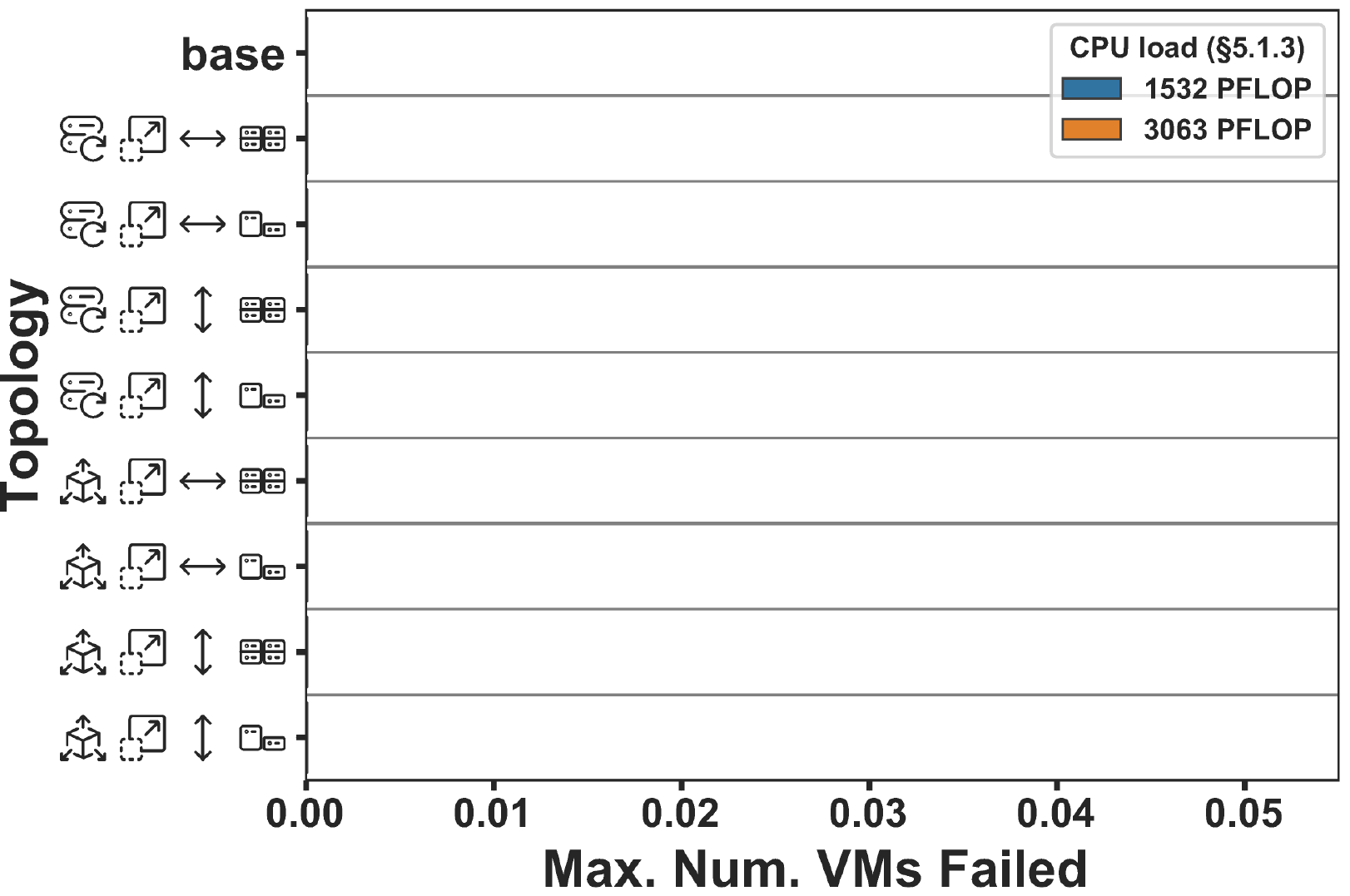}}%
    \caption{Performance of different horizontally and vertically expanded topologies, compared across workloads. For a legend of topologies, see Table~\ref{tab:experiment-overview}.}
    \label{fig:full:hor-ver:3}
\end{figure*}

\begin{figure*}
    \centering
    \subfloat[Requested CPU cycles\label{fig:full:hor-ver:summary:requested}]{\includegraphics[width=0.5\linewidth]{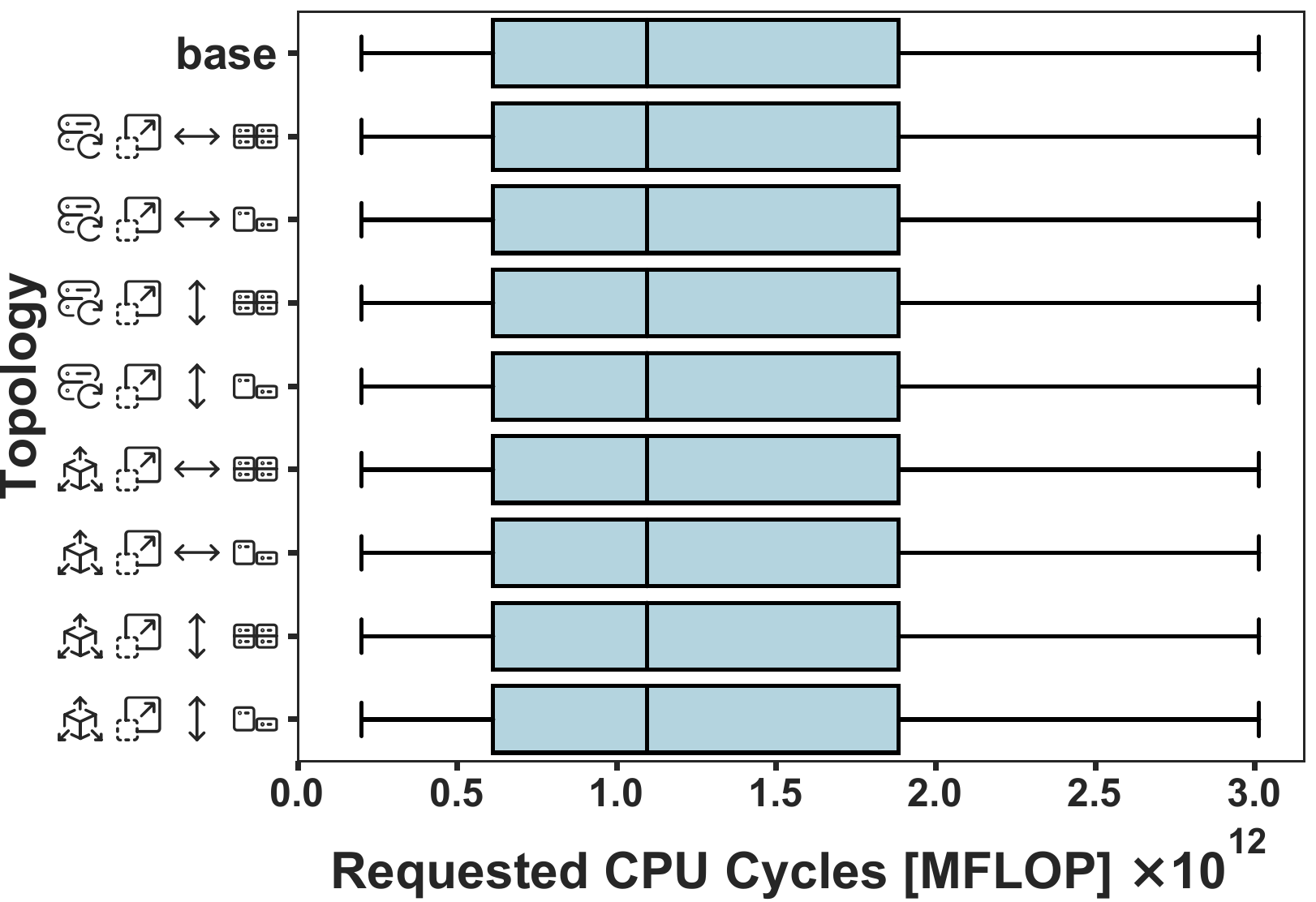}}%
    \subfloat[Granted CPU cycles\label{fig:full:hor-ver:summary:granted}]{\includegraphics[width=0.5\linewidth]{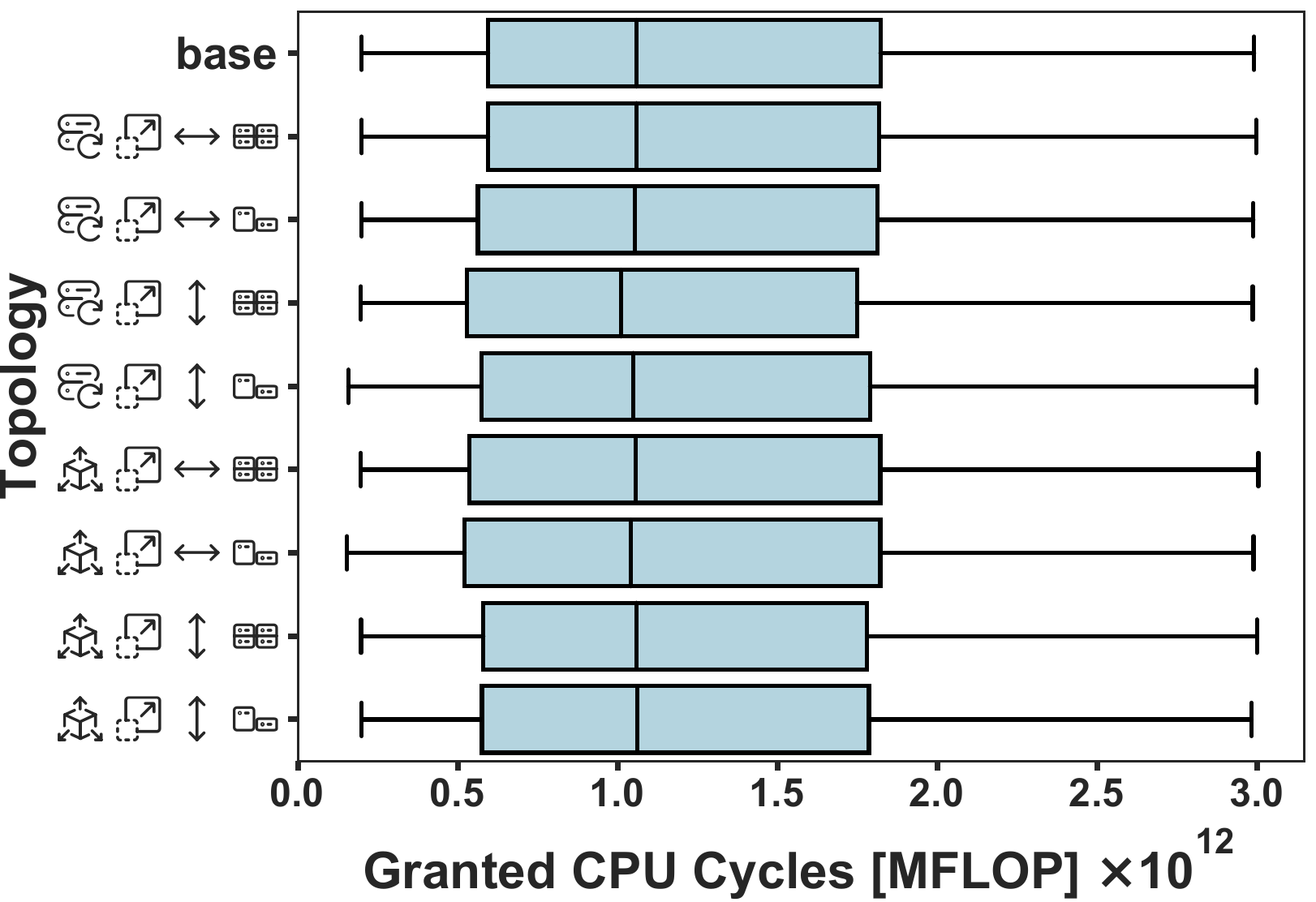}}\\
    \subfloat[Overcommitted CPU cycles\label{fig:full:hor-ver:summary:overcommitted}]{\includegraphics[width=0.5\linewidth]{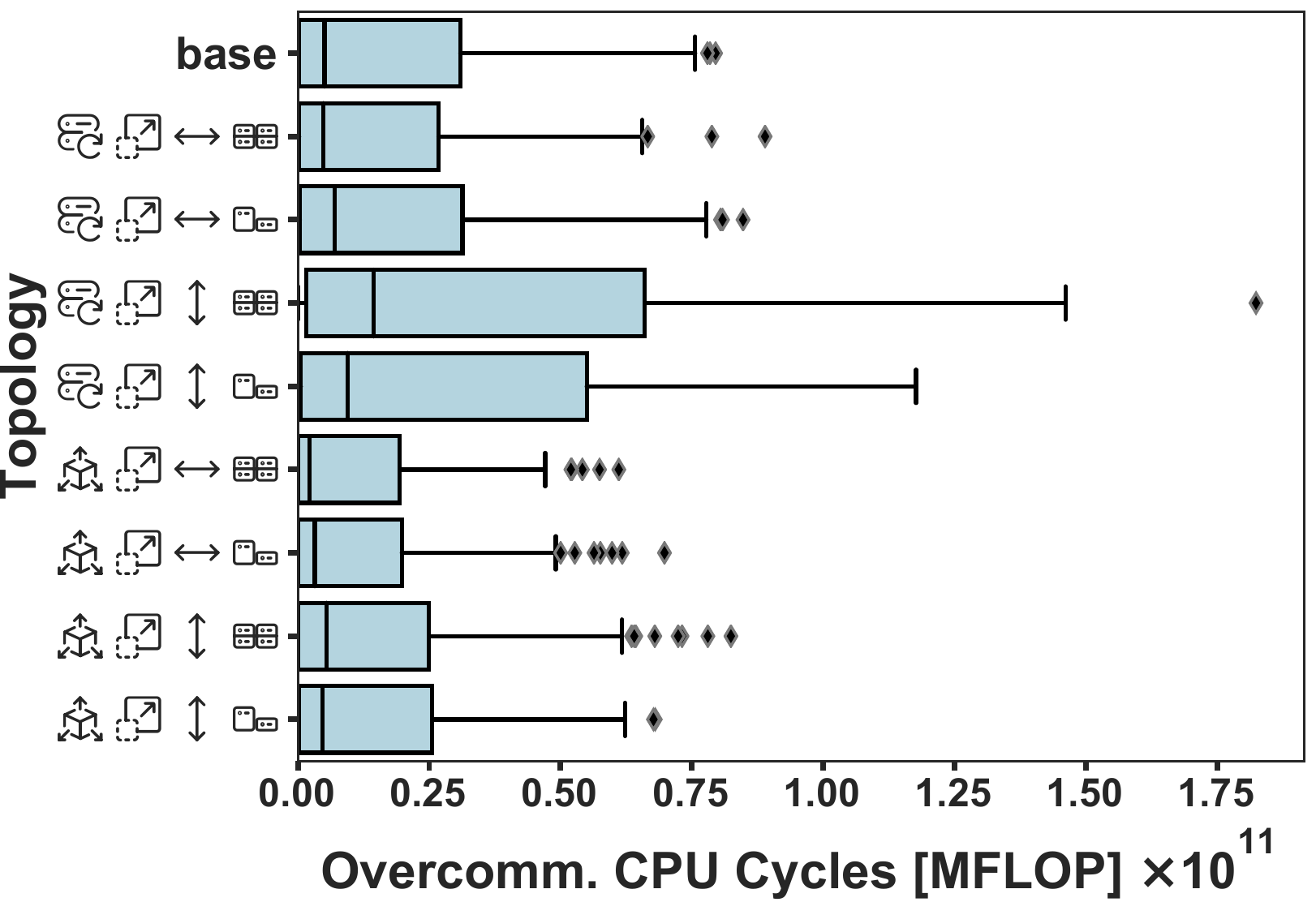}}%
    \subfloat[Interfered CPU cycles\label{fig:full:hor-ver:summary:interfered}]{\includegraphics[width=0.5\linewidth]{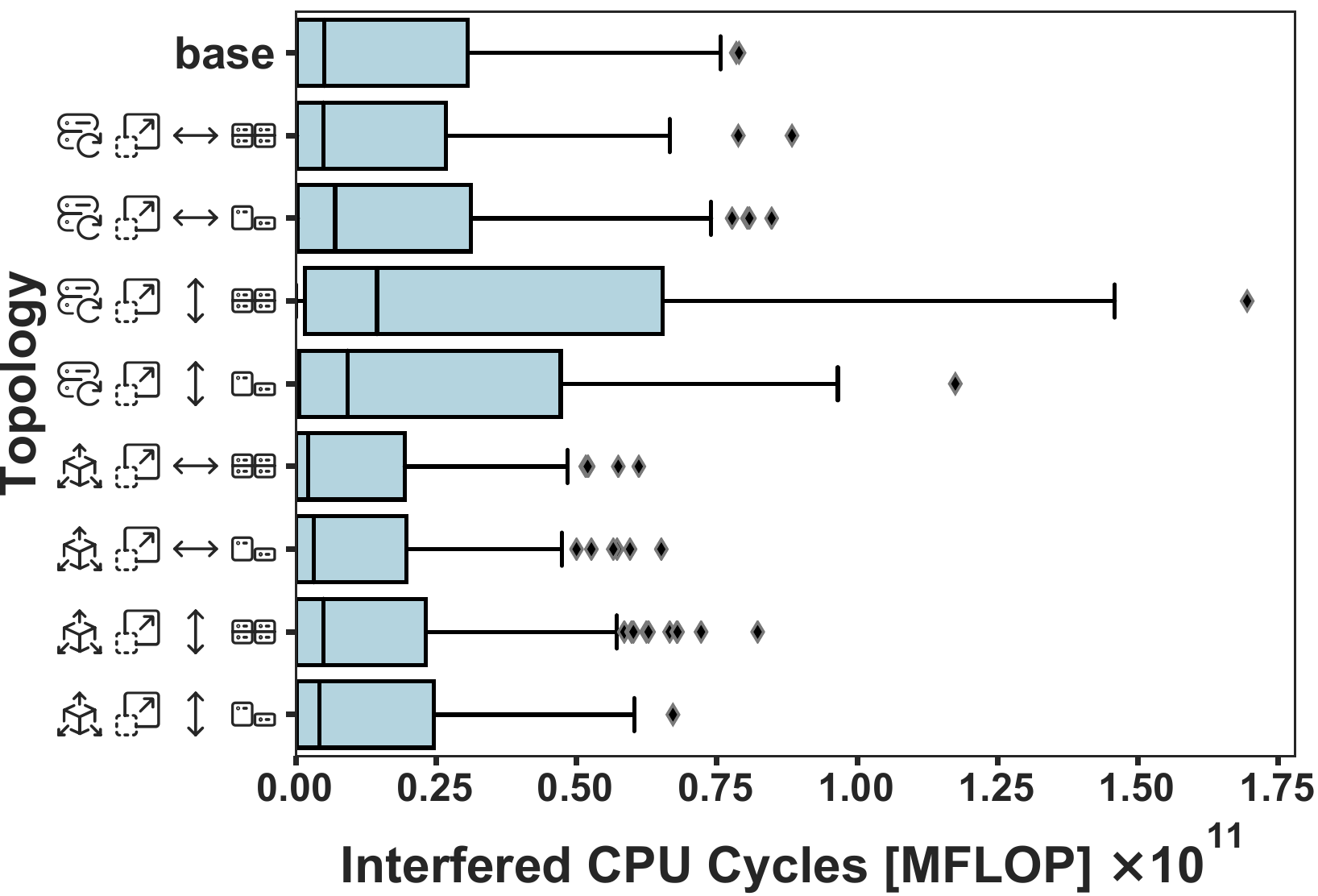}}\\
    \subfloat[Total power consumption\label{fig:full:hor-ver:summary:power}]{\includegraphics[width=0.5\linewidth]{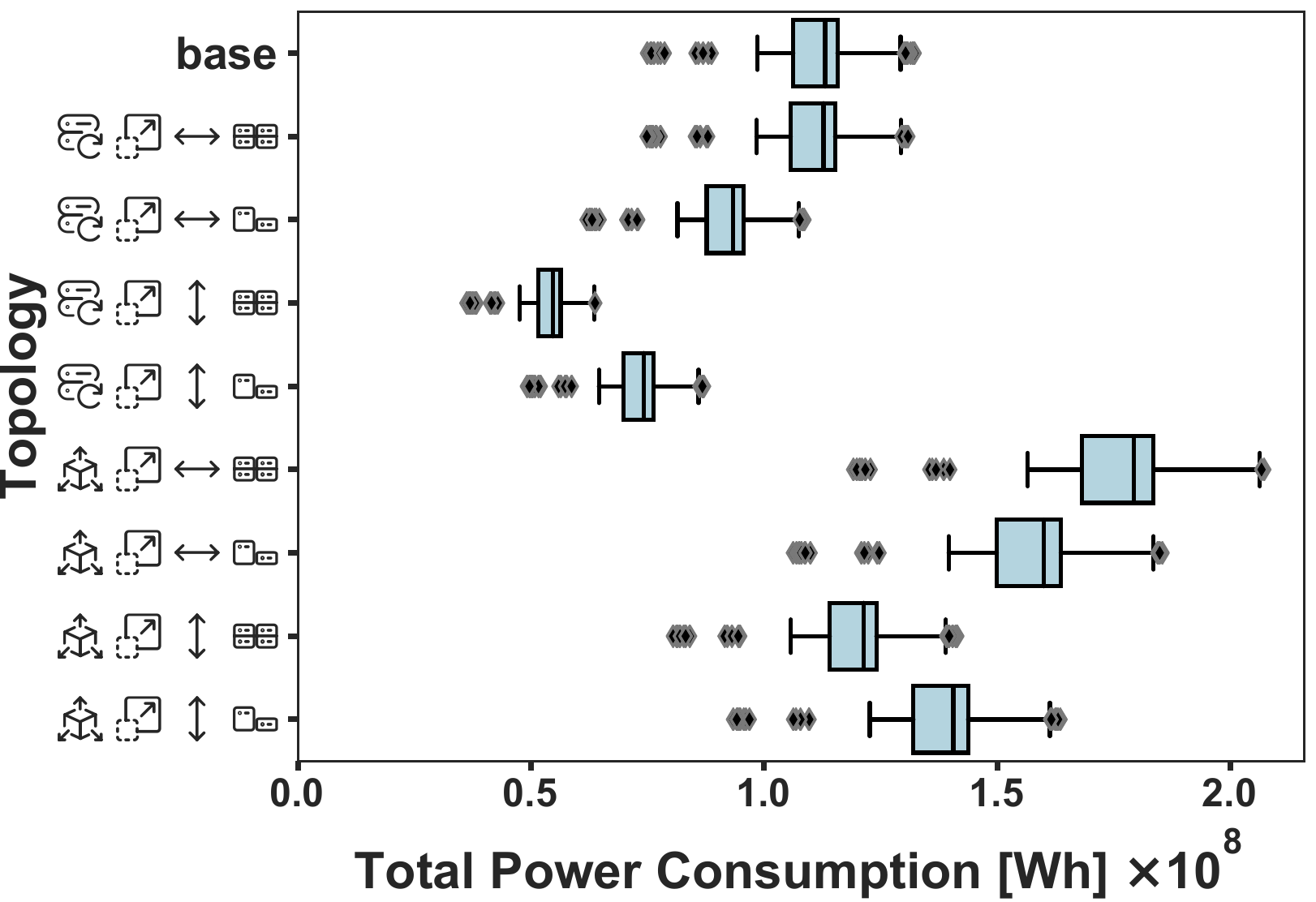}}%
    \subfloat[Total number of time slices in which a \gls{VM} is failed, aggregated across \glspl{VM}\label{fig:full:hor-ver:summary:failures:vms}]{\includegraphics[width=0.5\linewidth]{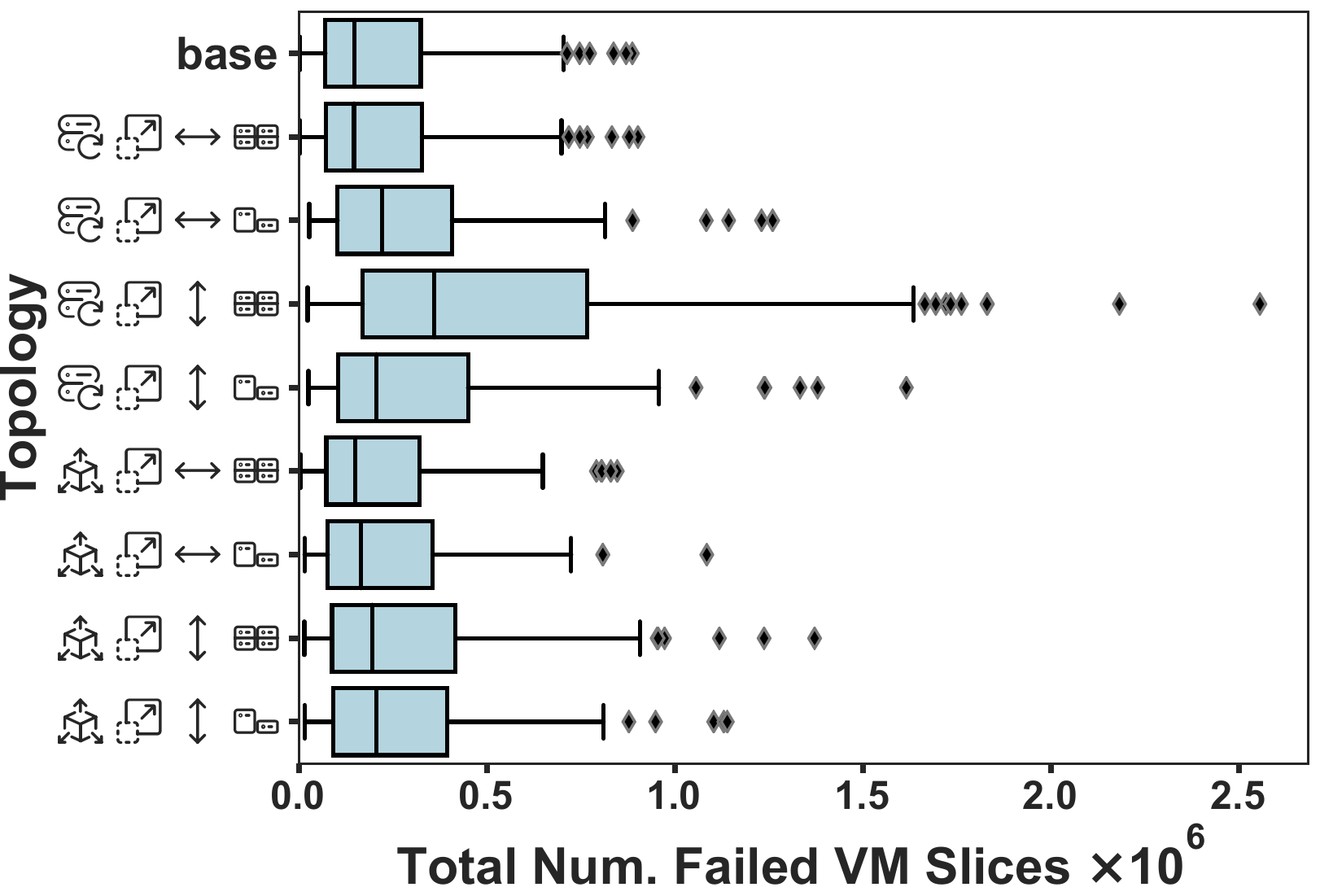}}\\
    \caption{Performance of different horizontally and vertically expanded topologies, compared across workloads. Results aggregated across the full set of workloads, including workloads not displayed in the more detailed figure. For a legend of topologies, see Table~\ref{tab:experiment-overview}. Continued in Figure~\ref{fig:full:hor-ver:summary:2}.}
    \label{fig:full:hor-ver:summary:1}
\end{figure*}

\begin{figure*}
    \subfloat[Mean CPU usage\label{fig:full:hor-ver:summary:cpu-usage}]{\includegraphics[width=0.5\linewidth]{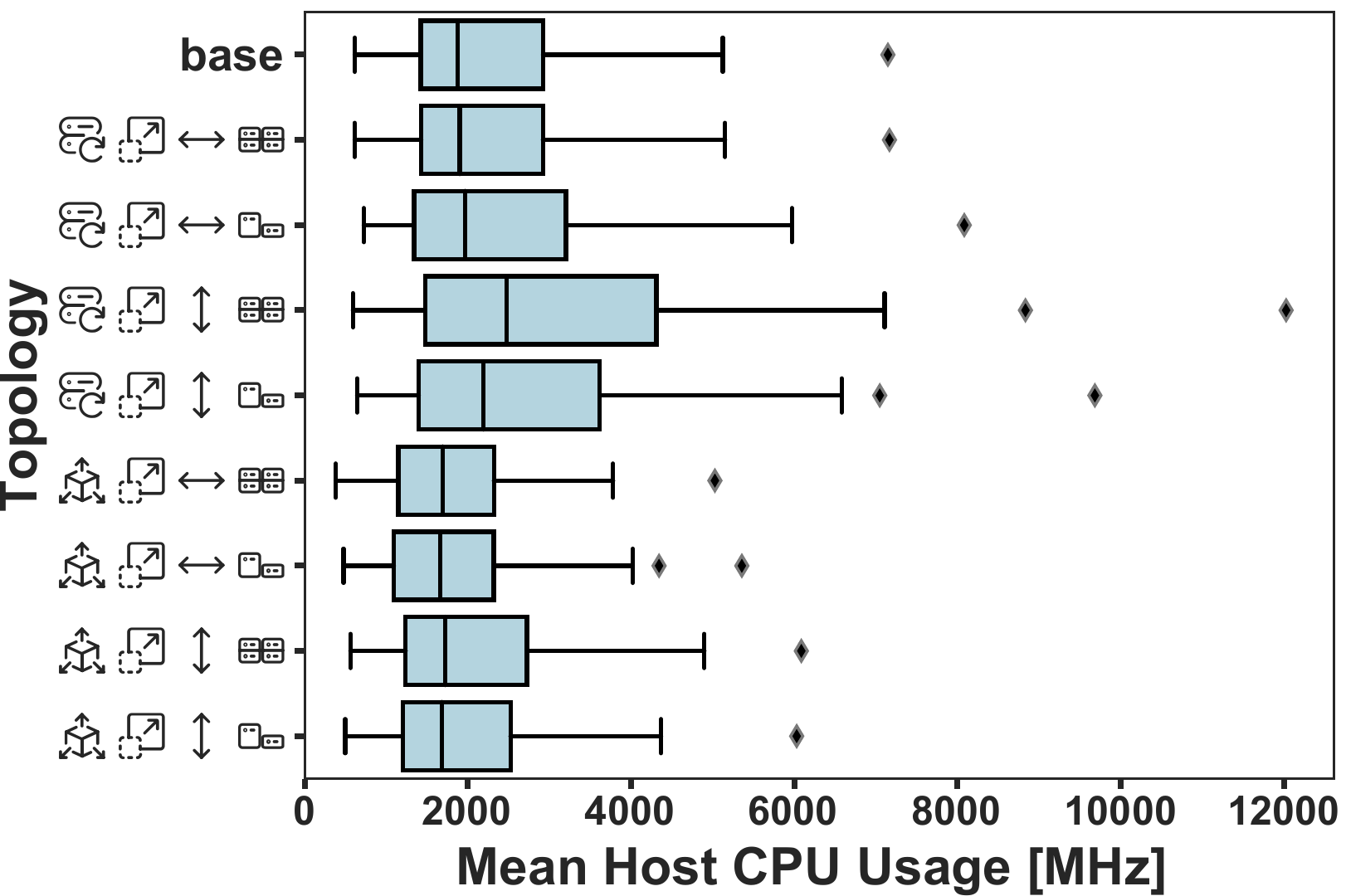}}%
    \subfloat[Mean CPU demand\label{fig:full:hor-ver:summary:cpu-demand}]{\includegraphics[width=0.5\linewidth]{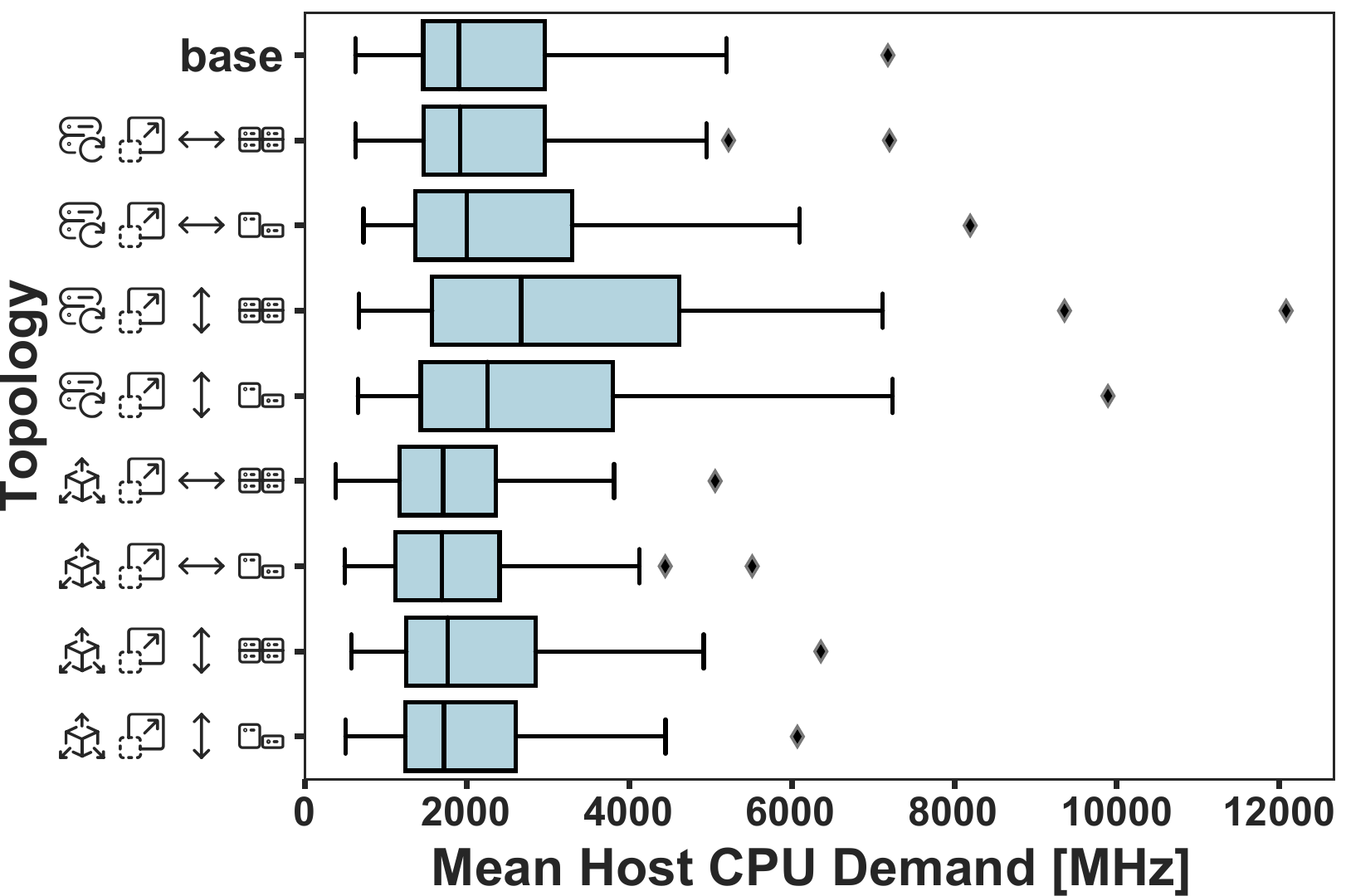}}\\
    \subfloat[Mean number of \glspl{VM} per host\label{fig:full:hor-ver:summary:mean-vm-count}]{\includegraphics[width=0.5\linewidth]{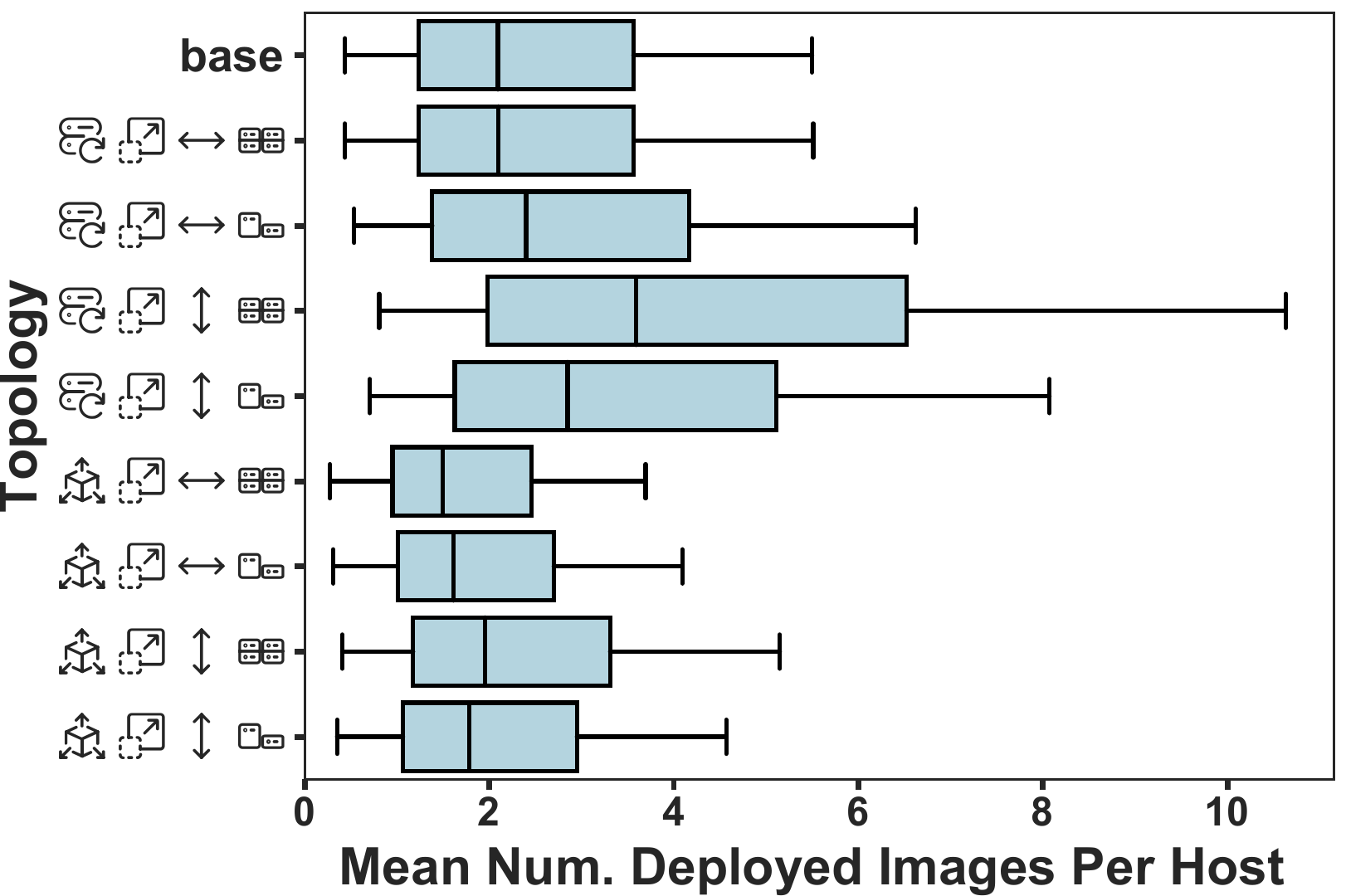}}%
    \subfloat[Max number of \glspl{VM} per host\label{fig:full:hor-ver:summary:max-vm-count}]{\includegraphics[width=0.5\linewidth]{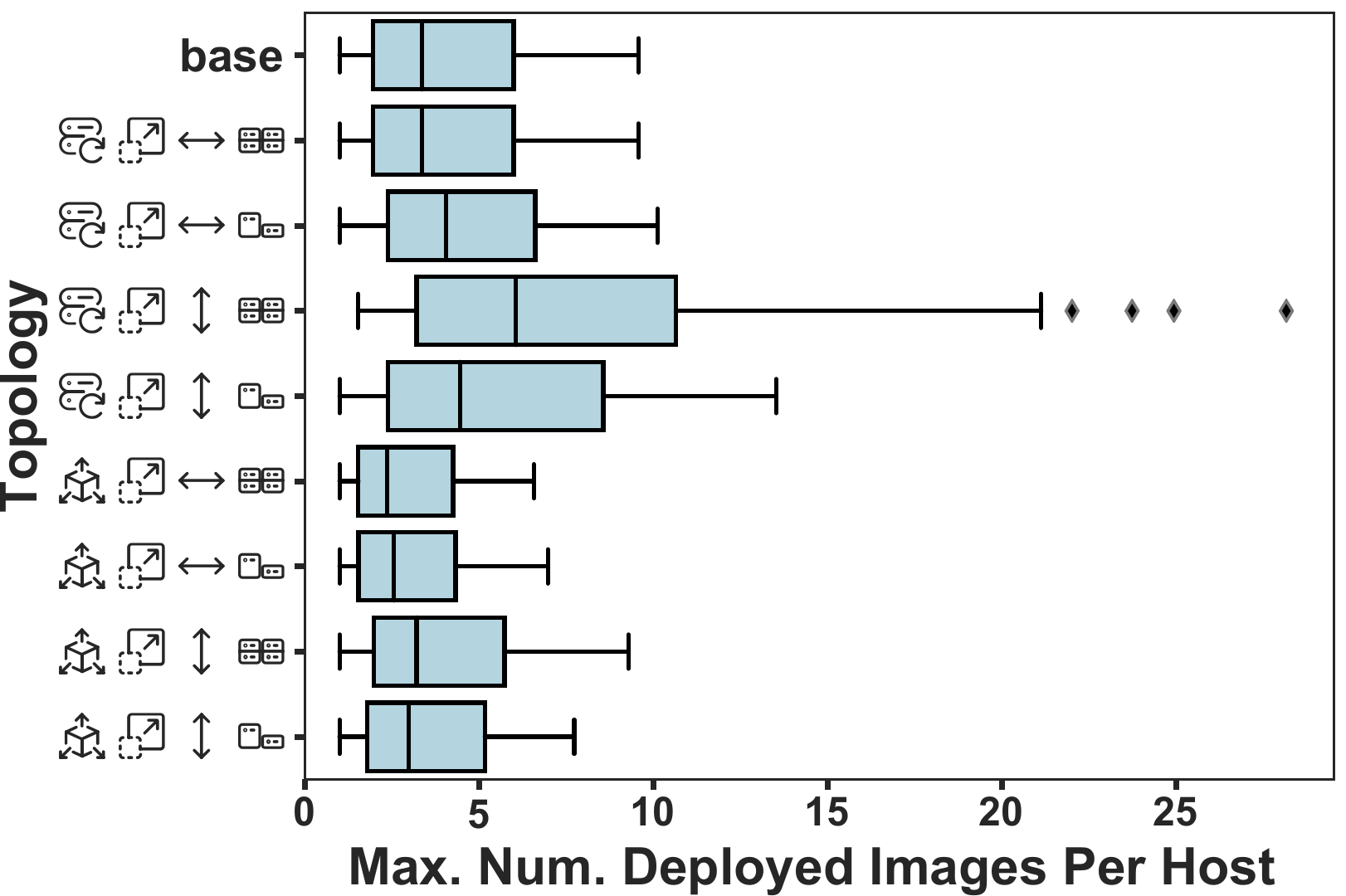}}\\
    \caption{Performance of different horizontally and vertically expanded topologies, compared across workloads. Results aggregated across the full set of workloads, including workloads not displayed in the more detailed figure. For a legend of topologies, see Table~\ref{tab:experiment-overview}. Continued in Figure~\ref{fig:full:hor-ver:summary:3}.}
    \label{fig:full:hor-ver:summary:2}
\end{figure*}

\begin{figure*}
    \subfloat[Total VMs Submitted\label{fig:full:hor-ver:summary:vms-submitted}]{\includegraphics[width=0.5\linewidth]{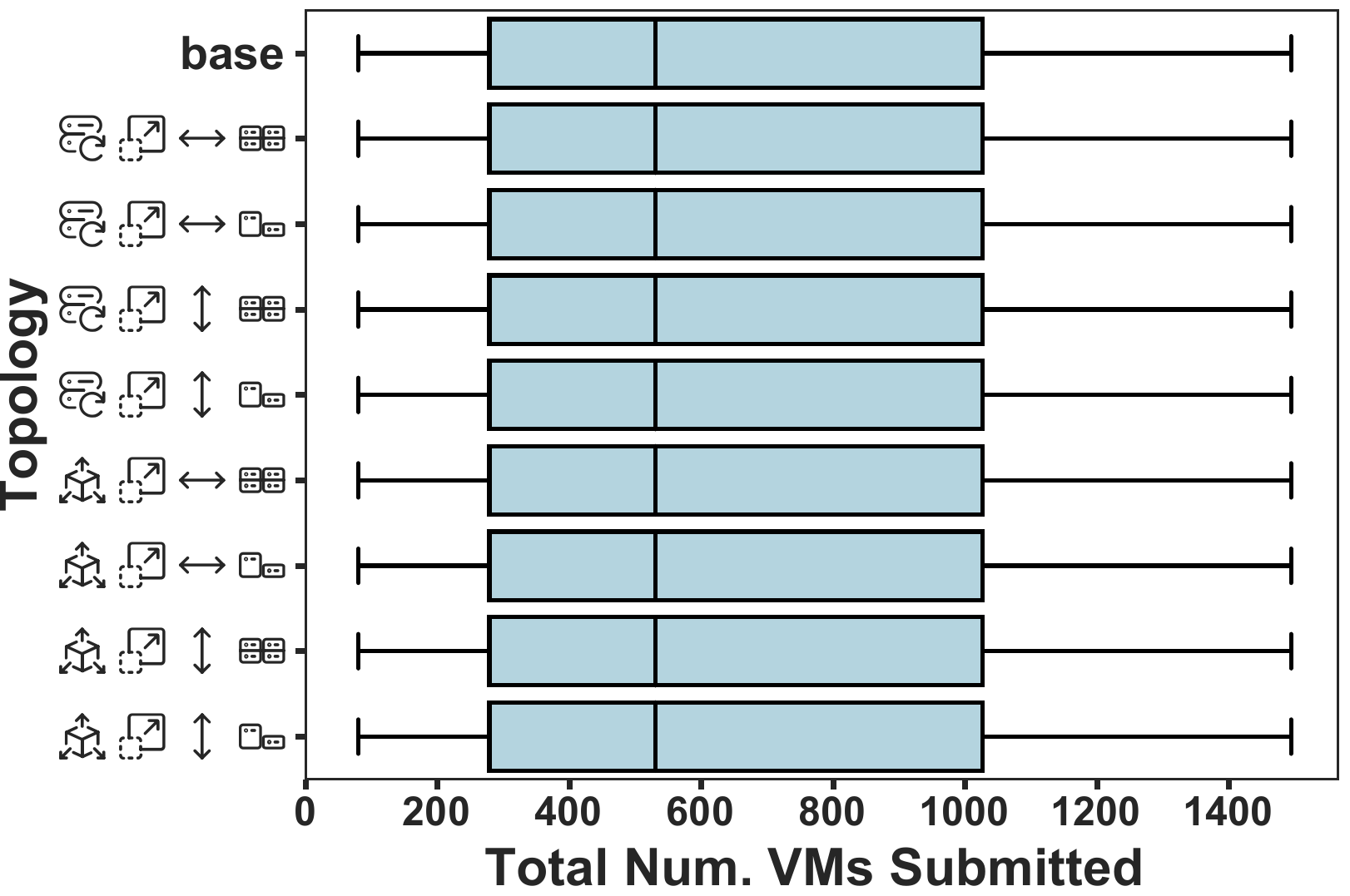}}%
    \subfloat[Total VMs Queued\label{fig:full:hor-ver:summary:vms-queued}]{\includegraphics[width=0.5\linewidth]{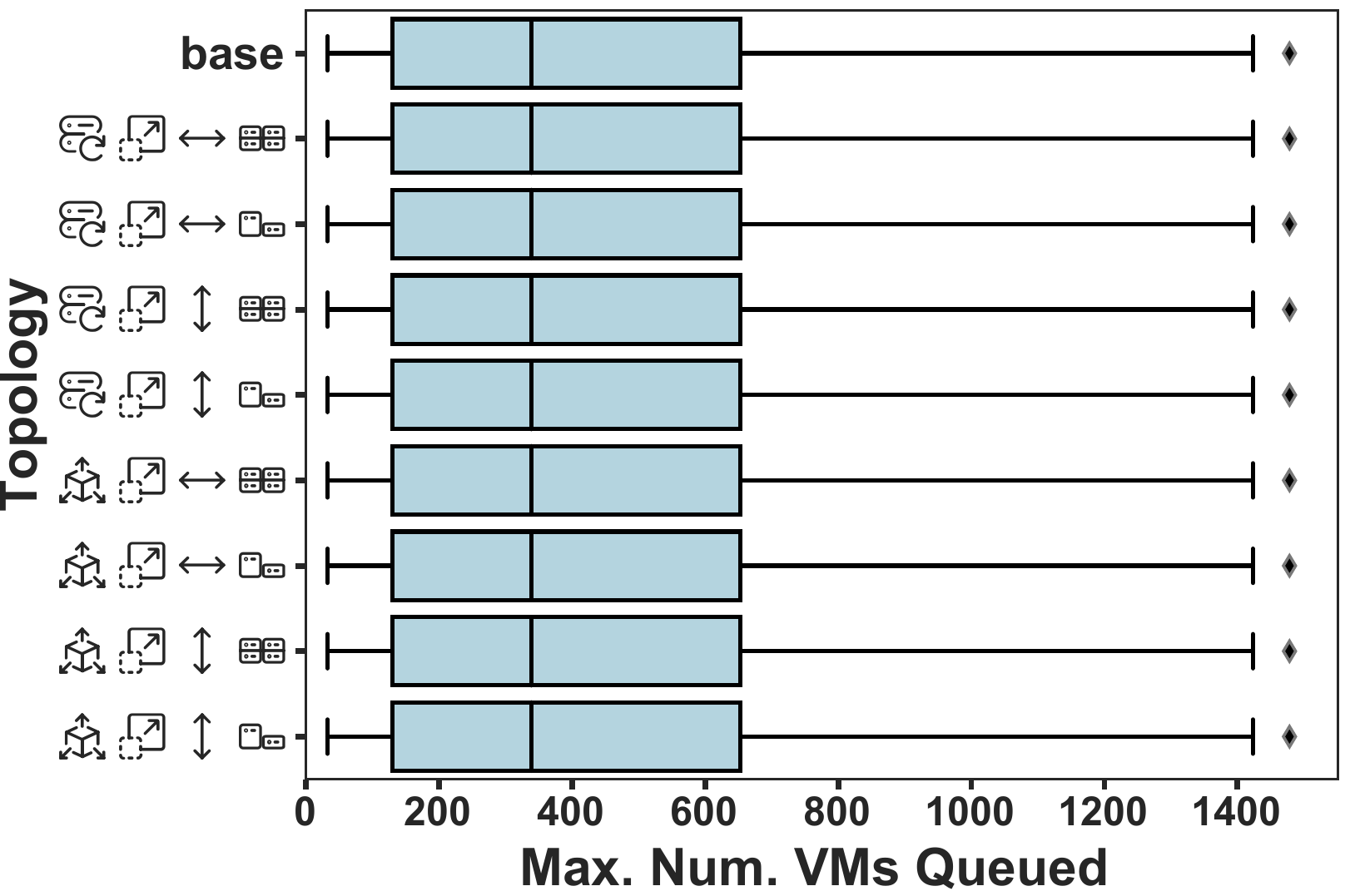}}\\
    \subfloat[Total VMs Finished\label{fig:full:hor-ver:summary:vms-finished}]{\includegraphics[width=0.5\linewidth]{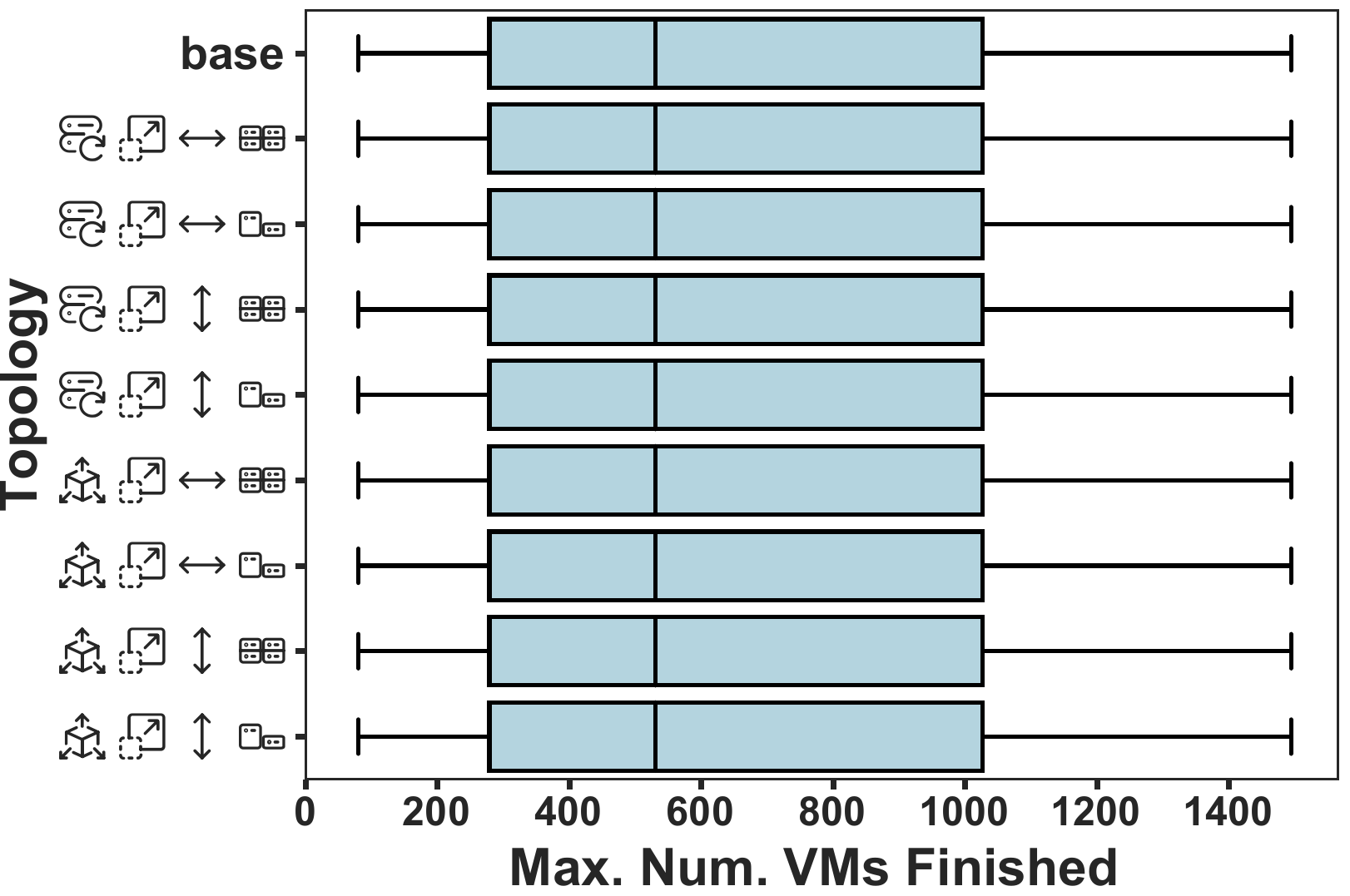}}%
    \subfloat[Total VMs Failed\label{fig:full:hor-ver:summary:vms-failed}]{\includegraphics[width=0.5\linewidth]{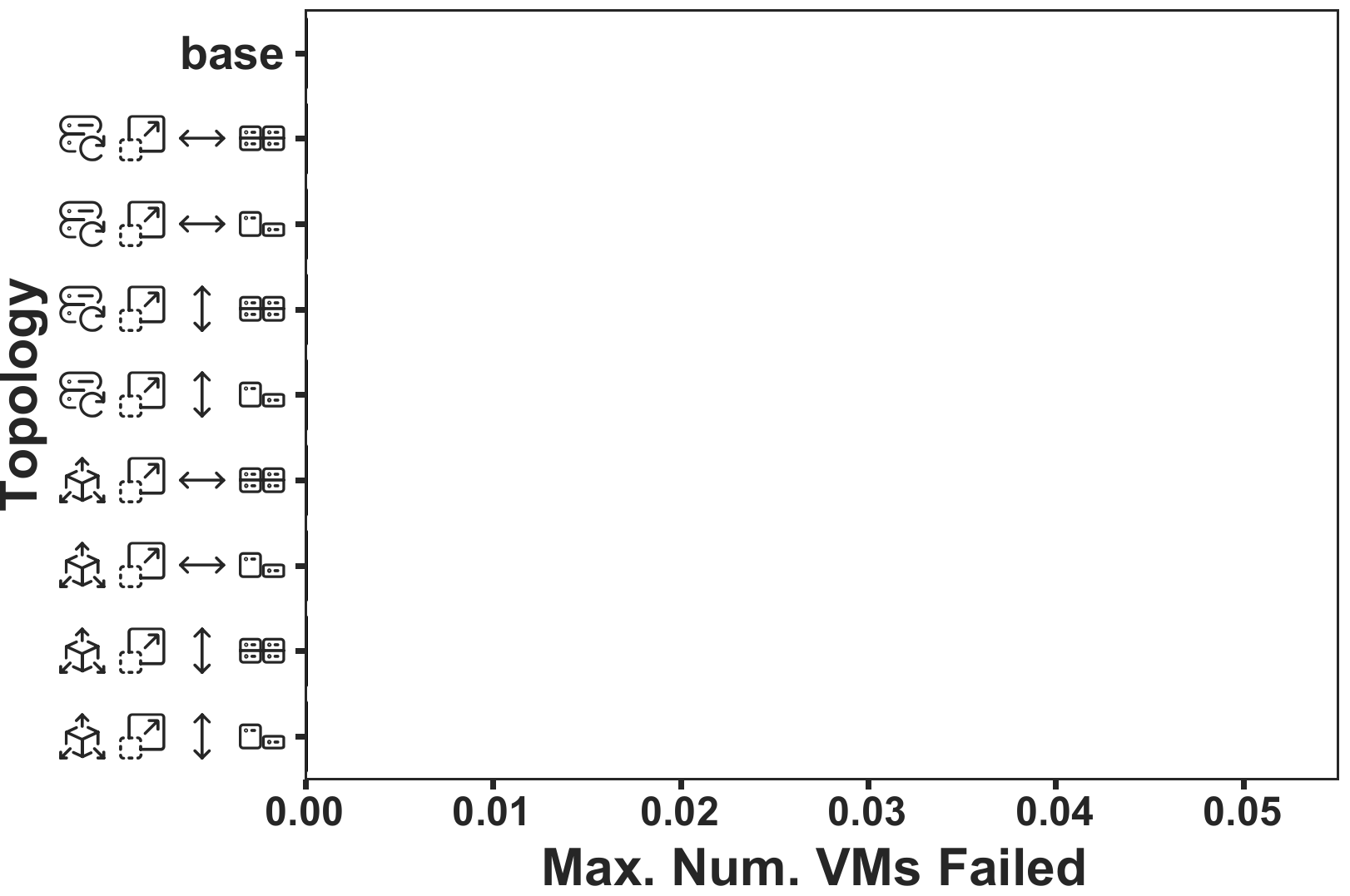}}%
    \caption{Performance of different horizontally and vertically expanded topologies, compared across workloads. Results aggregated across the full set of workloads, including workloads not displayed in the more detailed figure. For a legend of topologies, see Table~\ref{tab:experiment-overview}.}
    \label{fig:full:hor-ver:summary:3}
\end{figure*}

\begin{figure*}
    \centering
    \subfloat[Requested CPU cycles\label{fig:full:more-vel:requested}]{\includegraphics[width=0.5\linewidth]{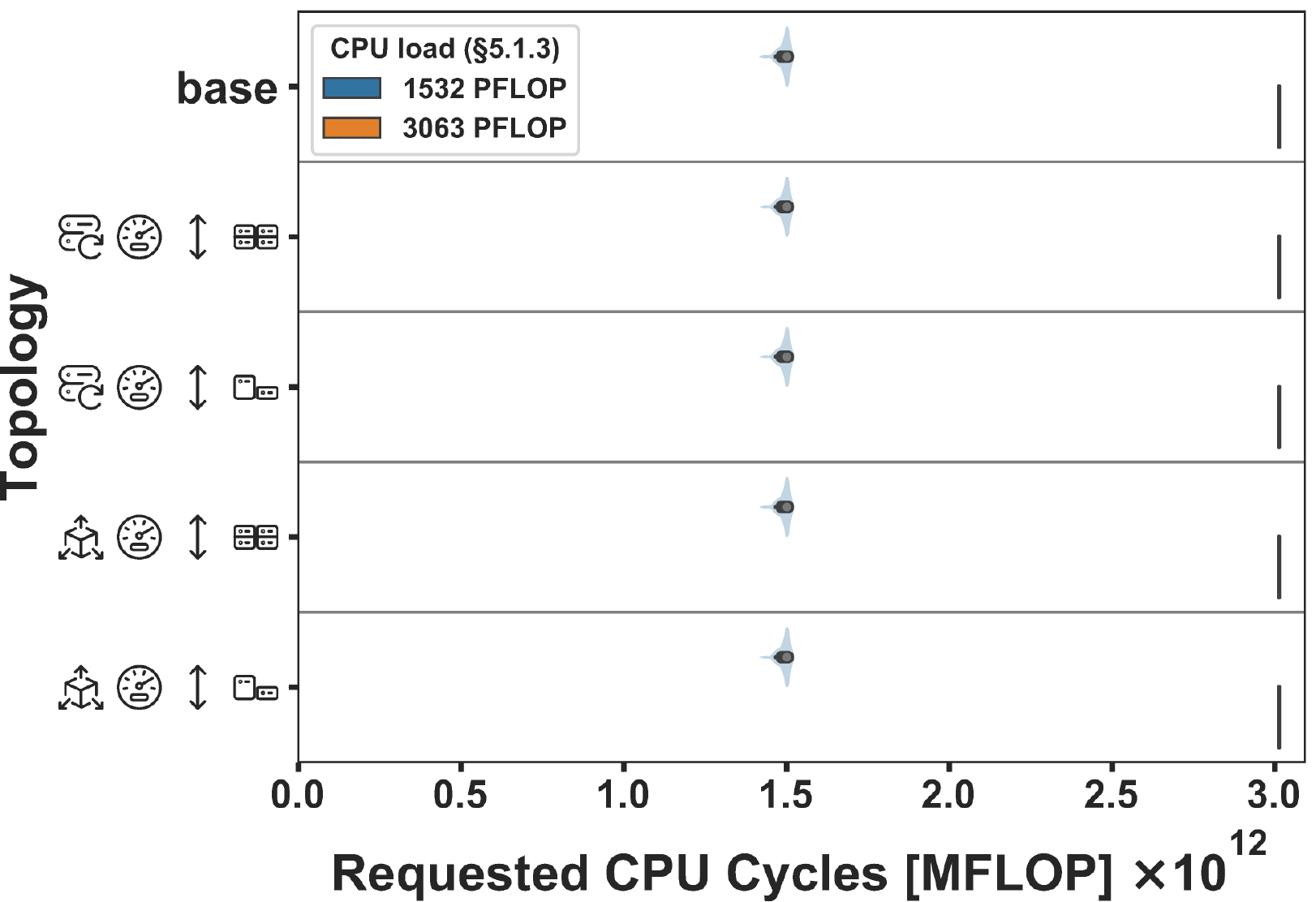}}%
    \subfloat[Granted CPU cycles\label{fig:full:more-vel:granted}]{\includegraphics[width=0.5\linewidth]{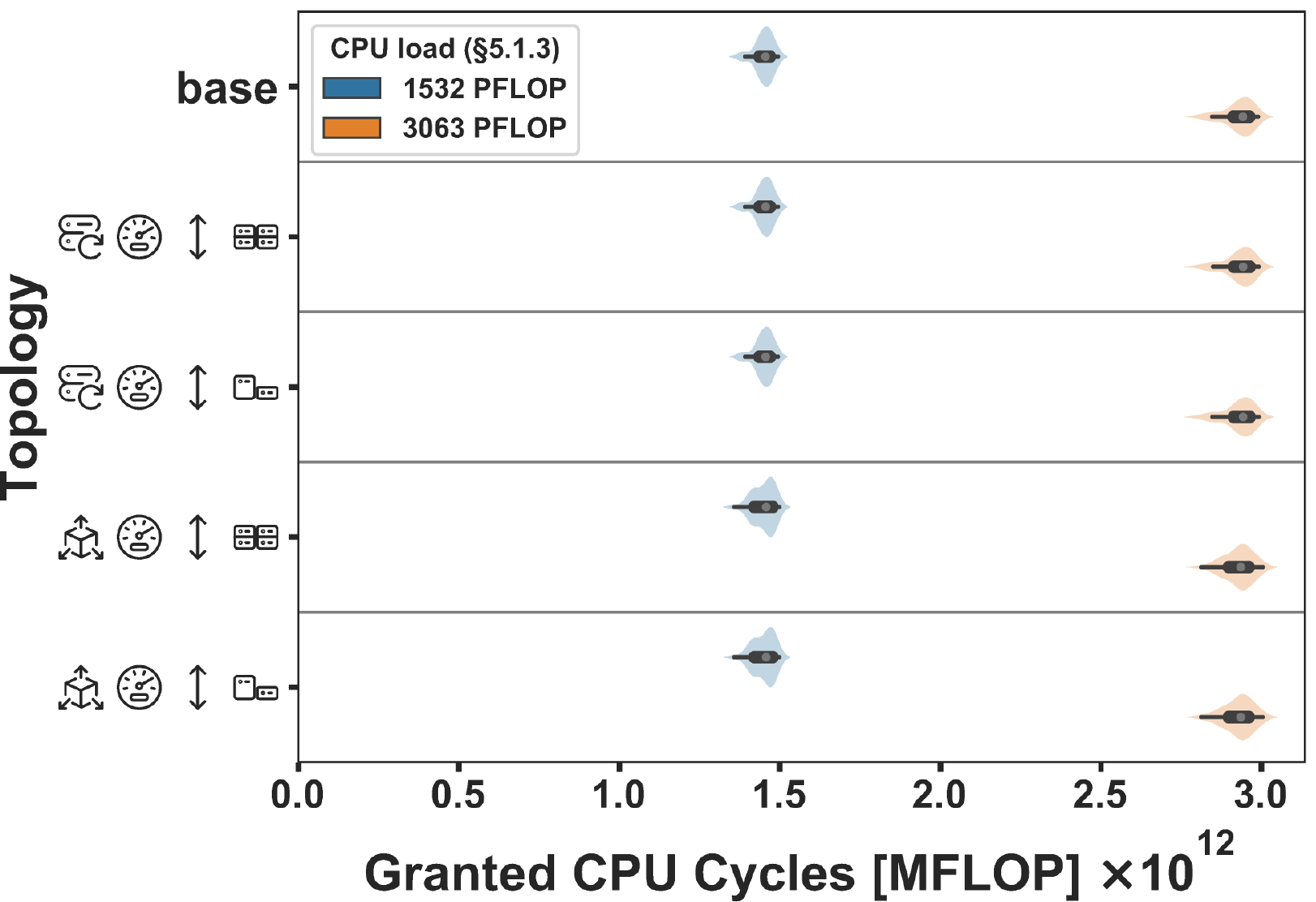}}\\
    \subfloat[Overcommitted CPU cycles\label{fig:full:more-vel:overcommitted}]{\includegraphics[width=0.5\linewidth]{figures/plots/more_velocity_total_overcommitted_burst.pdf}}%
    \subfloat[Interfered CPU cycles\label{fig:full:more-vel:interfered}]{\includegraphics[width=0.5\linewidth]{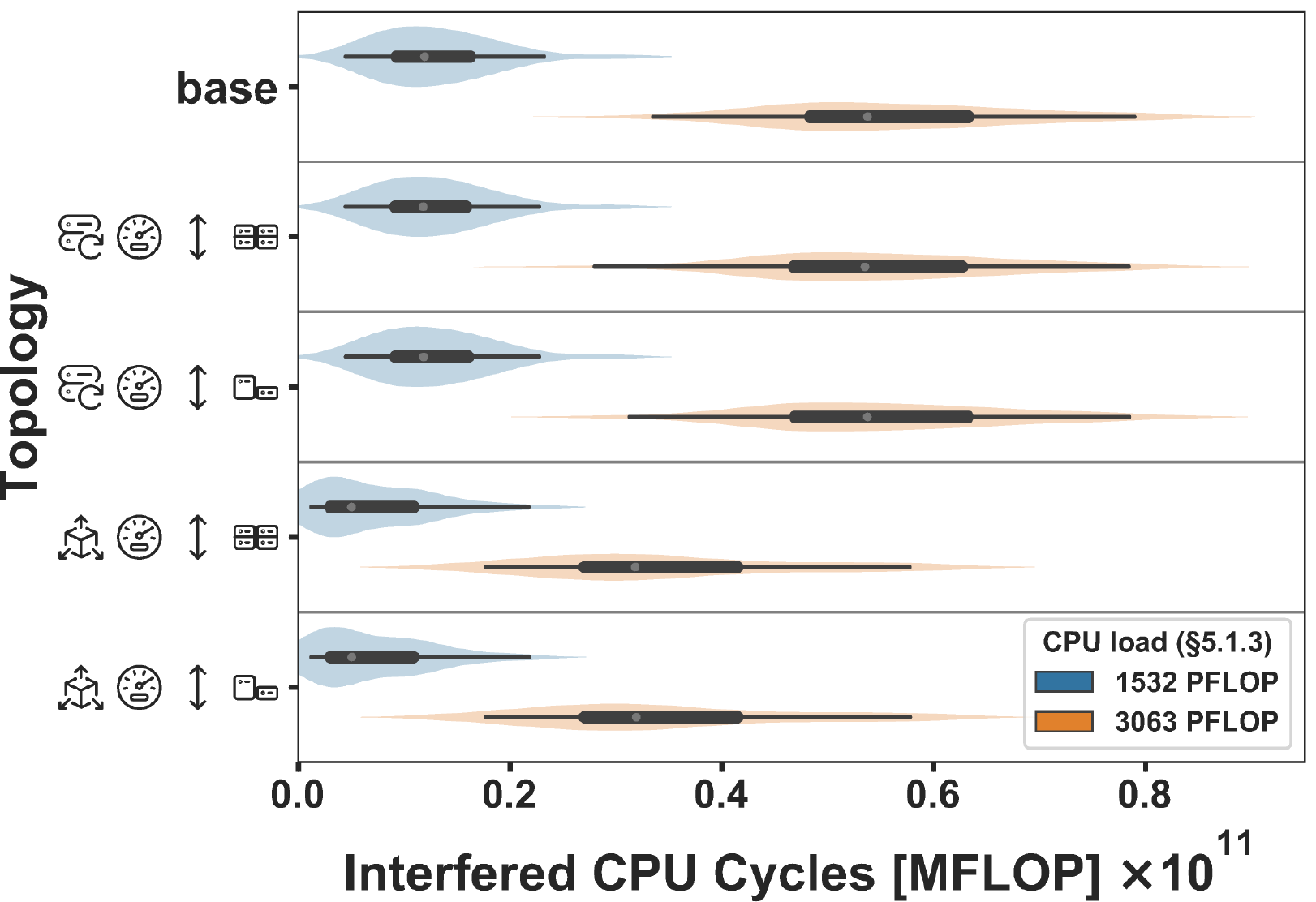}}\\
    \subfloat[Total power consumption\label{fig:full:more-vel:power}]{\includegraphics[width=0.5\linewidth]{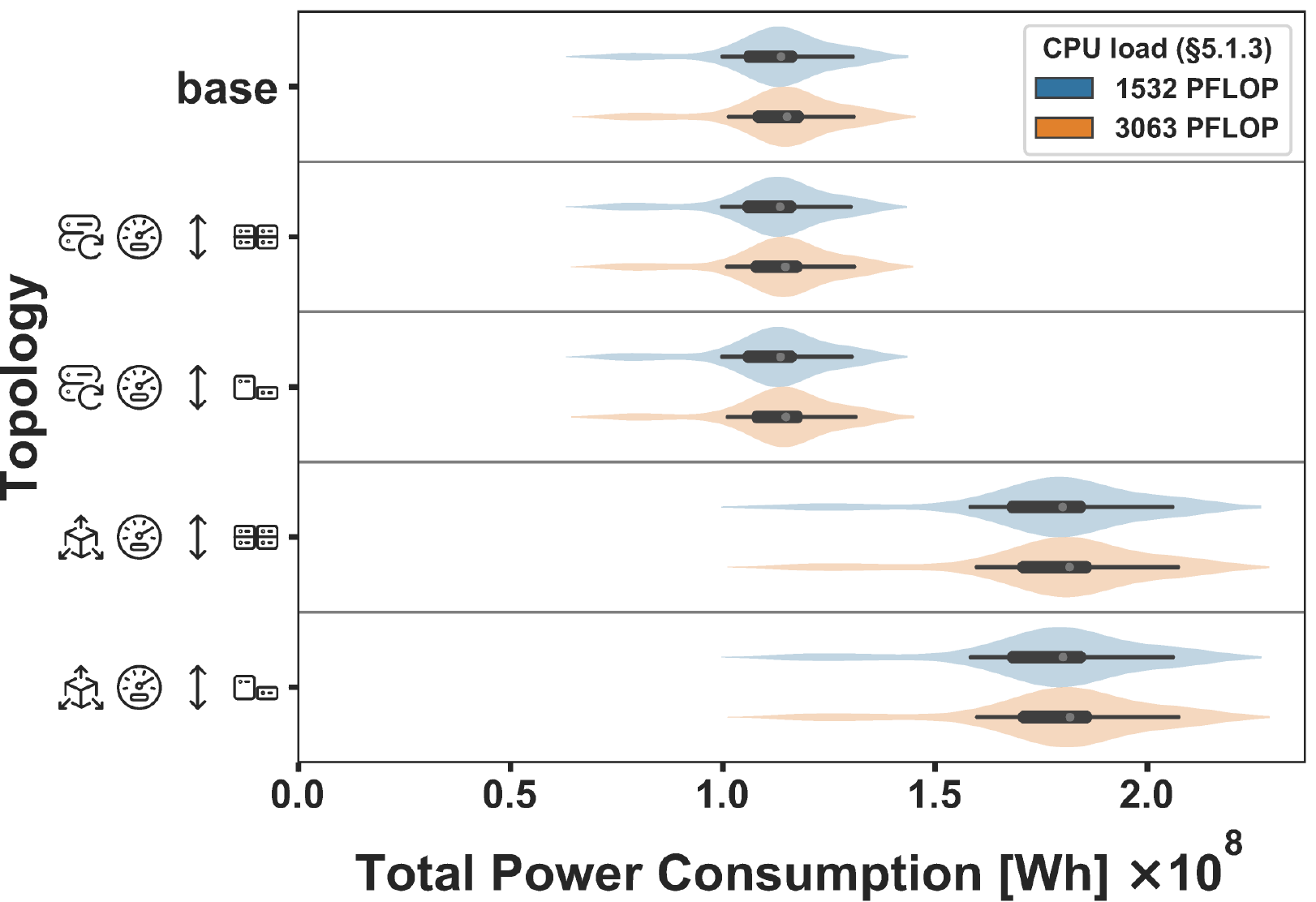}}%
    \subfloat[Total number of time slices in which a \gls{VM} is failed, aggregated across \glspl{VM}\label{fig:full:more-vel:failures:vms}]{\includegraphics[width=0.5\linewidth]{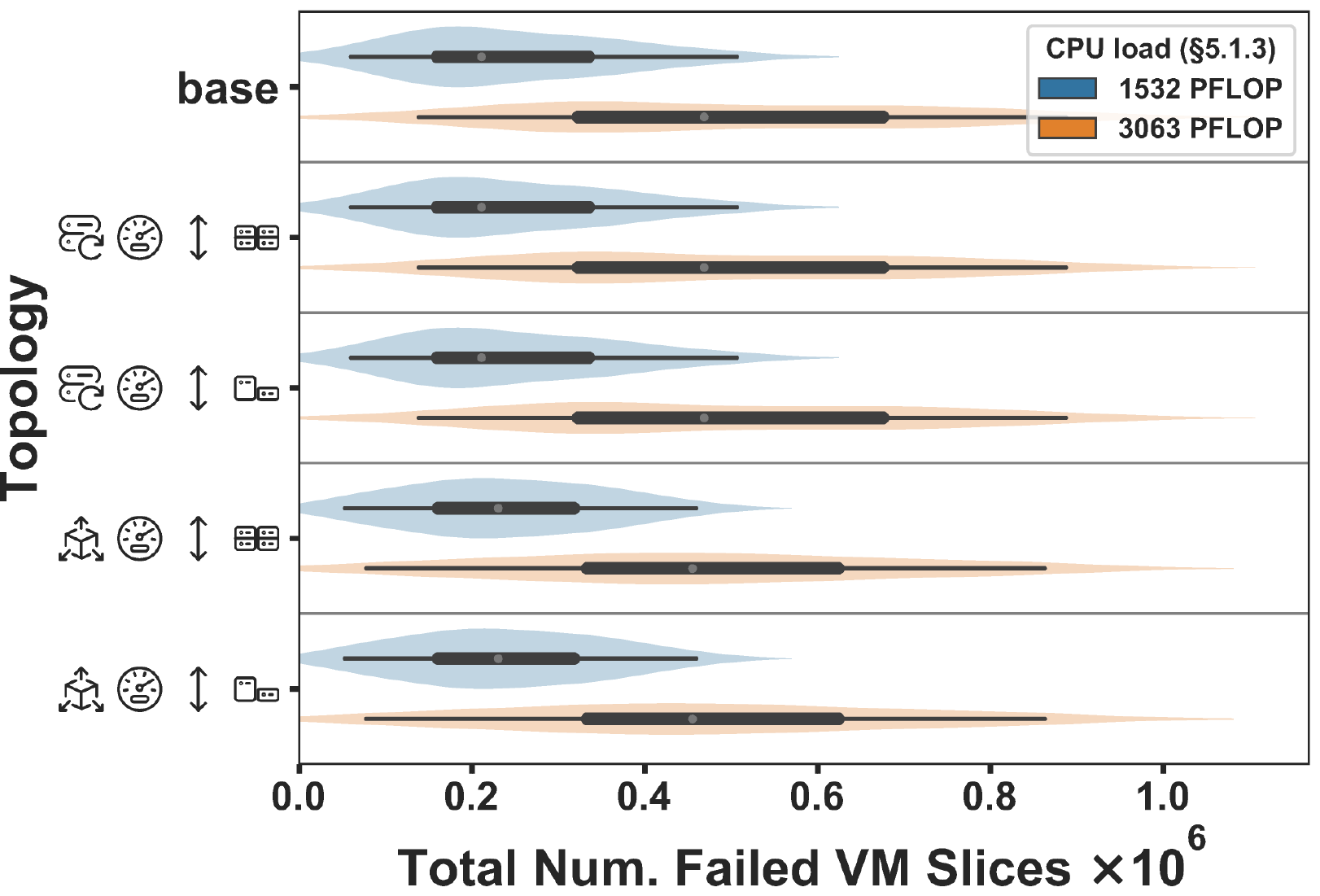}}\\
    \caption{Performance of different topologies expanded on velocity, compared across workloads. For a legend of topologies, see Table~\ref{tab:experiment-overview}. Continued in Figure~\ref{fig:full:more-vel:2}.}
    \label{fig:full:more-vel:1}
\end{figure*}

\begin{figure*}
    \subfloat[Mean CPU usage\label{fig:full:more-vel:cpu-usage}]{\includegraphics[width=0.5\linewidth]{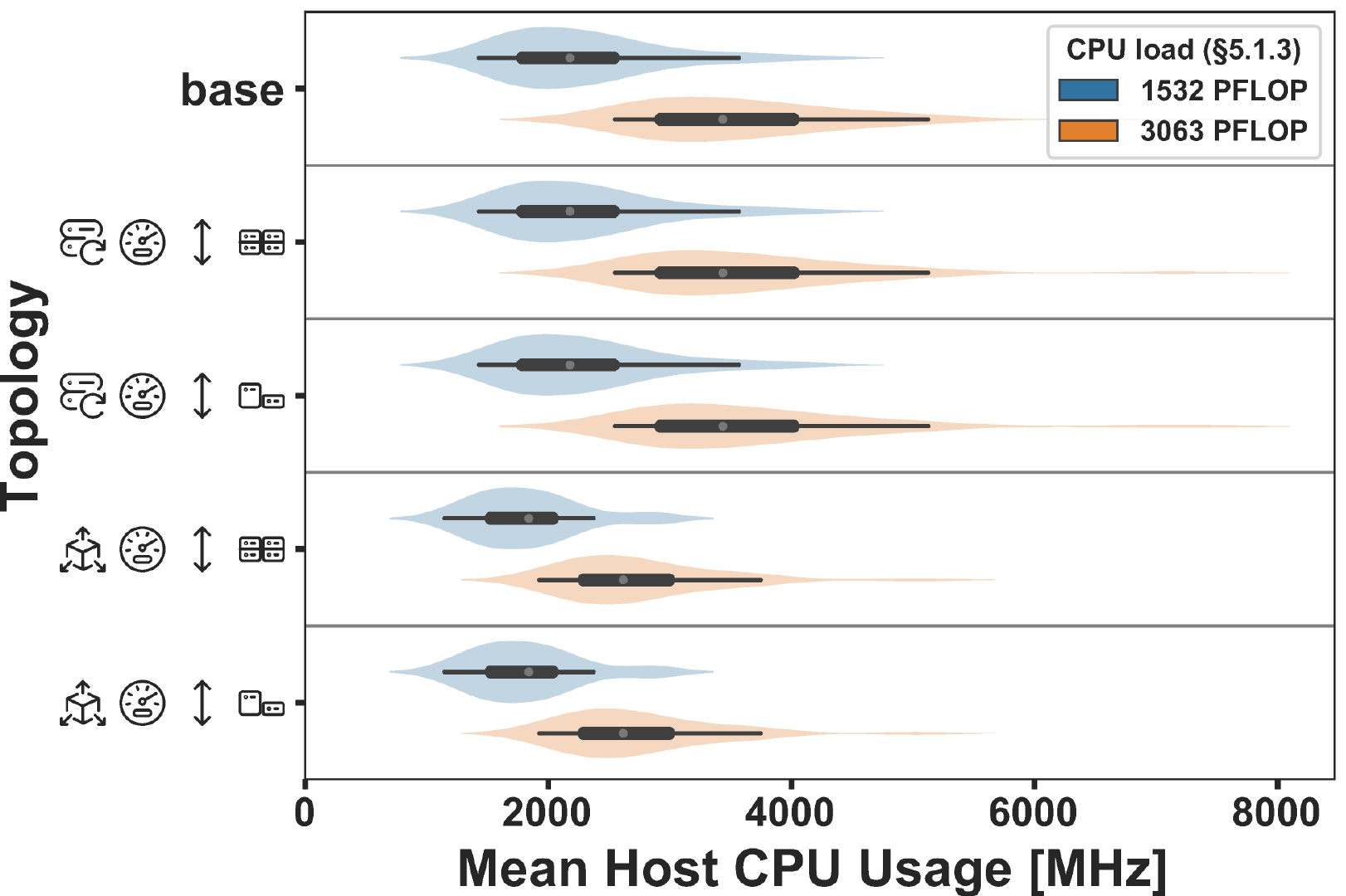}}%
    \subfloat[Mean CPU demand\label{fig:full:more-vel:cpu-demand}]{\includegraphics[width=0.5\linewidth]{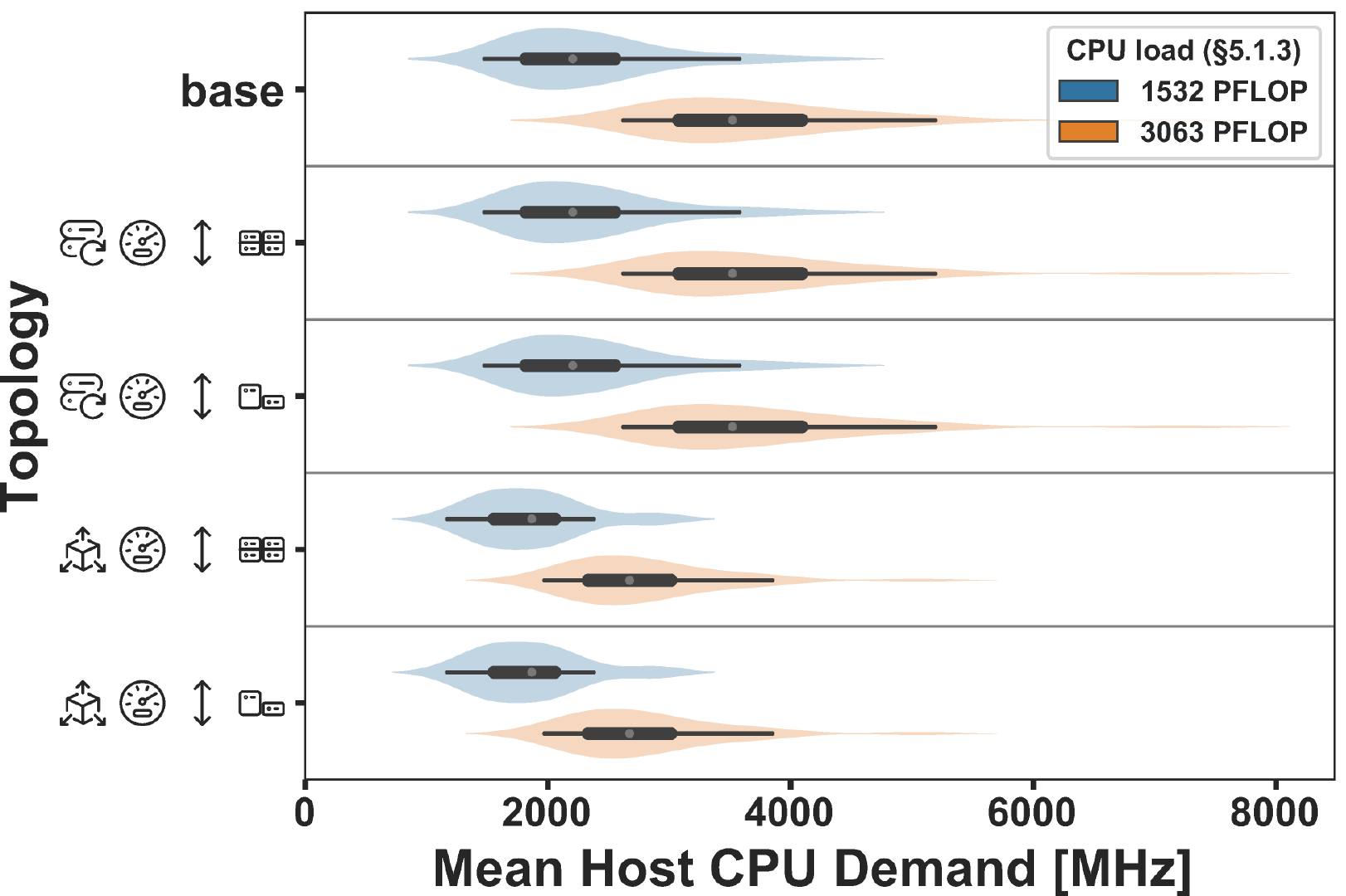}}\\
    \subfloat[Mean number of \glspl{VM} per host\label{fig:full:more-vel:mean-vm-count}]{\includegraphics[width=0.5\linewidth]{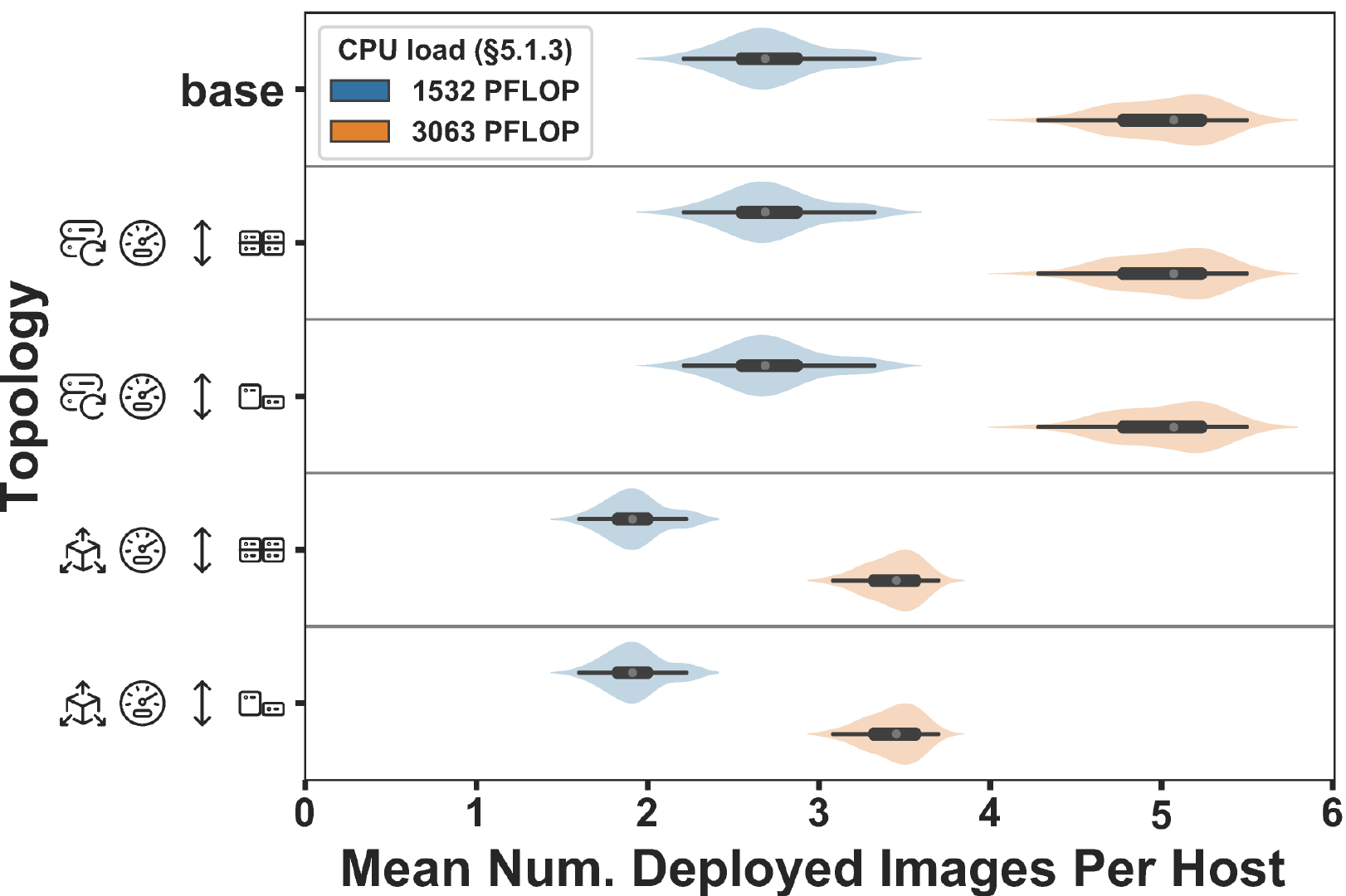}}%
    \subfloat[Max number of \glspl{VM} per host\label{fig:full:more-vel:max-vm-count}]{\includegraphics[width=0.5\linewidth]{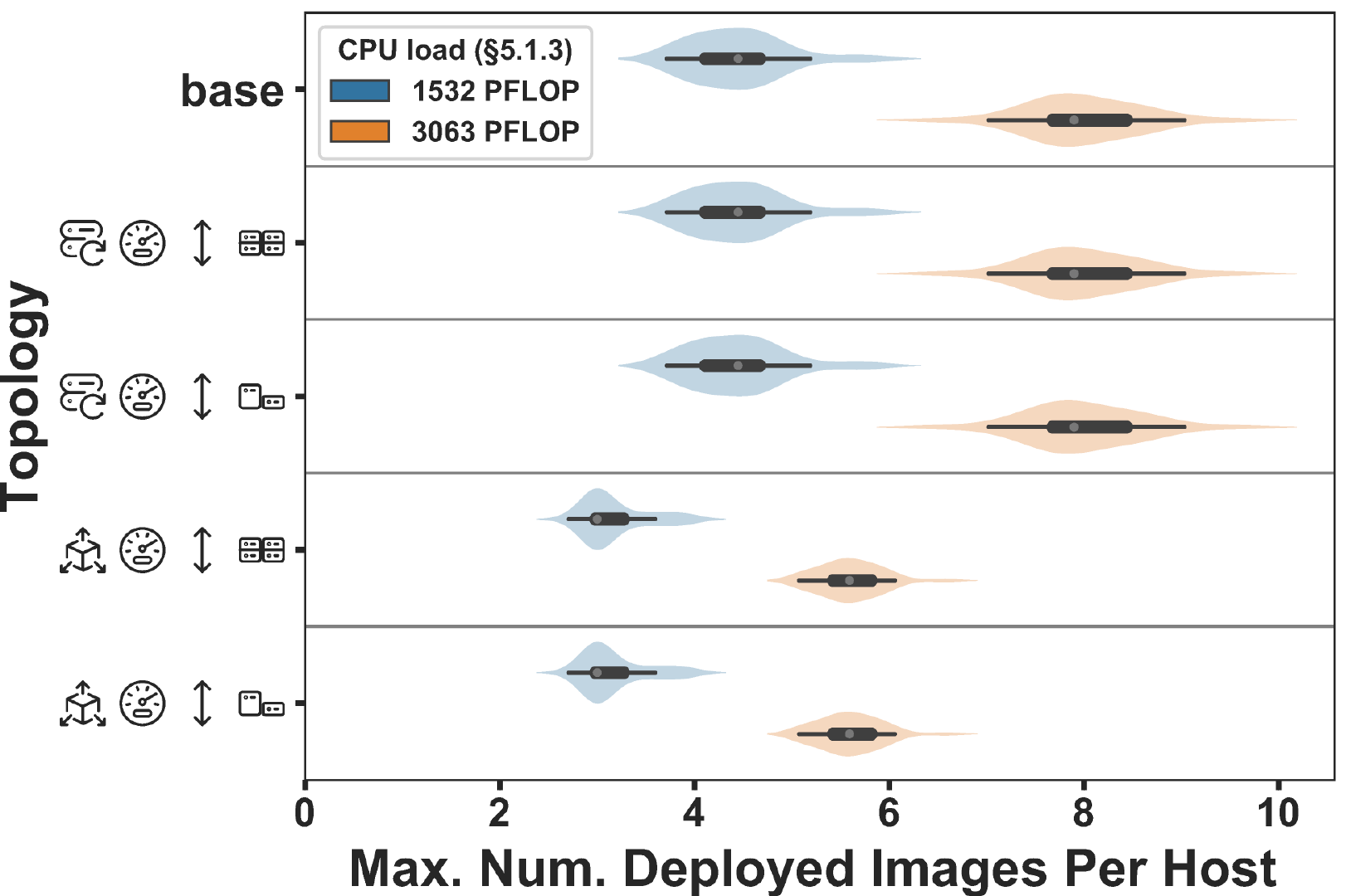}}\\
    \caption{Performance of different topologies expanded on velocity, compared across workloads. For a legend of topologies, see Table~\ref{tab:experiment-overview}. Continued in Figure~\ref{fig:full:more-vel:3}.}
    \label{fig:full:more-vel:2}
\end{figure*}

\begin{figure*}
    \subfloat[Total VMs Submitted\label{fig:full:more-vel:vms-submitted}]{\includegraphics[width=0.5\linewidth]{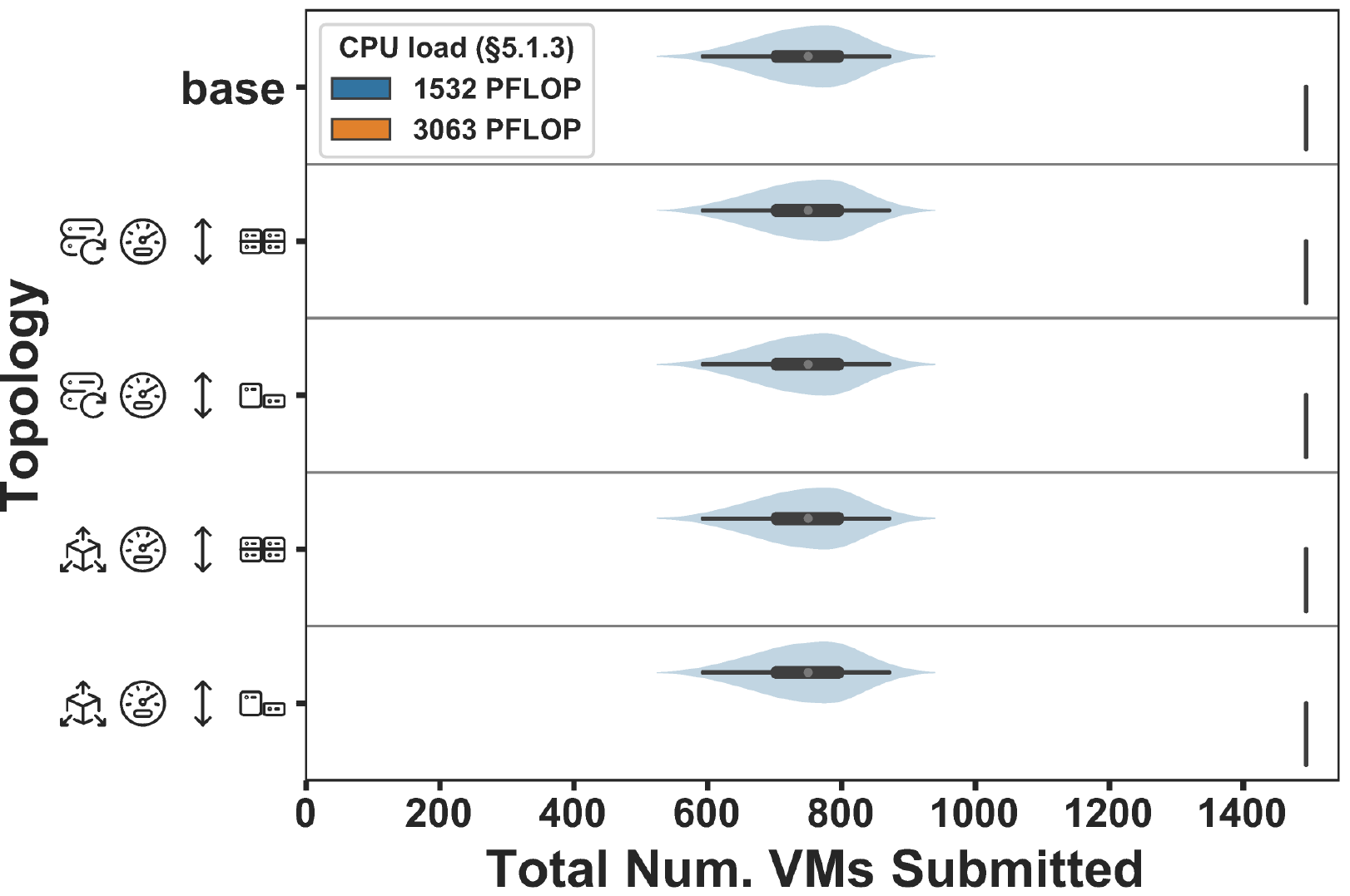}}%
    \subfloat[Total VMs Queued\label{fig:full:more-vel:vms-queued}]{\includegraphics[width=0.5\linewidth]{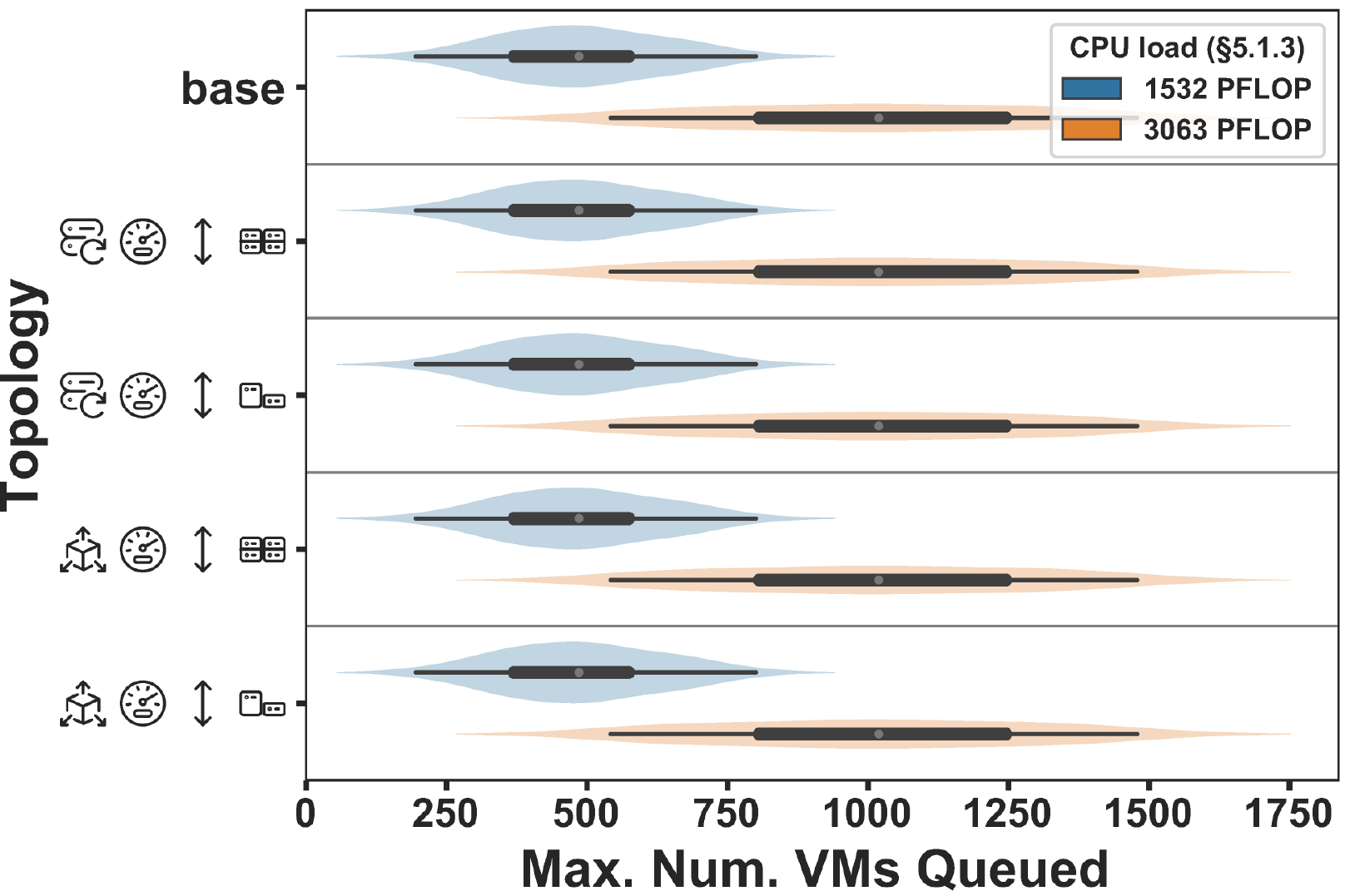}}\\
    \subfloat[Total VMs Finished\label{fig:full:more-vel:vms-finished}]{\includegraphics[width=0.5\linewidth]{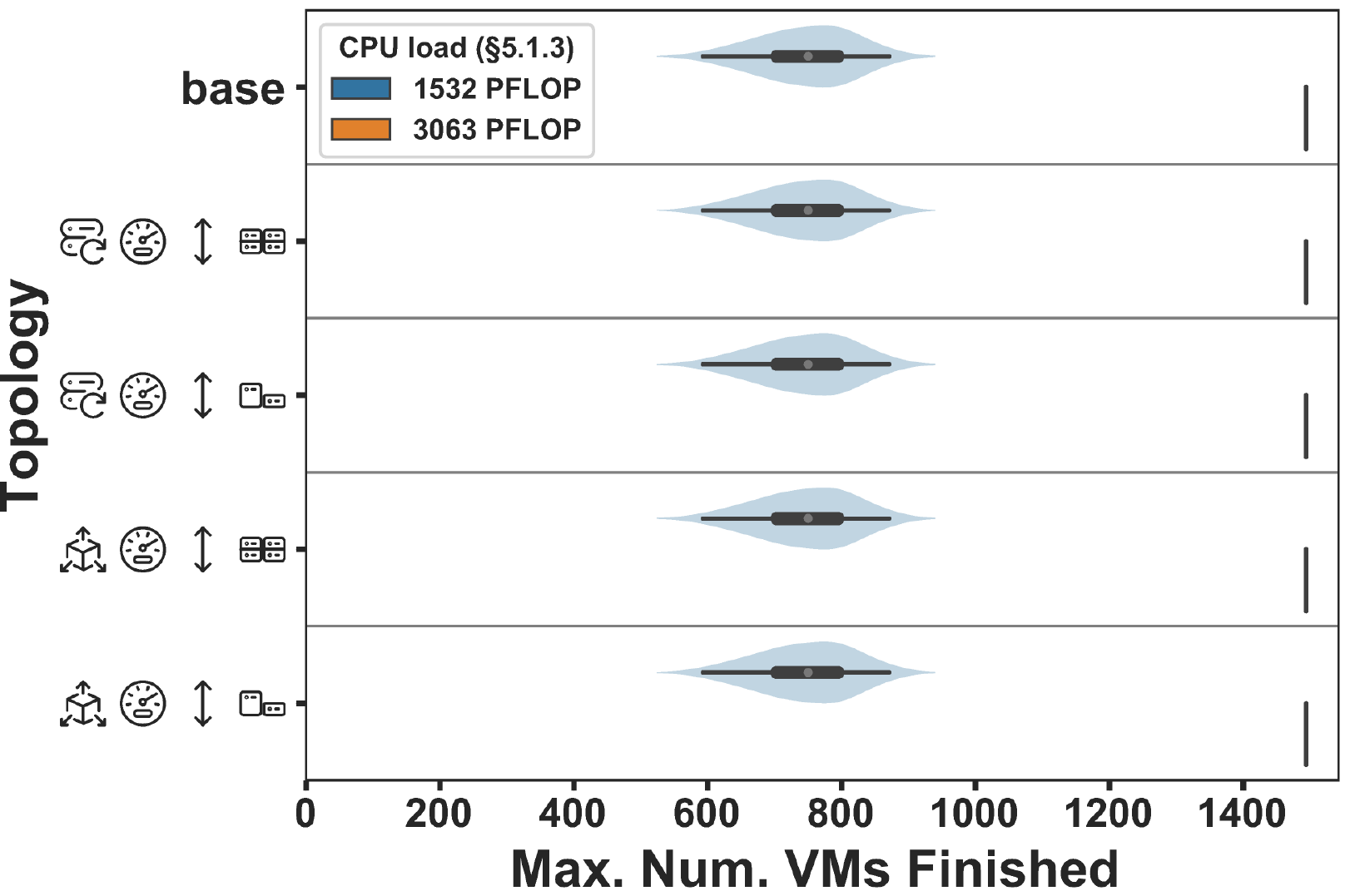}}%
    \subfloat[Total VMs Failed\label{fig:full:more-vel:vms-failed}]{\includegraphics[width=0.5\linewidth]{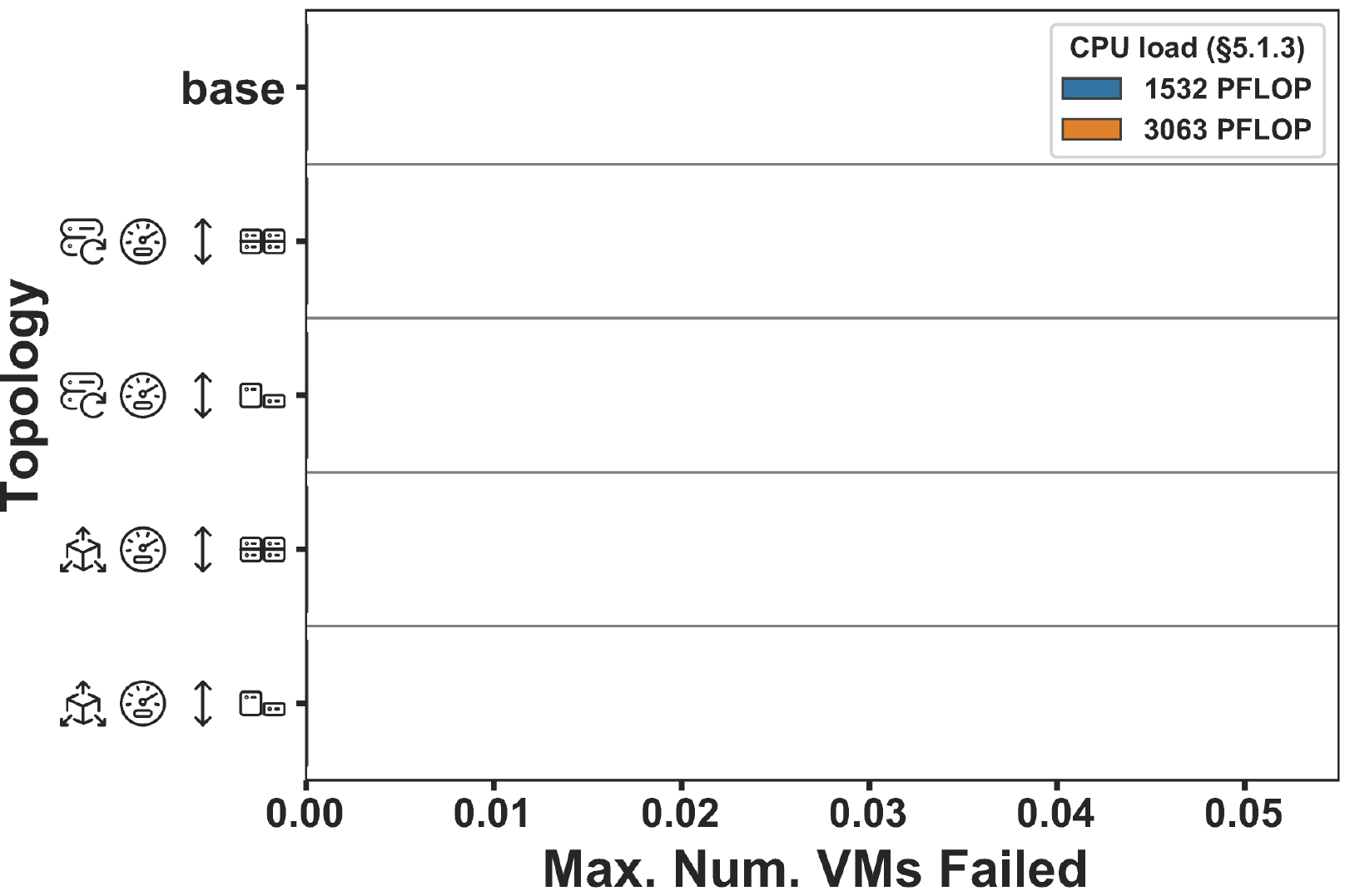}}%
    \caption{Performance of different topologies expanded on velocity, compared across workloads. For a legend of topologies, see Table~\ref{tab:experiment-overview}.}
    \label{fig:full:more-vel:3}
\end{figure*}

\begin{figure*}
    \centering
    \subfloat[Requested CPU cycles\label{fig:full:more-vel:summary:requested}]{\includegraphics[width=0.5\linewidth]{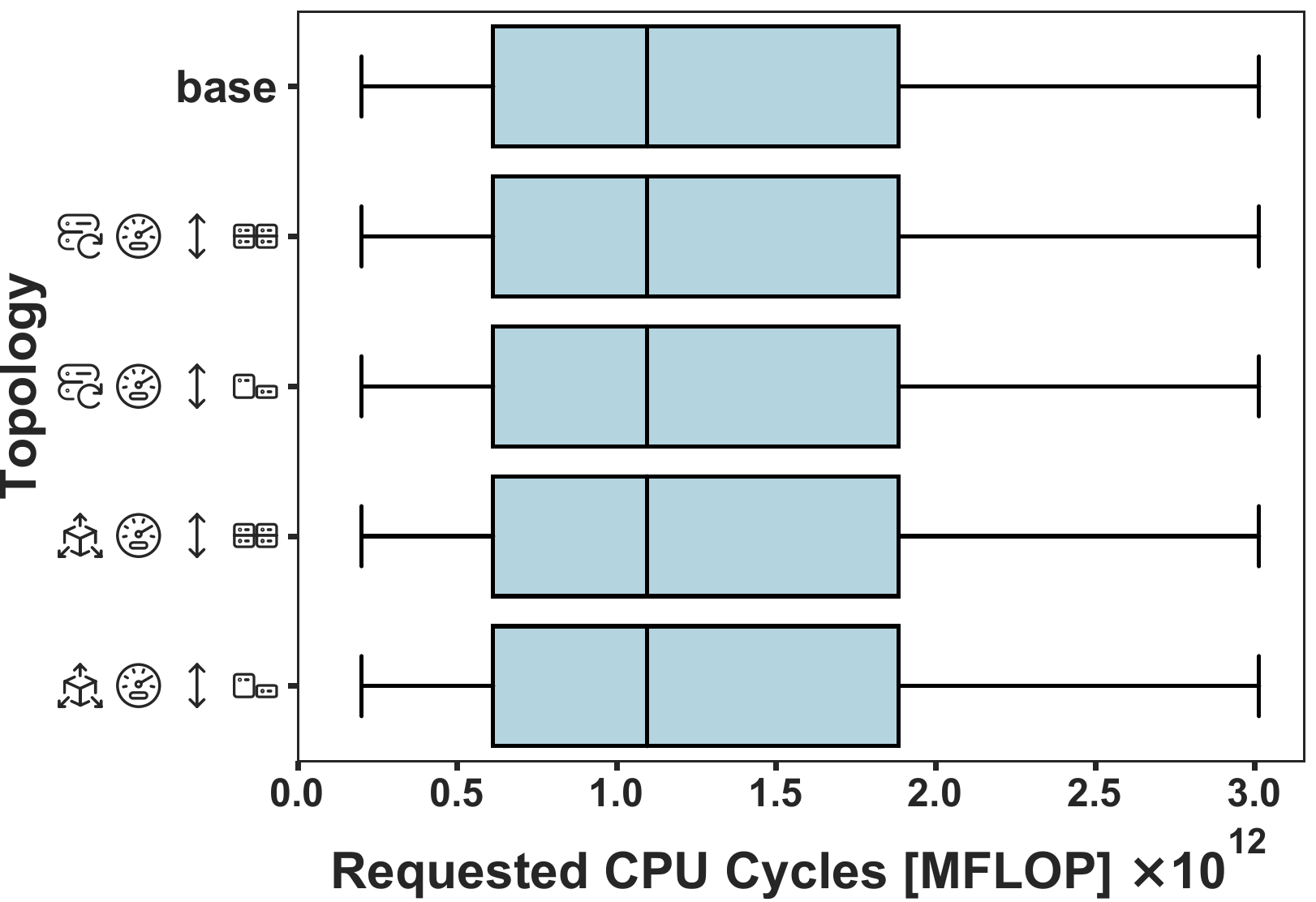}}%
    \subfloat[Granted CPU cycles\label{fig:full:more-vel:summary:granted}]{\includegraphics[width=0.5\linewidth]{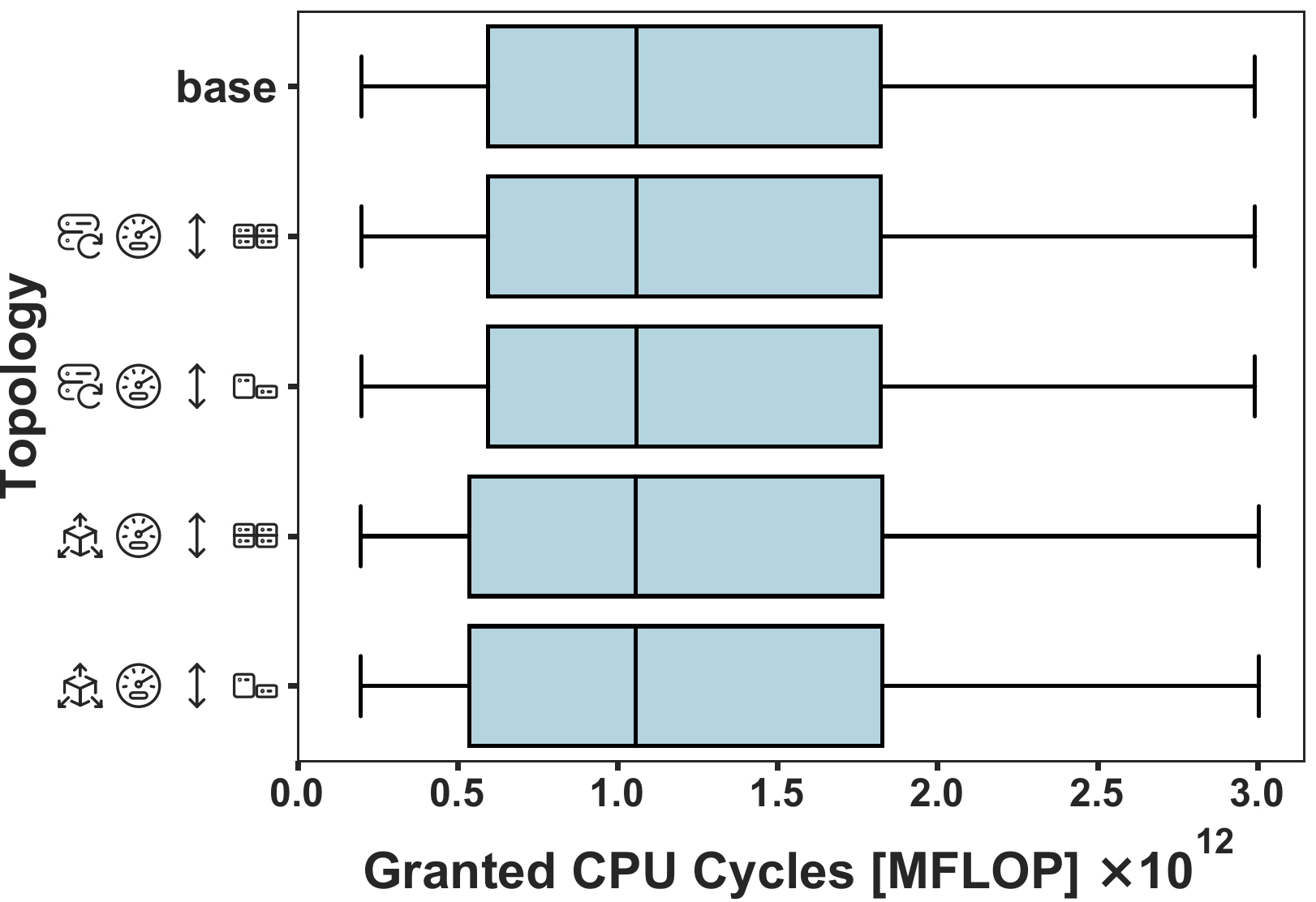}}\\
    \subfloat[Overcommitted CPU cycles\label{fig:full:more-vel:summary:overcommitted}]{\includegraphics[width=0.5\linewidth]{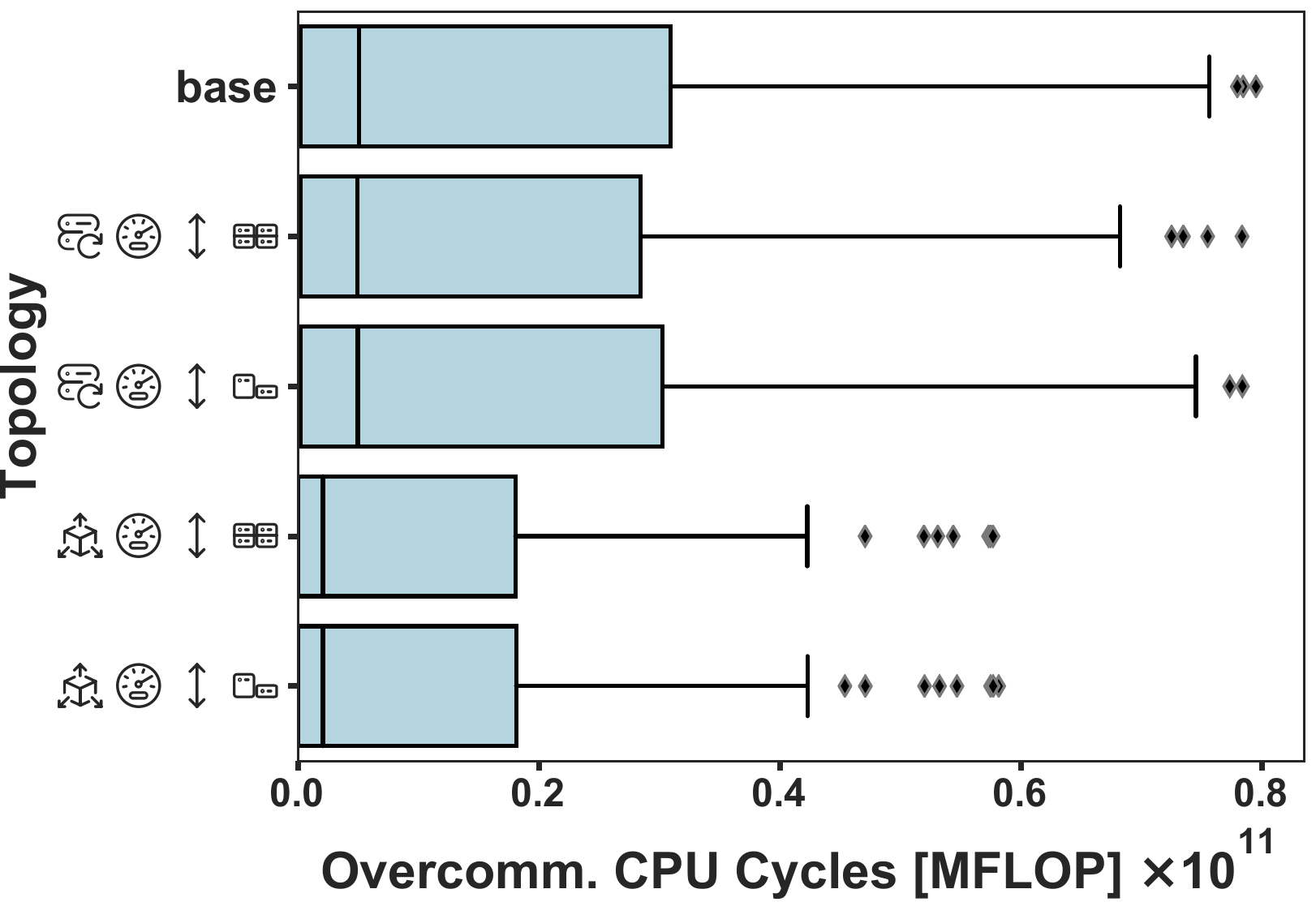}}%
    \subfloat[Interfered CPU cycles\label{fig:full:more-vel:summary:interfered}]{\includegraphics[width=0.5\linewidth]{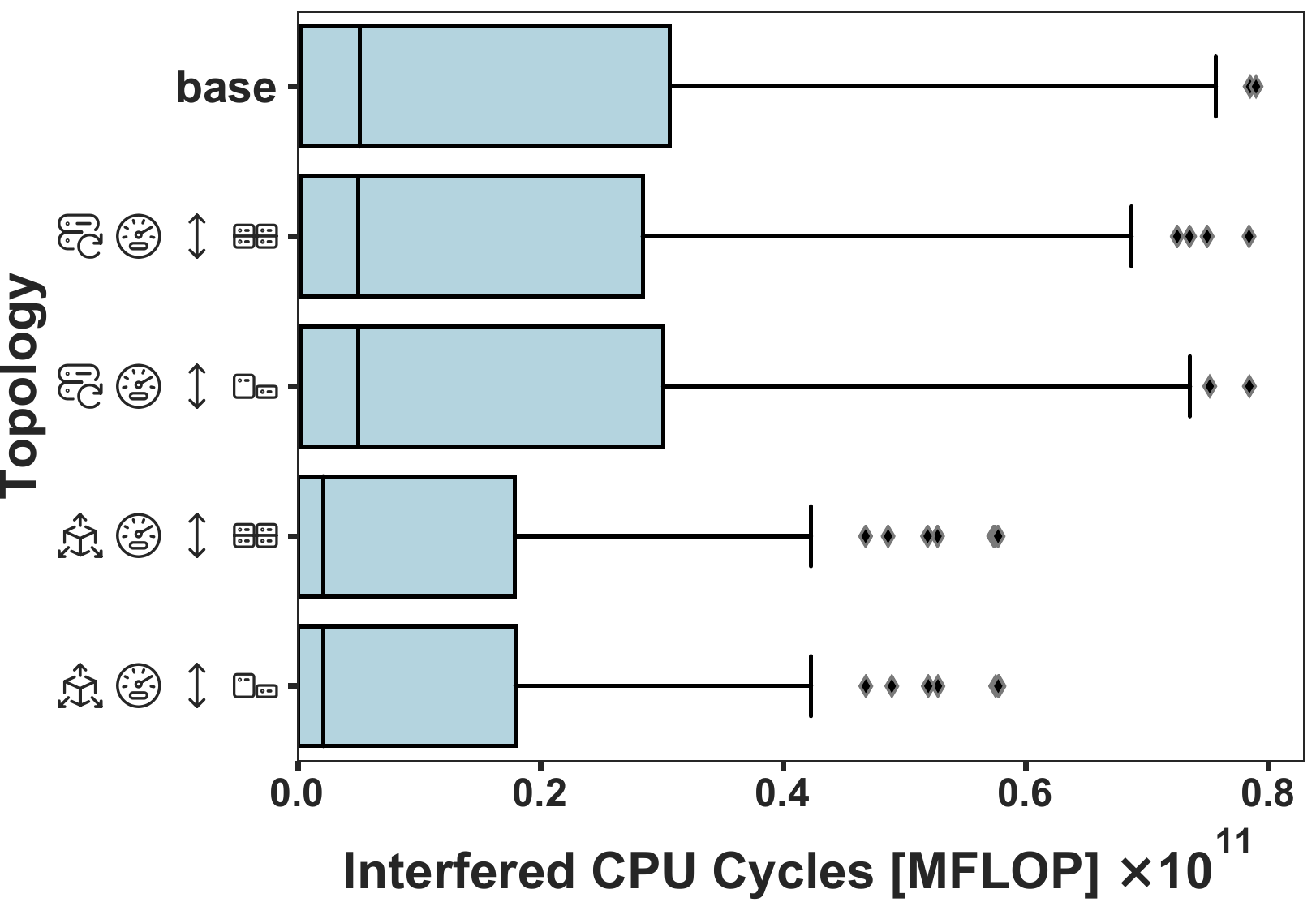}}\\
    \subfloat[Total power consumption\label{fig:full:more-vel:summary:power}]{\includegraphics[width=0.5\linewidth]{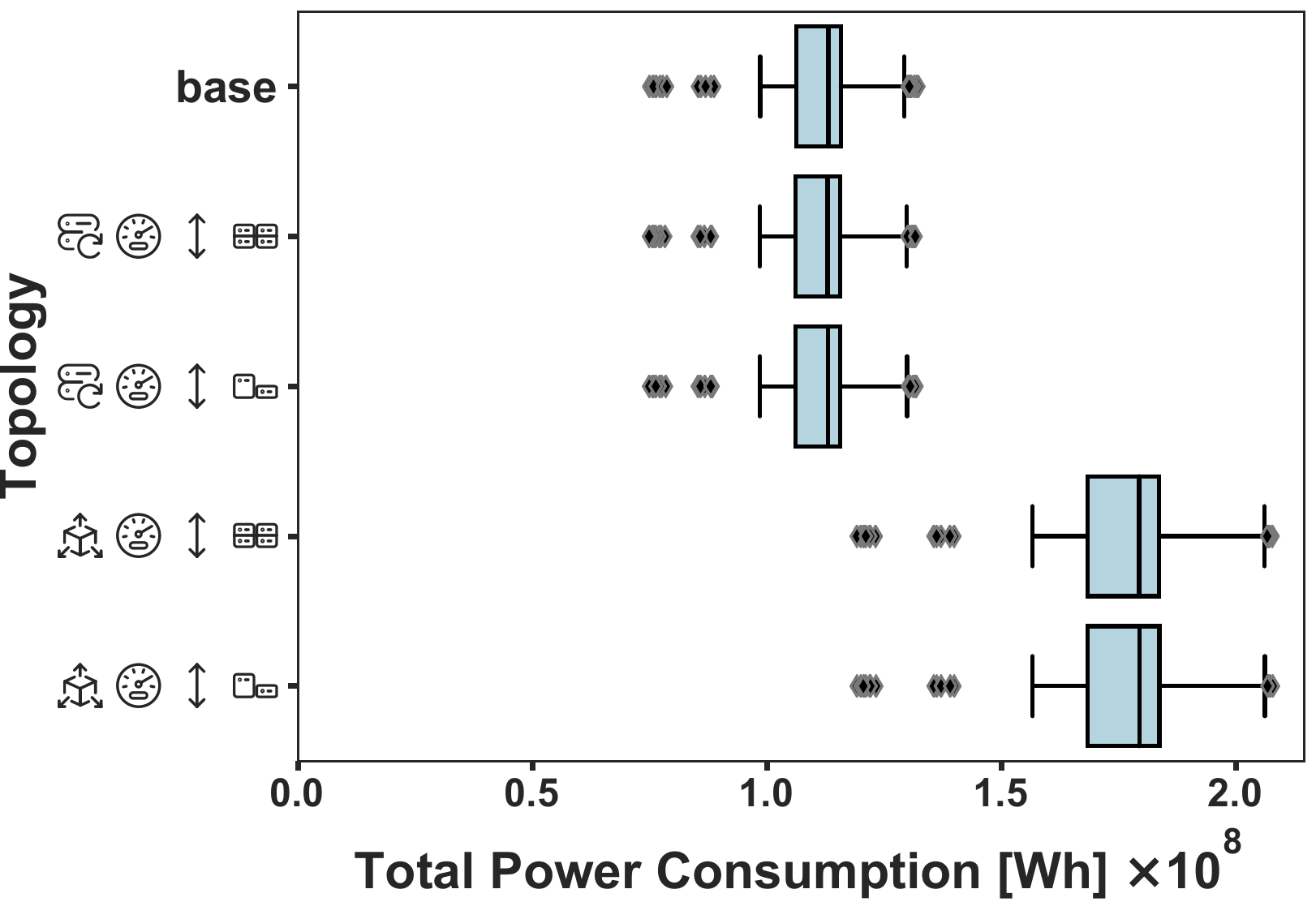}}%
    \subfloat[Total number of time slices in which a \gls{VM} is failed, aggregated across \glspl{VM}\label{fig:full:more-vel:summary:failures:vms}]{\includegraphics[width=0.5\linewidth]{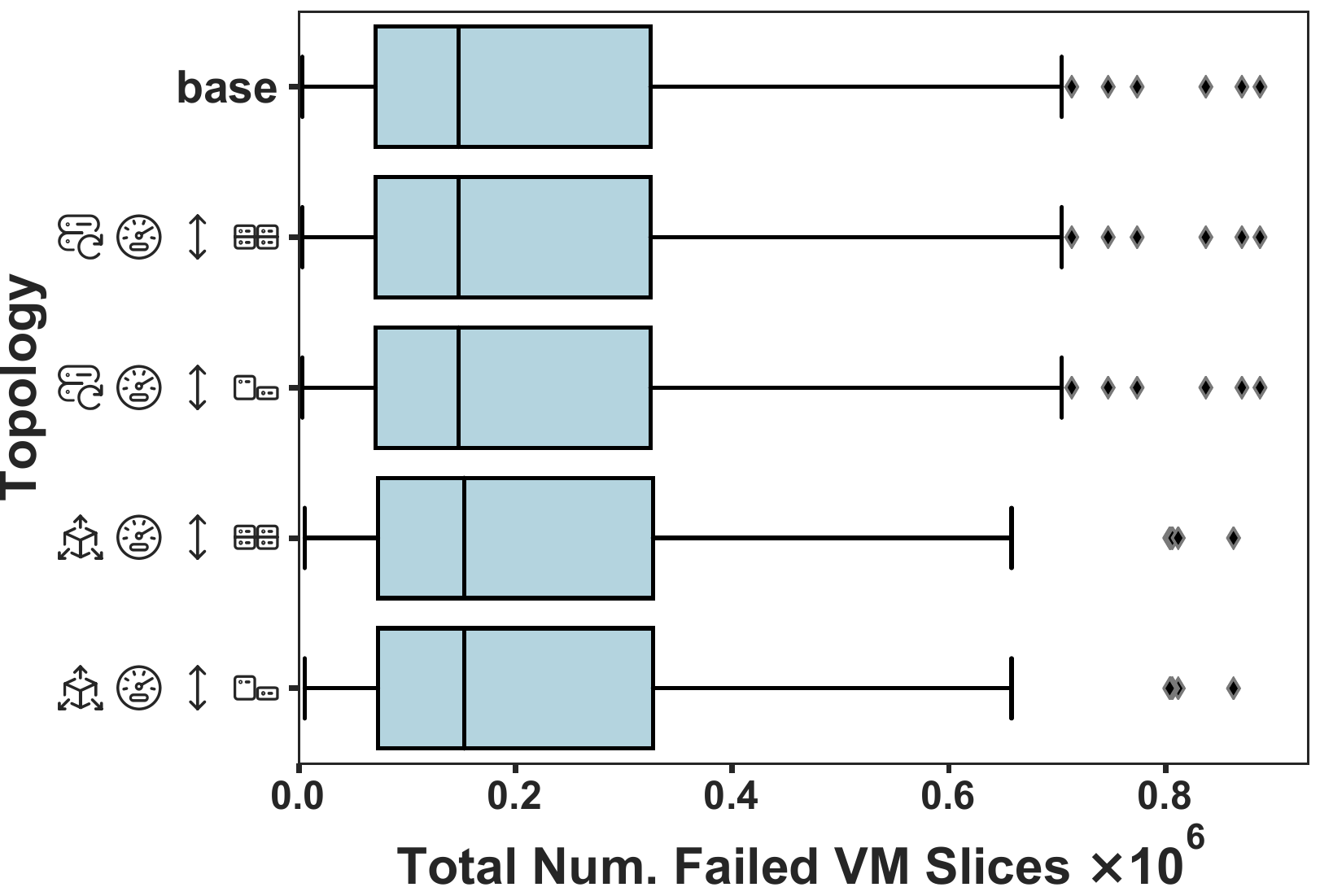}}\\
    \caption{Performance of different topologies expanded on velocity, compared across workloads. Results aggregated across the full set of workloads, including workloads not displayed in the more detailed figure. For a legend of topologies, see Table~\ref{tab:experiment-overview}. Continued in Figure~\ref{fig:full:more-vel:summary:2}.}
    \label{fig:full:more-vel:summary:1}
\end{figure*}

\begin{figure*}
    \subfloat[Mean CPU usage\label{fig:full:more-vel:summary:cpu-usage}]{\includegraphics[width=0.5\linewidth]{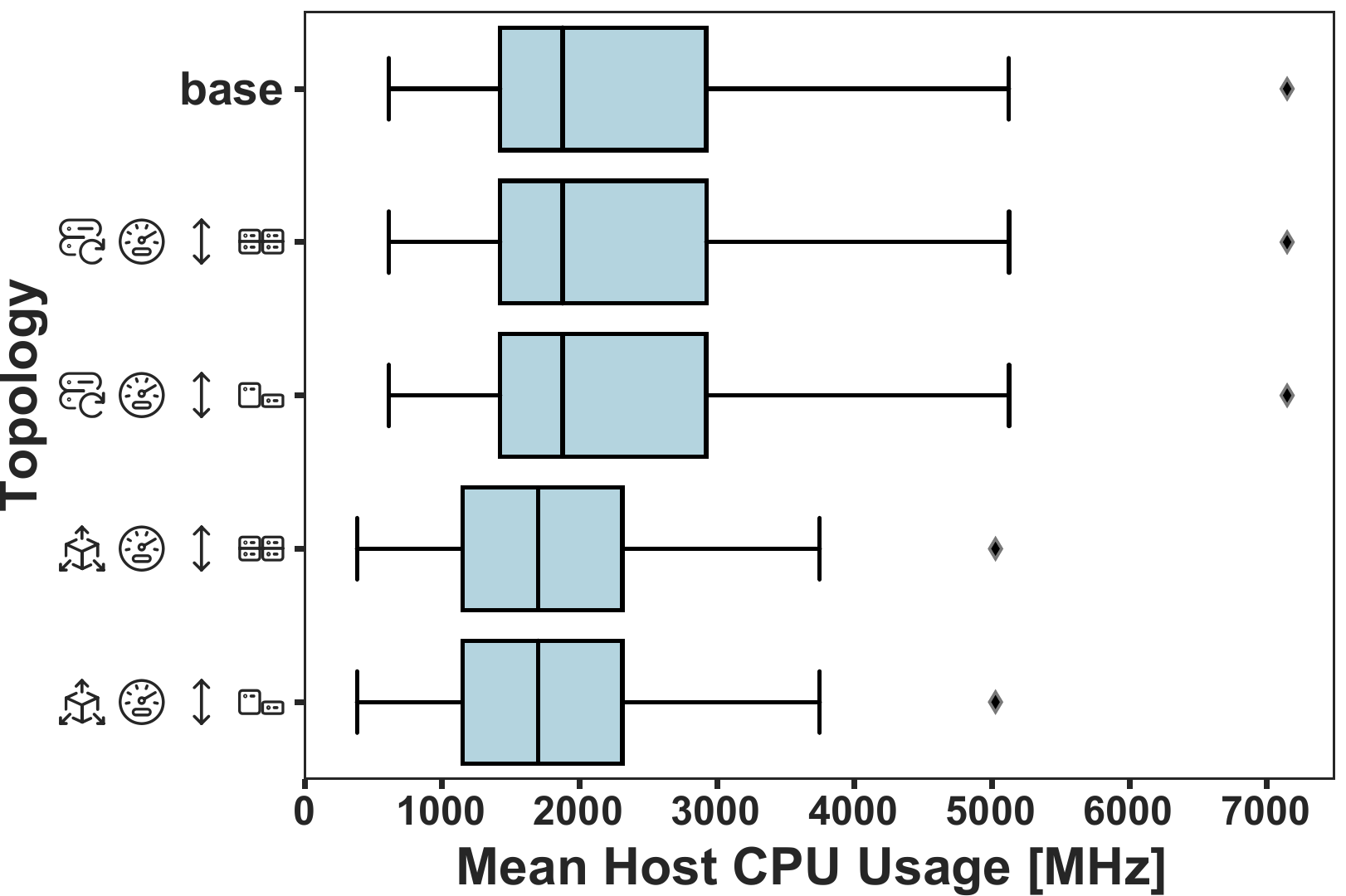}}%
    \subfloat[Mean CPU demand\label{fig:full:more-vel:summary:cpu-demand}]{\includegraphics[width=0.5\linewidth]{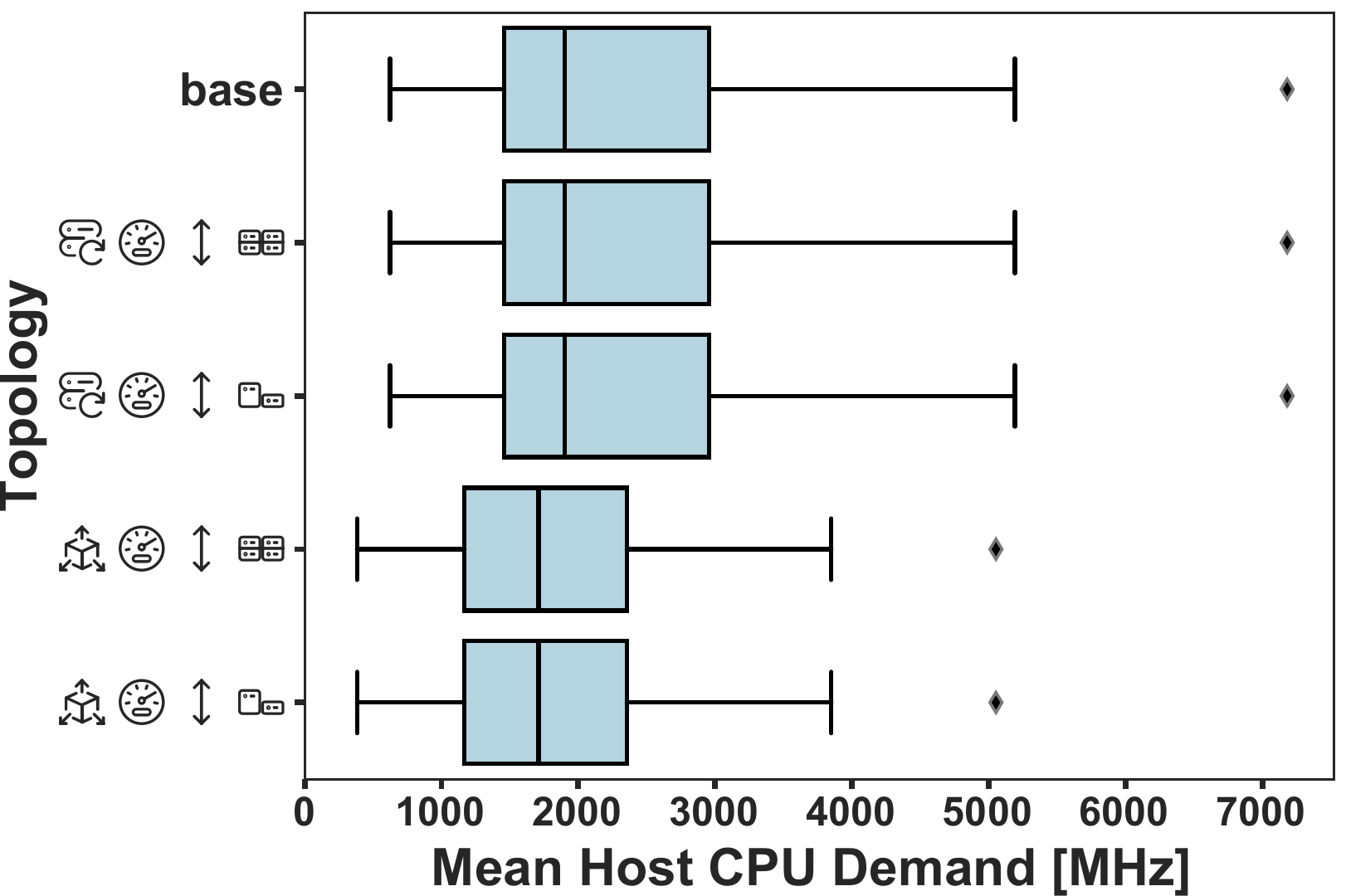}}\\
    \subfloat[Mean number of \glspl{VM} per host\label{fig:full:more-vel:summary:mean-vm-count}]{\includegraphics[width=0.5\linewidth]{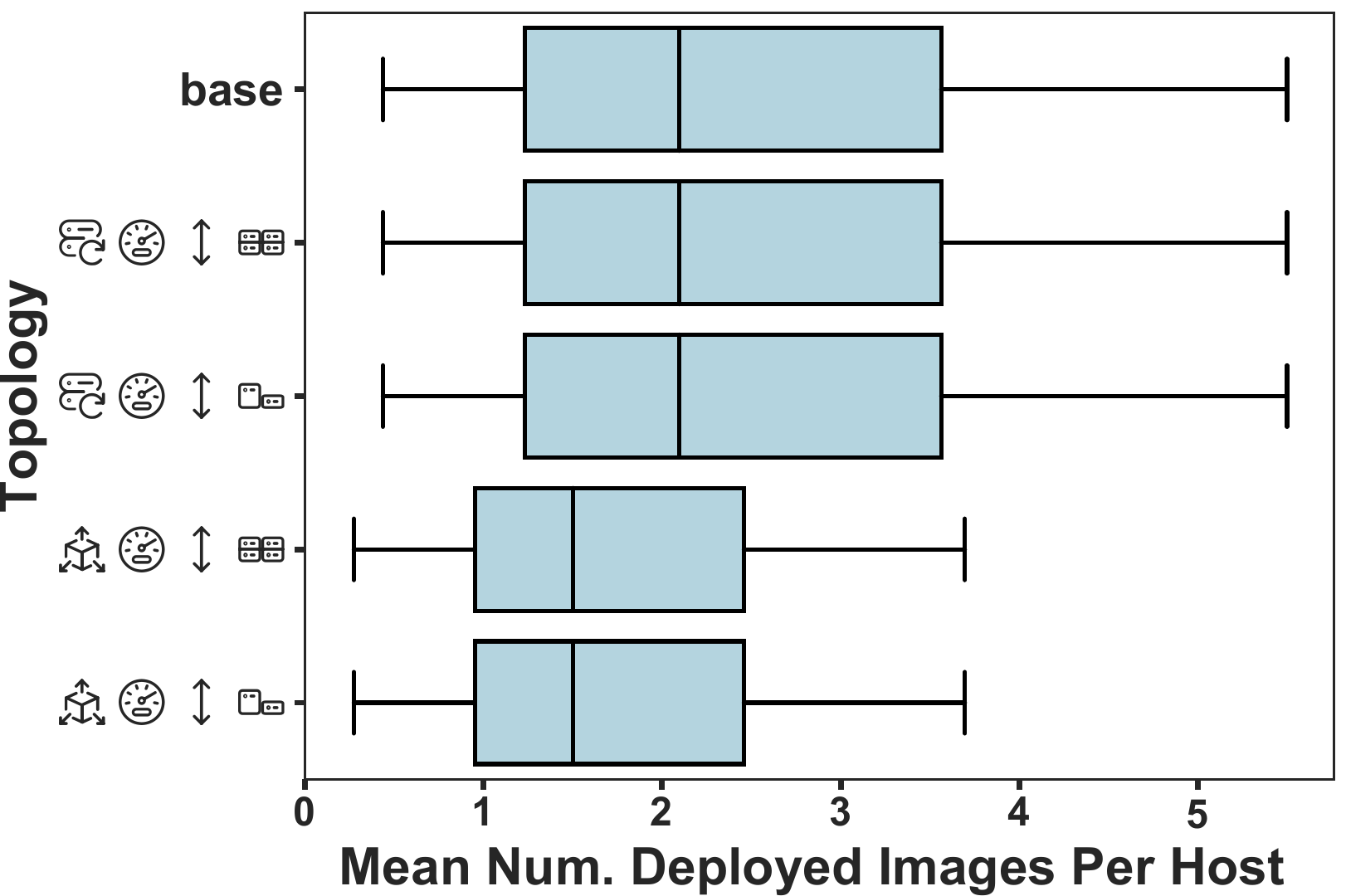}}%
    \subfloat[Max number of \glspl{VM} per host\label{fig:full:more-vel:summary:max-vm-count}]{\includegraphics[width=0.5\linewidth]{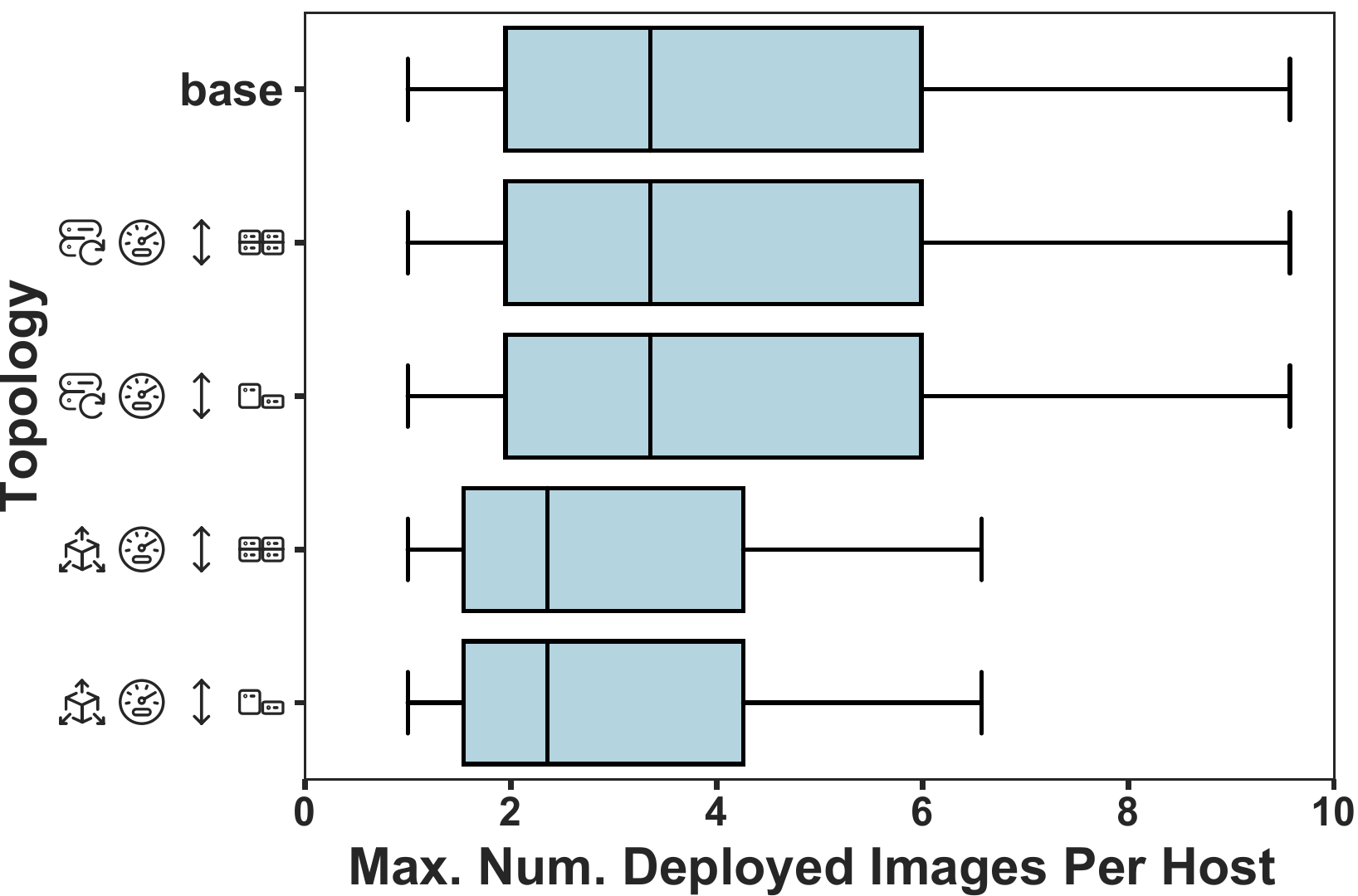}}\\
    \caption{Performance of different topologies expanded on velocity, compared across workloads. Results aggregated across the full set of workloads, including workloads not displayed in the more detailed figure. For a legend of topologies, see Table~\ref{tab:experiment-overview}. Continued in Figure~\ref{fig:full:more-vel:summary:3}.}
    \label{fig:full:more-vel:summary:2}
\end{figure*}

\begin{figure*}
    \subfloat[Total VMs Submitted\label{fig:full:more-vel:summary:vms-submitted}]{\includegraphics[width=0.5\linewidth]{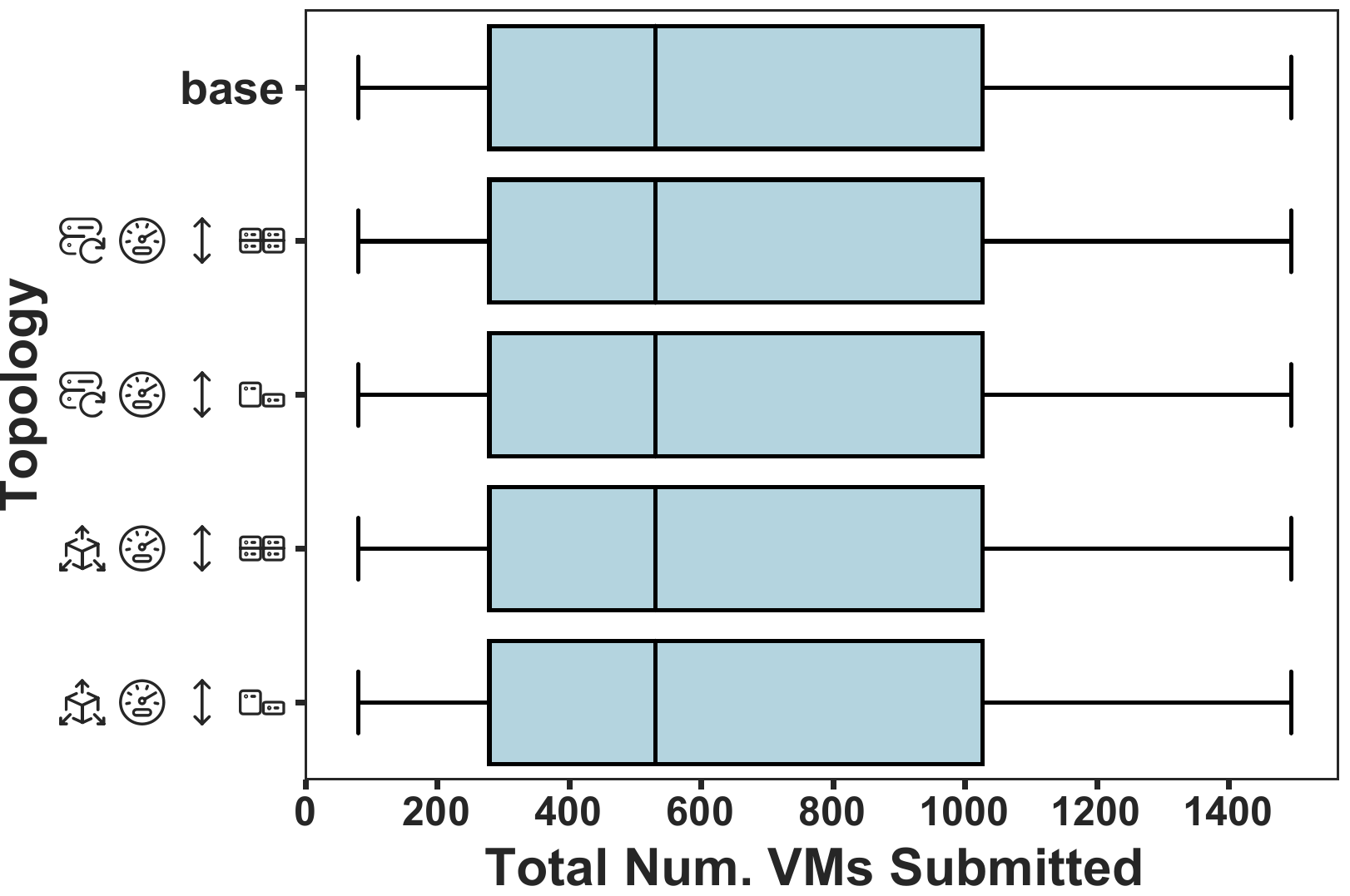}}%
    \subfloat[Total VMs Queued\label{fig:full:more-vel:summary:vms-queued}]{\includegraphics[width=0.5\linewidth]{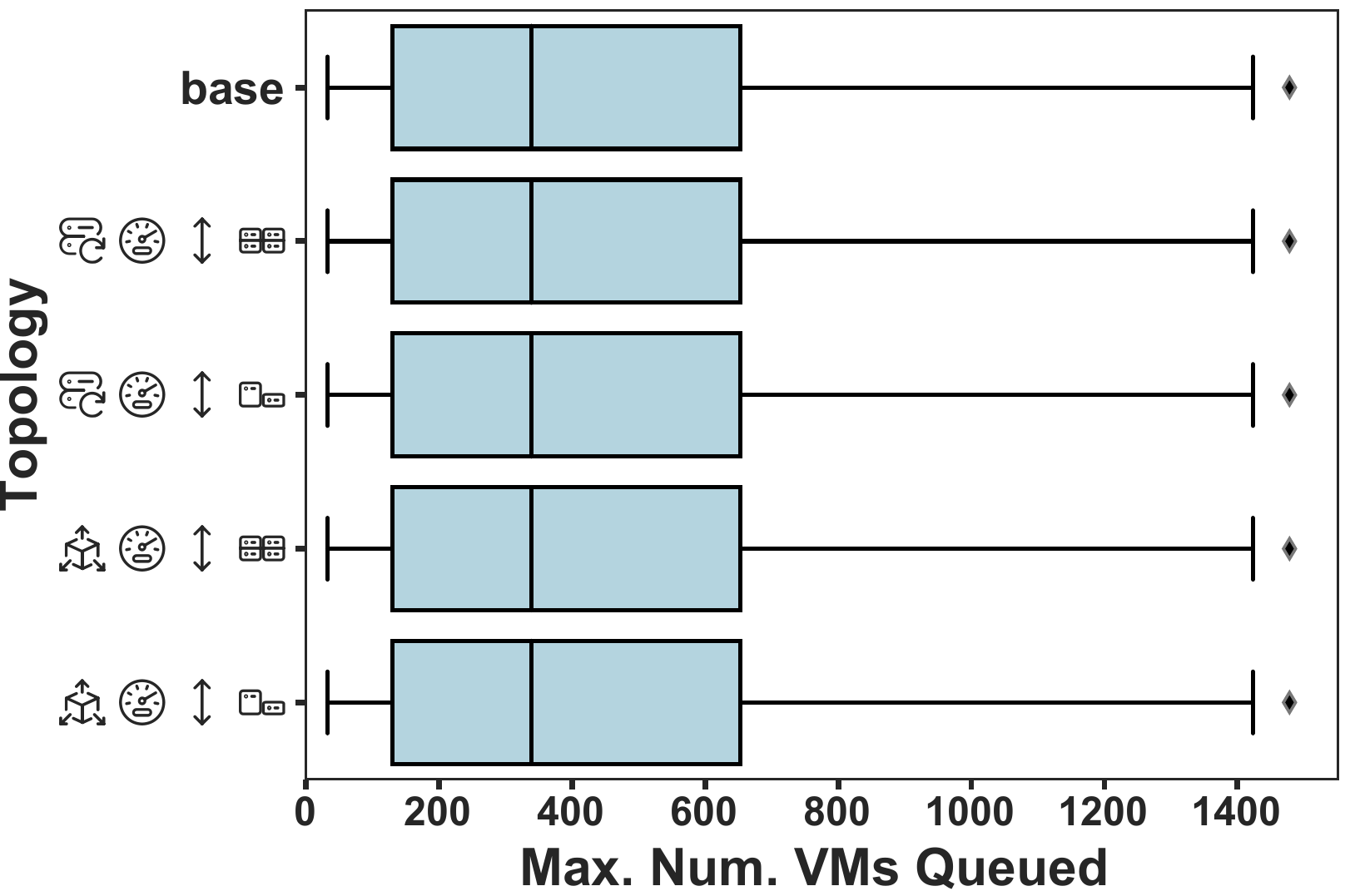}}\\
    \subfloat[Total VMs Finished\label{fig:full:more-vel:summary:vms-finished}]{\includegraphics[width=0.5\linewidth]{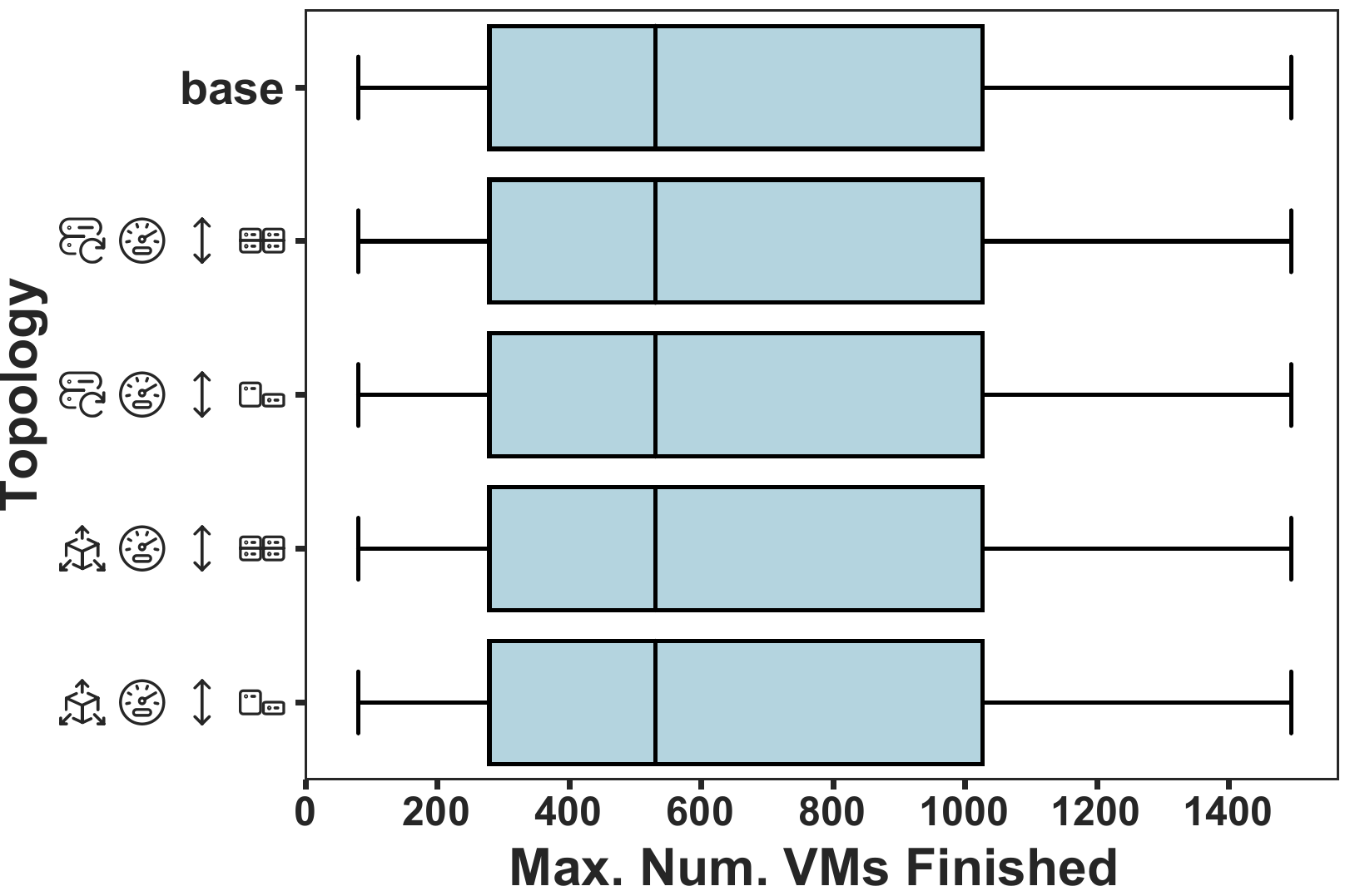}}%
    \subfloat[Total VMs Failed\label{fig:full:more-vel:summary:vms-failed}]{\includegraphics[width=0.5\linewidth]{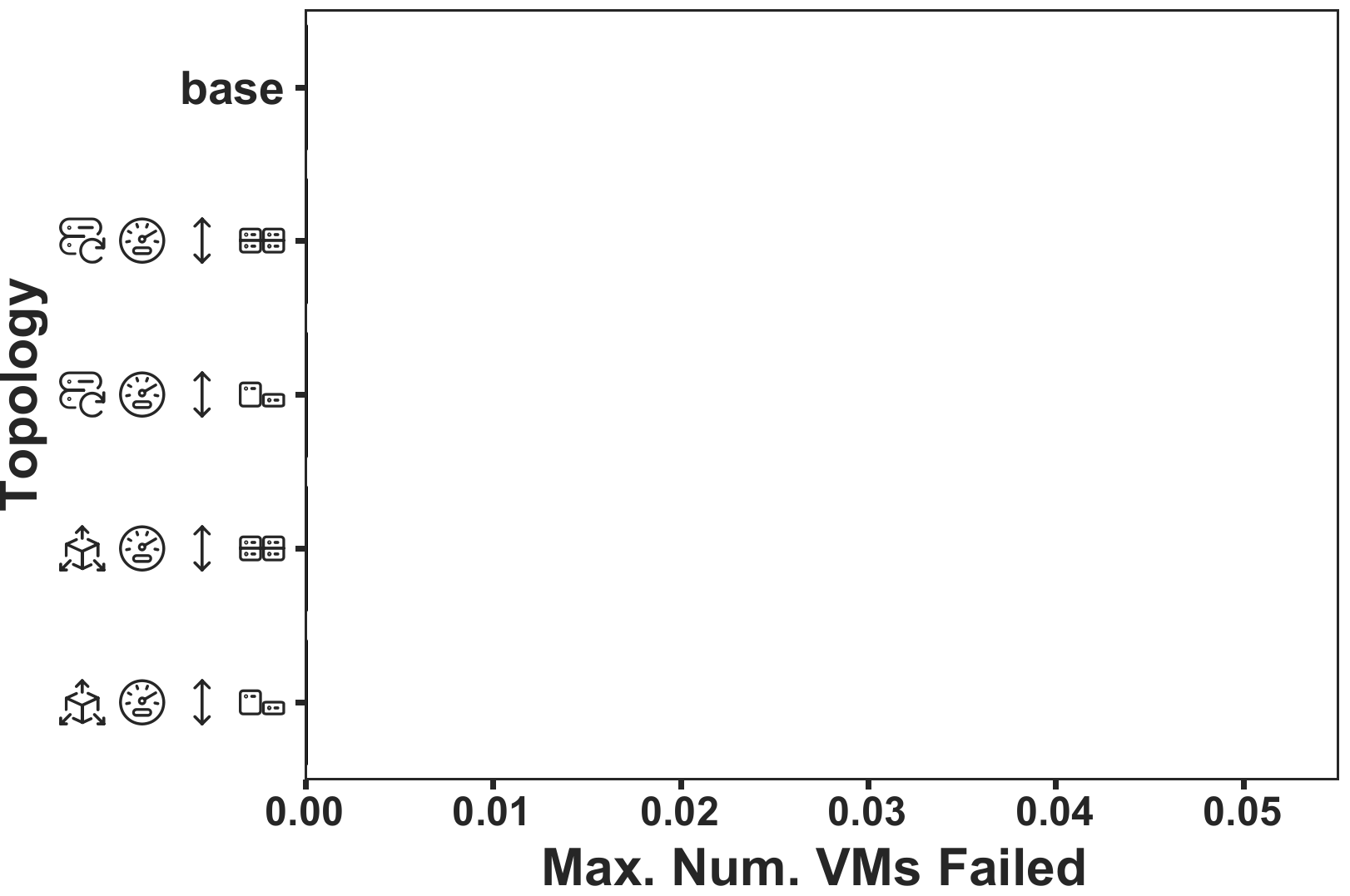}}%
    \caption{Performance of different topologies expanded on velocity, compared across workloads. Results aggregated across the full set of workloads, including workloads not displayed in the more detailed figure. For a legend of topologies, see Table~\ref{tab:experiment-overview}.}
    \label{fig:full:more-vel:summary:3}
\end{figure*}

\begin{figure*}
    \subfloat[Total VMs Submitted\label{fig:full:phenomena:vms-submitted}]{\includegraphics[width=0.5\linewidth]{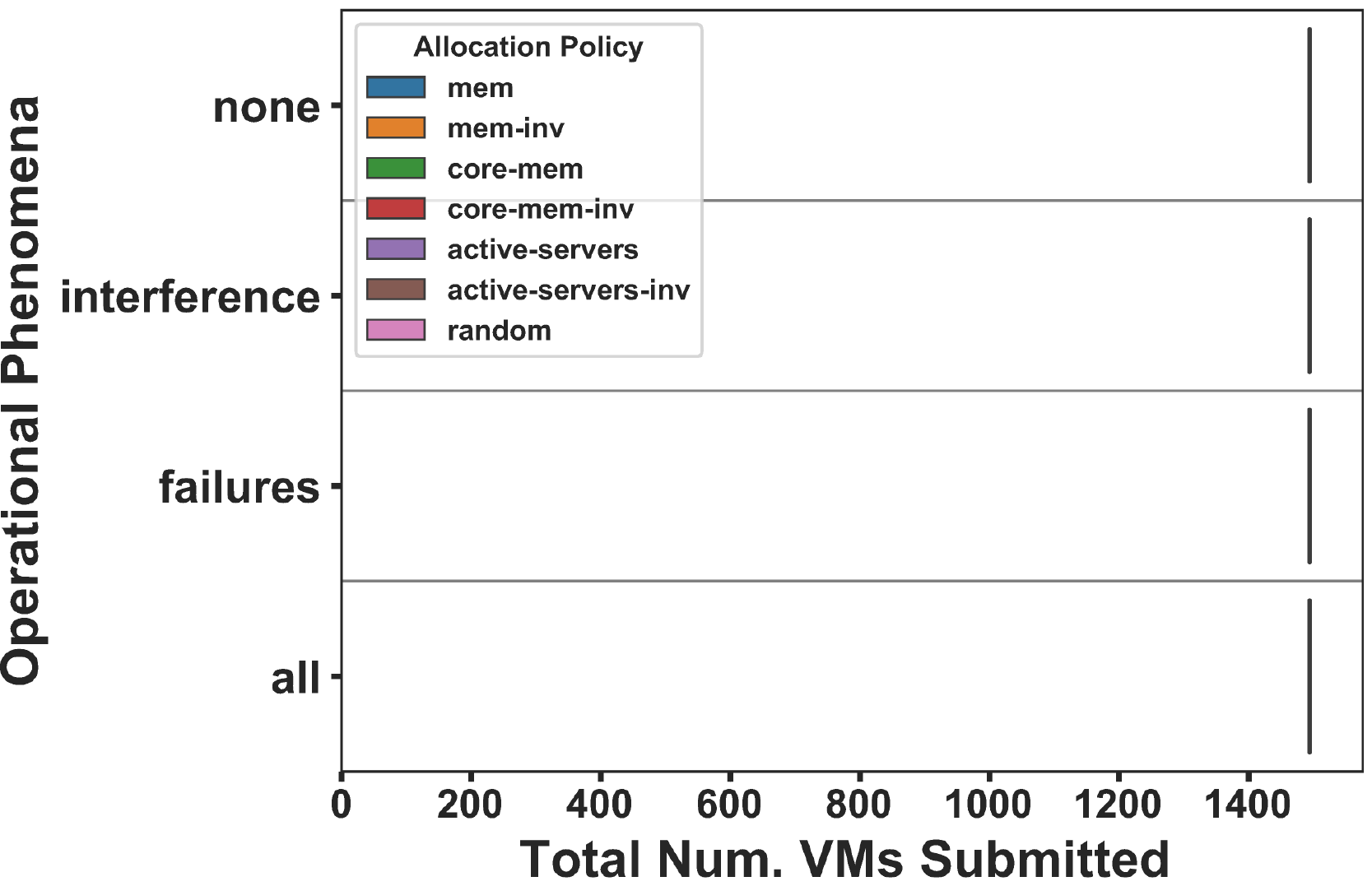}}%
    \subfloat[Total VMs Queued\label{fig:full:phenomena:vms-queued}]{\includegraphics[width=0.5\linewidth]{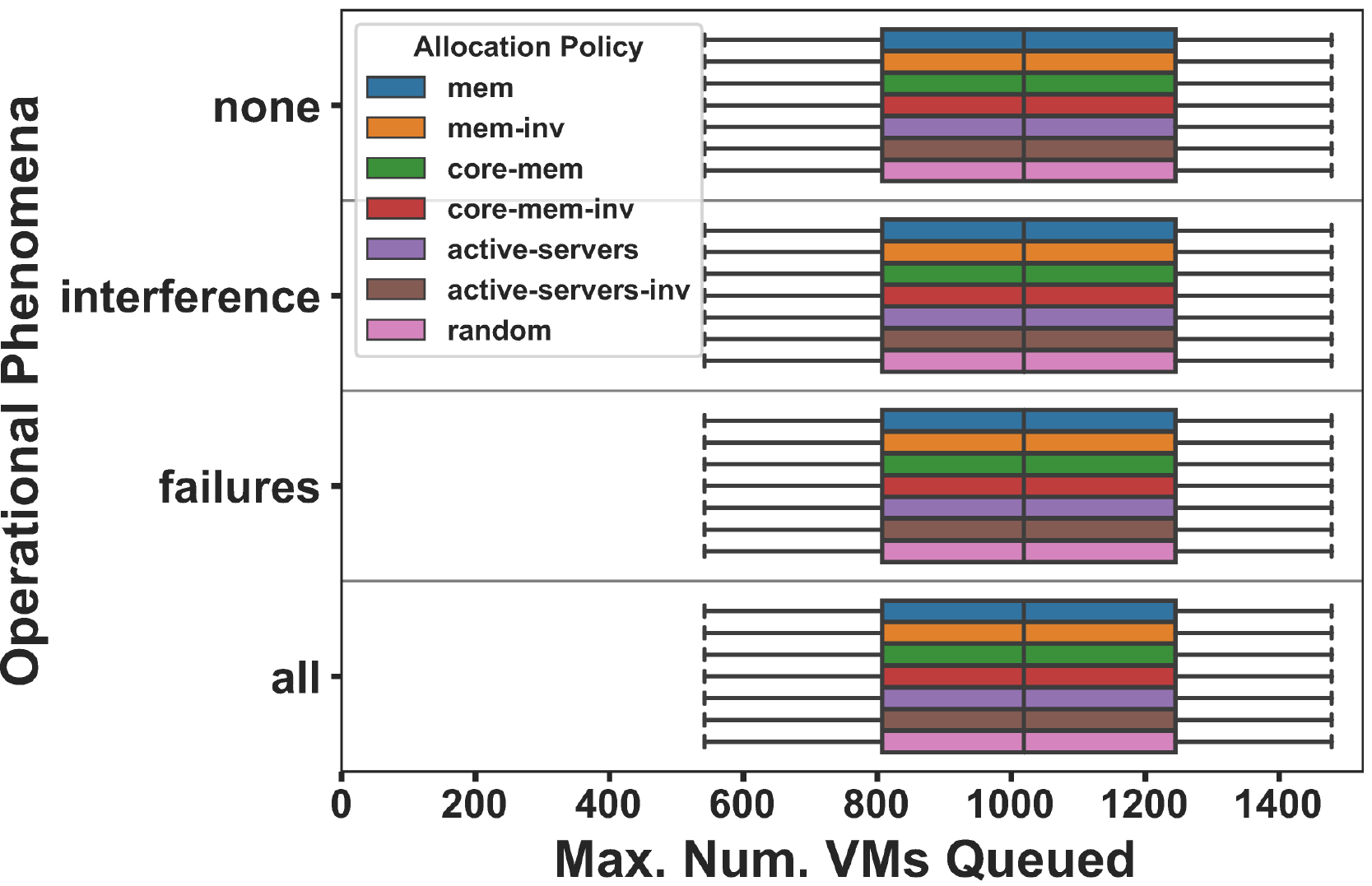}}\\
    \subfloat[Total VMs Finished\label{fig:full:phenomena:vms-finished}]{\includegraphics[width=0.5\linewidth]{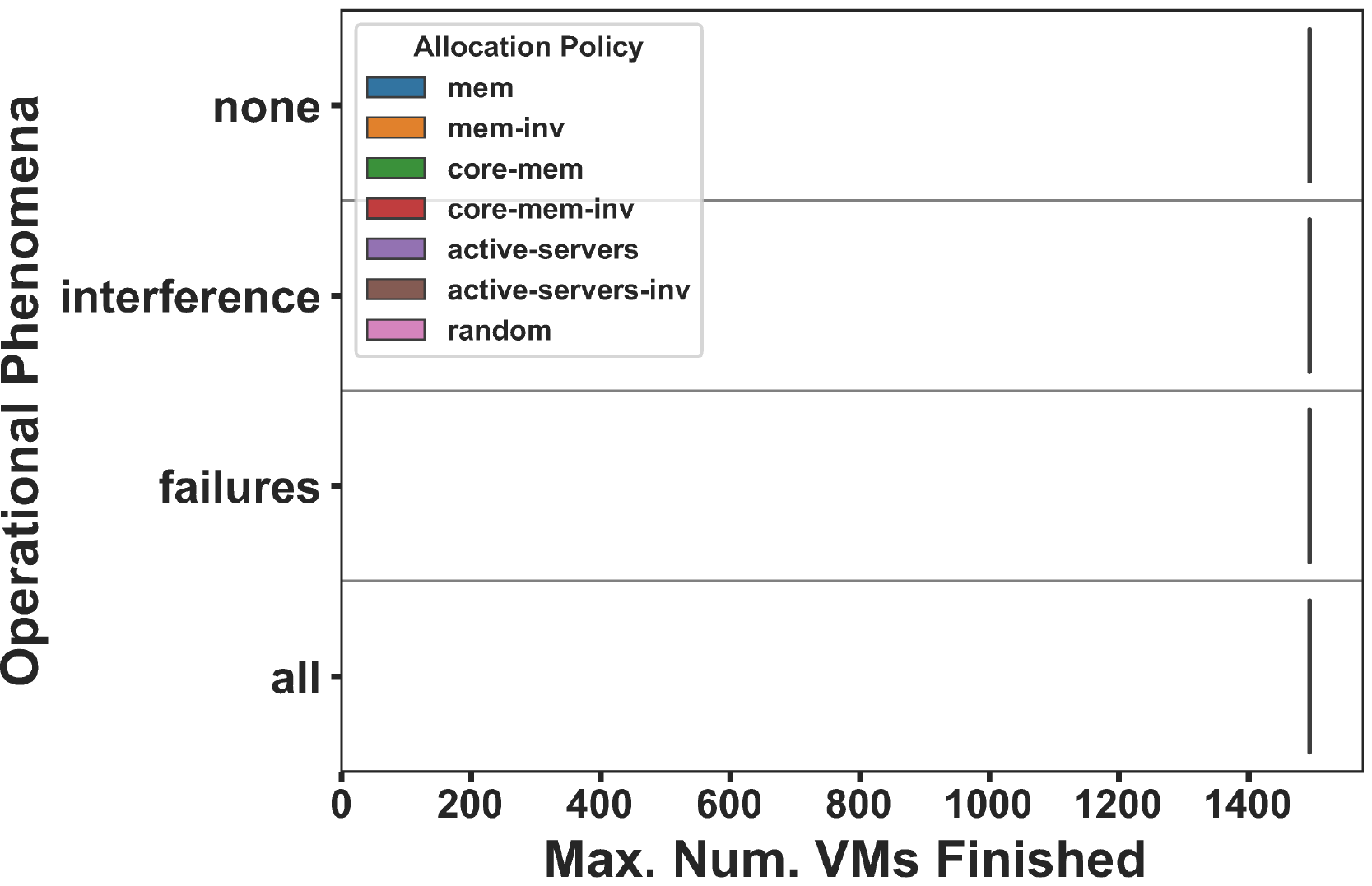}}%
    \subfloat[Total VMs Failed\label{fig:full:phenomena:vms-failed}]{\includegraphics[width=0.5\linewidth]{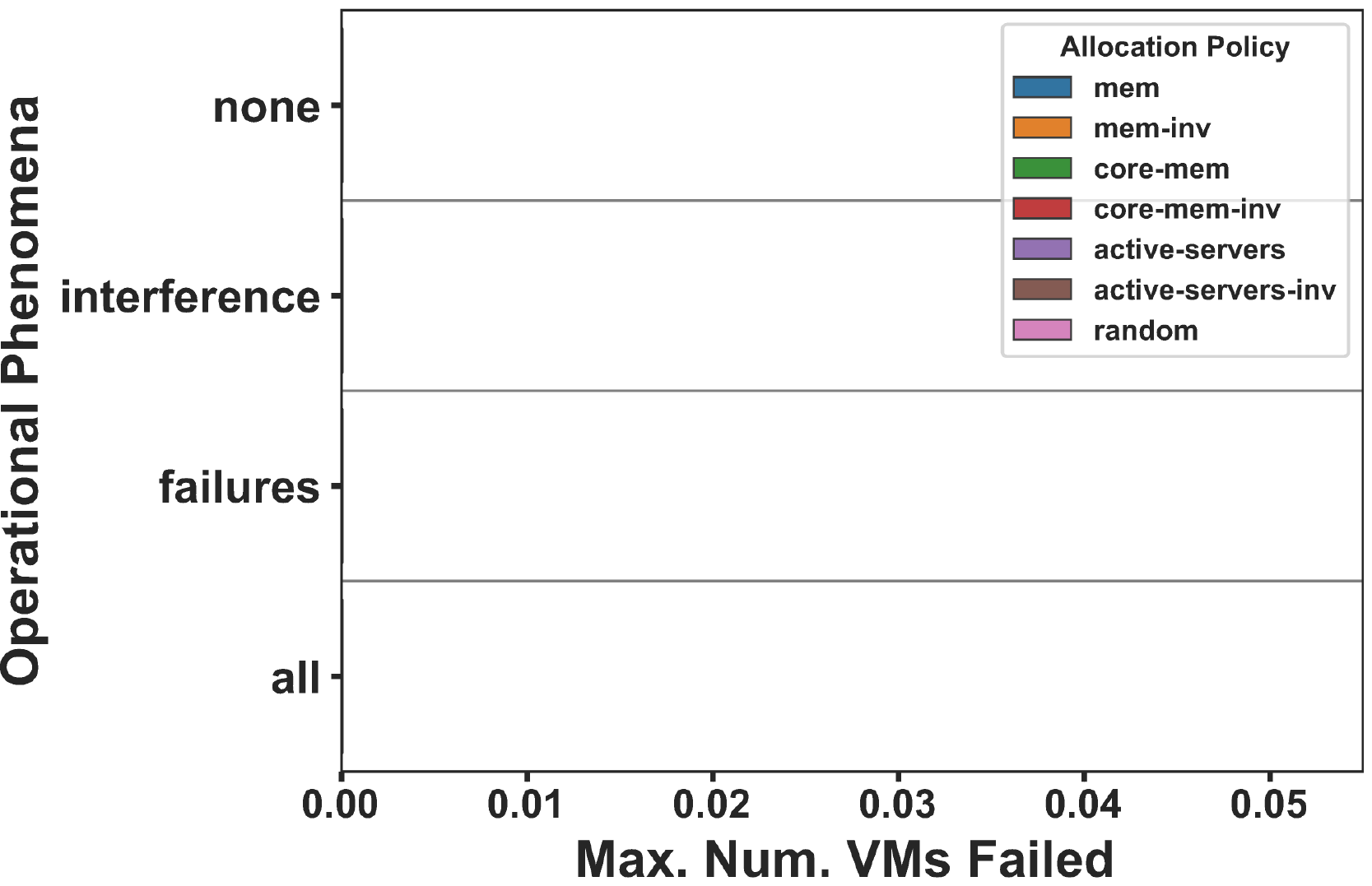}}%
    \caption{Impact of operational phenomena and different allocation policies on the base topology. For a legend of topologies, see Table~\ref{tab:experiment-overview}.}
    \label{fig:full:phenomena:3}
\end{figure*}

\begin{figure*}
    \centering
    \subfloat[Requested CPU cycles\label{fig:full:phenomena:summary:requested}]{\includegraphics[width=0.5\linewidth]{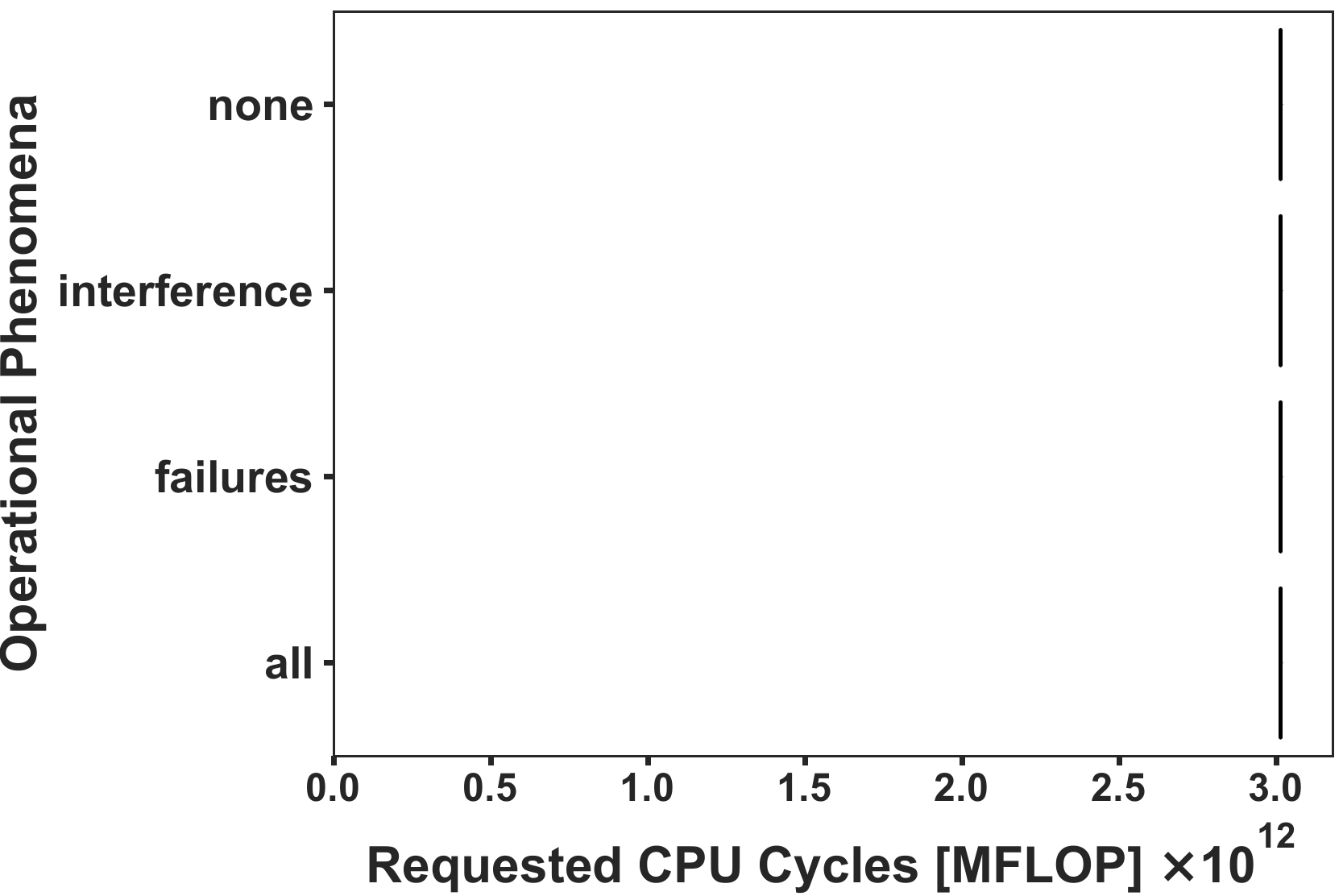}}%
    \subfloat[Granted CPU cycles\label{fig:full:phenomena:summary:granted}]{\includegraphics[width=0.5\linewidth]{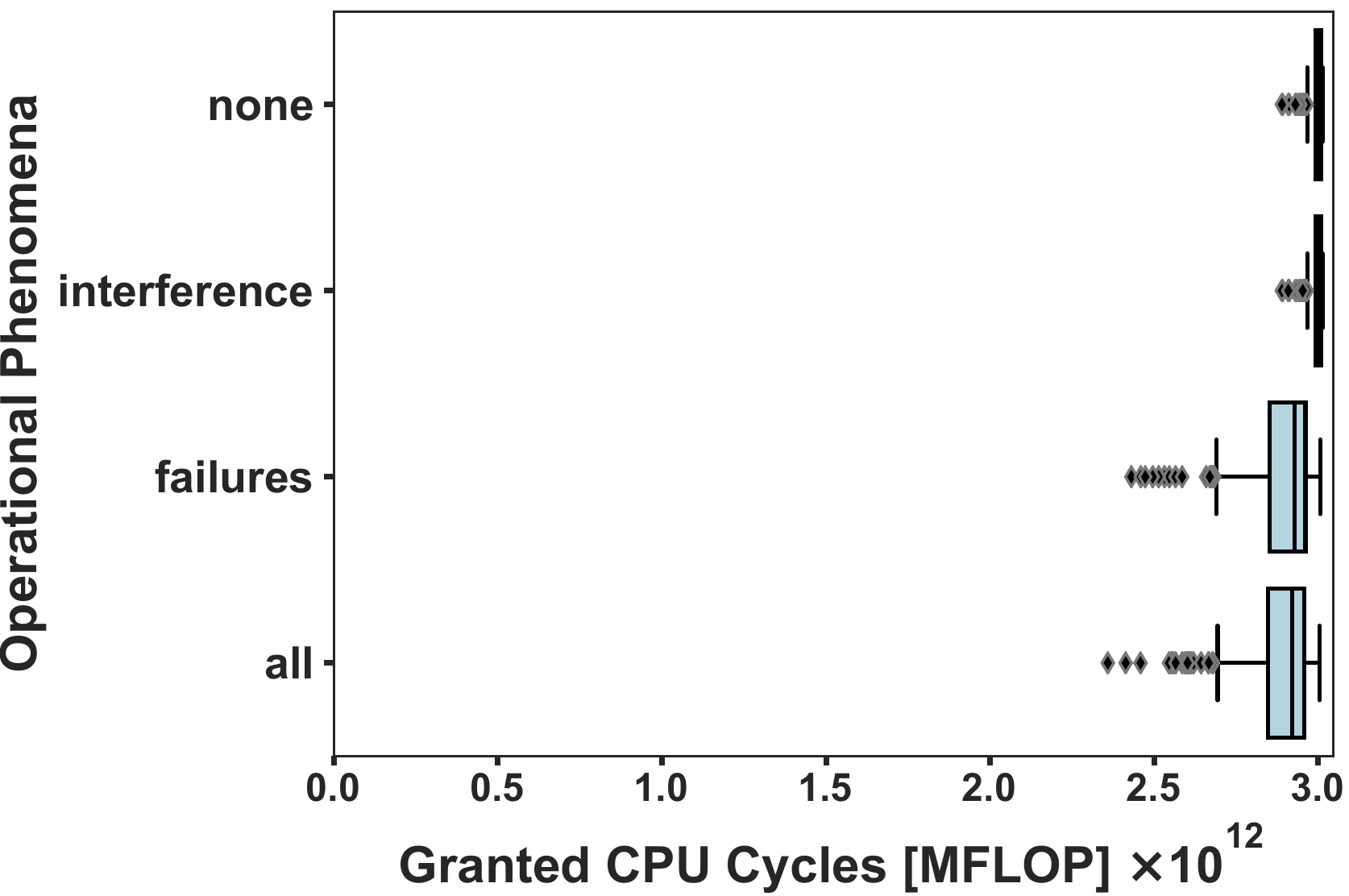}}\\
    \subfloat[Overcommitted CPU cycles\label{fig:full:phenomena:summary:overcommitted}]{\includegraphics[width=0.5\linewidth]{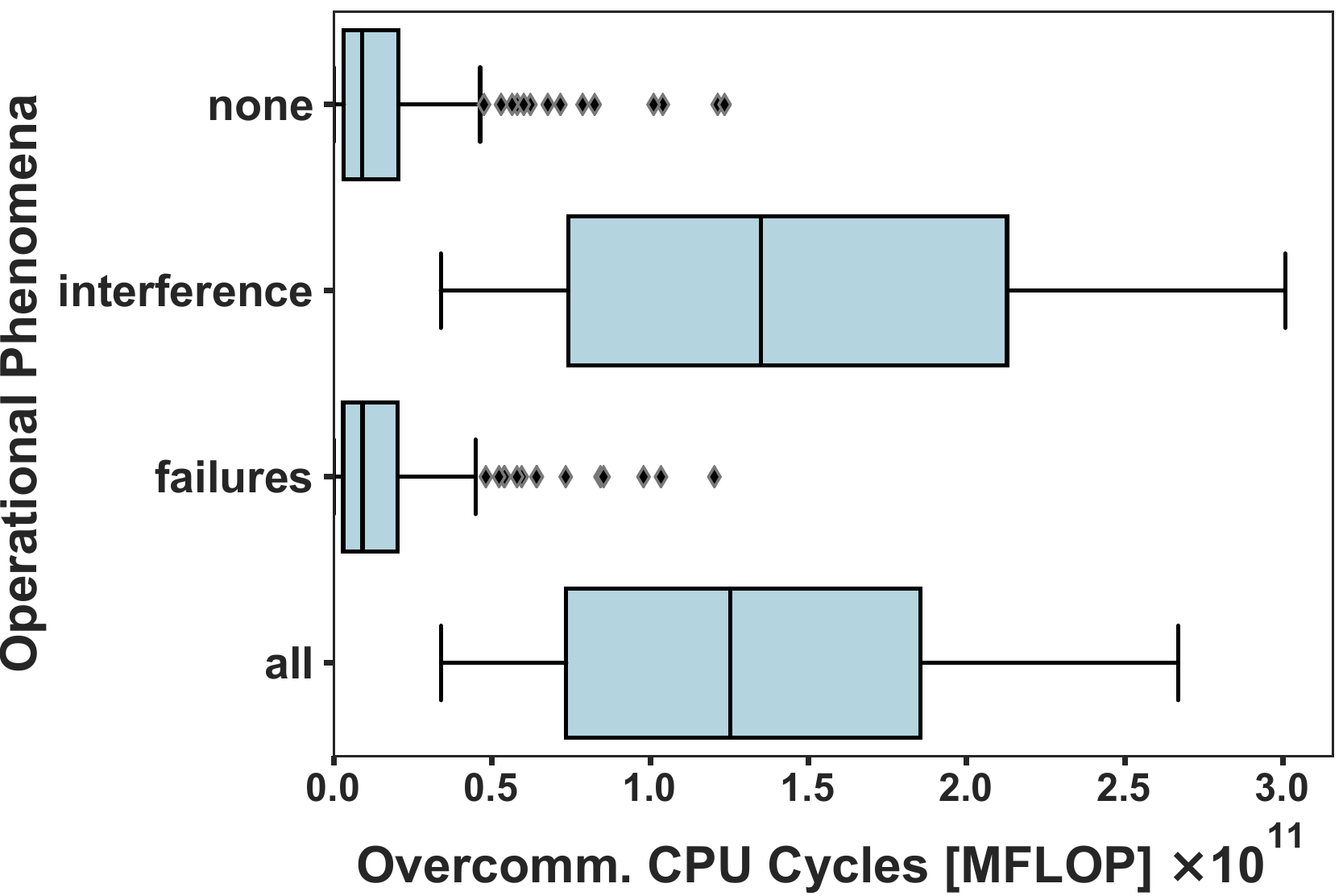}}%
    \subfloat[Interfered CPU cycles\label{fig:full:phenomena:summary:interfered}]{\includegraphics[width=0.5\linewidth]{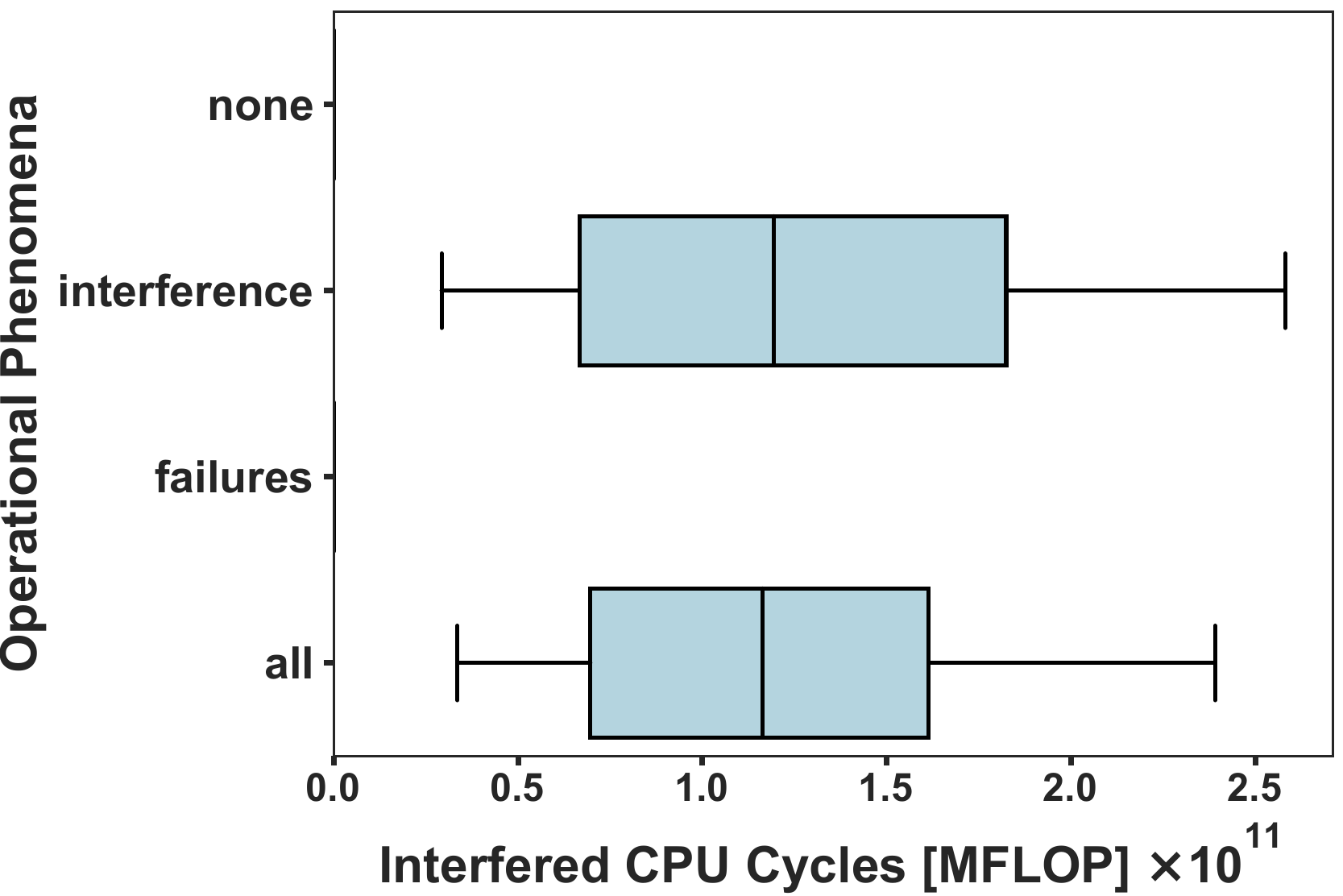}}\\
    \subfloat[Total power consumption\label{fig:full:phenomena:summary:power}]{\includegraphics[width=0.5\linewidth]{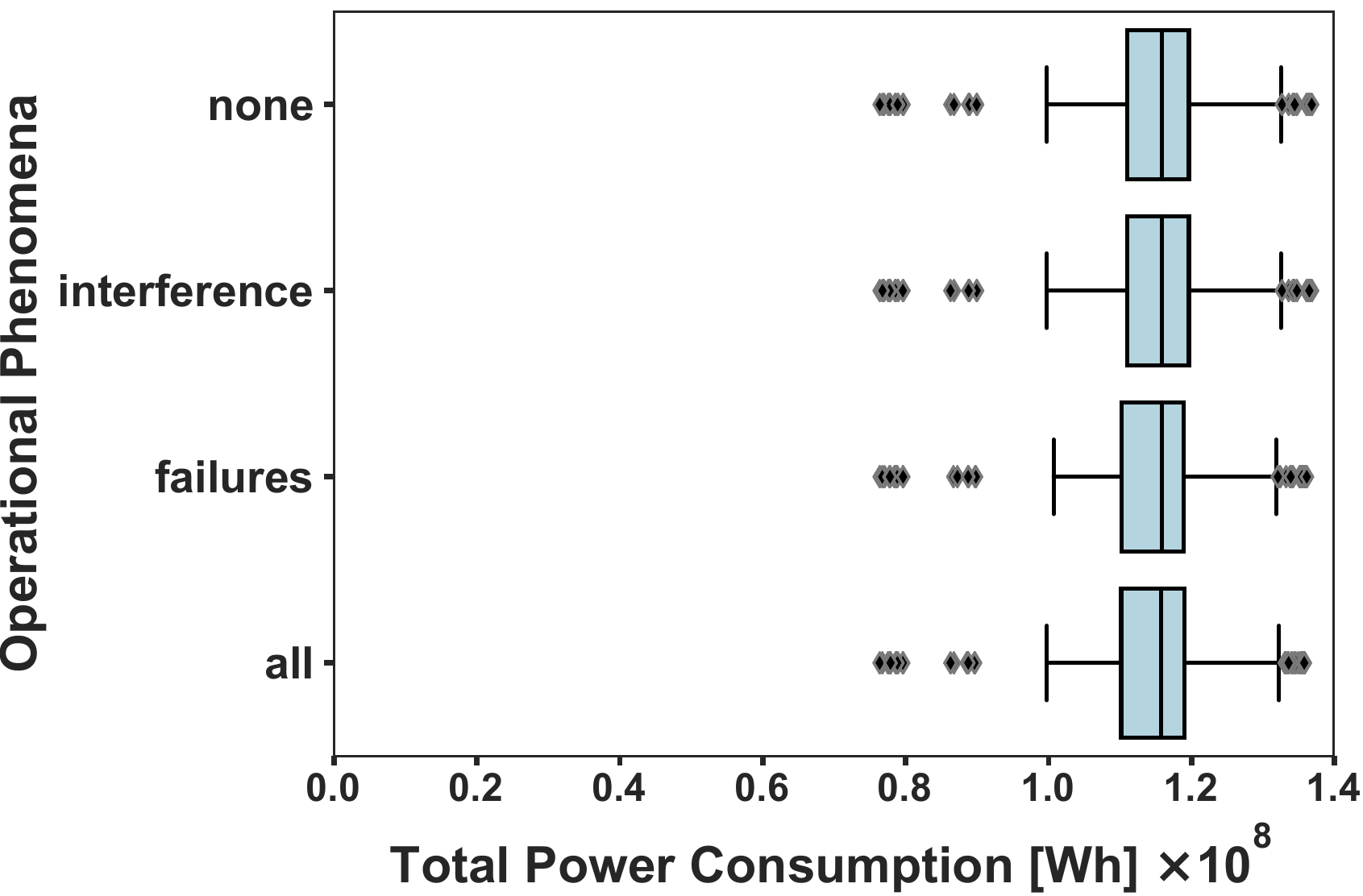}}%
    \subfloat[Total number of time slices in which a \gls{VM} is failed, aggregated across \glspl{VM}\label{fig:full:phenomena:summary:failures:vms}]{\includegraphics[width=0.5\linewidth]{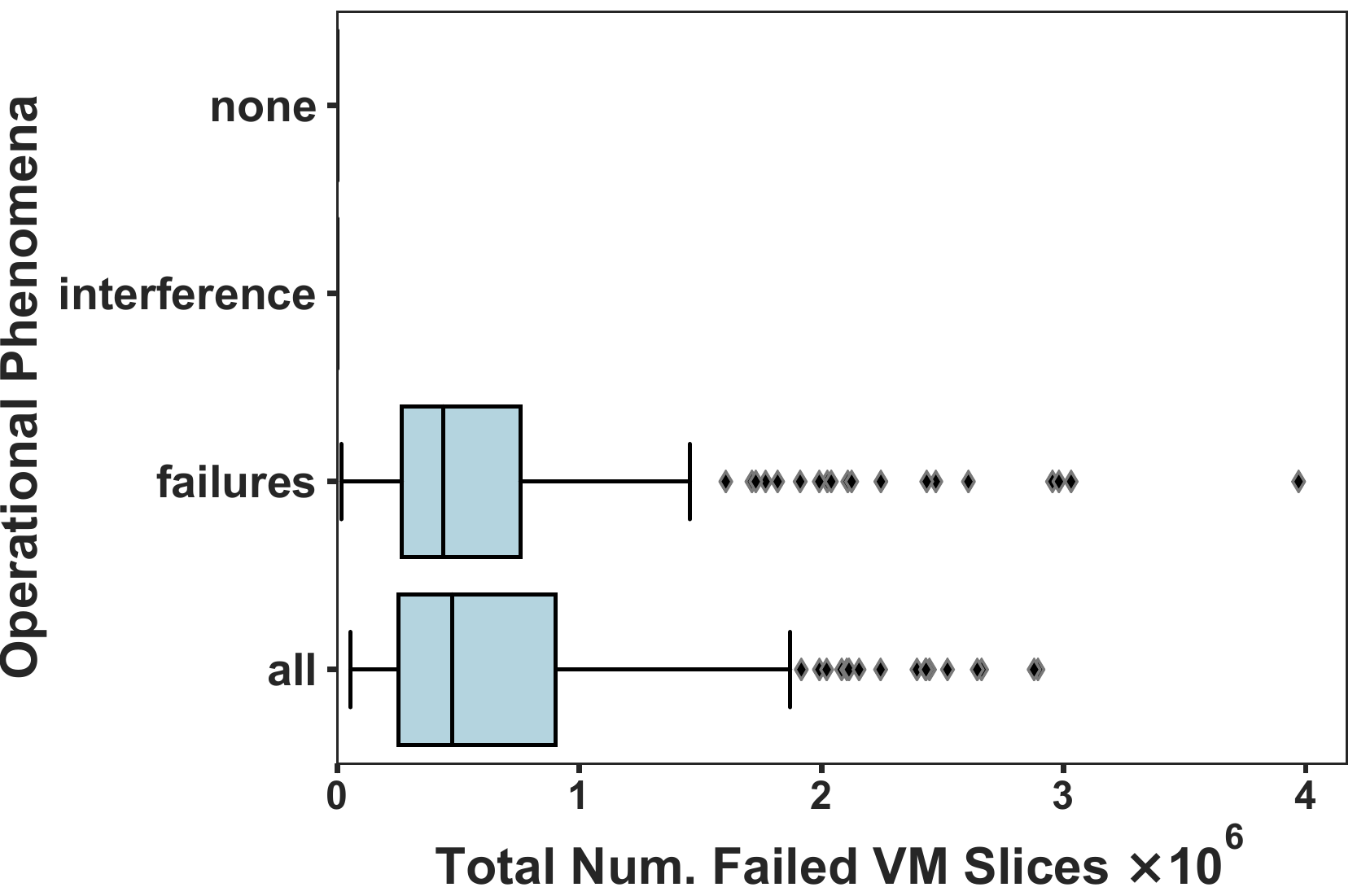}}\\
    \caption{Impact of operational phenomena and different allocation policies on the base topology. For a legend of topologies, see Table~\ref{tab:experiment-overview}. Continued in Figure~\ref{fig:full:phenomena:summary:2}.}
    \label{fig:full:phenomena:summary:1}
\end{figure*}

\begin{figure*}
    \subfloat[Mean CPU usage\label{fig:full:phenomena:summary:cpu-usage}]{\includegraphics[width=0.5\linewidth]{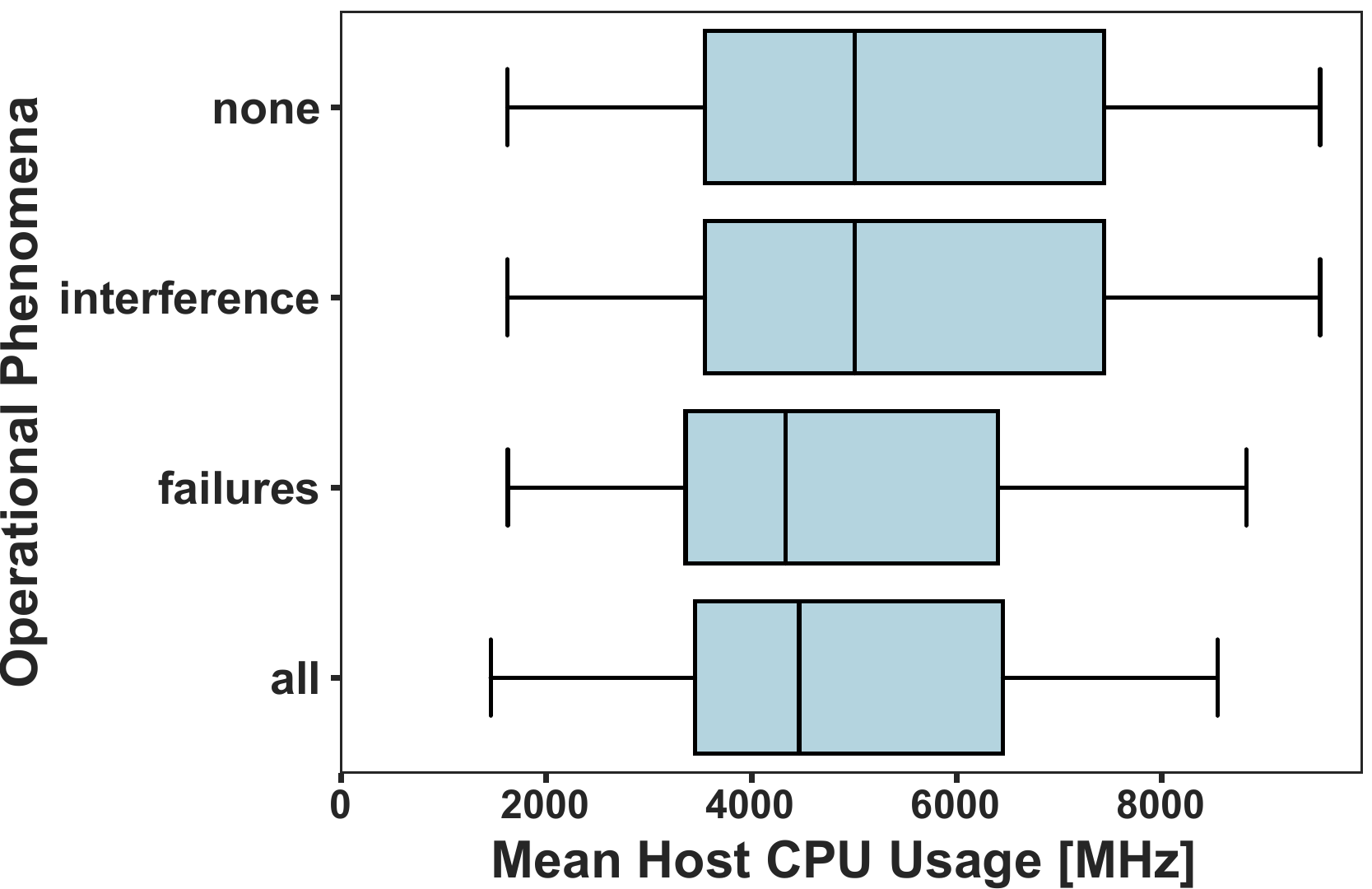}}%
    \subfloat[Mean CPU demand\label{fig:full:phenomena:summary:cpu-demand}]{\includegraphics[width=0.5\linewidth]{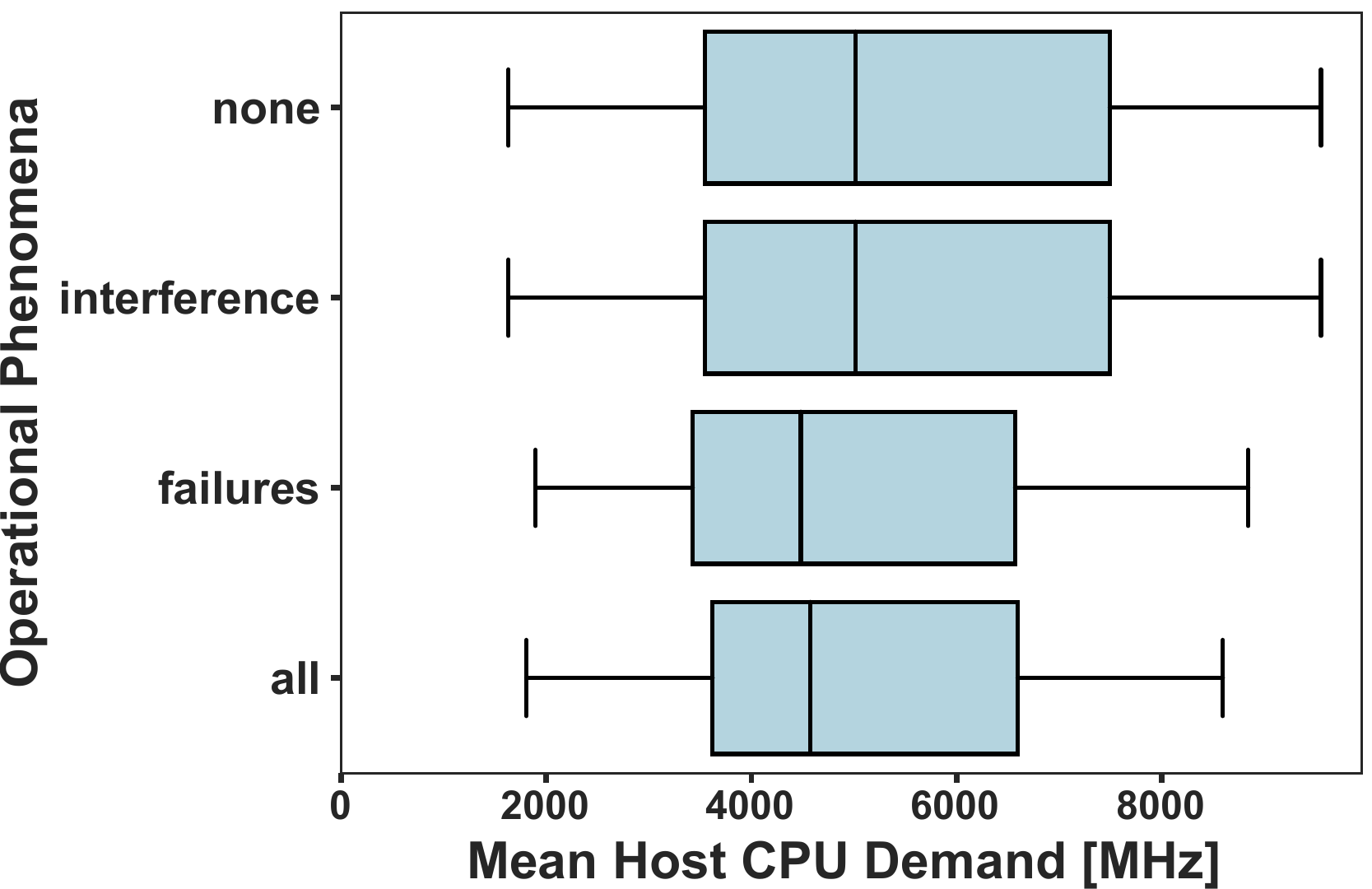}}\\
    \subfloat[Mean number of \glspl{VM} per host\label{fig:full:phenomena:summary:mean-vm-count}]{\includegraphics[width=0.5\linewidth]{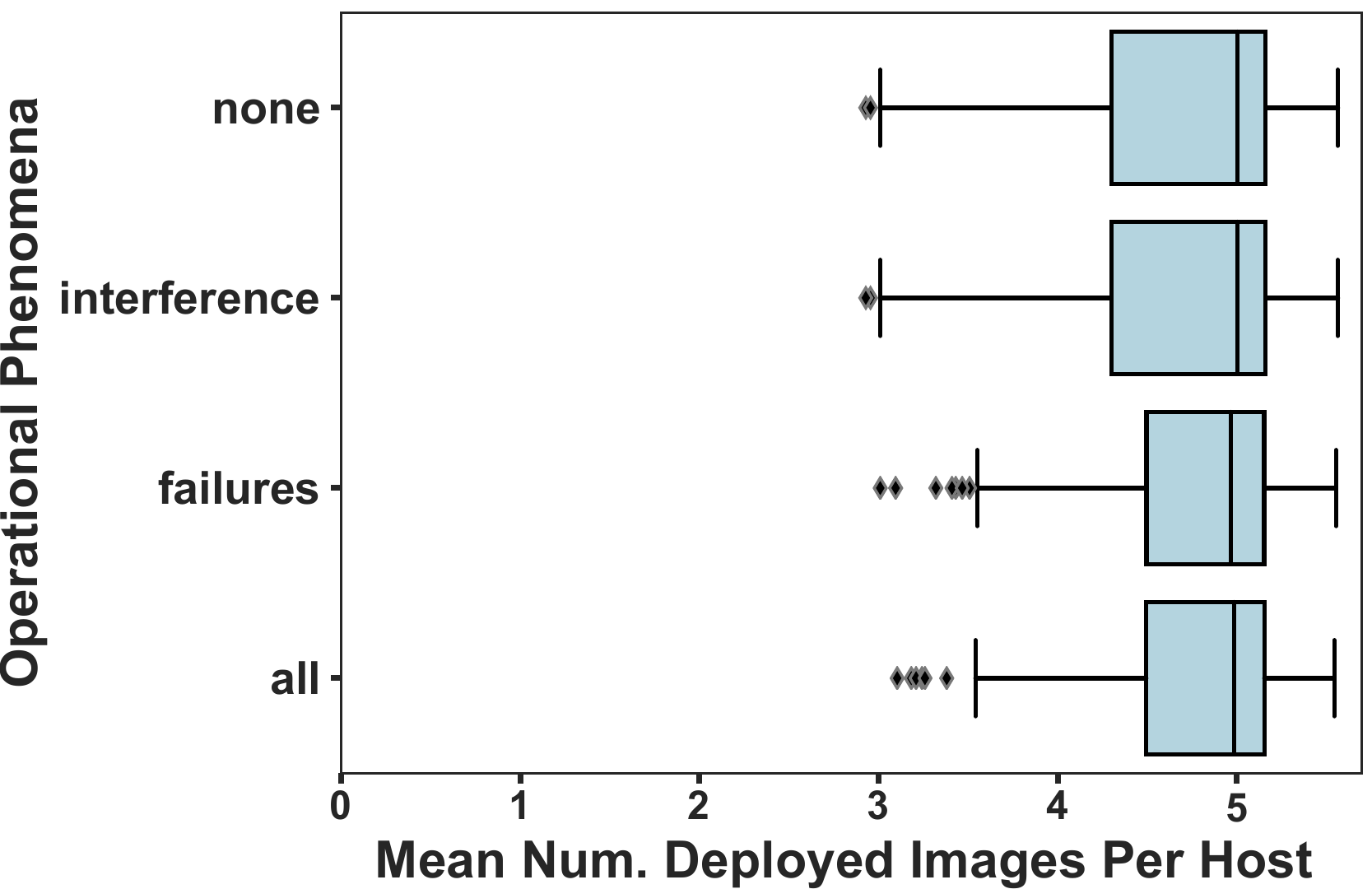}}%
    \subfloat[Max number of \glspl{VM} per host\label{fig:full:phenomena:summary:max-vm-count}]{\includegraphics[width=0.5\linewidth]{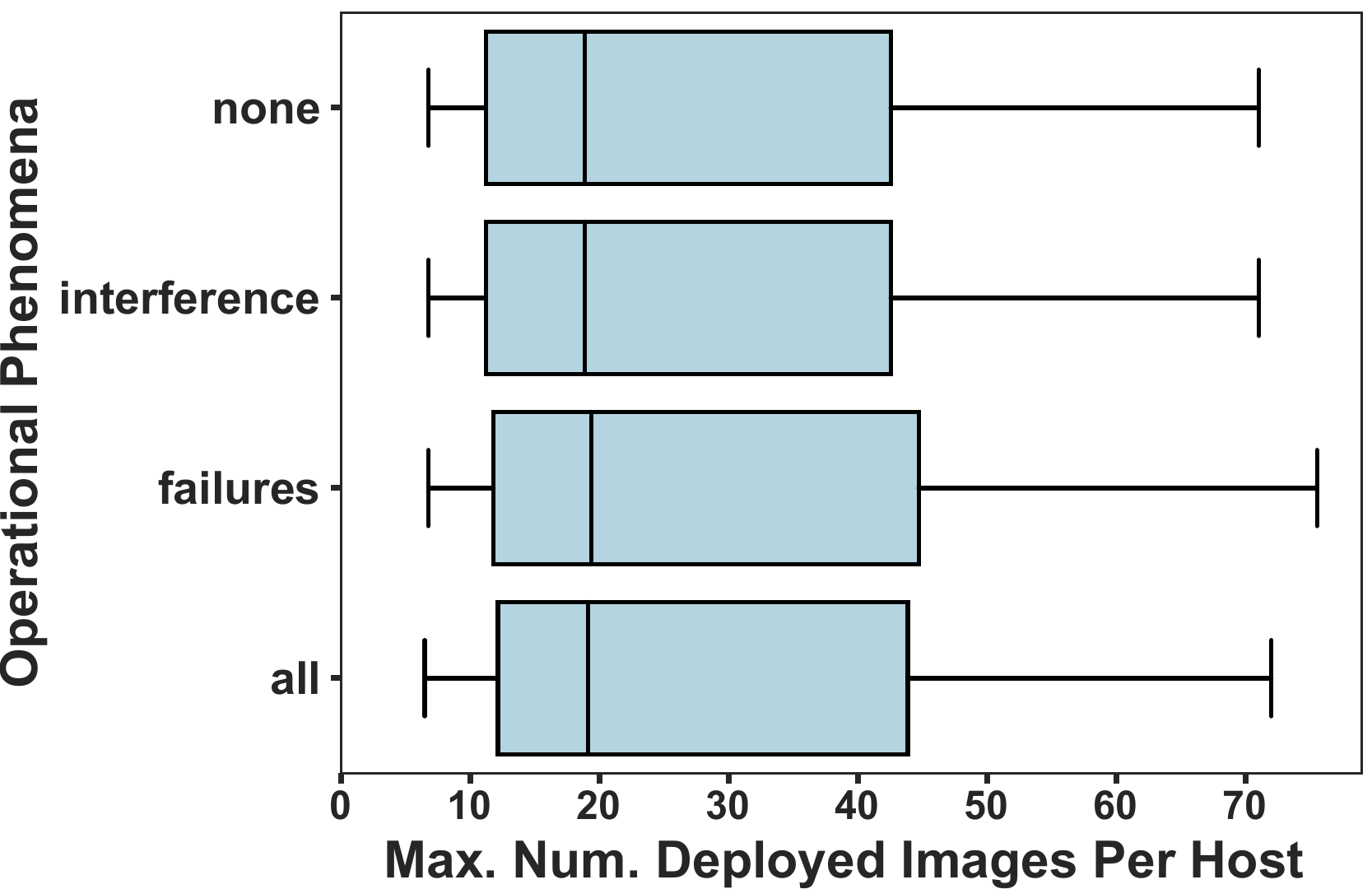}}\\
    \caption{Impact of operational phenomena and different allocation policies on the base topology. For a legend of topologies, see Table~\ref{tab:experiment-overview}. Continued in Figure~\ref{fig:full:phenomena:summary:3}.}
    \label{fig:full:phenomena:summary:2}
\end{figure*}

\begin{figure*}
    \subfloat[Total VMs Submitted\label{fig:full:phenomena:summary:vms-submitted}]{\includegraphics[width=0.5\linewidth]{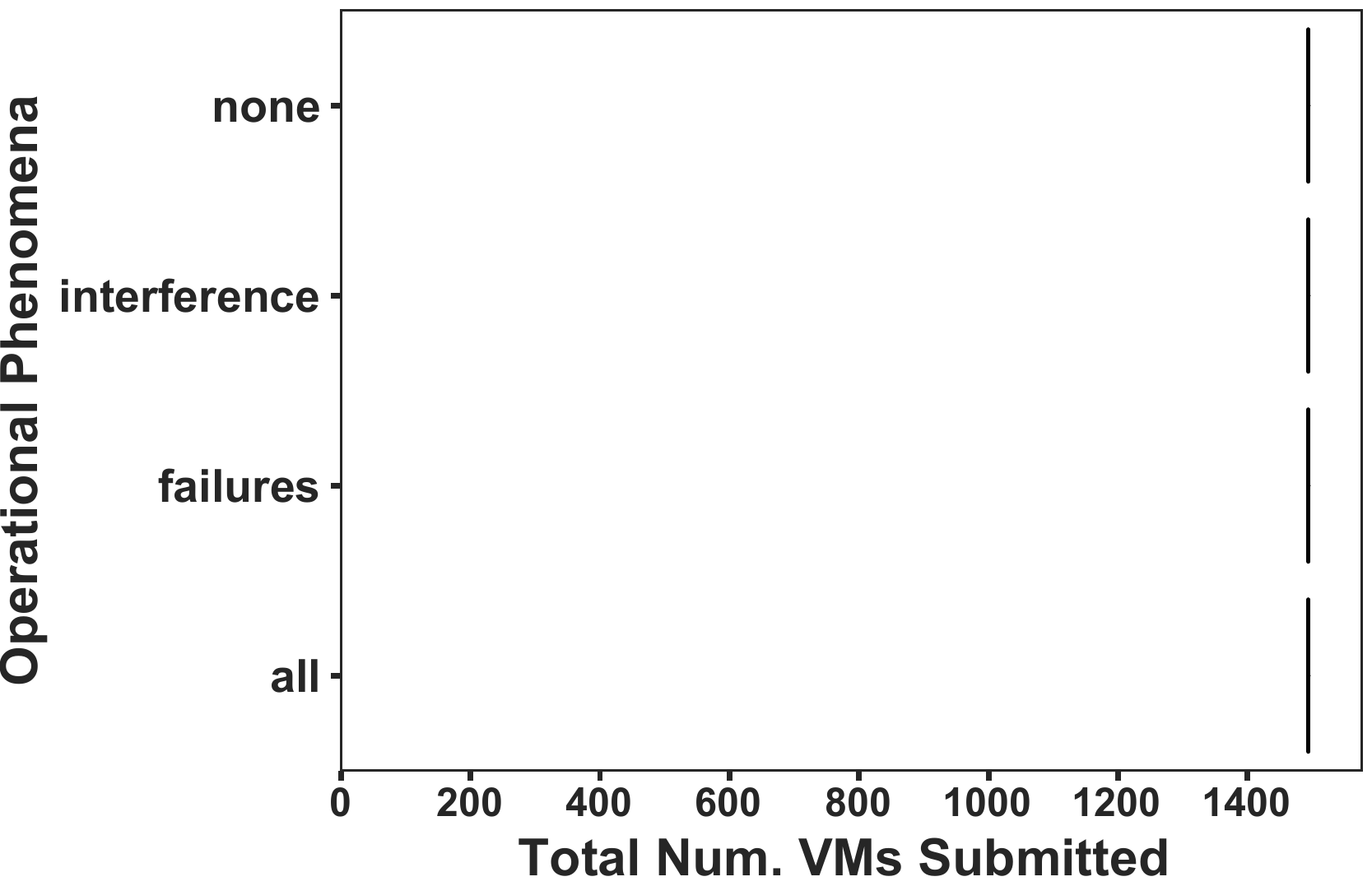}}%
    \subfloat[Total VMs Queued\label{fig:full:phenomena:summary:vms-queued}]{\includegraphics[width=0.5\linewidth]{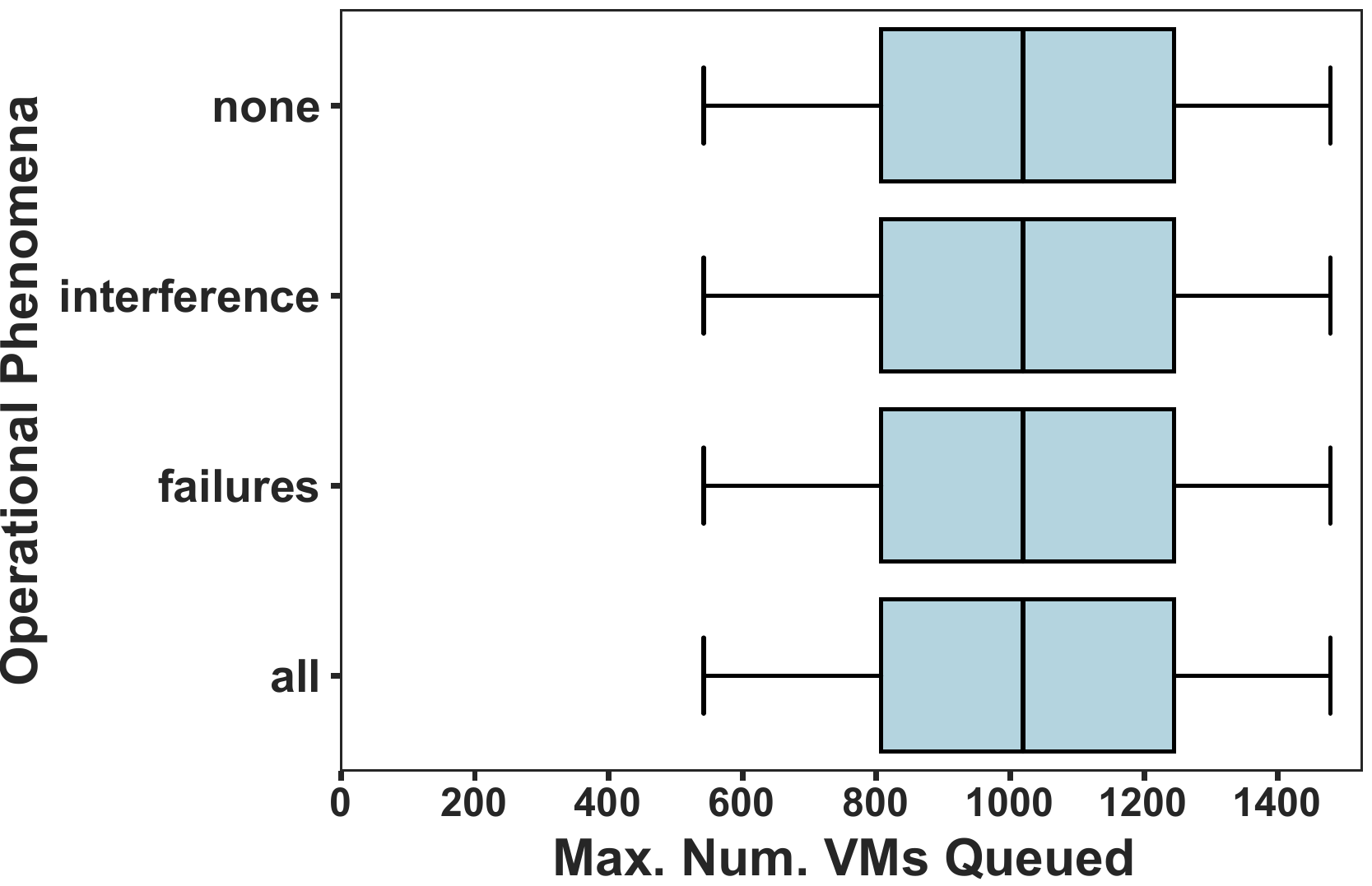}}\\
    \subfloat[Total VMs Finished\label{fig:full:phenomena:summary:vms-finished}]{\includegraphics[width=0.5\linewidth]{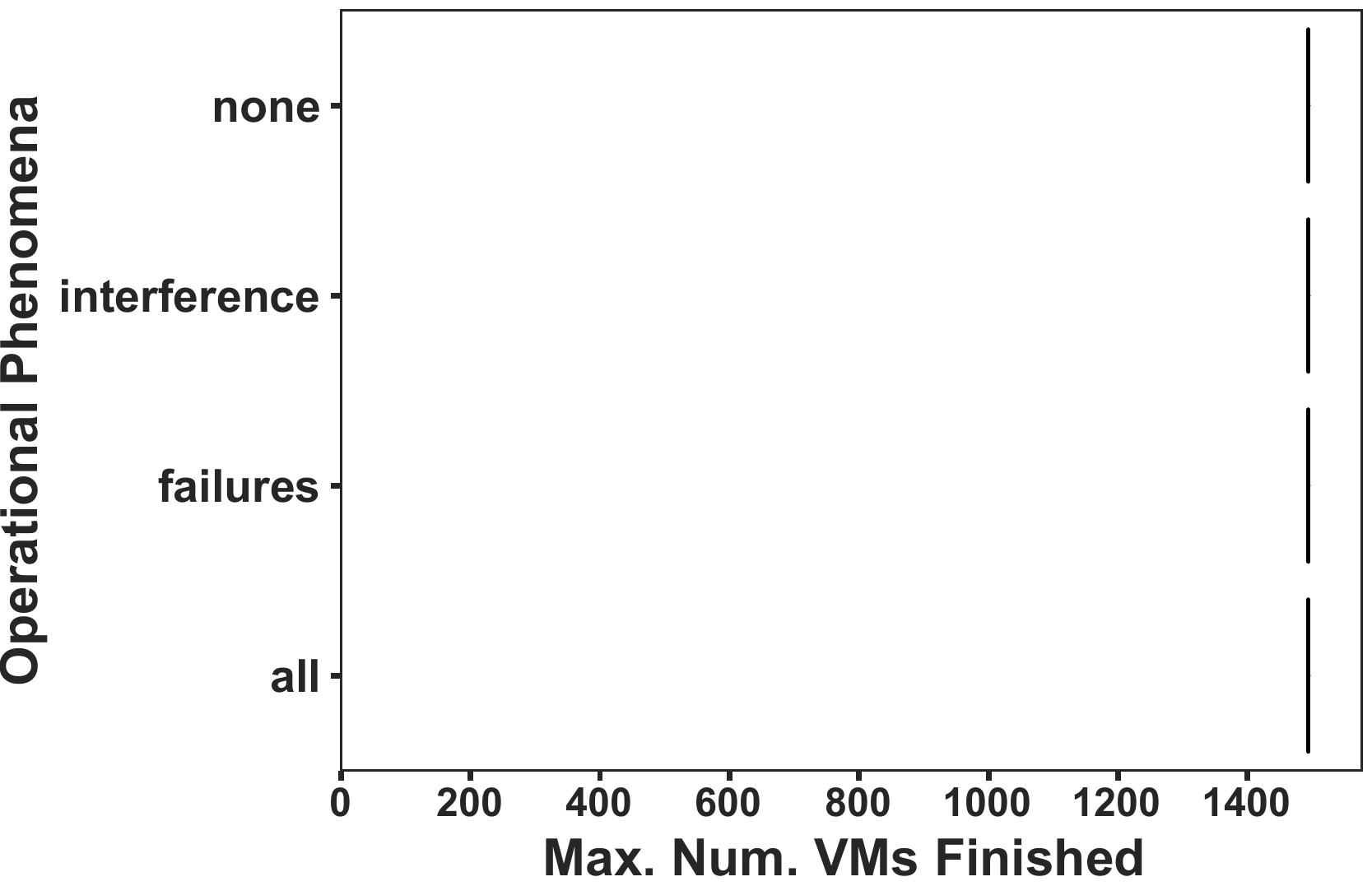}}%
    \subfloat[Total VMs Failed\label{fig:full:phenomena:summary:vms-failed}]{\includegraphics[width=0.5\linewidth]{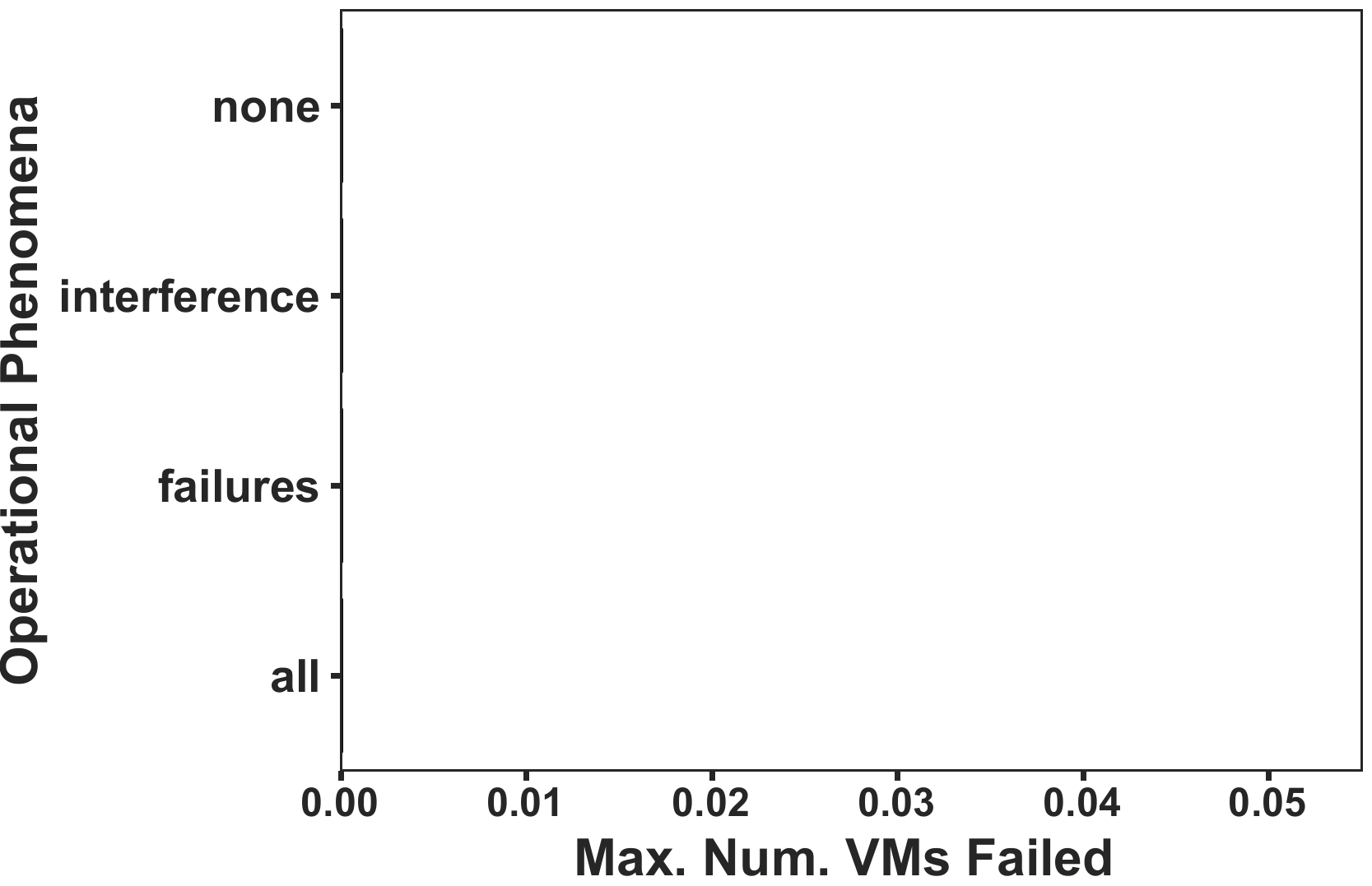}}%
    \caption{Impact of operational phenomena and different allocation policies on the base topology. For a legend of topologies, see Table~\ref{tab:experiment-overview}.}
    \label{fig:full:phenomena:summary:3}
\end{figure*}

\begin{figure*}
    \subfloat[Mean CPU usage\label{fig:full:composite:cpu-usage}]{\includegraphics[width=0.5\linewidth]{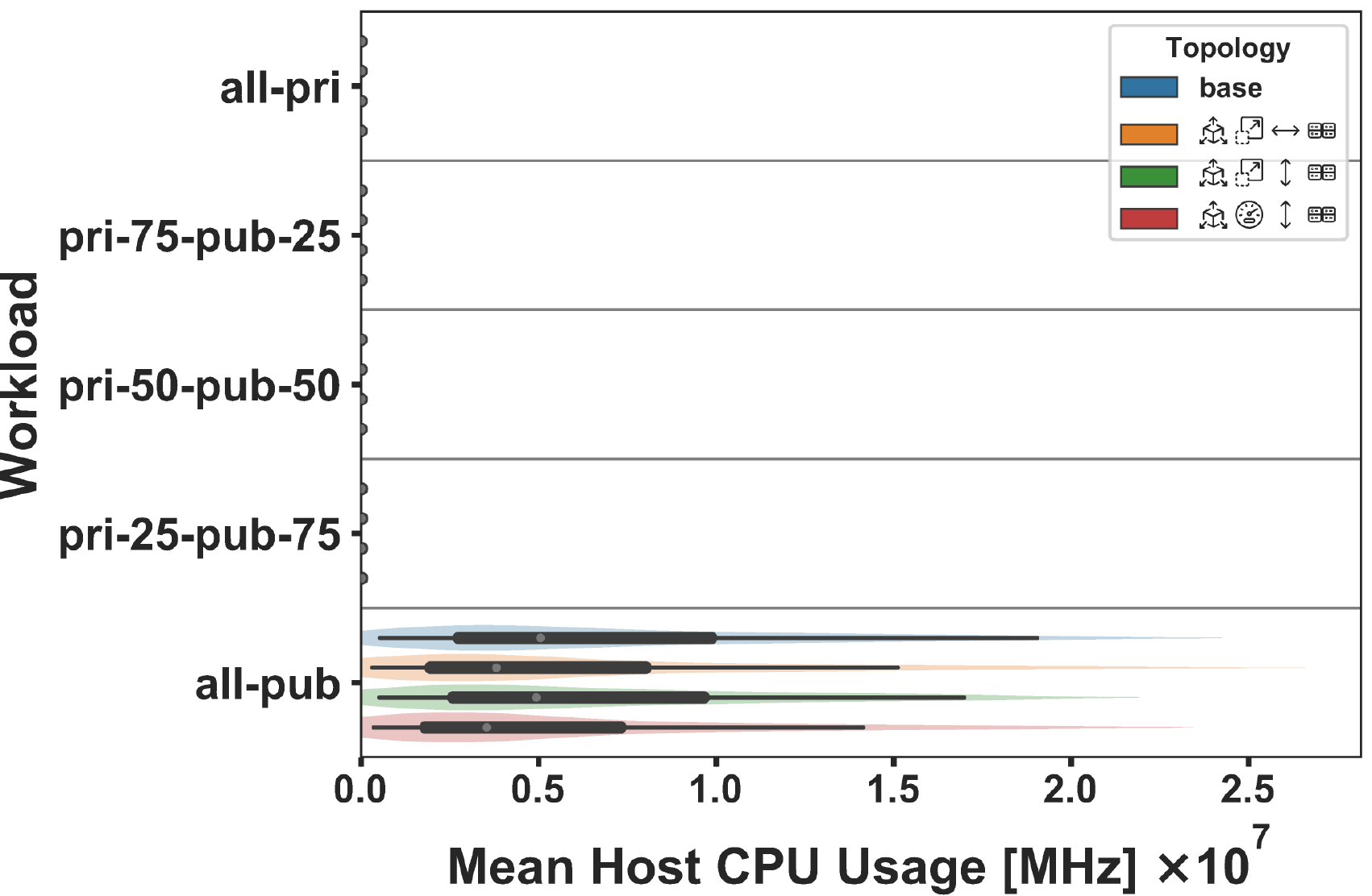}}%
    \subfloat[Mean CPU demand\label{fig:full:composite:cpu-demand}]{\includegraphics[width=0.5\linewidth]{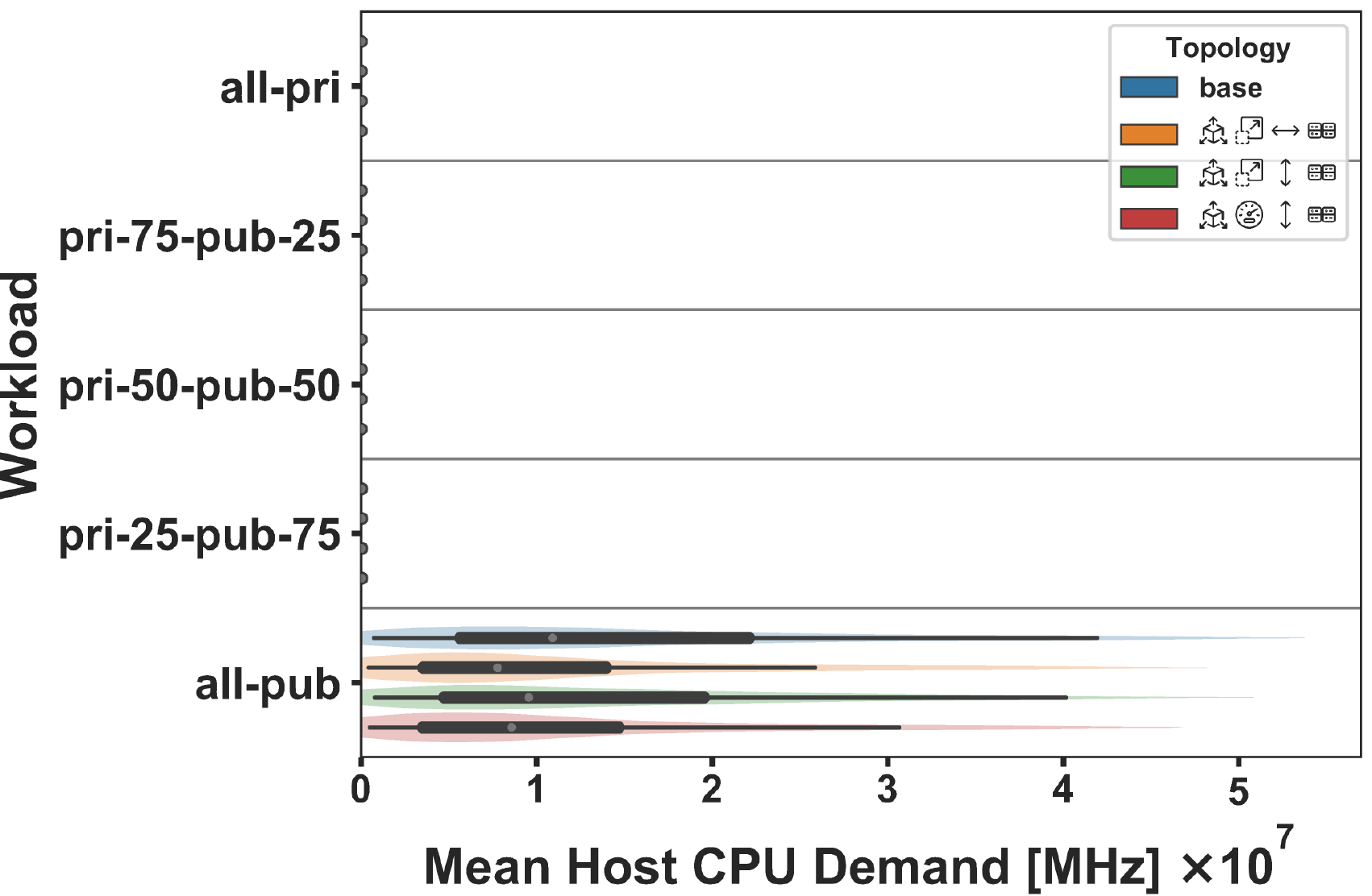}}\\
    \subfloat[Mean number of \glspl{VM} per host\label{fig:full:composite:mean-vm-count}]{\includegraphics[width=0.5\linewidth]{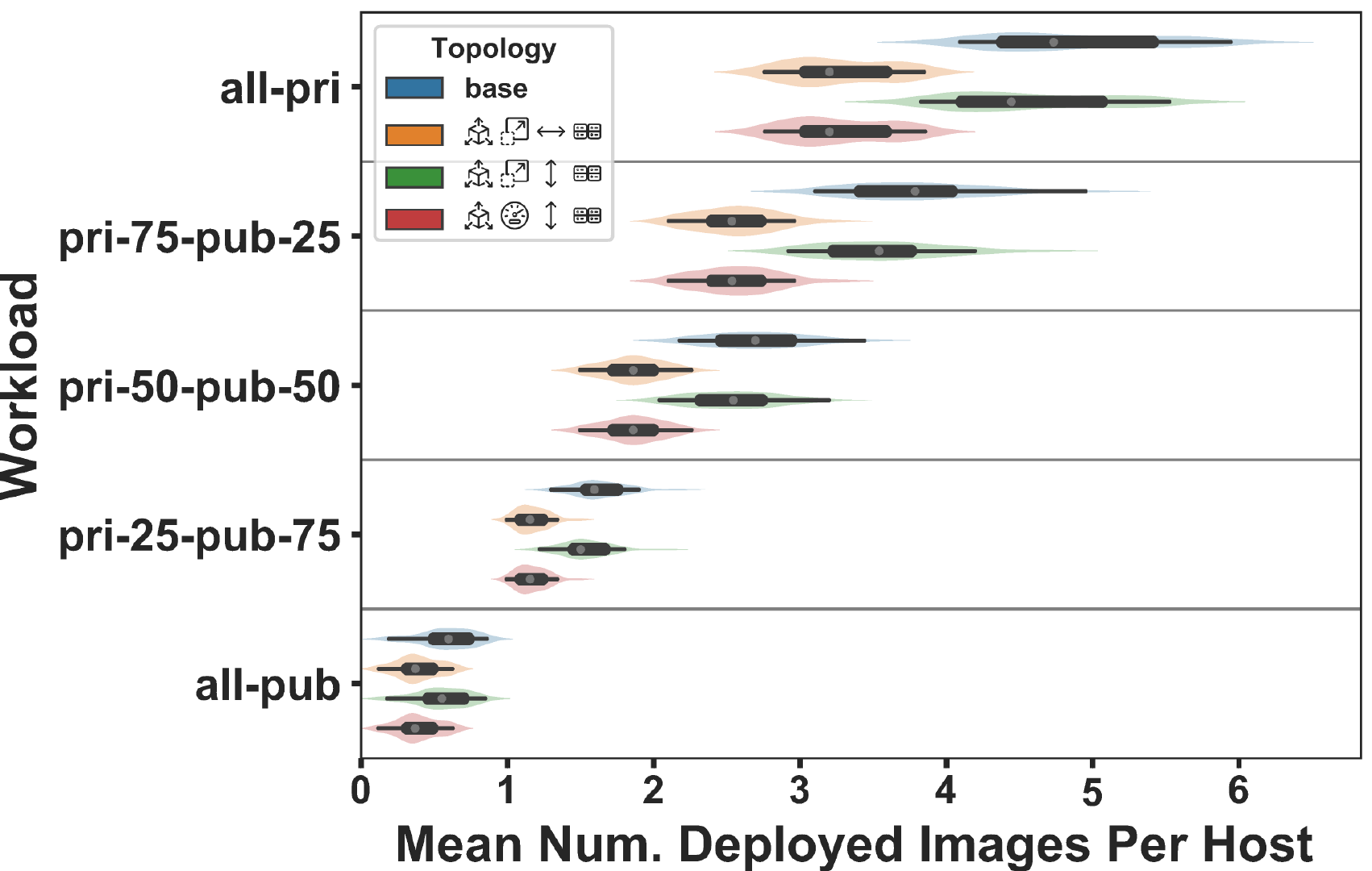}}%
    \subfloat[Max number of \glspl{VM} per host\label{fig:full:composite:max-vm-count}]{\includegraphics[width=0.5\linewidth]{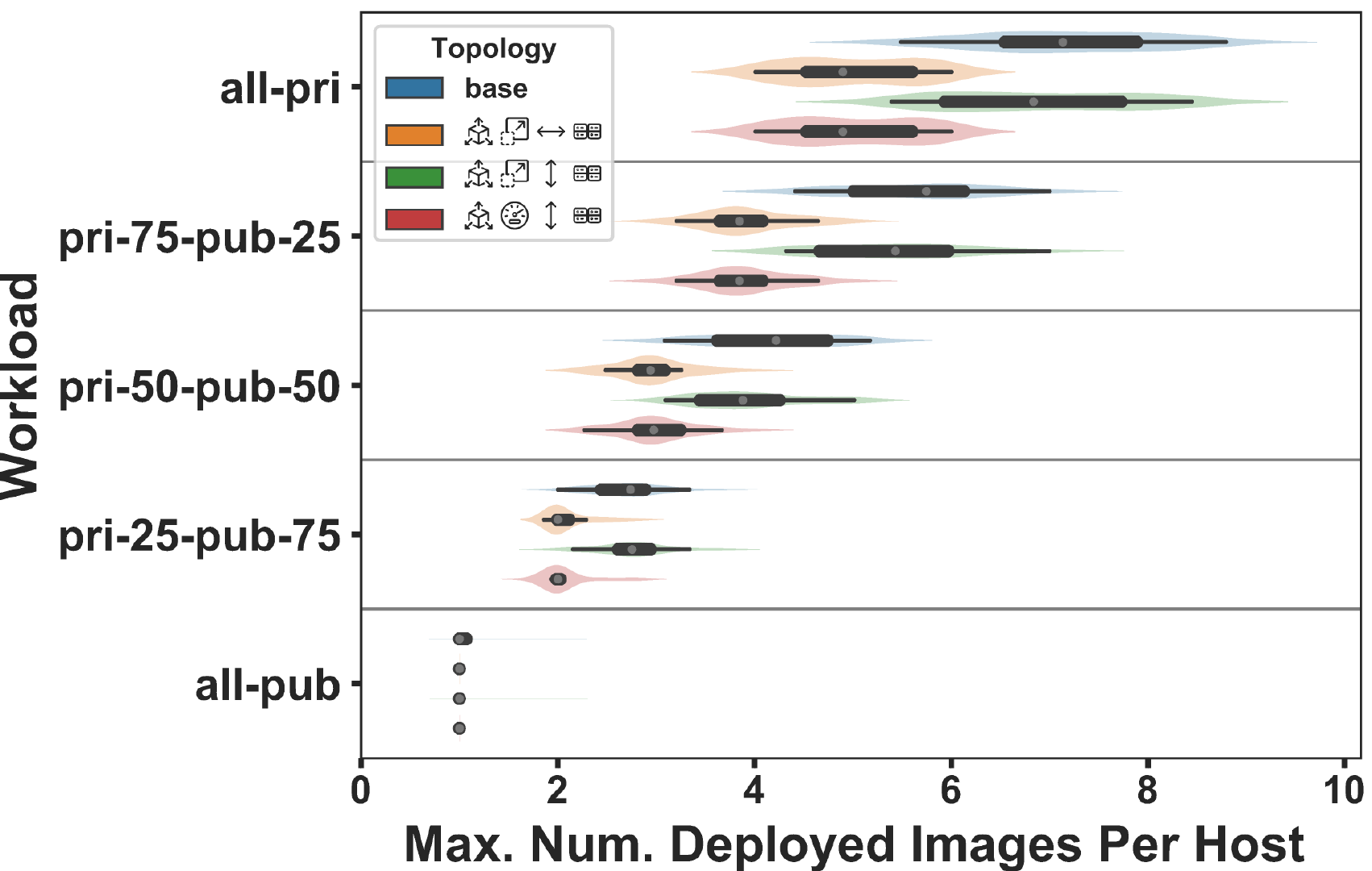}}\\
    \caption{Impact of a composite workload (consisting of private and public workloads) on different topologies. For a legend of topologies, see Table~\ref{tab:experiment-overview}. Continued in Figure~\ref{fig:full:composite:3}.}
    \label{fig:full:composite:2}
\end{figure*}

\begin{figure*}
    \subfloat[Total VMs Submitted\label{fig:full:composite:vms-submitted}]{\includegraphics[width=0.5\linewidth]{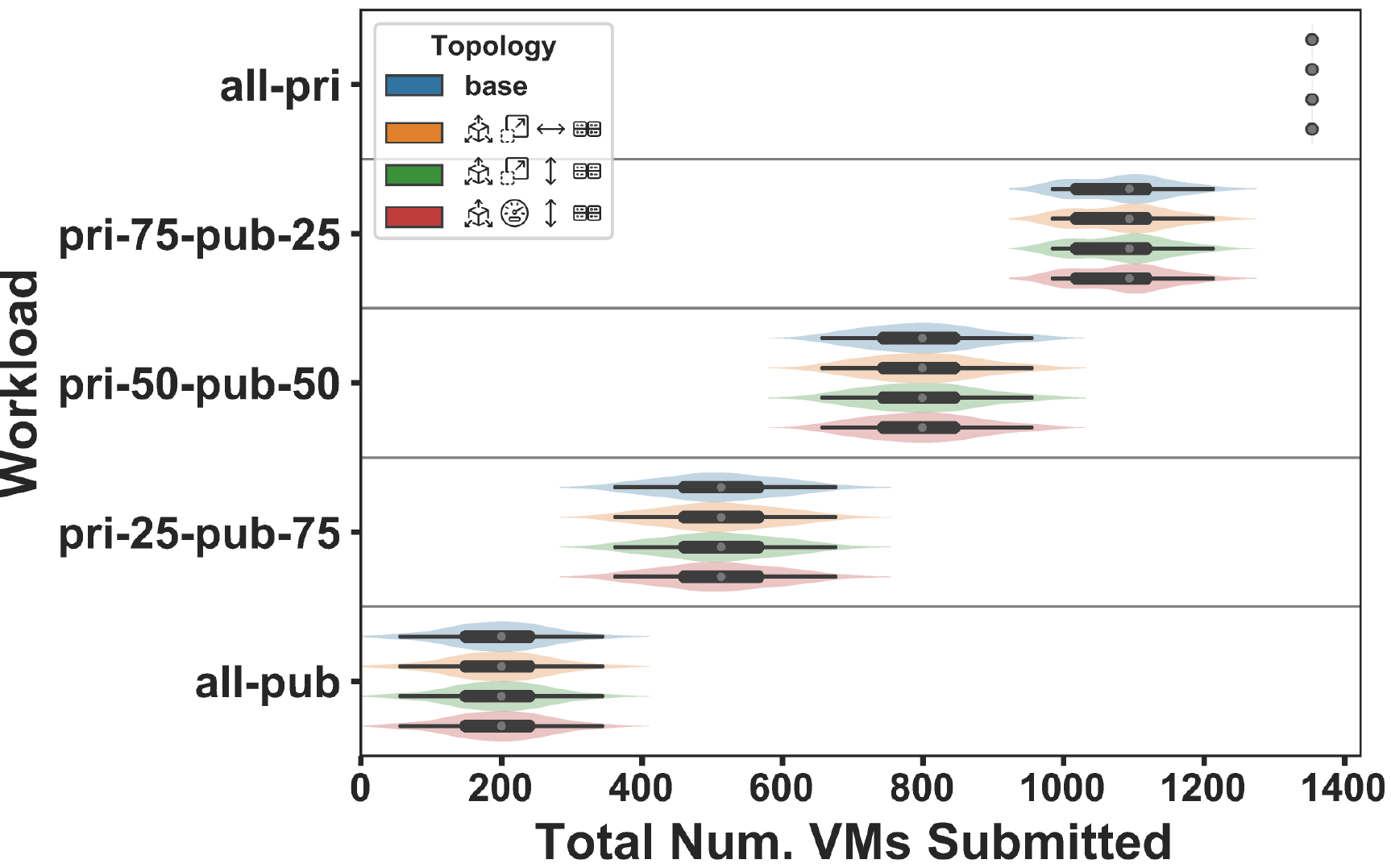}}%
    \subfloat[Total VMs Queued\label{fig:full:composite:vms-queued}]{\includegraphics[width=0.5\linewidth]{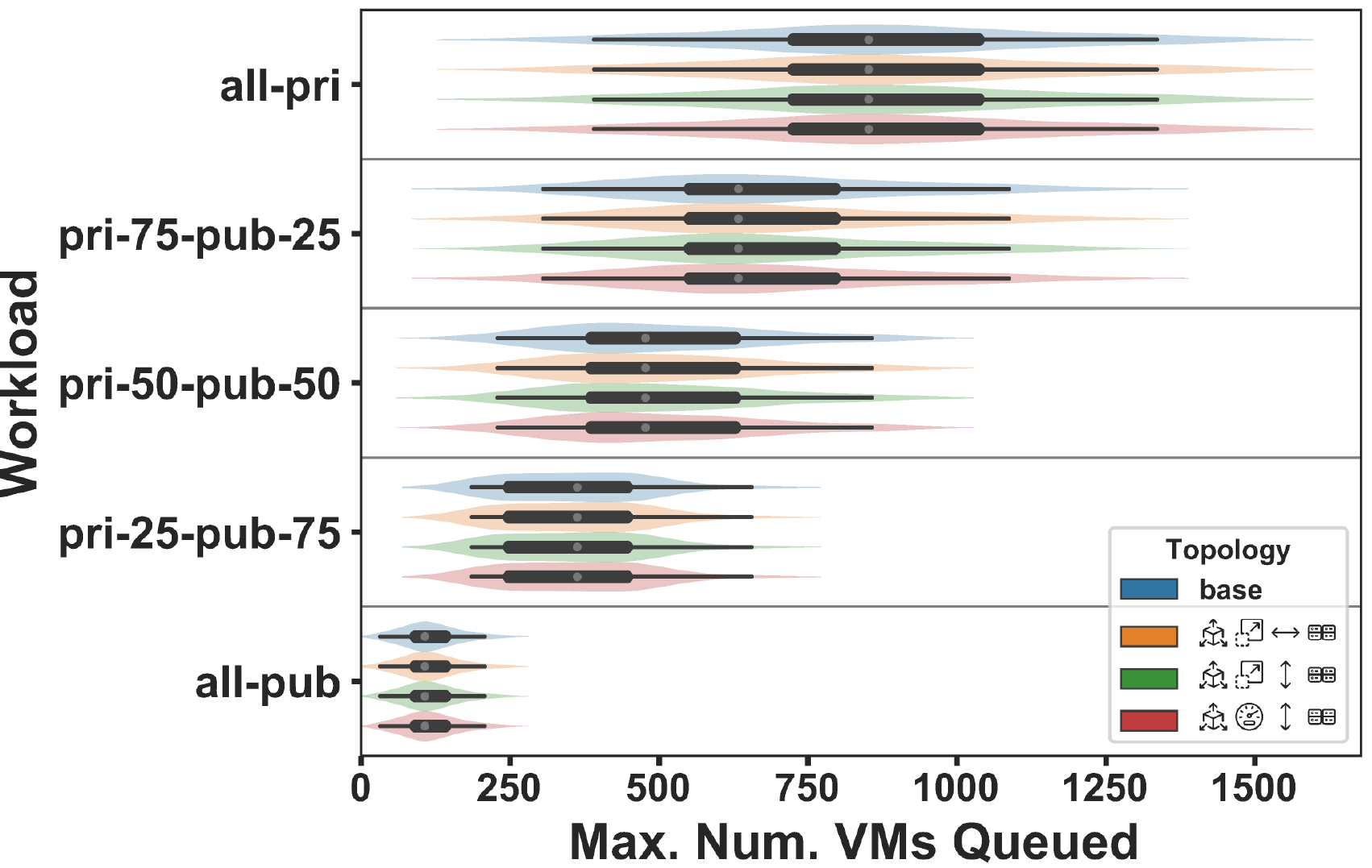}}\\
    \subfloat[Total VMs Finished\label{fig:full:composite:vms-finished}]{\includegraphics[width=0.5\linewidth]{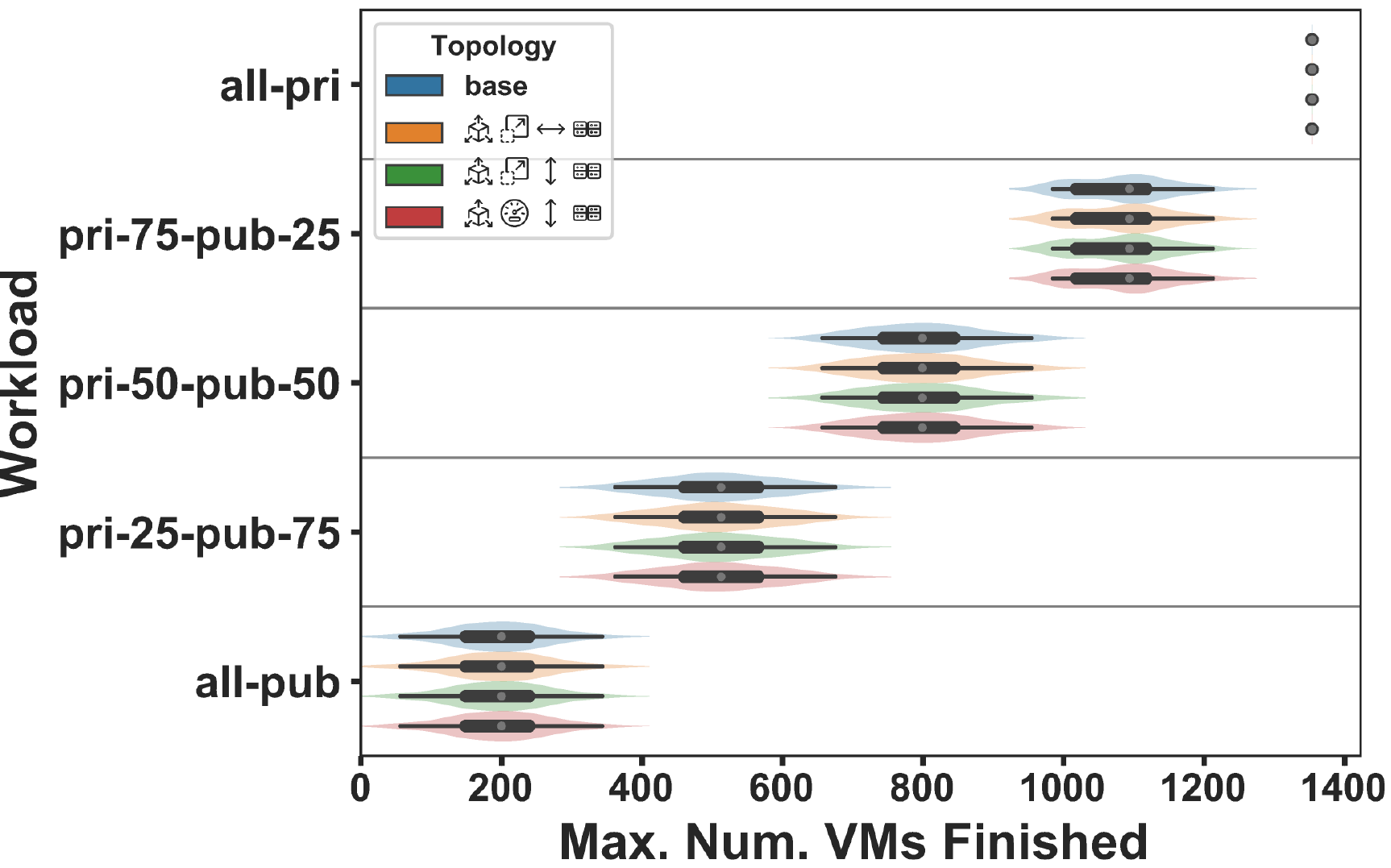}}%
    \subfloat[Total VMs Failed\label{fig:full:composite:vms-failed}]{\includegraphics[width=0.5\linewidth]{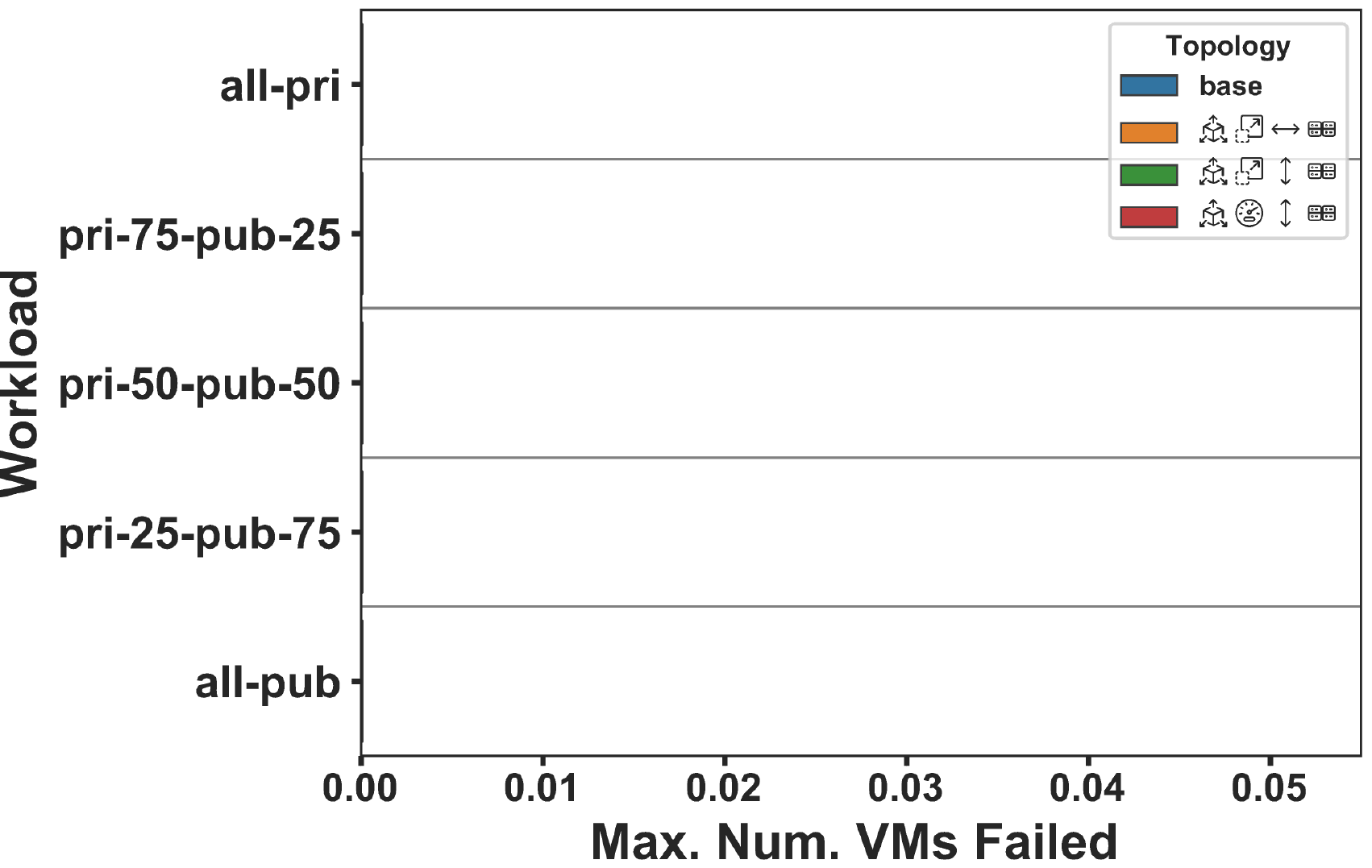}}%
    \caption{Impact of a composite workload (consisting of private and public workloads) on different topologies. For a legend of topologies, see Table~\ref{tab:experiment-overview}.}
    \label{fig:full:composite:3}
\end{figure*}

\begin{figure*}
    \centering
    \subfloat[Requested CPU cycles\label{fig:full:composite:summary:requested}]{\includegraphics[width=0.5\linewidth]{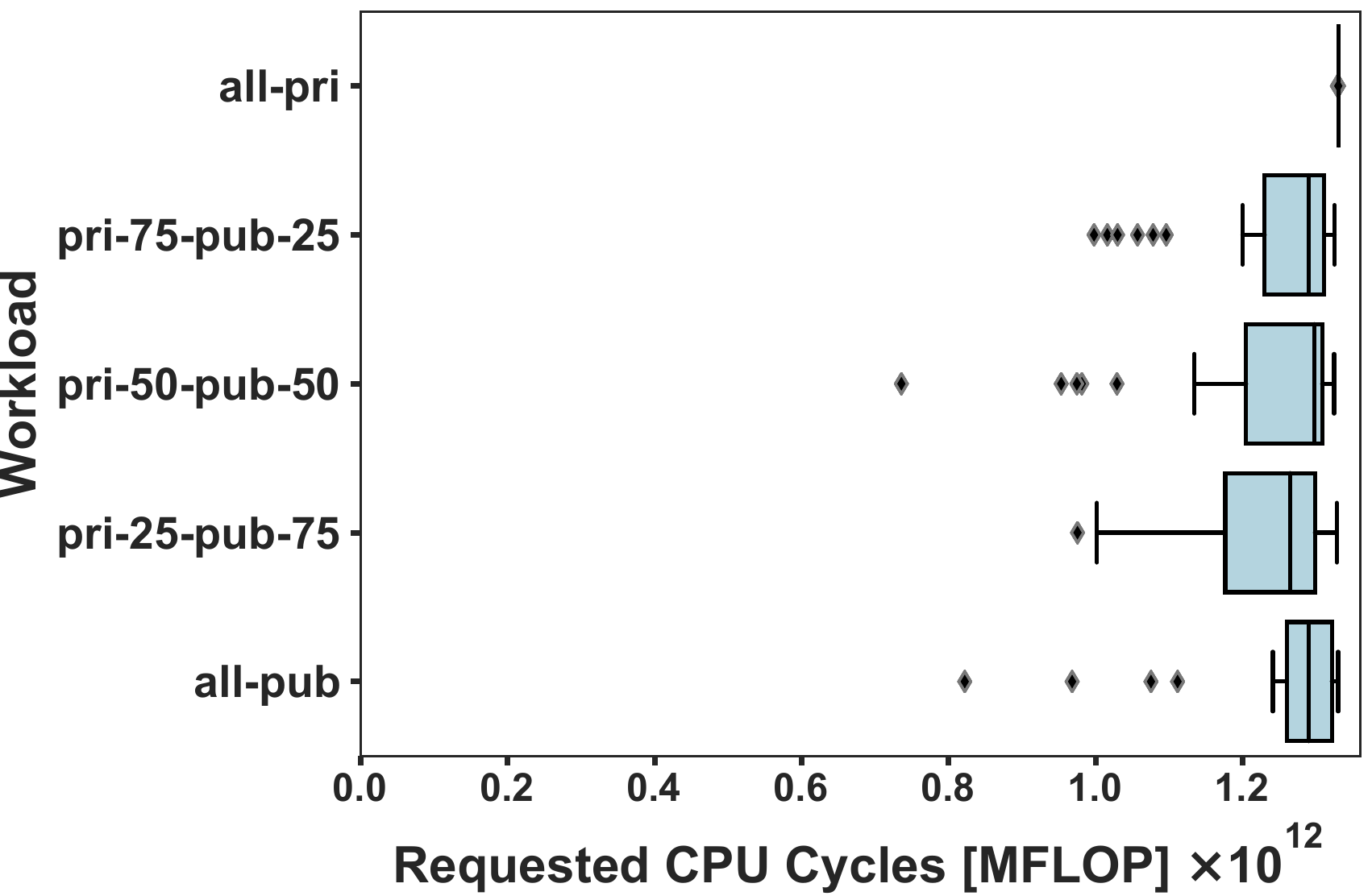}}%
    \subfloat[Granted CPU cycles\label{fig:full:composite:summary:granted}]{\includegraphics[width=0.5\linewidth]{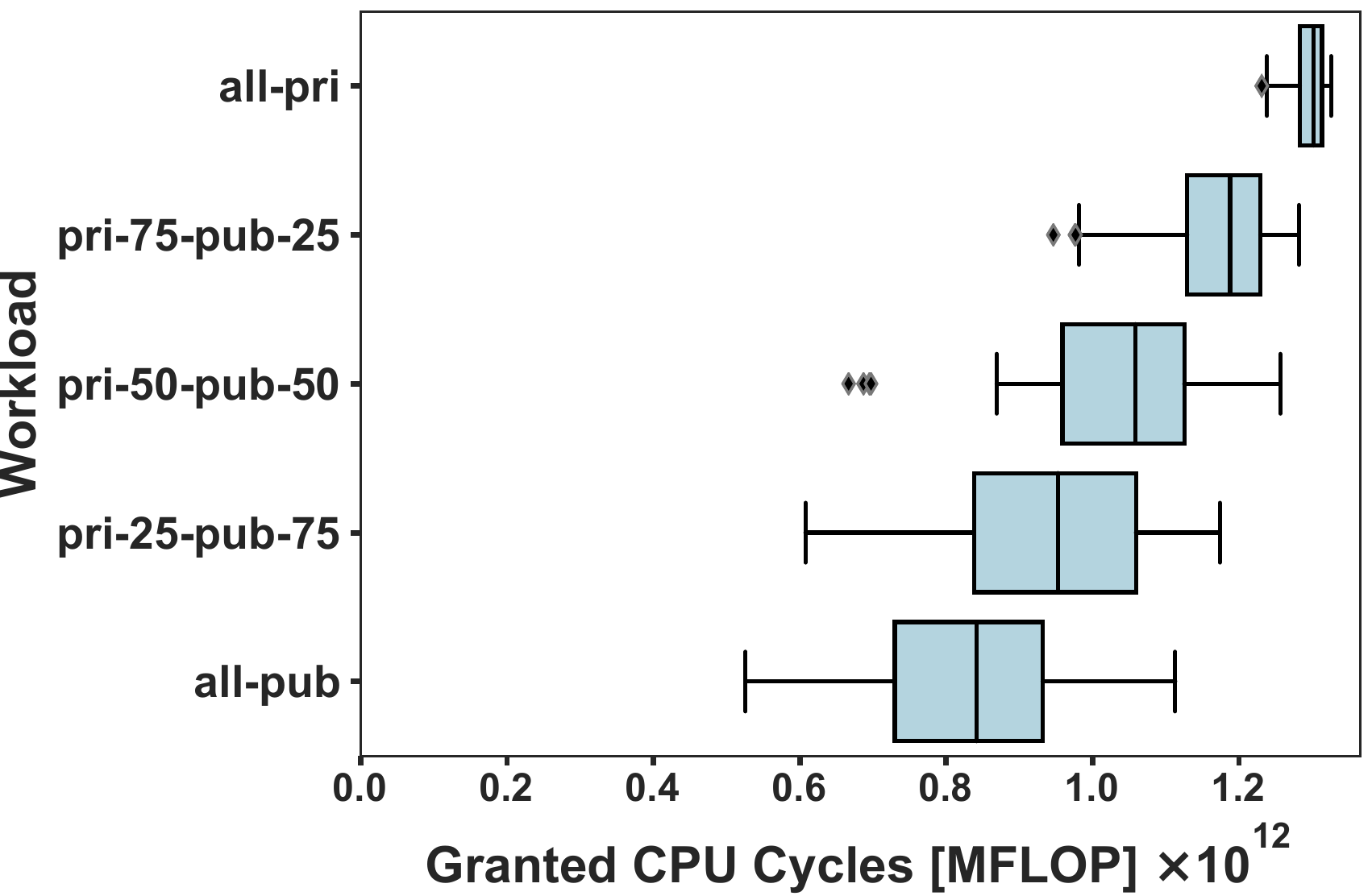}}\\
    \subfloat[Overcommitted CPU cycles\label{fig:full:composite:summary:overcommitted}]{\includegraphics[width=0.5\linewidth]{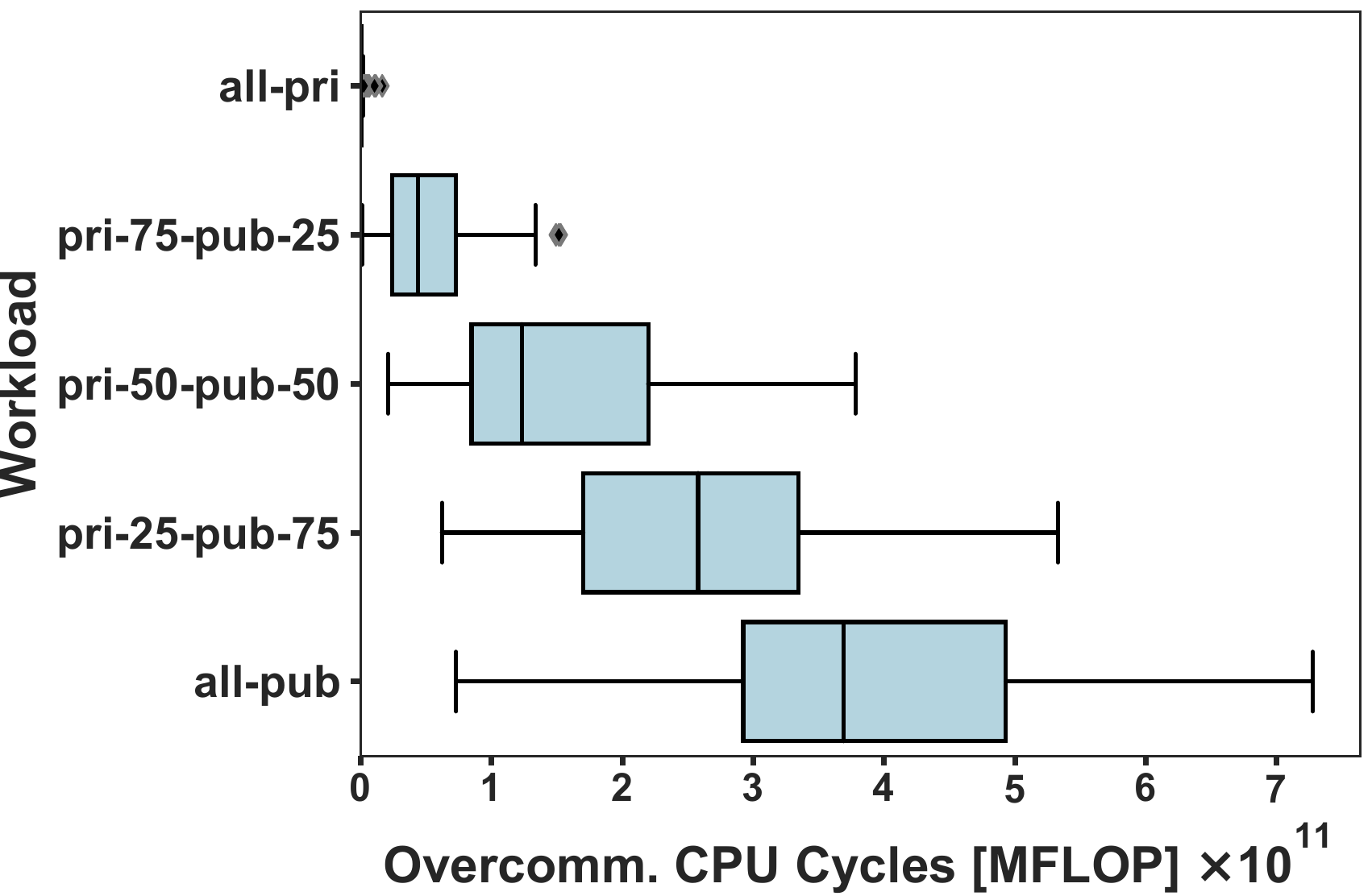}}%
    \subfloat[Interfered CPU cycles\label{fig:full:composite:summary:interfered}]{\includegraphics[width=0.5\linewidth]{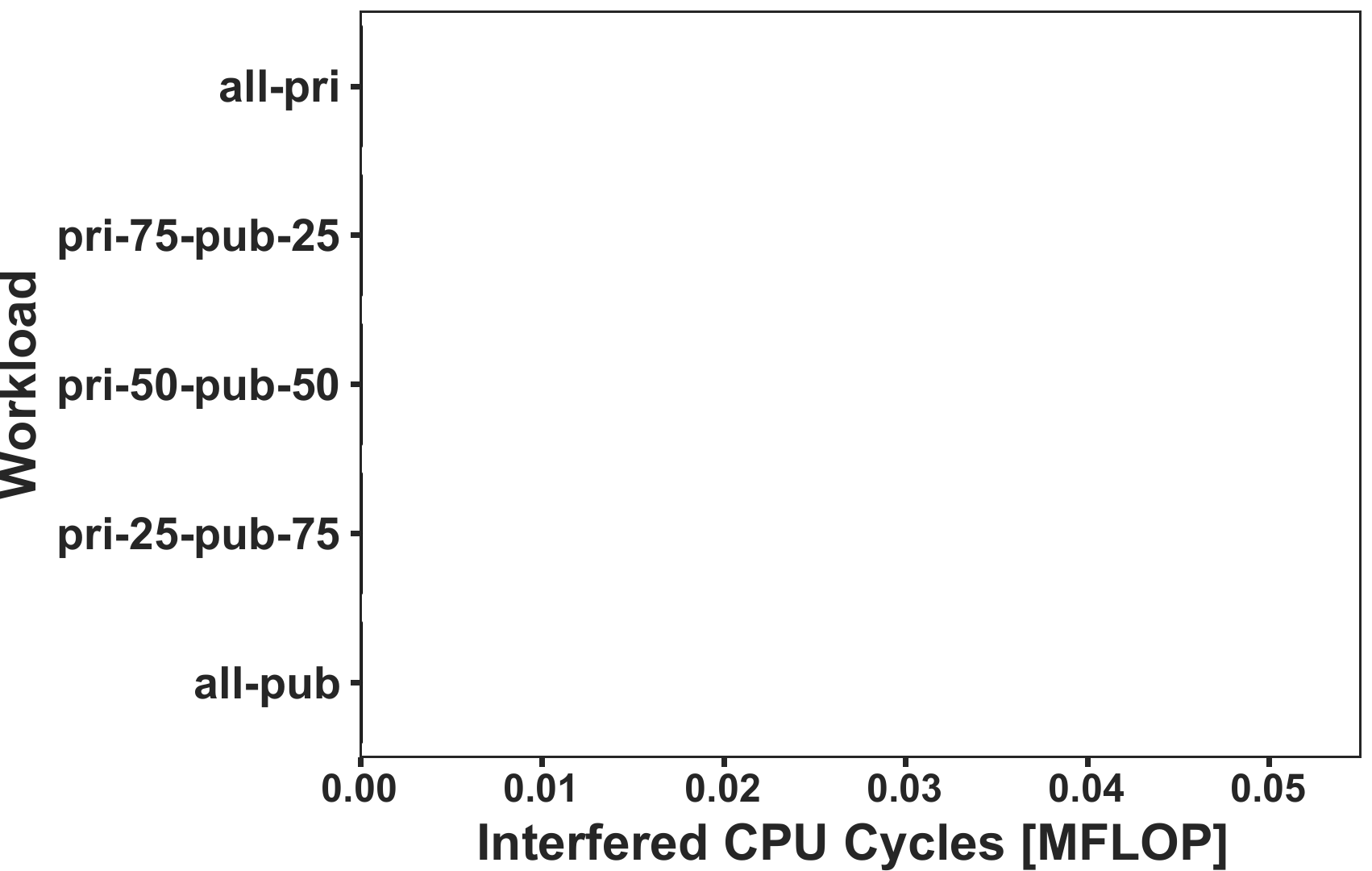}}\\
    \subfloat[Total power consumption\label{fig:full:composite:summary:power}]{\includegraphics[width=0.5\linewidth]{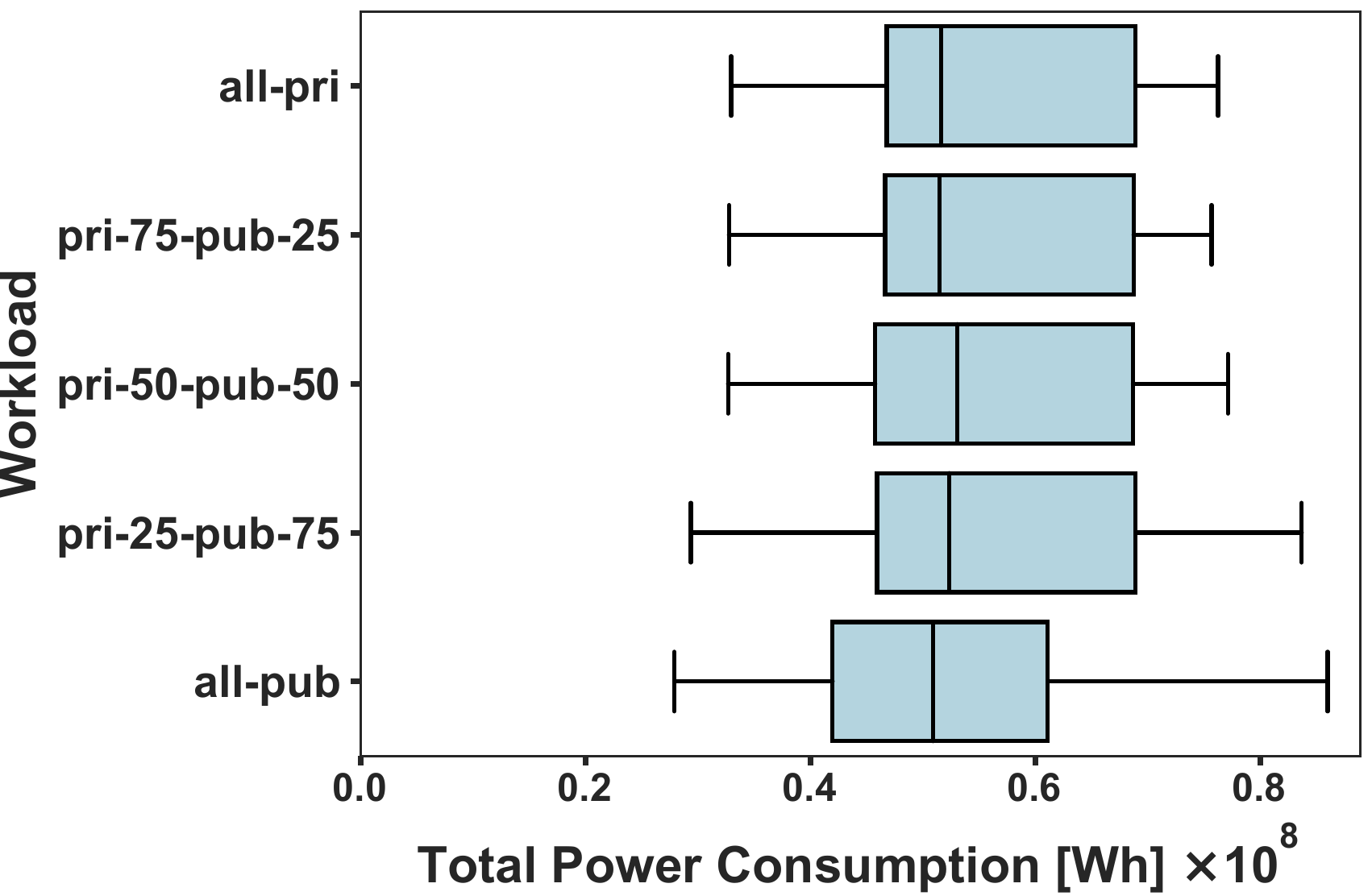}}%
    \subfloat[Total number of time slices in which a \gls{VM} is failed, aggregated across \glspl{VM}\label{fig:full:composite:summary:failures:vms}]{\includegraphics[width=0.5\linewidth]{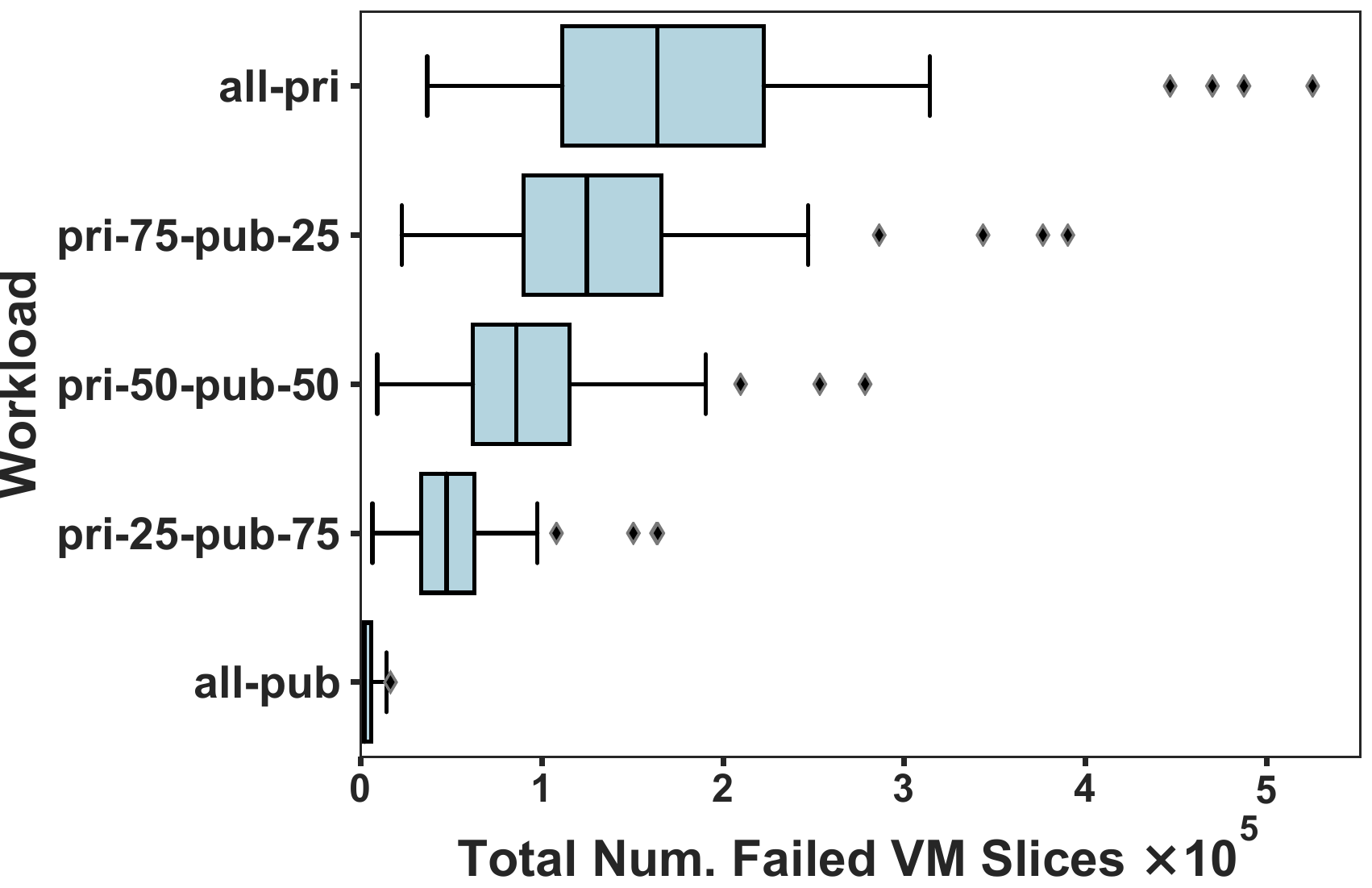}}\\
    \caption{Impact of a composite workload (consisting of private and public workloads) on different topologies. For a legend of topologies, see Table~\ref{tab:experiment-overview}. Continued in Figure~\ref{fig:full:composite:summary:2}.}
    \label{fig:full:composite:summary:1}
\end{figure*}

\begin{figure*}
    \subfloat[Mean CPU usage\label{fig:full:composite:summary:cpu-usage}]{\includegraphics[width=0.5\linewidth]{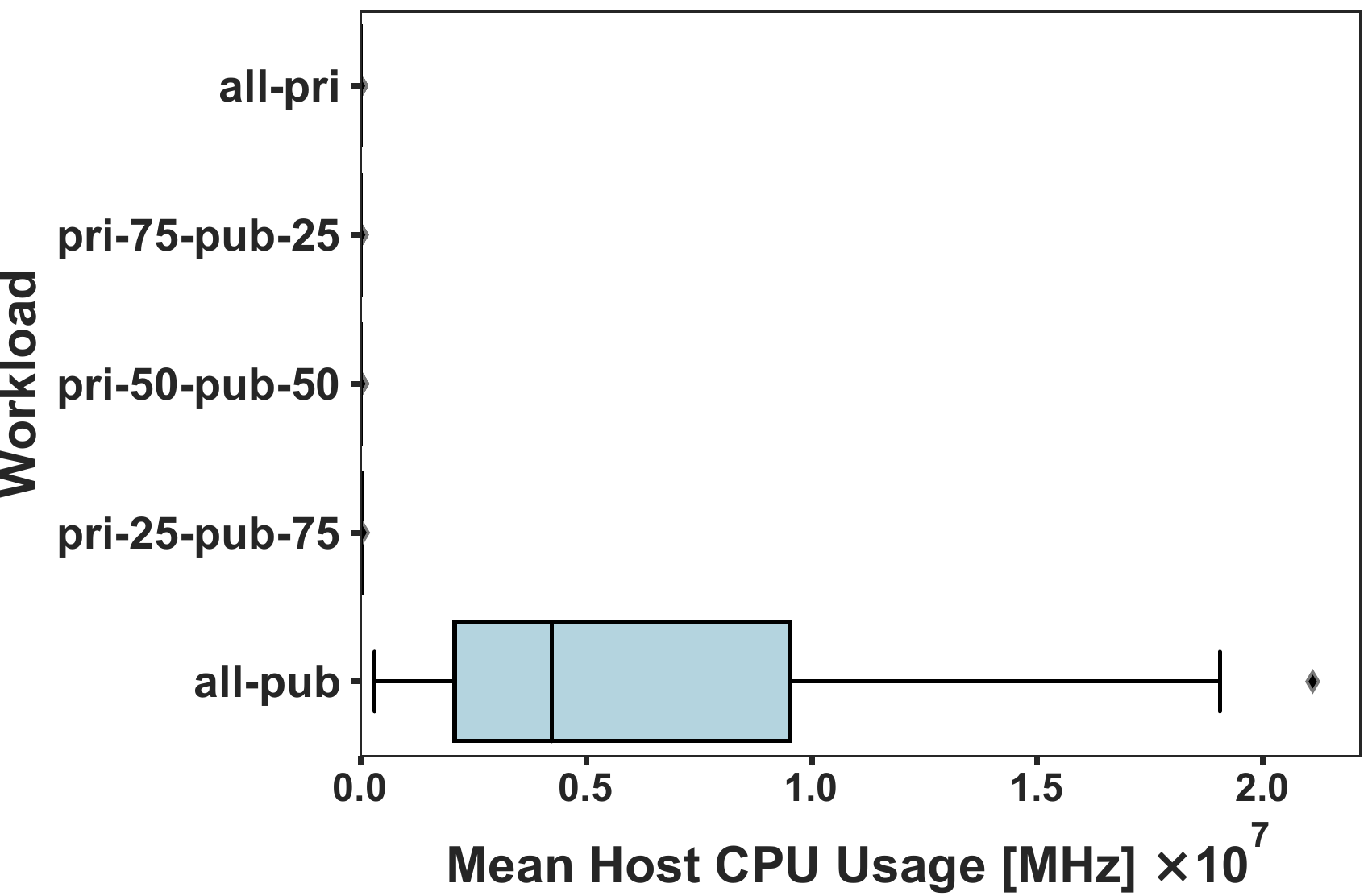}}%
    \subfloat[Mean CPU demand\label{fig:full:composite:summary:cpu-demand}]{\includegraphics[width=0.5\linewidth]{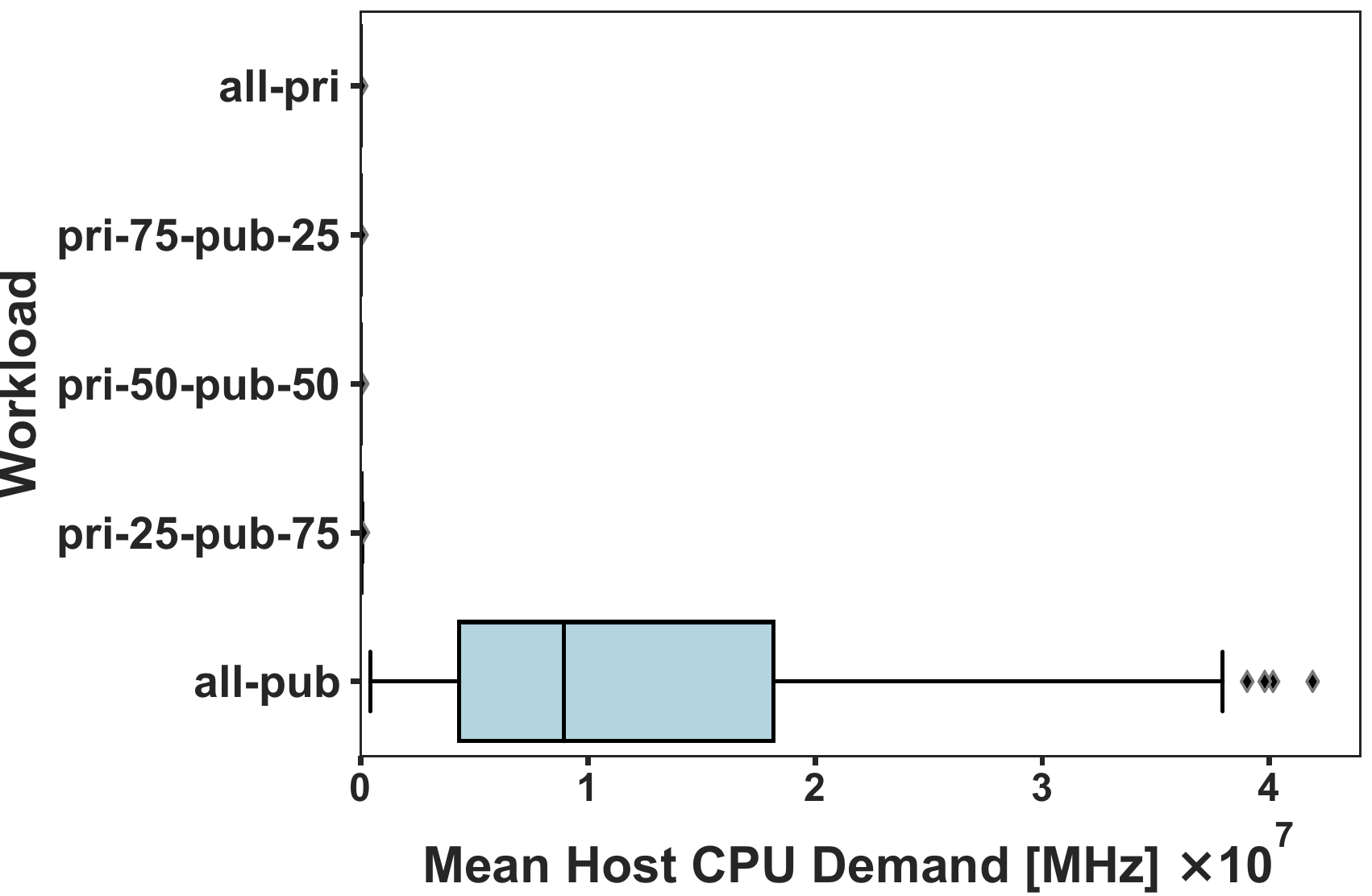}}\\
    \subfloat[Mean number of \glspl{VM} per host\label{fig:full:composite:summary:mean-vm-count}]{\includegraphics[width=0.5\linewidth]{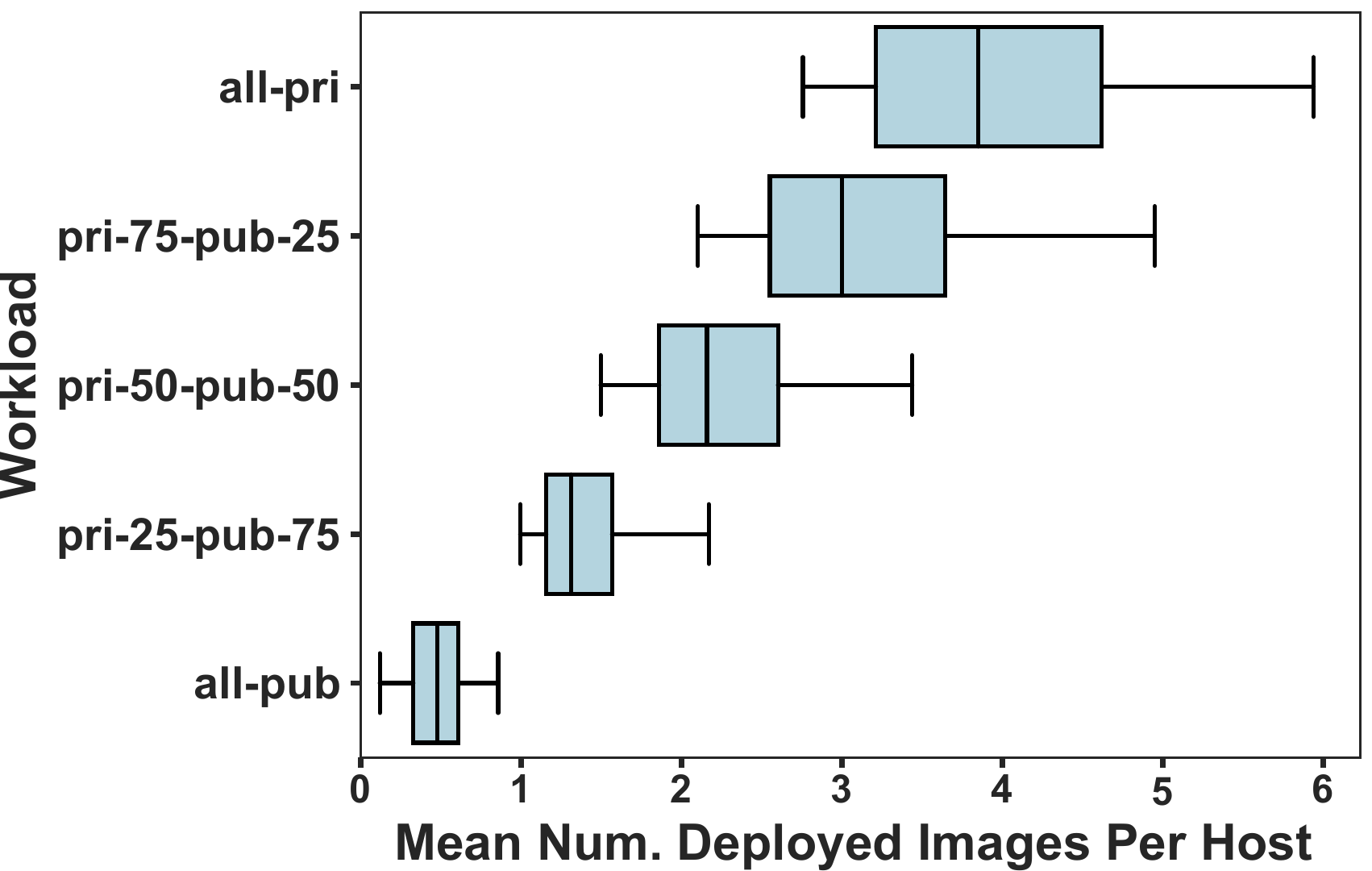}}%
    \subfloat[Max number of \glspl{VM} per host\label{fig:full:composite:summary:max-vm-count}]{\includegraphics[width=0.5\linewidth]{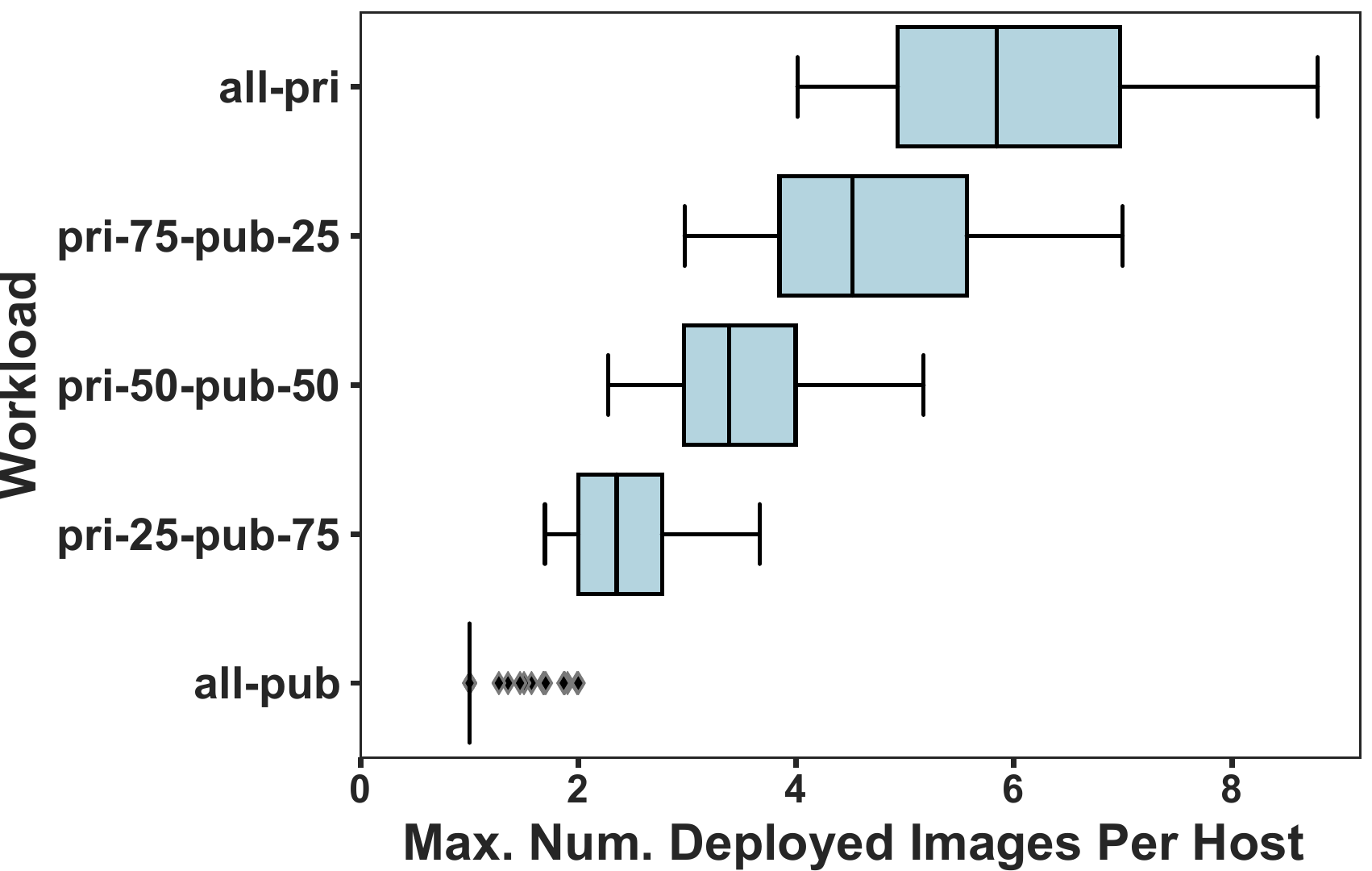}}\\
    \caption{Impact of a composite workload (consisting of private and public workloads) on different topologies. For a legend of topologies, see Table~\ref{tab:experiment-overview}. Continued in Figure~\ref{fig:full:composite:summary:3}.}
    \label{fig:full:composite:summary:2}
\end{figure*}

\begin{figure*}
    \subfloat[Total VMs Submitted\label{fig:full:composite:summary:vms-submitted}]{\includegraphics[width=0.5\linewidth]{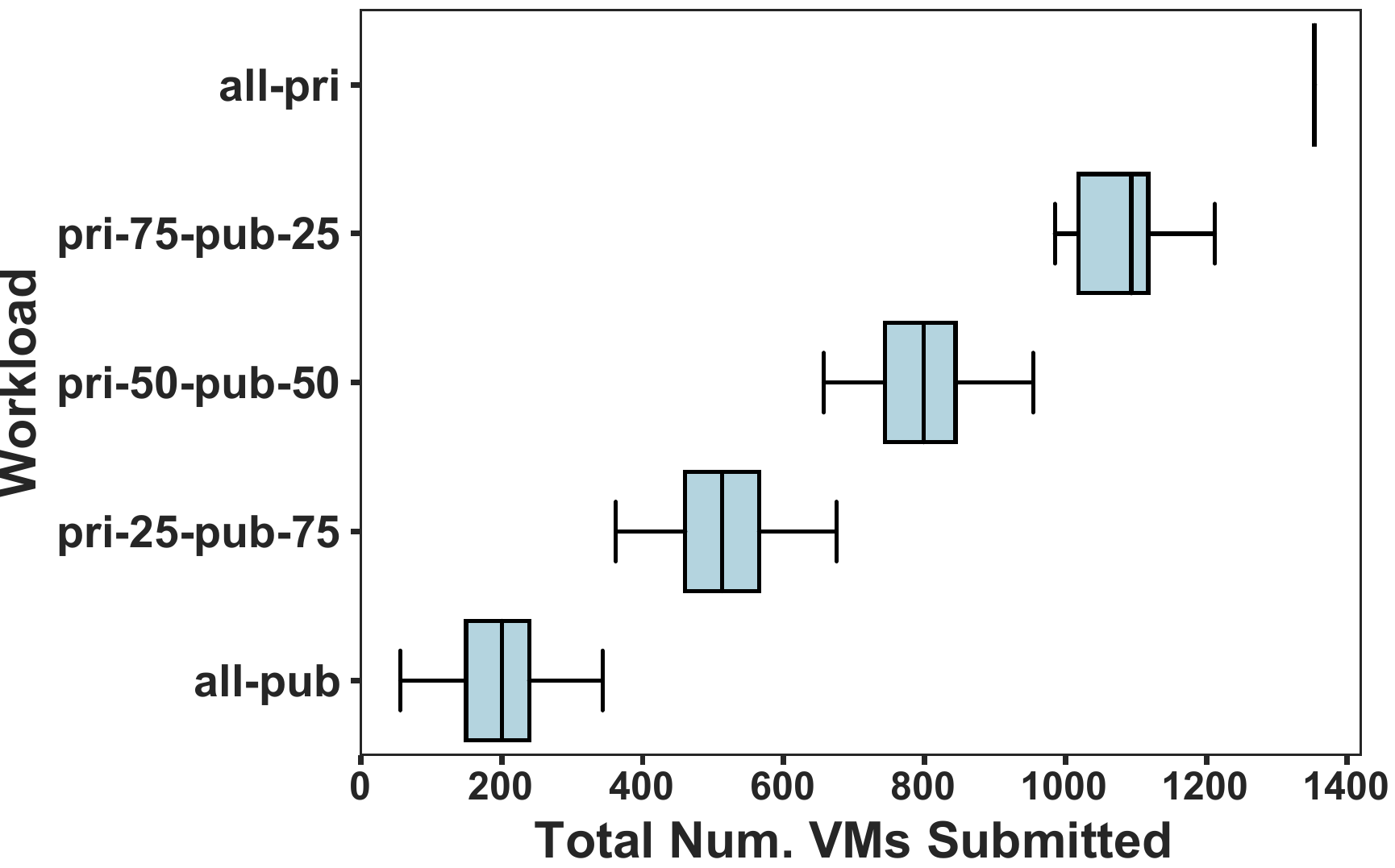}}%
    \subfloat[Total VMs Queued\label{fig:full:composite:summary:vms-queued}]{\includegraphics[width=0.5\linewidth]{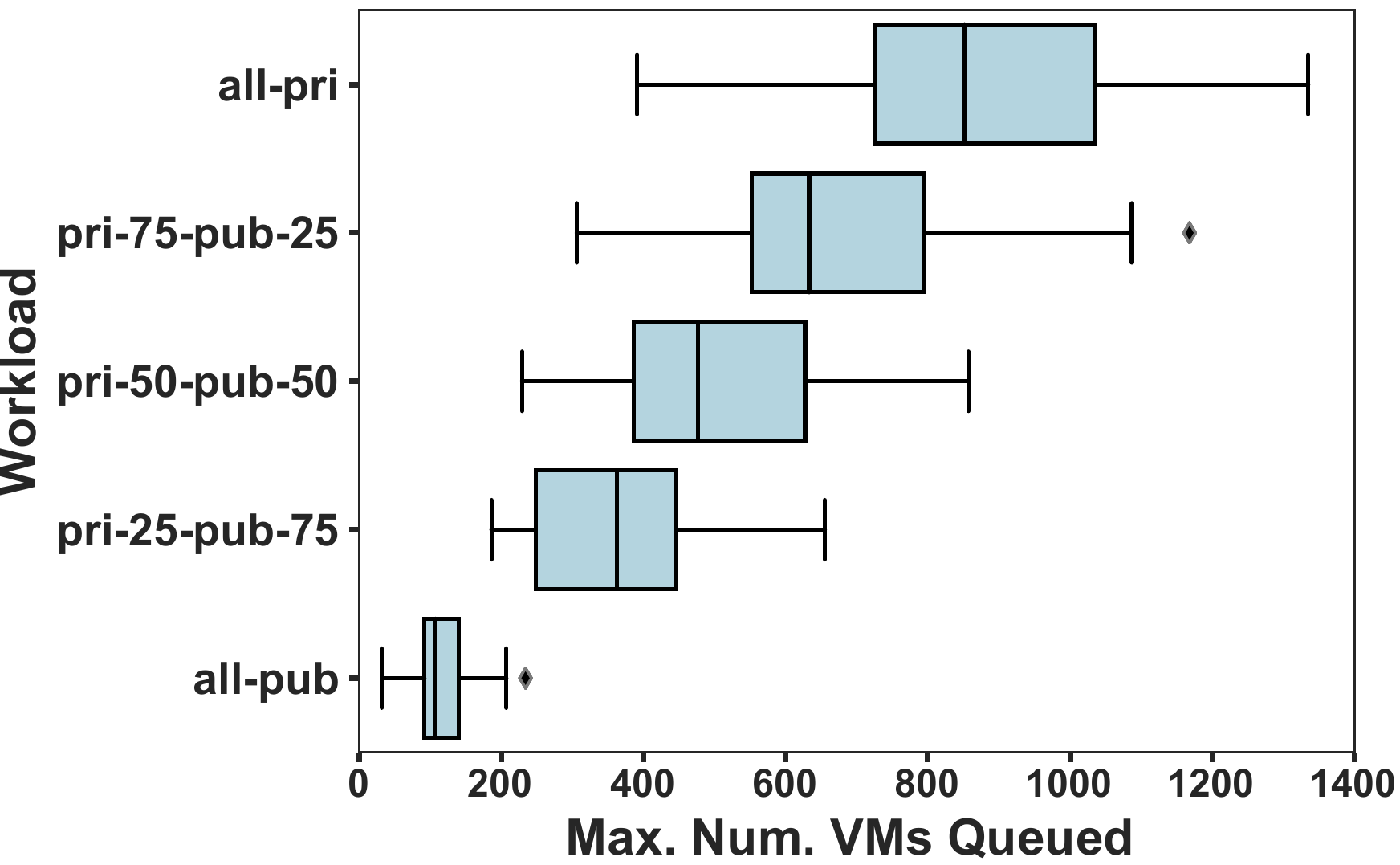}}\\
    \subfloat[Total VMs Finished\label{fig:full:composite:summary:vms-finished}]{\includegraphics[width=0.5\linewidth]{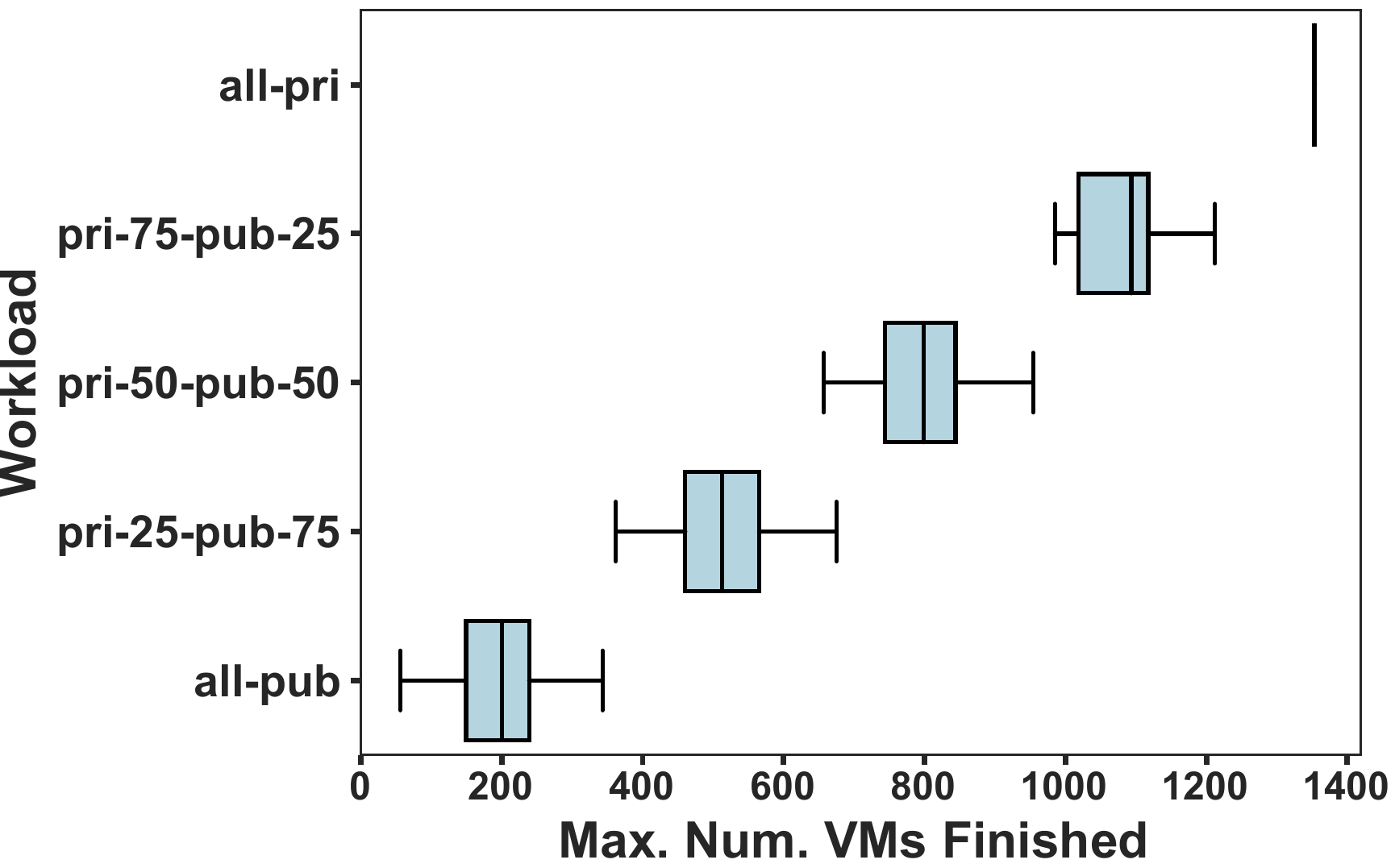}}%
    \subfloat[Total VMs Failed\label{fig:full:composite:summary:vms-failed}]{\includegraphics[width=0.5\linewidth]{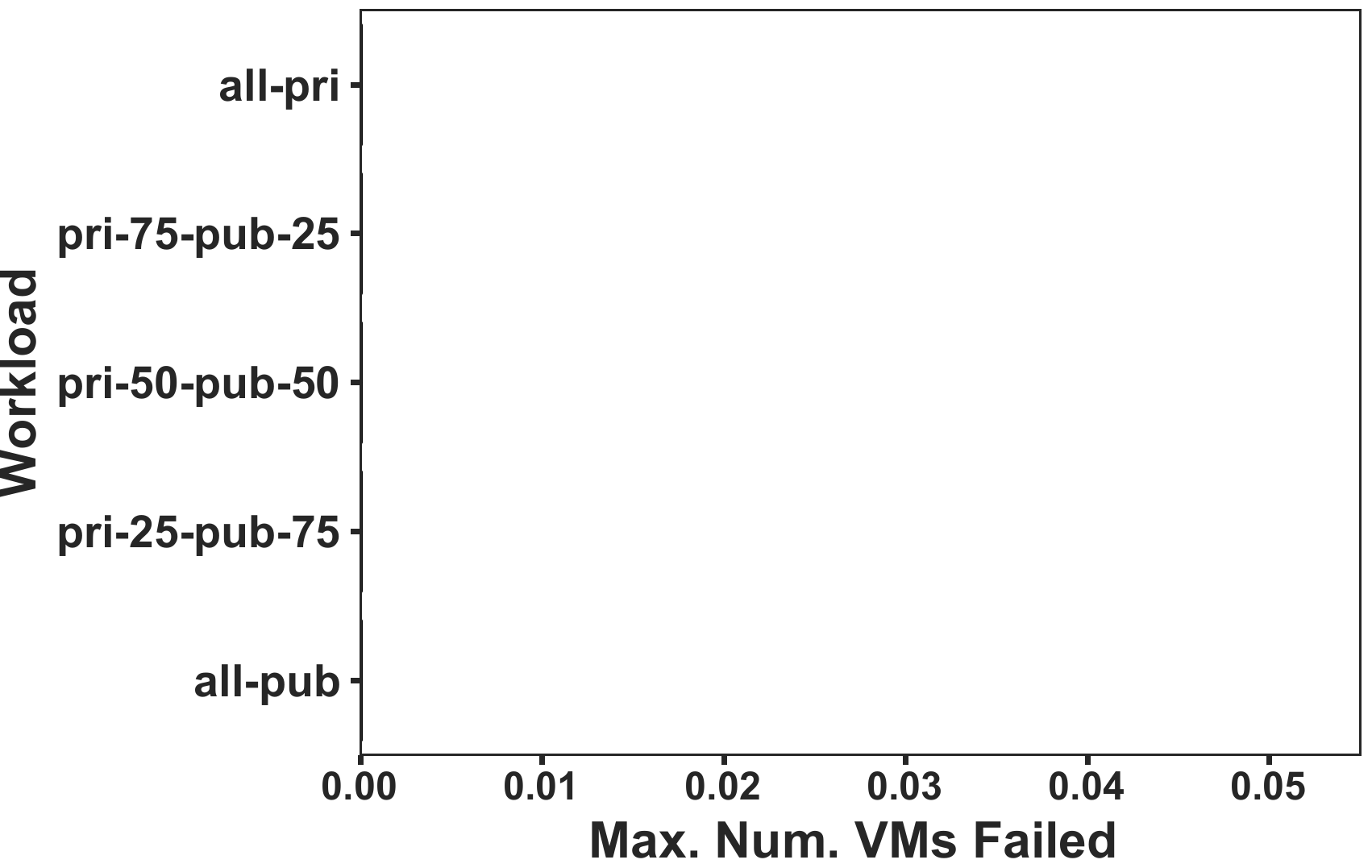}}%
    \caption{Impact of a composite workload (consisting of private and public workloads) on different topologies. For a legend of topologies, see Table~\ref{tab:experiment-overview}.}
    \label{fig:full:composite:summary:3}
\end{figure*}

\begin{figure*}
    \centering
    \subfloat[Requested CPU cycles\label{fig:full:replay:requested}]{\includegraphics[width=0.5\linewidth]{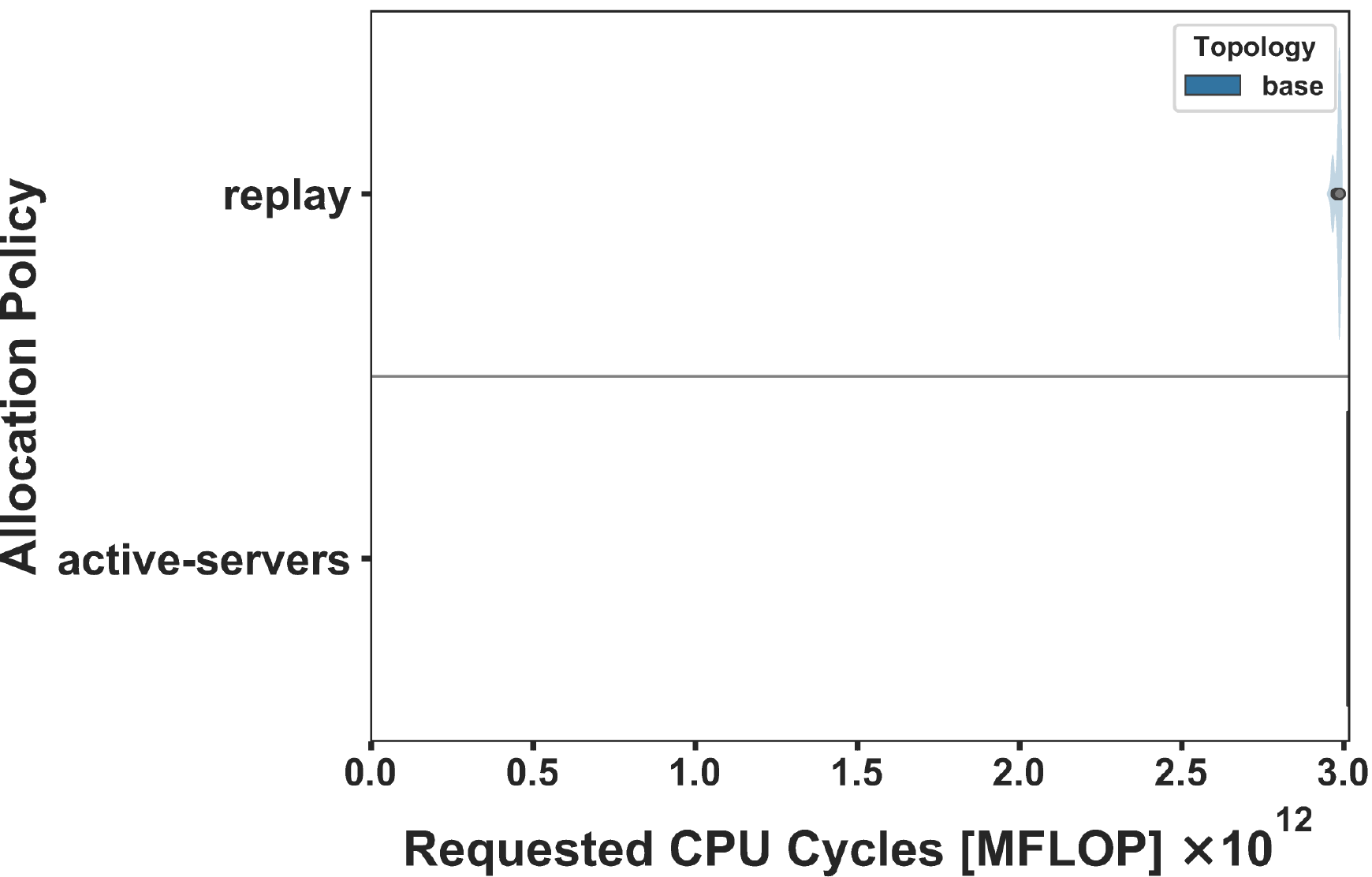}}%
    \subfloat[Granted CPU cycles\label{fig:full:replay:granted}]{\includegraphics[width=0.5\linewidth]{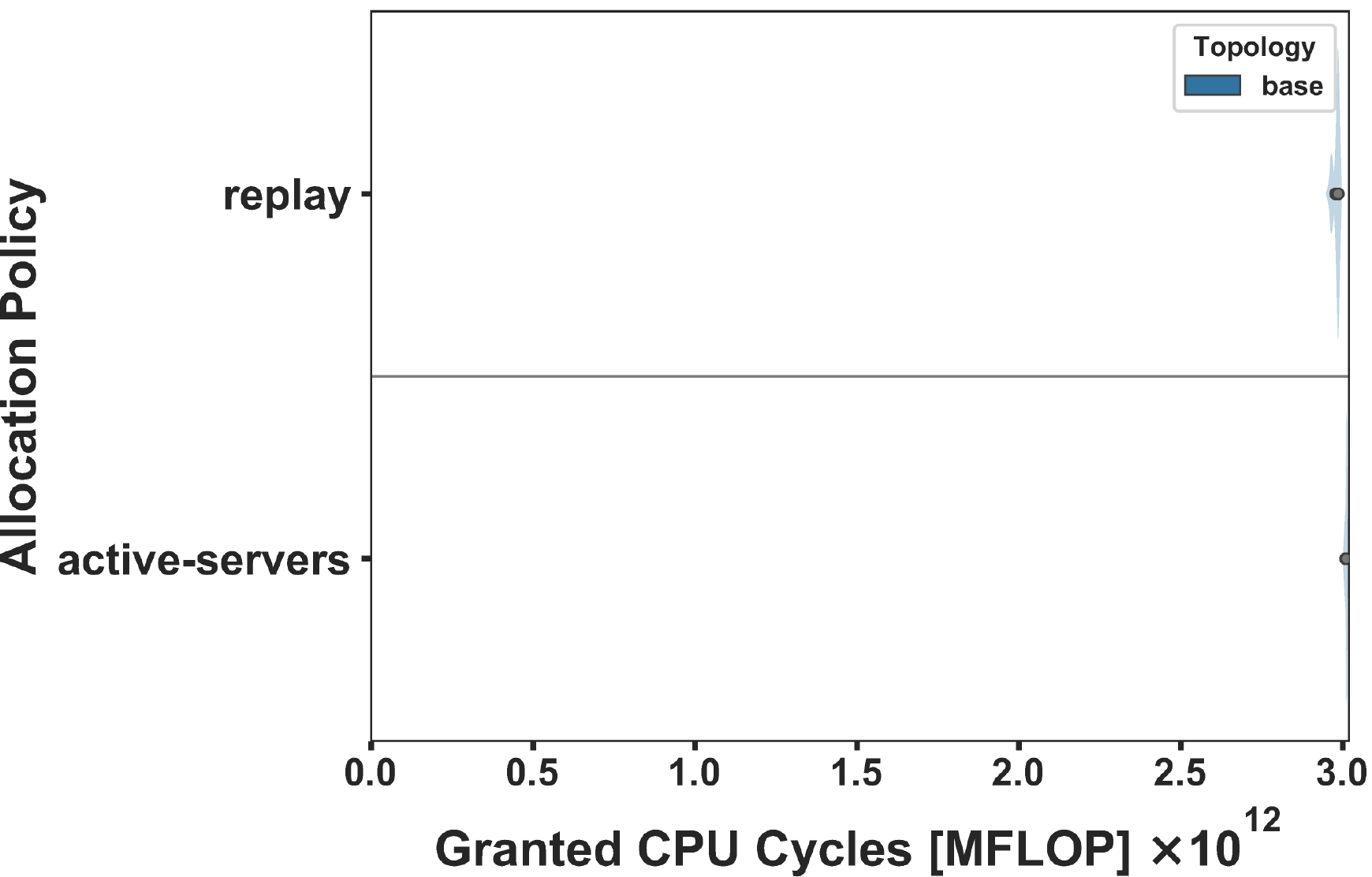}}\\
    \subfloat[Overcommitted CPU cycles\label{fig:full:replay:overcommitted}]{\includegraphics[width=0.5\linewidth]{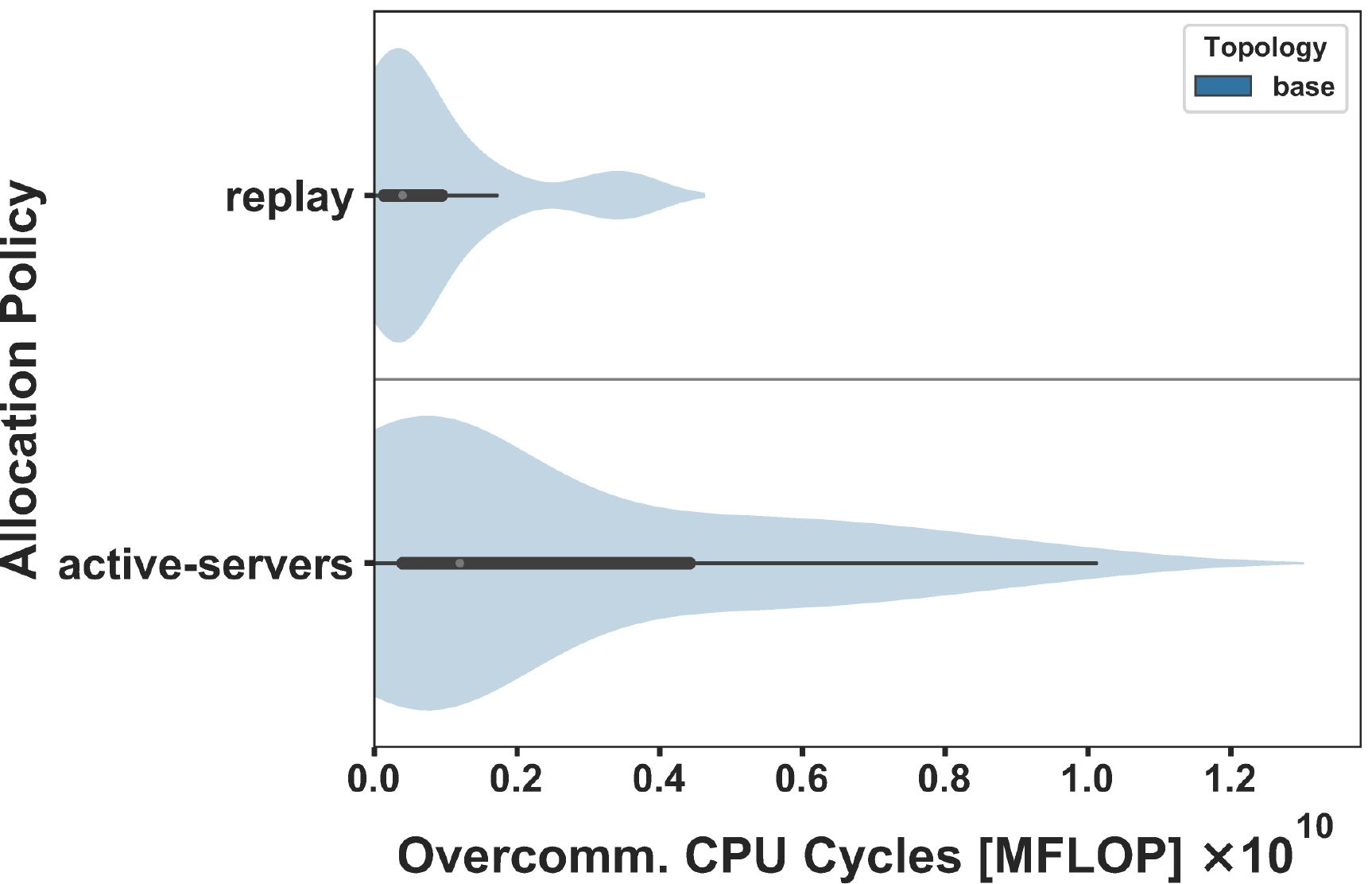}}%
    \subfloat[Interfered CPU cycles\label{fig:full:replay:interfered}]{\includegraphics[width=0.5\linewidth]{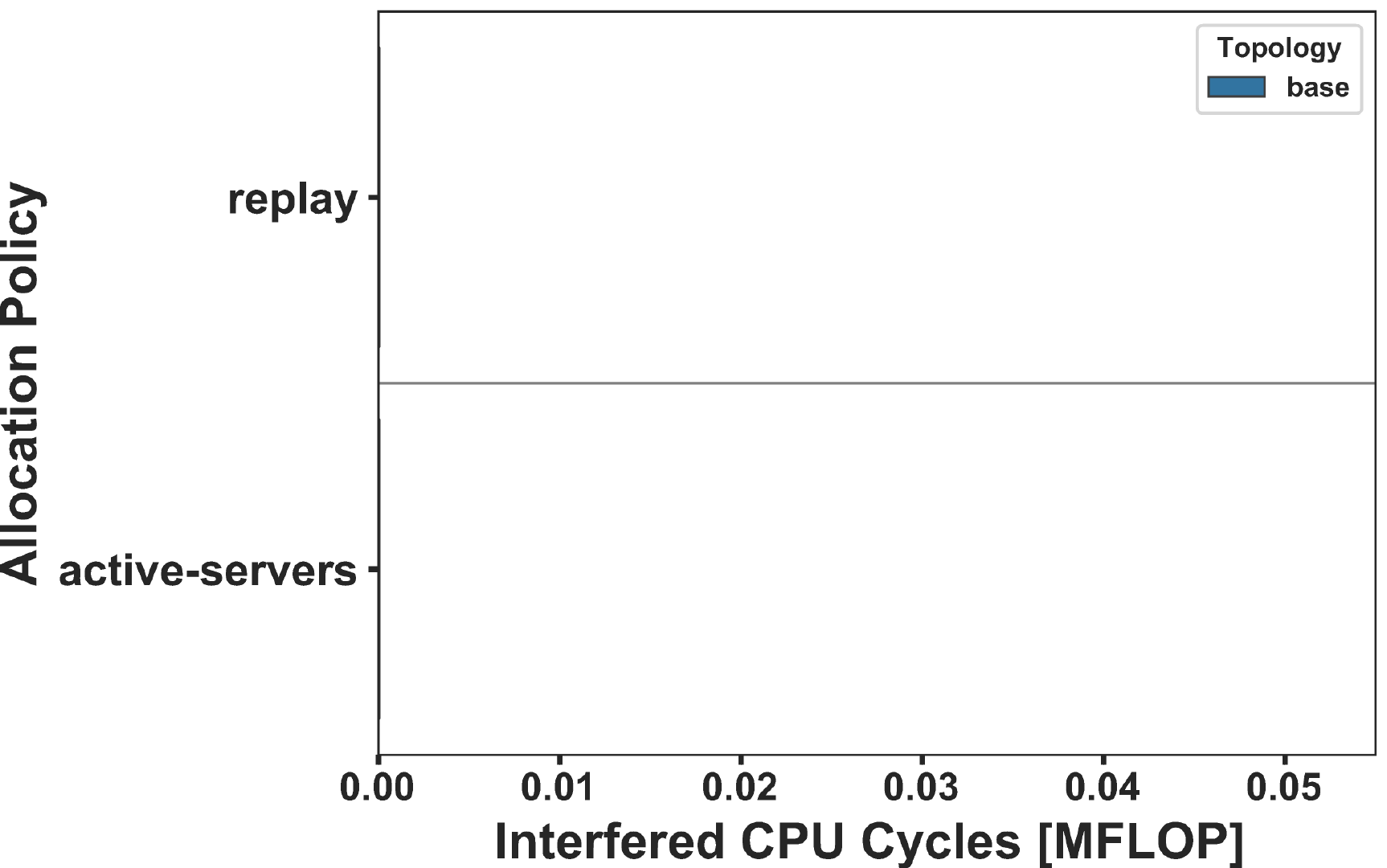}}\\
    \subfloat[Total power consumption\label{fig:full:replay:power}]{\includegraphics[width=0.5\linewidth]{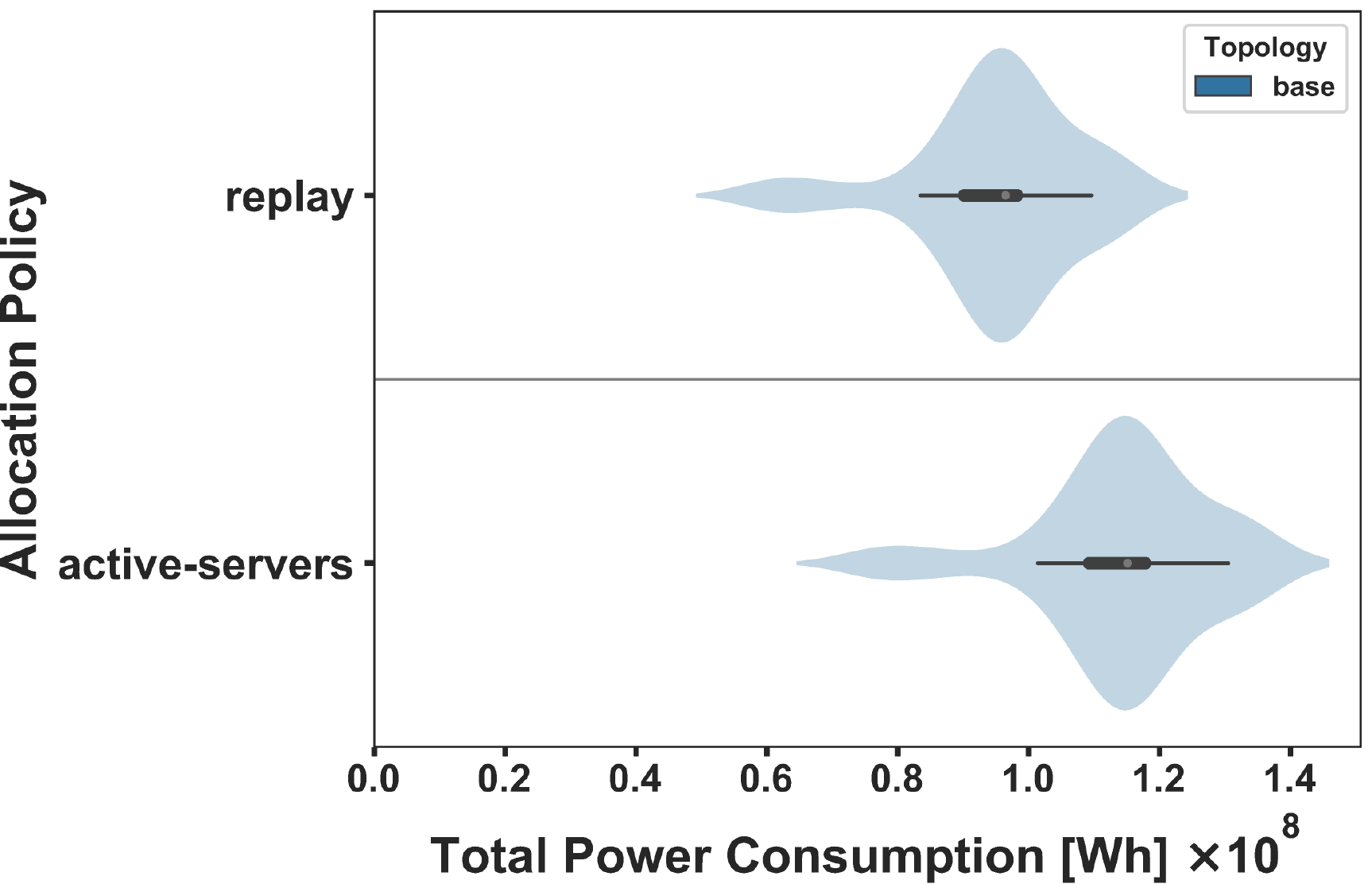}}%
    \subfloat[Total number of time slices in which a \gls{VM} is failed, aggregated across \glspl{VM}\label{fig:full:replay:failures:vms}]{\includegraphics[width=0.5\linewidth]{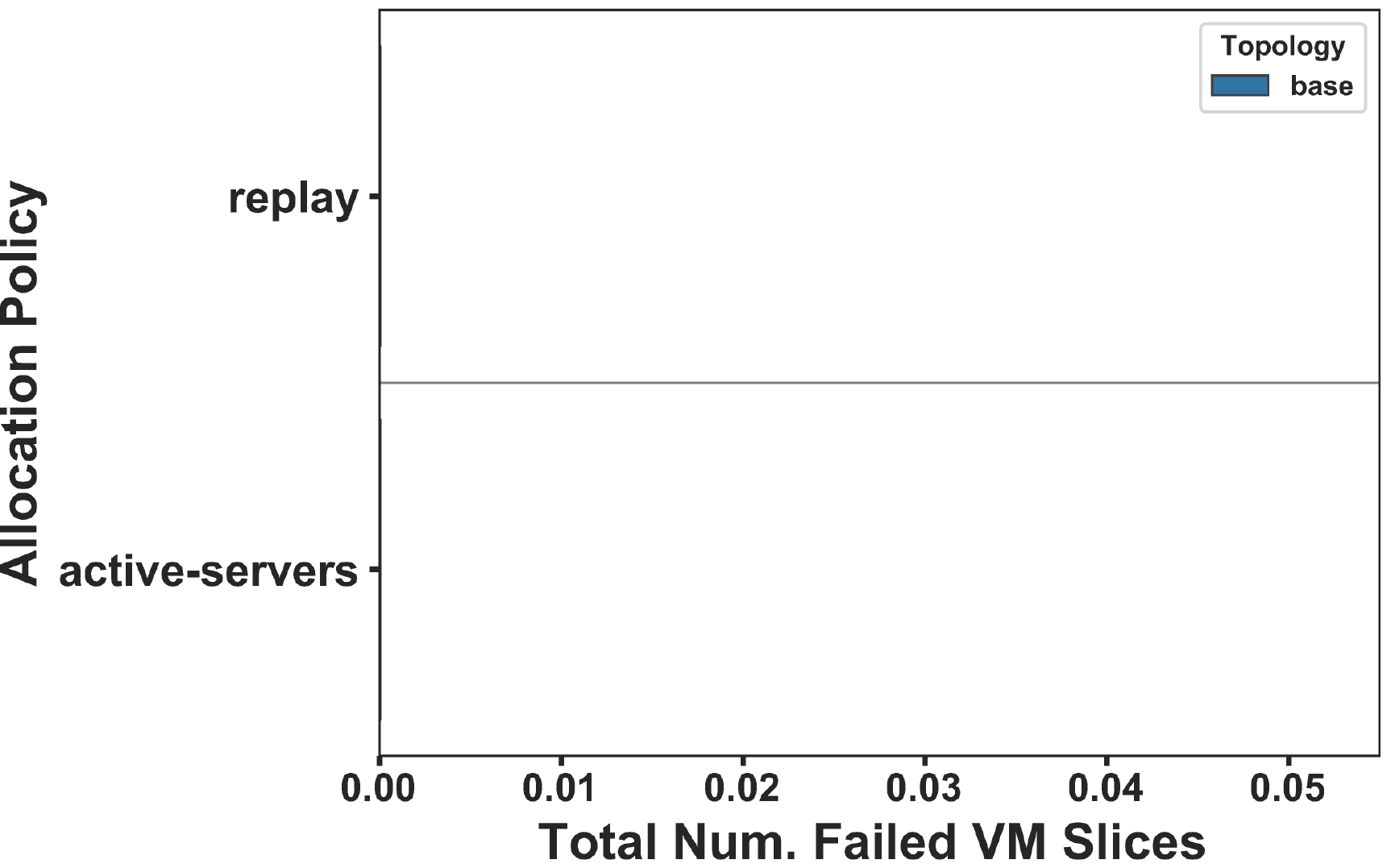}}\\
    \caption{Validation with a replay policy, copying the exact cluster assignment of the original deployment. For a legend of topologies, see Table~\ref{tab:experiment-overview}. Continued in Figure~\ref{fig:full:replay:2}.}
    \label{fig:full:replay:1}
\end{figure*}

\begin{figure*}
    \subfloat[Mean CPU usage\label{fig:full:replay:cpu-usage}]{\includegraphics[width=0.5\linewidth]{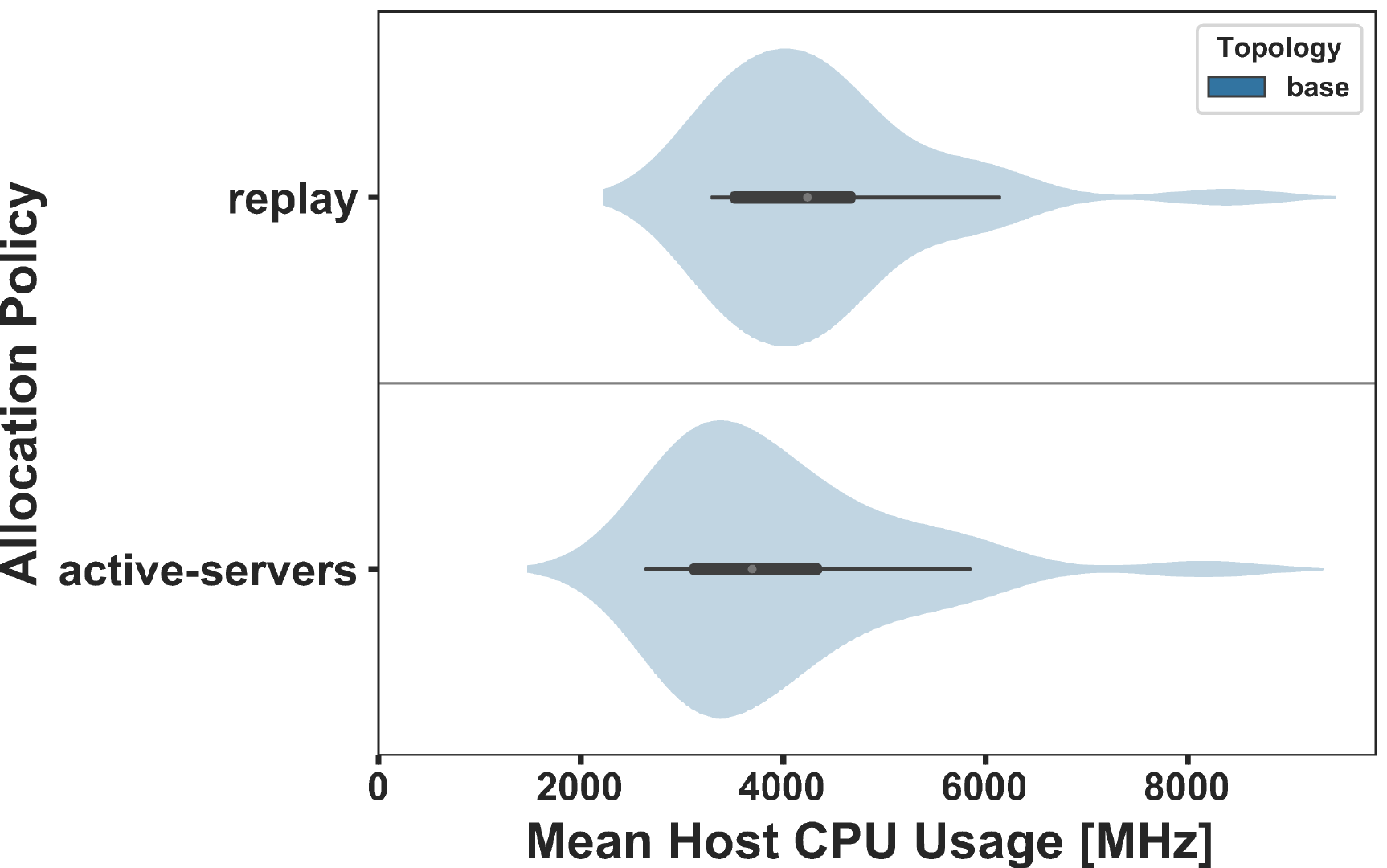}}%
    \subfloat[Mean CPU demand\label{fig:full:replay:cpu-demand}]{\includegraphics[width=0.5\linewidth]{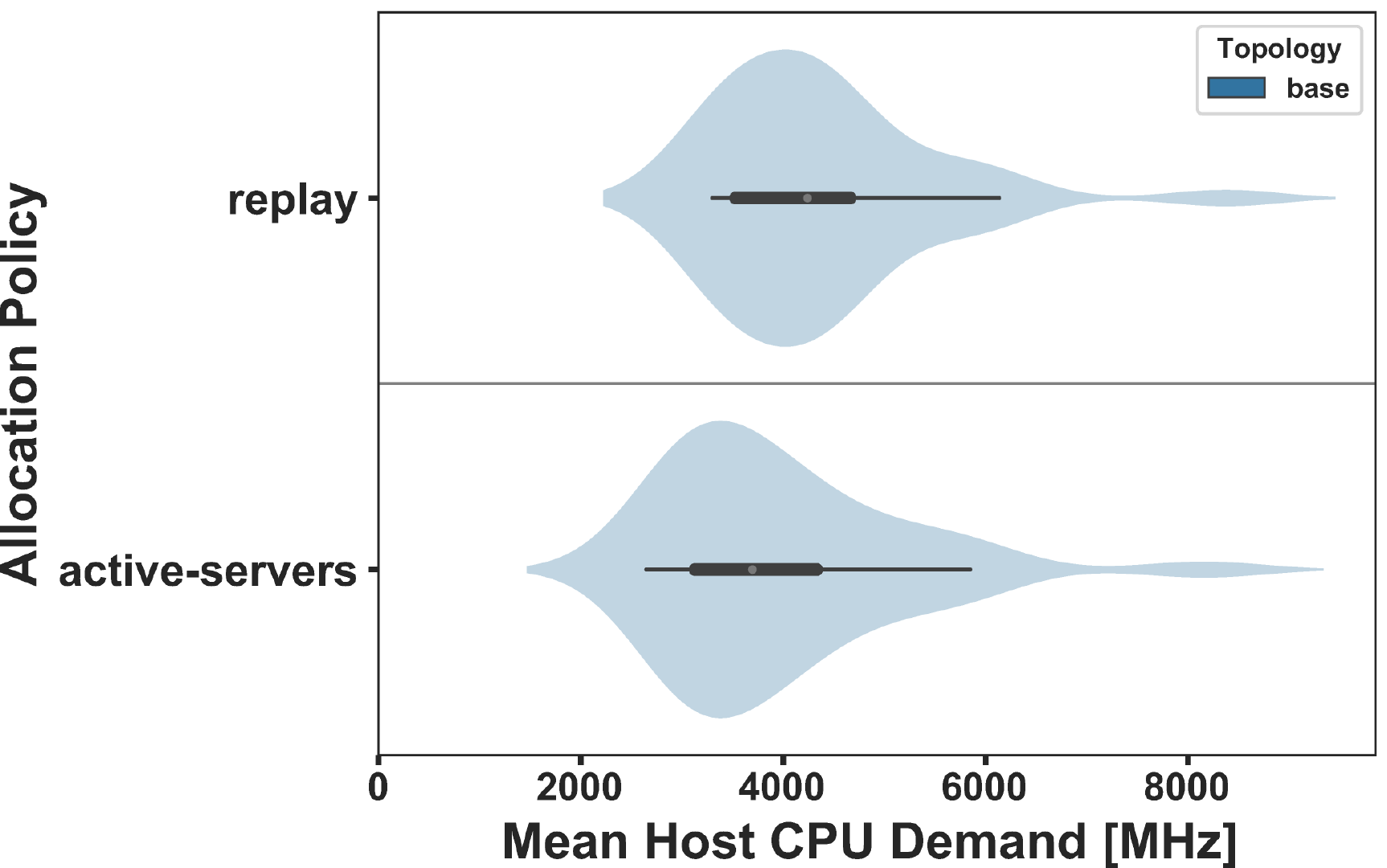}}\\
    \subfloat[Mean number of \glspl{VM} per host\label{fig:full:replay:mean-vm-count}]{\includegraphics[width=0.5\linewidth]{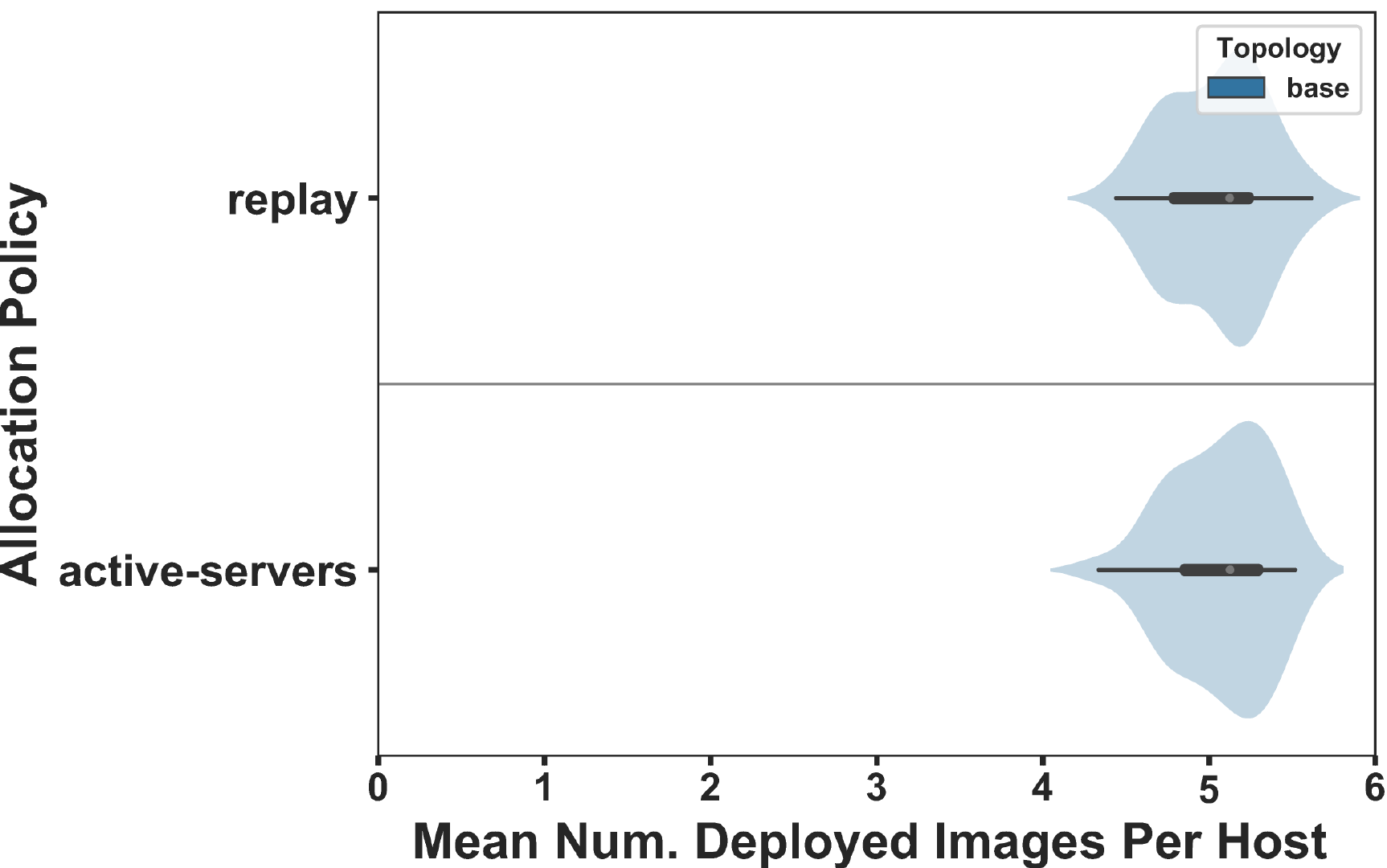}}%
    \subfloat[Max number of \glspl{VM} per host\label{fig:full:replay:max-vm-count}]{\includegraphics[width=0.5\linewidth]{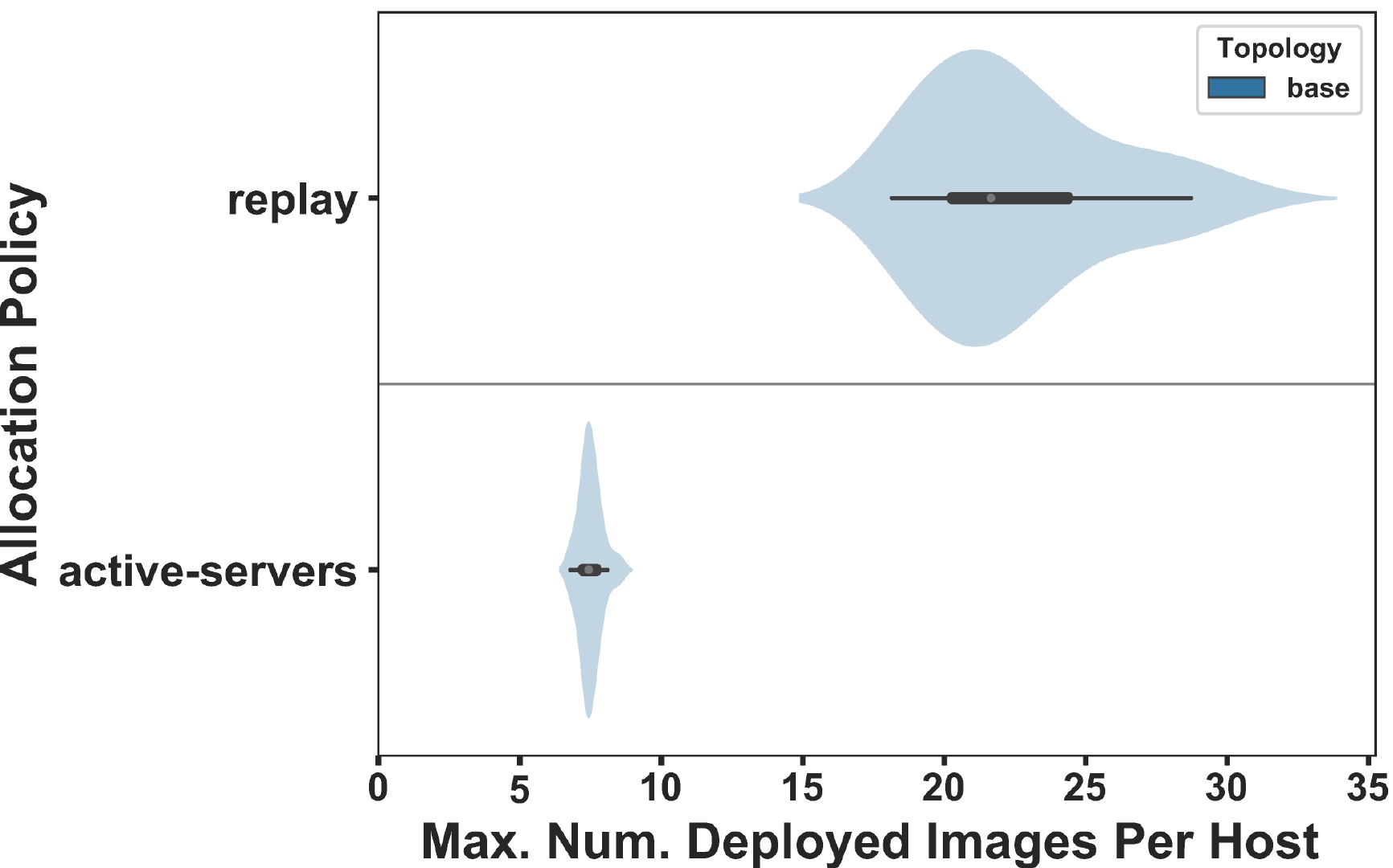}}\\
    \caption{Validation with a replay policy, copying the exact cluster assignment of the original deployment. For a legend of topologies, see Table~\ref{tab:experiment-overview}. Continued in Figure~\ref{fig:full:replay:3}.}
    \label{fig:full:replay:2}
\end{figure*}

\begin{figure*}
    \subfloat[Total VMs Submitted\label{fig:full:replay:vms-submitted}]{\includegraphics[width=0.5\linewidth]{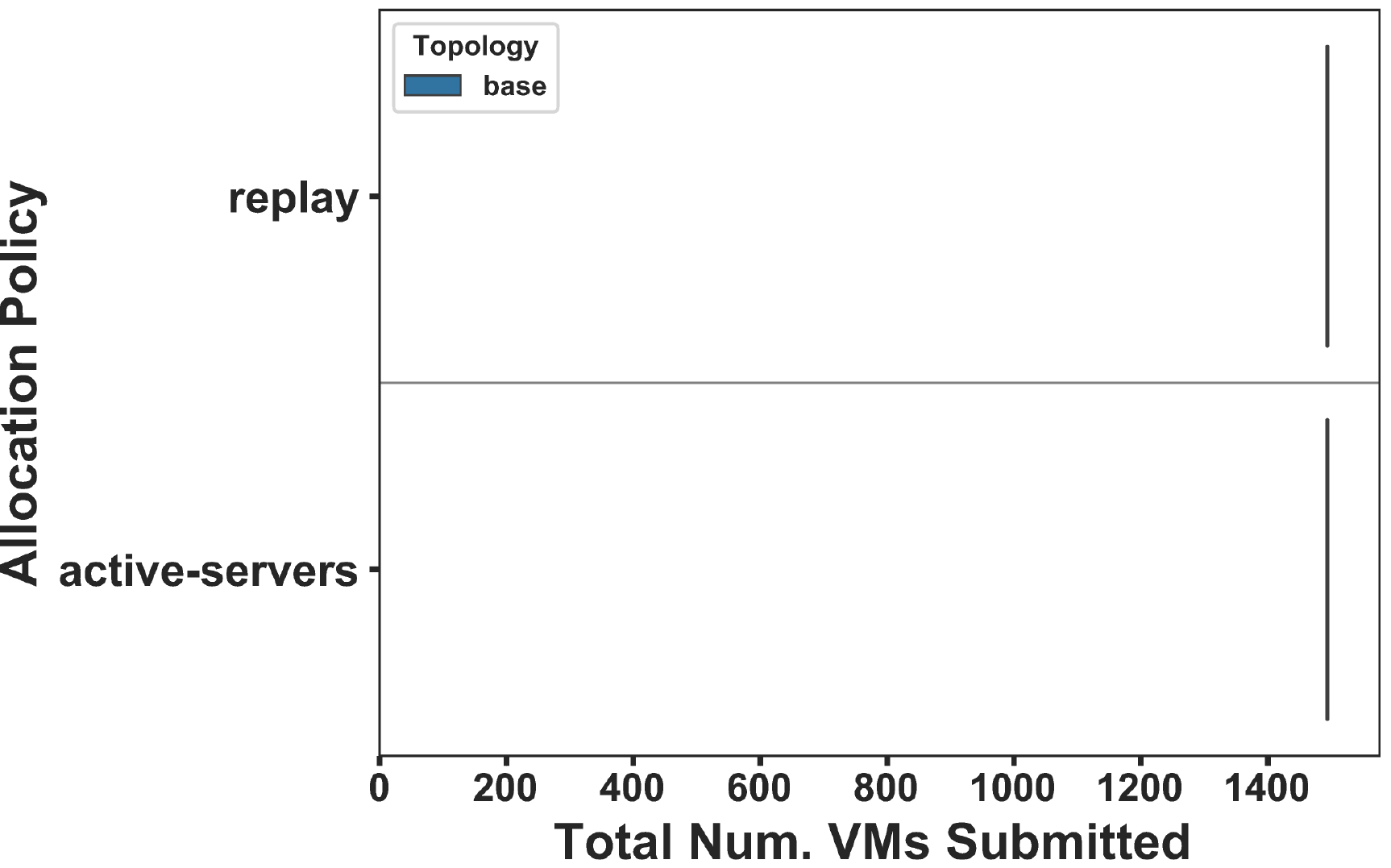}}%
    \subfloat[Total VMs Queued\label{fig:full:replay:vms-queued}]{\includegraphics[width=0.5\linewidth]{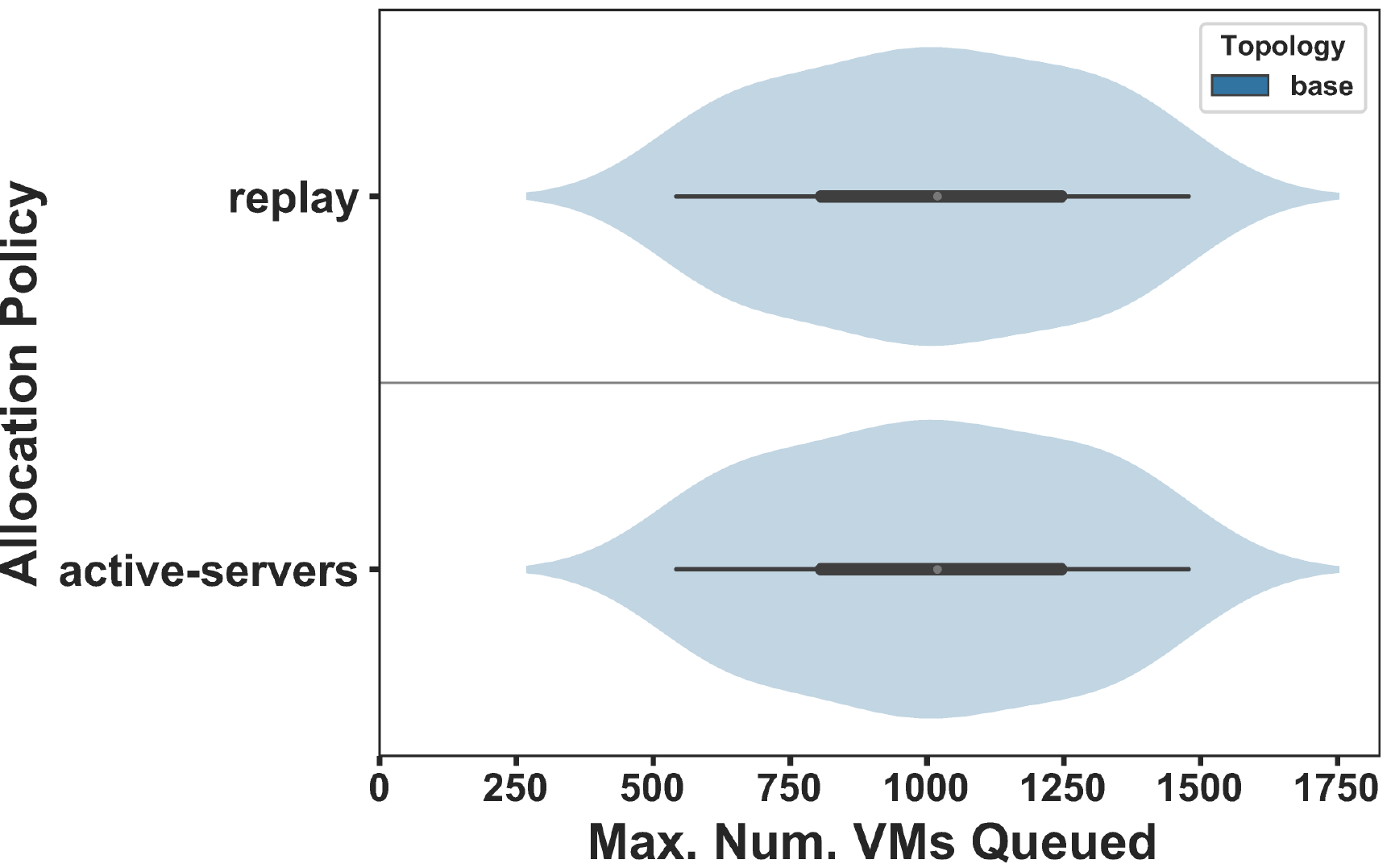}}\\
    \subfloat[Total VMs Finished\label{fig:full:replay:vms-finished}]{\includegraphics[width=0.5\linewidth]{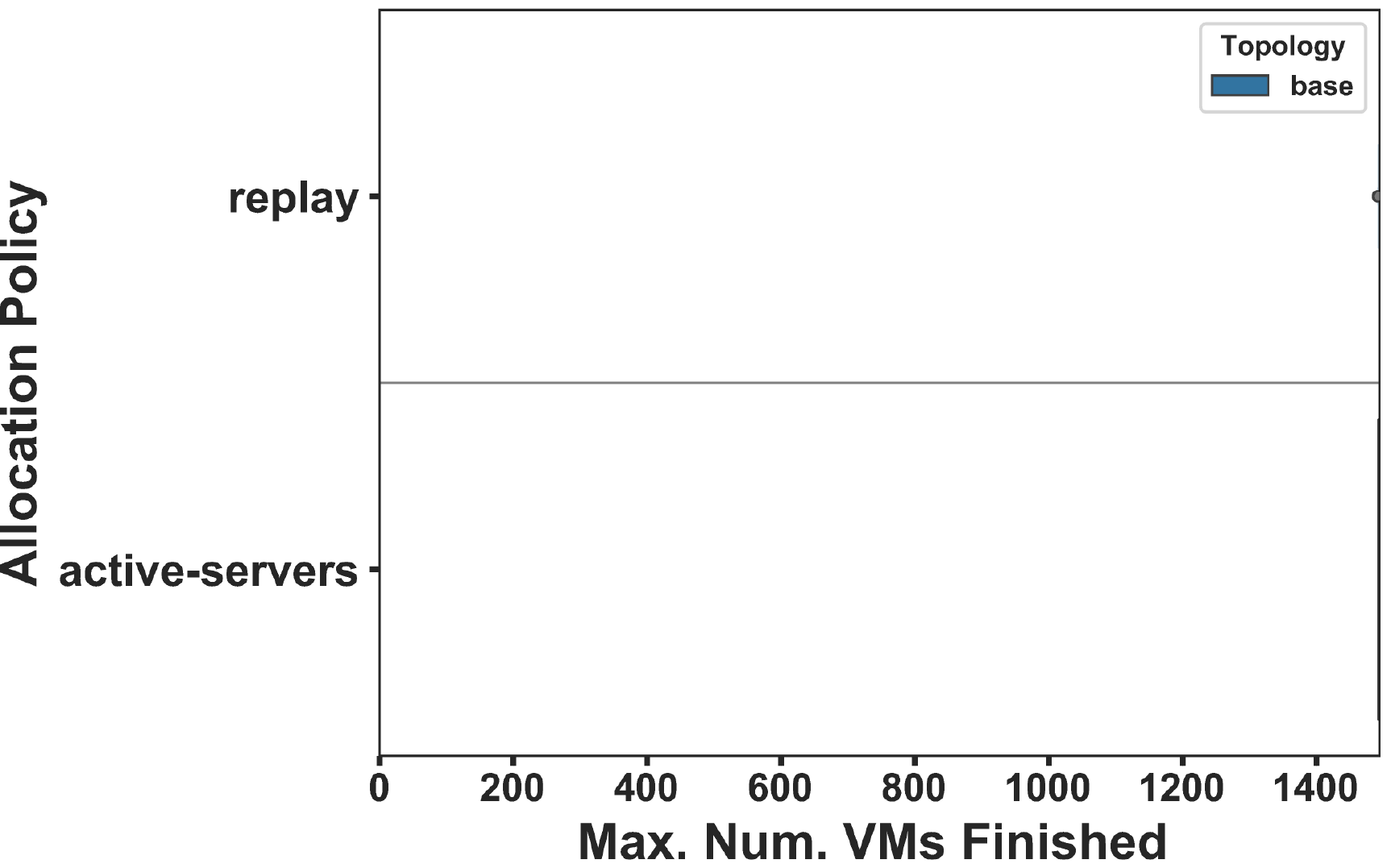}}%
    \subfloat[Total VMs Failed\label{fig:full:replay:vms-failed}]{\includegraphics[width=0.5\linewidth]{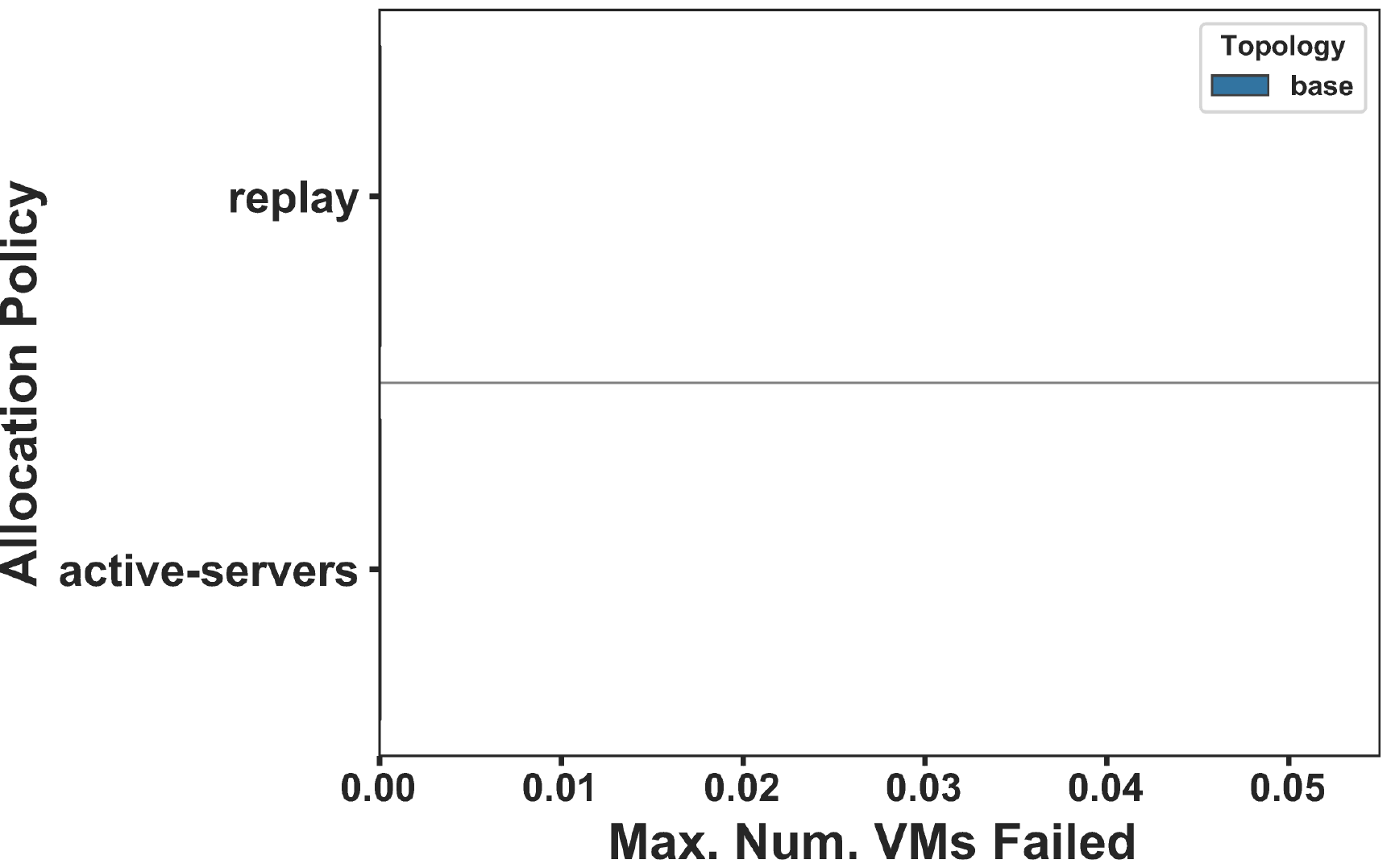}}%
    \caption{Validation with a replay policy, copying the exact cluster assignment of the original deployment. For a legend of topologies, see Table~\ref{tab:experiment-overview}.}
    \label{fig:full:replay:3}
\end{figure*}

\clearpage
    \section{Full Tabular Results} \label{sec:full-tabular-results}

In the tables below, we list the full results of two of the most important metrics (\S\ref{sec:exp:setup:metrics}) for each experiment: overcommission and power consumption.

\clearpage
\onecolumn
    
\begin{longtable}{rrrrrrrr}
\caption{Results of experiment 1, with key metrics and their mean and standard deviation.}\\
\toprule
                                    Topology &    Workload & Op. Phen. &   Alloc. Policy & \multicolumn{2}{l}{Overcommission [MFLOP]} & \multicolumn{2}{l}{Power Consumption [Wh]} \\
                                             &  &  &  &                       median &       std &                       median &       std \\
\midrule
\endhead
\midrule
\multicolumn{8}{r}{{Continued on next page}} \\
\midrule
\endfoot

\bottomrule
\endlastfoot
                                        base &   306 PFLOP &       all &  active-servers &                            0 & 7.653e+07 &                    1.106e+08 & 1.343e+07 \\
                                        base &   766 PFLOP &       all &  active-servers &                    1.059e+09 & 1.514e+09 &                    1.128e+08 & 1.288e+07 \\
                                        base &  1532 PFLOP &       all &  active-servers &                    1.192e+10 & 5.622e+09 &                    1.137e+08 &  1.32e+07 \\
                                        base &  3063 PFLOP &       all &  active-servers &                    5.691e+10 & 1.131e+10 &                    1.151e+08 &  1.32e+07 \\
 \iconRep{} \iconVol{} \iconHor{} \iconHom{} &   306 PFLOP &       all &  active-servers &                            0 & 7.655e+07 &                    1.105e+08 & 1.341e+07 \\
 \iconRep{} \iconVol{} \iconHor{} \iconHom{} &   766 PFLOP &       all &  active-servers &                    8.603e+08 & 1.492e+09 &                    1.125e+08 & 1.285e+07 \\
 \iconRep{} \iconVol{} \iconHor{} \iconHom{} &  1532 PFLOP &       all &  active-servers &                    1.124e+10 & 5.613e+09 &                    1.131e+08 & 1.314e+07 \\
 \iconRep{} \iconVol{} \iconHor{} \iconHom{} &  3063 PFLOP &       all &  active-servers &                    5.261e+10 & 1.212e+10 &                    1.142e+08 & 1.315e+07 \\
 \iconRep{} \iconVol{} \iconHor{} \iconHet{} &   306 PFLOP &       all &  active-servers &                    7.288e+05 & 2.182e+08 &                    9.139e+07 & 1.109e+07 \\
 \iconRep{} \iconVol{} \iconHor{} \iconHet{} &   766 PFLOP &       all &  active-servers &                    1.678e+09 & 3.129e+09 &                    9.311e+07 & 1.064e+07 \\
 \iconRep{} \iconVol{} \iconHor{} \iconHet{} &  1532 PFLOP &       all &  active-servers &                    1.263e+10 & 6.379e+09 &                    9.365e+07 & 1.087e+07 \\
 \iconRep{} \iconVol{} \iconHor{} \iconHet{} &  3063 PFLOP &       all &  active-servers &                    6.172e+10 & 1.377e+10 &                    9.466e+07 & 1.086e+07 \\
 \iconRep{} \iconVol{} \iconVer{} \iconHom{} &   306 PFLOP &       all &  active-servers &                    1.466e+08 & 1.348e+09 &                    5.327e+07 & 6.447e+06 \\
 \iconRep{} \iconVol{} \iconVer{} \iconHom{} &   766 PFLOP &       all &  active-servers &                    5.407e+09 & 4.606e+09 &                    5.443e+07 & 6.197e+06 \\
 \iconRep{} \iconVol{} \iconVer{} \iconHom{} &  1532 PFLOP &       all &  active-servers &                    2.661e+10 & 1.035e+10 &                    5.498e+07 & 6.326e+06 \\
 \iconRep{} \iconVol{} \iconVer{} \iconHom{} &  3063 PFLOP &       all &  active-servers &                    1.054e+11 & 2.293e+10 &                    5.585e+07 & 6.351e+06 \\
 \iconRep{} \iconVol{} \iconVer{} \iconHet{} &   306 PFLOP &       all &  active-servers &                    4.594e+07 & 4.369e+08 &                    7.239e+07 & 8.774e+06 \\
 \iconRep{} \iconVol{} \iconVer{} \iconHet{} &   766 PFLOP &       all &  active-servers &                    3.174e+09 & 3.518e+09 &                    7.388e+07 & 8.423e+06 \\
 \iconRep{} \iconVol{} \iconVer{} \iconHet{} &  1532 PFLOP &       all &  active-servers &                    1.915e+10 & 8.997e+09 &                    7.455e+07 & 8.615e+06 \\
 \iconRep{} \iconVol{} \iconVer{} \iconHet{} &  3063 PFLOP &       all &  active-servers &                    8.054e+10 & 1.323e+10 &                    7.575e+07 & 8.638e+06 \\
 \iconExp{} \iconVol{} \iconHor{} \iconHom{} &   306 PFLOP &       all &  active-servers &                            0 &         0 &                    1.757e+08 & 2.133e+07 \\
 \iconExp{} \iconVol{} \iconHor{} \iconHom{} &   766 PFLOP &       all &  active-servers &                    2.351e+08 & 2.375e+09 &                     1.79e+08 & 2.048e+07 \\
 \iconExp{} \iconVol{} \iconHor{} \iconHom{} &  1532 PFLOP &       all &  active-servers &                     5.07e+09 & 5.265e+09 &                    1.798e+08 & 2.093e+07 \\
 \iconExp{} \iconVol{} \iconHor{} \iconHom{} &  3063 PFLOP &       all &  active-servers &                    3.311e+10 & 1.195e+10 &                    1.813e+08 & 2.095e+07 \\
 \iconExp{} \iconVol{} \iconHor{} \iconHet{} &   306 PFLOP &       all &  active-servers &                            0 & 6.543e+04 &                    1.567e+08 & 1.904e+07 \\
 \iconExp{} \iconVol{} \iconHor{} \iconHet{} &   766 PFLOP &       all &  active-servers &                    3.597e+08 & 3.238e+09 &                    1.597e+08 & 1.826e+07 \\
 \iconExp{} \iconVol{} \iconHor{} \iconHet{} &  1532 PFLOP &       all &  active-servers &                    6.354e+09 & 5.345e+09 &                    1.605e+08 & 1.866e+07 \\
 \iconExp{} \iconVol{} \iconHor{} \iconHet{} &  3063 PFLOP &       all &  active-servers &                     3.74e+10 & 1.284e+10 &                    1.618e+08 & 1.868e+07 \\
 \iconExp{} \iconVol{} \iconVer{} \iconHom{} &   306 PFLOP &       all &  active-servers &                            0 & 4.804e+06 &                    1.187e+08 & 1.442e+07 \\
 \iconExp{} \iconVol{} \iconVer{} \iconHom{} &   766 PFLOP &       all &  active-servers &                    1.265e+09 & 2.186e+09 &                     1.21e+08 & 1.383e+07 \\
 \iconExp{} \iconVol{} \iconVer{} \iconHom{} &  1532 PFLOP &       all &  active-servers &                    1.098e+10 &  5.14e+09 &                    1.218e+08 & 1.412e+07 \\
 \iconExp{} \iconVol{} \iconVer{} \iconHom{} &  3063 PFLOP &       all &  active-servers &                    5.072e+10 & 1.248e+10 &                    1.232e+08 & 1.414e+07 \\
 \iconExp{} \iconVol{} \iconVer{} \iconHet{} &   306 PFLOP &       all &  active-servers &                            0 & 3.796e+07 &                    1.377e+08 & 1.673e+07 \\
 \iconExp{} \iconVol{} \iconVer{} \iconHet{} &   766 PFLOP &       all &  active-servers &                    8.684e+08 & 1.667e+09 &                    1.403e+08 & 1.606e+07 \\
 \iconExp{} \iconVol{} \iconVer{} \iconHet{} &  1532 PFLOP &       all &  active-servers &                    9.854e+09 & 5.417e+09 &                    1.411e+08 & 1.641e+07 \\
 \iconExp{} \iconVol{} \iconVer{} \iconHet{} &  3063 PFLOP &       all &  active-servers &                    4.787e+10 & 1.087e+10 &                    1.425e+08 & 1.644e+07 \\
\end{longtable}

\clearpage
\twocolumn

\clearpage
\onecolumn
    
\begin{longtable}{rrrrrrrr}
\caption{Results of experiment 2, with key metrics and their mean and standard deviation.}\\
\toprule
                                    Topology &    Workload & Op. Phen. &   Alloc. Policy & \multicolumn{2}{l}{Overcommission [MFLOP]} & \multicolumn{2}{l}{Power Consumption [Wh]} \\
                                             &  &  &  &                       median &       std &                       median &       std \\
\midrule
\endhead
\midrule
\multicolumn{8}{r}{{Continued on next page}} \\
\midrule
\endfoot

\bottomrule
\endlastfoot
                                        base &   306 PFLOP &       all &  active-servers &                            0 & 7.653e+07 &                    1.106e+08 & 1.343e+07 \\
                                        base &   766 PFLOP &       all &  active-servers &                    1.059e+09 & 1.514e+09 &                    1.128e+08 & 1.288e+07 \\
                                        base &  1532 PFLOP &       all &  active-servers &                    1.192e+10 & 5.622e+09 &                    1.137e+08 &  1.32e+07 \\
                                        base &  3063 PFLOP &       all &  active-servers &                    5.691e+10 & 1.131e+10 &                    1.151e+08 &  1.32e+07 \\
 \iconRep{} \iconVel{} \iconVer{} \iconHom{} &   306 PFLOP &       all &  active-servers &                            0 & 7.656e+07 &                    1.105e+08 & 1.342e+07 \\
 \iconRep{} \iconVel{} \iconVer{} \iconHom{} &   766 PFLOP &       all &  active-servers &                    1.059e+09 & 1.511e+09 &                    1.126e+08 & 1.286e+07 \\
 \iconRep{} \iconVel{} \iconVer{} \iconHom{} &  1532 PFLOP &       all &  active-servers &                    1.179e+10 & 5.547e+09 &                    1.135e+08 & 1.317e+07 \\
 \iconRep{} \iconVel{} \iconVer{} \iconHom{} &  3063 PFLOP &       all &  active-servers &                    5.504e+10 & 1.134e+10 &                    1.147e+08 & 1.318e+07 \\
 \iconRep{} \iconVel{} \iconVer{} \iconHet{} &   306 PFLOP &       all &  active-servers &                            0 & 7.657e+07 &                    1.106e+08 & 1.343e+07 \\
 \iconRep{} \iconVel{} \iconVer{} \iconHet{} &   766 PFLOP &       all &  active-servers &                    1.059e+09 & 1.514e+09 &                    1.127e+08 & 1.287e+07 \\
 \iconRep{} \iconVel{} \iconVer{} \iconHet{} &  1532 PFLOP &       all &  active-servers &                    1.183e+10 & 5.608e+09 &                    1.135e+08 & 1.318e+07 \\
 \iconRep{} \iconVel{} \iconVer{} \iconHet{} &  3063 PFLOP &       all &  active-servers &                    5.527e+10 & 1.131e+10 &                    1.149e+08 & 1.319e+07 \\
 \iconExp{} \iconVel{} \iconVer{} \iconHom{} &   306 PFLOP &       all &  active-servers &                            0 &         0 &                    1.758e+08 & 2.135e+07 \\
 \iconExp{} \iconVel{} \iconVer{} \iconHom{} &   766 PFLOP &       all &  active-servers &                    2.803e+08 & 2.374e+09 &                    1.791e+08 &  2.05e+07 \\
 \iconExp{} \iconVel{} \iconVer{} \iconHom{} &  1532 PFLOP &       all &  active-servers &                     5.02e+09 & 5.349e+09 &                      1.8e+08 & 2.096e+07 \\
 \iconExp{} \iconVel{} \iconVer{} \iconHom{} &  3063 PFLOP &       all &  active-servers &                    3.298e+10 & 1.203e+10 &                    1.816e+08 & 2.097e+07 \\
 \iconExp{} \iconVel{} \iconVer{} \iconHet{} &   306 PFLOP &       all &  active-servers &                            0 & 4.703e+04 &                    1.758e+08 & 2.135e+07 \\
 \iconExp{} \iconVel{} \iconVer{} \iconHet{} &   766 PFLOP &       all &  active-servers &                    2.803e+08 & 2.374e+09 &                    1.792e+08 &  2.05e+07 \\
 \iconExp{} \iconVel{} \iconVer{} \iconHet{} &  1532 PFLOP &       all &  active-servers &                    5.043e+09 & 5.352e+09 &                    1.801e+08 & 2.097e+07 \\
 \iconExp{} \iconVel{} \iconVer{} \iconHet{} &  3063 PFLOP &       all &  active-servers &                    3.423e+10 & 1.221e+10 &                    1.817e+08 & 2.098e+07 \\
\end{longtable}

\clearpage
\twocolumn

\clearpage
\onecolumn
    
\begin{longtable}{rrrrrrrr}
\caption{Results of experiment 3, with key metrics and their mean and standard deviation.}\\
\toprule
Topology &    Workload &     Op. Phen. &       Alloc. Policy & \multicolumn{2}{l}{Overcommission [MFLOP]} & \multicolumn{2}{l}{Power Consumption [Wh]} \\
         &  &  &  &                       median &       std &                       median &       std \\
\midrule
\endhead
\midrule
\multicolumn{8}{r}{{Continued on next page}} \\
\midrule
\endfoot

\bottomrule
\endlastfoot
    base &   306 PFLOP &          none &                 mem &                            0 &         0 &                    1.102e+08 & 1.335e+07 \\
    base &   306 PFLOP &          none &             mem-inv &                    5.662e+07 & 3.309e+09 &                    1.108e+08 & 1.347e+07 \\
    base &   306 PFLOP &          none &            core-mem &                            0 &         0 &                    1.102e+08 & 1.335e+07 \\
    base &   306 PFLOP &          none &        core-mem-inv &                    8.319e+07 & 2.172e+09 &                    1.106e+08 & 1.344e+07 \\
    base &   306 PFLOP &          none &      active-servers &                            0 & 6.543e+04 &                    1.106e+08 & 1.346e+07 \\
    base &   306 PFLOP &          none &  active-servers-inv &                    1.026e+08 & 2.163e+09 &                    1.104e+08 & 1.342e+07 \\
    base &   306 PFLOP &          none &              random &                            0 & 7.099e+04 &                    1.105e+08 & 1.345e+07 \\
    base &   306 PFLOP &  interference &                 mem &                    1.951e+09 & 1.786e+09 &                    1.102e+08 & 1.335e+07 \\
    base &   306 PFLOP &  interference &             mem-inv &                     8.42e+09 & 5.136e+09 &                    1.108e+08 & 1.347e+07 \\
    base &   306 PFLOP &  interference &            core-mem &                    1.222e+09 & 1.487e+09 &                    1.102e+08 & 1.335e+07 \\
    base &   306 PFLOP &  interference &        core-mem-inv &                     1.09e+10 & 3.629e+09 &                    1.106e+08 & 1.344e+07 \\
    base &   306 PFLOP &  interference &      active-servers &                            0 & 1.573e+06 &                    1.106e+08 & 1.346e+07 \\
    base &   306 PFLOP &  interference &  active-servers-inv &                    1.203e+10 & 6.329e+09 &                    1.104e+08 & 1.342e+07 \\
    base &   306 PFLOP &  interference &              random &                     1.36e+08 & 1.069e+09 &                    1.105e+08 & 1.345e+07 \\
    base &   306 PFLOP &      failures &                 mem &                            0 &         0 &                    1.102e+08 & 1.335e+07 \\
    base &   306 PFLOP &      failures &             mem-inv &                    9.672e+07 & 2.497e+09 &                    1.107e+08 & 1.346e+07 \\
    base &   306 PFLOP &      failures &            core-mem &                            0 &         0 &                    1.102e+08 & 1.335e+07 \\
    base &   306 PFLOP &      failures &        core-mem-inv &                    1.045e+08 & 2.164e+09 &                    1.106e+08 & 1.344e+07 \\
    base &   306 PFLOP &      failures &      active-servers &                            0 & 4.703e+04 &                    1.106e+08 & 1.343e+07 \\
    base &   306 PFLOP &      failures &  active-servers-inv &                    1.026e+08 & 2.165e+09 &                    1.104e+08 & 1.342e+07 \\
    base &   306 PFLOP &      failures &              random &                            0 & 3.939e+07 &                    1.105e+08 & 1.341e+07 \\
    base &   306 PFLOP &           all &                 mem &                    1.423e+09 & 1.917e+09 &                    1.102e+08 & 1.335e+07 \\
    base &   306 PFLOP &           all &             mem-inv &                    8.388e+09 & 5.075e+09 &                    1.107e+08 & 1.346e+07 \\
    base &   306 PFLOP &           all &            core-mem &                    1.502e+09 & 2.269e+09 &                    1.102e+08 & 1.335e+07 \\
    base &   306 PFLOP &           all &        core-mem-inv &                    1.079e+10 & 3.501e+09 &                    1.106e+08 & 1.343e+07 \\
    base &   306 PFLOP &           all &      active-servers &                            0 & 7.653e+07 &                    1.106e+08 & 1.343e+07 \\
    base &   306 PFLOP &           all &  active-servers-inv &                    1.148e+10 & 5.756e+09 &                    1.104e+08 & 1.342e+07 \\
    base &   306 PFLOP &           all &              random &                    2.613e+08 & 1.797e+09 &                    1.105e+08 & 1.341e+07 \\
    base &   766 PFLOP &          none &                 mem &                            0 &         0 &                    1.121e+08 & 1.279e+07 \\
    base &   766 PFLOP &          none &             mem-inv &                    1.734e+09 & 5.145e+09 &                    1.137e+08 & 1.306e+07 \\
    base &   766 PFLOP &          none &            core-mem &                            0 &         0 &                    1.121e+08 & 1.279e+07 \\
    base &   766 PFLOP &          none &        core-mem-inv &                    7.918e+08 & 3.135e+09 &                    1.135e+08 & 1.301e+07 \\
    base &   766 PFLOP &          none &      active-servers &                            0 & 1.242e+05 &                    1.128e+08 & 1.293e+07 \\
    base &   766 PFLOP &          none &  active-servers-inv &                    2.481e+09 & 8.611e+09 &                     1.13e+08 & 1.296e+07 \\
    base &   766 PFLOP &          none &              random &                            0 & 2.897e+06 &                     1.13e+08 & 1.294e+07 \\
    base &   766 PFLOP &  interference &                 mem &                    1.901e+10 & 6.425e+09 &                    1.121e+08 & 1.279e+07 \\
    base &   766 PFLOP &  interference &             mem-inv &                     2.65e+10 & 9.494e+09 &                    1.137e+08 & 1.306e+07 \\
    base &   766 PFLOP &  interference &            core-mem &                    1.838e+10 & 7.025e+09 &                    1.121e+08 & 1.279e+07 \\
    base &   766 PFLOP &  interference &        core-mem-inv &                    2.905e+10 & 7.354e+09 &                    1.135e+08 & 1.301e+07 \\
    base &   766 PFLOP &  interference &      active-servers &                    5.298e+08 & 1.934e+09 &                    1.128e+08 & 1.293e+07 \\
    base &   766 PFLOP &  interference &  active-servers-inv &                    4.045e+10 &  1.41e+10 &                     1.13e+08 & 1.296e+07 \\
    base &   766 PFLOP &  interference &              random &                    2.085e+09 & 3.228e+09 &                     1.13e+08 & 1.294e+07 \\
    base &   766 PFLOP &      failures &                 mem &                            0 & 2.956e+07 &                    1.121e+08 & 1.279e+07 \\
    base &   766 PFLOP &      failures &             mem-inv &                    1.499e+09 & 3.789e+09 &                    1.135e+08 & 1.301e+07 \\
    base &   766 PFLOP &      failures &            core-mem &                            0 & 8.401e+06 &                    1.121e+08 & 1.279e+07 \\
    base &   766 PFLOP &      failures &        core-mem-inv &                    1.077e+09 & 3.287e+09 &                    1.132e+08 & 1.298e+07 \\
    base &   766 PFLOP &      failures &      active-servers &                            0 & 3.987e+07 &                    1.127e+08 & 1.289e+07 \\
    base &   766 PFLOP &      failures &  active-servers-inv &                    2.134e+09 & 8.564e+09 &                    1.128e+08 & 1.295e+07 \\
    base &   766 PFLOP &      failures &              random &                            0 &  1.11e+08 &                    1.128e+08 & 1.289e+07 \\
    base &   766 PFLOP &           all &                 mem &                    1.859e+10 & 6.785e+09 &                    1.121e+08 & 1.278e+07 \\
    base &   766 PFLOP &           all &             mem-inv &                    2.603e+10 & 7.593e+09 &                    1.135e+08 & 1.303e+07 \\
    base &   766 PFLOP &           all &            core-mem &                    2.069e+10 & 6.969e+09 &                    1.121e+08 & 1.278e+07 \\
    base &   766 PFLOP &           all &        core-mem-inv &                    2.878e+10 & 7.465e+09 &                    1.134e+08 & 1.298e+07 \\
    base &   766 PFLOP &           all &      active-servers &                    1.059e+09 & 1.514e+09 &                    1.128e+08 & 1.288e+07 \\
    base &   766 PFLOP &           all &  active-servers-inv &                    3.754e+10 &  1.48e+10 &                    1.129e+08 & 1.295e+07 \\
    base &   766 PFLOP &           all &              random &                    3.214e+09 & 2.526e+09 &                    1.129e+08 & 1.289e+07 \\
    base &  1532 PFLOP &          none &                 mem &                    2.108e+06 & 1.128e+08 &                    1.124e+08 & 1.302e+07 \\
    base &  1532 PFLOP &          none &             mem-inv &                     5.12e+09 & 7.258e+09 &                    1.157e+08 &  1.36e+07 \\
    base &  1532 PFLOP &          none &            core-mem &                    1.023e+06 & 1.213e+09 &                    1.124e+08 & 1.303e+07 \\
    base &  1532 PFLOP &          none &        core-mem-inv &                    2.448e+09 &  4.48e+09 &                    1.155e+08 & 1.358e+07 \\
    base &  1532 PFLOP &          none &      active-servers &                      4.9e+05 & 4.845e+08 &                    1.137e+08 & 1.325e+07 \\
    base &  1532 PFLOP &          none &  active-servers-inv &                    1.242e+10 & 1.189e+10 &                    1.148e+08 & 1.347e+07 \\
    base &  1532 PFLOP &          none &              random &                     2.81e+06 & 1.536e+09 &                     1.14e+08 & 1.329e+07 \\
    base &  1532 PFLOP &  interference &                 mem &                    7.042e+10 & 1.328e+10 &                    1.124e+08 & 1.302e+07 \\
    base &  1532 PFLOP &  interference &             mem-inv &                    5.968e+10 & 1.333e+10 &                    1.157e+08 &  1.36e+07 \\
    base &  1532 PFLOP &  interference &            core-mem &                    7.529e+10 & 1.181e+10 &                    1.124e+08 & 1.303e+07 \\
    base &  1532 PFLOP &  interference &        core-mem-inv &                    5.868e+10 & 1.052e+10 &                    1.155e+08 & 1.358e+07 \\
    base &  1532 PFLOP &  interference &      active-servers &                    9.837e+09 & 4.374e+09 &                    1.137e+08 & 1.325e+07 \\
    base &  1532 PFLOP &  interference &  active-servers-inv &                    8.973e+10 & 1.804e+10 &                    1.148e+08 & 1.347e+07 \\
    base &  1532 PFLOP &  interference &              random &                    1.326e+10 &  6.45e+09 &                     1.14e+08 & 1.329e+07 \\
    base &  1532 PFLOP &      failures &                 mem &                      2.4e+07 & 4.724e+08 &                    1.124e+08 & 1.302e+07 \\
    base &  1532 PFLOP &      failures &             mem-inv &                    3.479e+09 & 6.233e+09 &                    1.154e+08 & 1.351e+07 \\
    base &  1532 PFLOP &      failures &            core-mem &                    6.982e+07 &  1.22e+09 &                    1.124e+08 & 1.302e+07 \\
    base &  1532 PFLOP &      failures &        core-mem-inv &                    2.127e+09 & 4.865e+09 &                    1.151e+08 & 1.349e+07 \\
    base &  1532 PFLOP &      failures &      active-servers &                    1.613e+06 &  1.18e+09 &                    1.137e+08 & 1.319e+07 \\
    base &  1532 PFLOP &      failures &  active-servers-inv &                     8.15e+09 & 1.143e+10 &                    1.142e+08 & 1.338e+07 \\
    base &  1532 PFLOP &      failures &              random &                    4.024e+06 & 2.341e+09 &                    1.139e+08 & 1.324e+07 \\
    base &  1532 PFLOP &           all &                 mem &                    7.815e+10 & 1.253e+10 &                    1.124e+08 & 1.302e+07 \\
    base &  1532 PFLOP &           all &             mem-inv &                    5.715e+10 & 9.696e+09 &                    1.152e+08 & 1.353e+07 \\
    base &  1532 PFLOP &           all &            core-mem &                    7.581e+10 & 1.721e+10 &                    1.124e+08 & 1.302e+07 \\
    base &  1532 PFLOP &           all &        core-mem-inv &                    5.885e+10 & 1.096e+10 &                    1.153e+08 & 1.349e+07 \\
    base &  1532 PFLOP &           all &      active-servers &                    1.192e+10 & 5.622e+09 &                    1.137e+08 &  1.32e+07 \\
    base &  1532 PFLOP &           all &  active-servers-inv &                    8.486e+10 & 1.848e+10 &                    1.145e+08 & 1.339e+07 \\
    base &  1532 PFLOP &           all &              random &                    1.432e+10 & 6.048e+09 &                    1.139e+08 & 1.323e+07 \\
    base &  3063 PFLOP &          none &                 mem &                    2.118e+10 &  1.58e+10 &                    1.132e+08 & 1.303e+07 \\
    base &  3063 PFLOP &          none &             mem-inv &                     8.24e+09 & 9.127e+09 &                    1.195e+08 & 1.433e+07 \\
    base &  3063 PFLOP &          none &            core-mem &                    1.254e+10 & 8.967e+09 &                    1.141e+08 & 1.322e+07 \\
    base &  3063 PFLOP &          none &        core-mem-inv &                    9.598e+09 & 1.125e+10 &                    1.195e+08 & 1.434e+07 \\
    base &  3063 PFLOP &          none &      active-servers &                    1.201e+09 & 2.916e+09 &                    1.151e+08 & 1.327e+07 \\
    base &  3063 PFLOP &          none &  active-servers-inv &                    3.632e+10 & 3.072e+10 &                    1.174e+08 &   1.4e+07 \\
    base &  3063 PFLOP &          none &              random &                    1.872e+09 & 4.763e+09 &                    1.156e+08 & 1.336e+07 \\
    base &  3063 PFLOP &  interference &                 mem &                    2.455e+11 & 2.342e+10 &                    1.132e+08 & 1.303e+07 \\
    base &  3063 PFLOP &  interference &             mem-inv &                    1.178e+11 & 1.717e+10 &                    1.195e+08 & 1.433e+07 \\
    base &  3063 PFLOP &  interference &            core-mem &                    1.944e+11 & 2.155e+10 &                    1.141e+08 & 1.322e+07 \\
    base &  3063 PFLOP &  interference &        core-mem-inv &                    1.208e+11 & 2.077e+10 &                    1.195e+08 & 1.434e+07 \\
    base &  3063 PFLOP &  interference &      active-servers &                    4.914e+10 &  1.23e+10 &                    1.151e+08 & 1.327e+07 \\
    base &  3063 PFLOP &  interference &  active-servers-inv &                    2.161e+11 & 3.762e+10 &                    1.174e+08 &   1.4e+07 \\
    base &  3063 PFLOP &  interference &              random &                    5.421e+10 & 1.603e+10 &                    1.156e+08 & 1.336e+07 \\
    base &  3063 PFLOP &      failures &                 mem &                    2.779e+10 & 1.701e+10 &                    1.135e+08 & 1.315e+07 \\
    base &  3063 PFLOP &      failures &             mem-inv &                    6.364e+09 & 8.511e+09 &                    1.188e+08 & 1.412e+07 \\
    base &  3063 PFLOP &      failures &            core-mem &                    8.847e+09 & 8.367e+09 &                    1.143e+08 & 1.317e+07 \\
    base &  3063 PFLOP &      failures &        core-mem-inv &                    6.153e+09 & 7.126e+09 &                    1.187e+08 & 1.413e+07 \\
    base &  3063 PFLOP &      failures &      active-servers &                    2.111e+09 & 4.357e+09 &                    1.151e+08 & 1.319e+07 \\
    base &  3063 PFLOP &      failures &  active-servers-inv &                    3.255e+10 & 2.869e+10 &                    1.168e+08 &  1.38e+07 \\
    base &  3063 PFLOP &      failures &              random &                    9.814e+08 &  3.34e+09 &                    1.156e+08 & 1.328e+07 \\
    base &  3063 PFLOP &           all &                 mem &                    2.217e+11 & 3.067e+10 &                    1.134e+08 & 1.332e+07 \\
    base &  3063 PFLOP &           all &             mem-inv &                    1.147e+11 & 1.401e+10 &                    1.188e+08 & 1.407e+07 \\
    base &  3063 PFLOP &           all &            core-mem &                    1.561e+11 & 2.525e+10 &                    1.144e+08 & 1.327e+07 \\
    base &  3063 PFLOP &           all &        core-mem-inv &                    1.216e+11 & 1.887e+10 &                    1.188e+08 & 1.405e+07 \\
    base &  3063 PFLOP &           all &      active-servers &                    5.691e+10 & 1.131e+10 &                    1.151e+08 &  1.32e+07 \\
    base &  3063 PFLOP &           all &  active-servers-inv &                    2.035e+11 & 2.452e+10 &                     1.17e+08 & 1.383e+07 \\
    base &  3063 PFLOP &           all &              random &                    5.885e+10 & 1.098e+10 &                    1.156e+08 & 1.325e+07 \\
\end{longtable}

\clearpage
\twocolumn

\clearpage
\onecolumn
    
\begin{longtable}{rrrrrrrr}
\caption{Results of experiment 4, with key metrics and their mean and standard deviation.}\\
\toprule
                                    Topology &       Workload & Op. Phen. &   Alloc. Policy & \multicolumn{2}{l}{Overcommission [MFLOP]} & \multicolumn{2}{l}{Power Consumption [Wh]} \\
                                             &  &  &  &                       median &       std &                       median &       std \\
\midrule
\endhead
\midrule
\multicolumn{8}{r}{{Continued on next page}} \\
\midrule
\endfoot

\bottomrule
\endlastfoot
                                        base &        all-pri &  failures &  active-servers &                    5.356e+08 & 2.428e+09 &                    4.382e+07 &  3.95e+06 \\
                                        base &  pri-75-pub-25 &  failures &  active-servers &                    4.324e+10 & 3.961e+10 &                    4.369e+07 & 4.089e+06 \\
                                        base &  pri-50-pub-50 &  failures &  active-servers &                    1.358e+11 & 9.609e+10 &                    4.353e+07 & 4.548e+06 \\
                                        base &  pri-25-pub-75 &  failures &  active-servers &                    2.838e+11 & 1.168e+11 &                    4.352e+07 & 6.325e+06 \\
                                        base &        all-pub &  failures &  active-servers &                    4.032e+11 & 1.541e+11 &                    3.988e+07 & 9.219e+06 \\
 \iconExp{} \iconVol{} \iconHor{} \iconHom{} &        all-pri &  failures &  active-servers &                    1.094e+07 & 5.261e+08 &                    6.885e+07 & 6.188e+06 \\
 \iconExp{} \iconVol{} \iconHor{} \iconHom{} &  pri-75-pub-25 &  failures &  active-servers &                    4.378e+10 & 3.267e+10 &                    6.872e+07 & 6.294e+06 \\
 \iconExp{} \iconVol{} \iconHor{} \iconHom{} &  pri-50-pub-50 &  failures &  active-servers &                    1.022e+11 &  9.59e+10 &                    6.862e+07 & 6.703e+06 \\
 \iconExp{} \iconVol{} \iconHor{} \iconHom{} &  pri-25-pub-75 &  failures &  active-servers &                    2.505e+11 & 1.061e+11 &                    6.911e+07 & 9.373e+06 \\
 \iconExp{} \iconVol{} \iconHor{} \iconHom{} &        all-pub &  failures &  active-servers &                    3.325e+11 & 1.444e+11 &                    6.102e+07 & 1.355e+07 \\
 \iconExp{} \iconVol{} \iconVer{} \iconHom{} &        all-pri &  failures &  active-servers &                    2.181e+08 & 3.535e+09 &                    4.694e+07 & 4.228e+06 \\
 \iconExp{} \iconVol{} \iconVer{} \iconHom{} &  pri-75-pub-25 &  failures &  active-servers &                    3.912e+10 & 3.111e+10 &                    4.684e+07 & 4.359e+06 \\
 \iconExp{} \iconVol{} \iconVer{} \iconHom{} &  pri-50-pub-50 &  failures &  active-servers &                     1.25e+11 & 1.008e+11 &                    4.667e+07 & 4.733e+06 \\
 \iconExp{} \iconVol{} \iconVer{} \iconHom{} &  pri-25-pub-75 &  failures &  active-servers &                    2.456e+11 & 1.116e+11 &                    4.664e+07 & 6.696e+06 \\
 \iconExp{} \iconVol{} \iconVer{} \iconHom{} &        all-pub &  failures &  active-servers &                    3.572e+11 & 1.698e+11 &                    4.242e+07 & 9.665e+06 \\
 \iconExp{} \iconVel{} \iconVer{} \iconHom{} &        all-pri &  failures &  active-servers &                    1.862e+07 & 5.246e+08 &                    6.901e+07 & 6.204e+06 \\
 \iconExp{} \iconVel{} \iconVer{} \iconHom{} &  pri-75-pub-25 &  failures &  active-servers &                    4.804e+10 & 3.466e+10 &                    6.883e+07 & 6.318e+06 \\
 \iconExp{} \iconVel{} \iconVer{} \iconHom{} &  pri-50-pub-50 &  failures &  active-servers &                    1.234e+11 &  9.76e+10 &                    6.872e+07 & 6.752e+06 \\
 \iconExp{} \iconVel{} \iconVer{} \iconHom{} &  pri-25-pub-75 &  failures &  active-servers &                    2.658e+11 & 1.131e+11 &                    6.922e+07 & 9.444e+06 \\
 \iconExp{} \iconVel{} \iconVer{} \iconHom{} &        all-pub &  failures &  active-servers &                    3.829e+11 & 1.559e+11 &                    6.125e+07 & 1.369e+07 \\
\end{longtable}

\clearpage
\twocolumn

\clearpage
\onecolumn
    
\begin{longtable}{rrrrrrrr}
\caption{Results of experiment 5, with key metrics and their mean and standard deviation.}\\
\toprule
Topology &    Workload & Op. Phen. &   Alloc. Policy & \multicolumn{2}{l}{Overcommission [MFLOP]} & \multicolumn{2}{l}{Power Consumption [Wh]} \\
         &  &  &  &                       median &       std &                       median &       std \\
\midrule
\endhead
\midrule
\multicolumn{8}{r}{{Continued on next page}} \\
\midrule
\endfoot

\bottomrule
\endlastfoot
    base &  3063 PFLOP &      none &          replay &                    393283398 & 1.076e+09 &                    9.646e+07 & 1.225e+07 \\
    base &  3063 PFLOP &      none &  active-servers &                   1200611704 & 2.916e+09 &                    1.151e+08 & 1.327e+07 \\
\end{longtable}

\clearpage
\twocolumn

\end{document}